\documentclass[preprint]{aastex61}
\pdfoutput=1 

\usepackage{graphicx}
\usepackage[space]{grffile}
\usepackage{latexsym}
\usepackage{textcomp}
\usepackage{multirow,booktabs}
\usepackage{amsfonts,amsmath,amssymb}
\usepackage{natbib}
\usepackage{url}

\usepackage[utf8]{inputenc}
\usepackage[ngerman,english]{babel}
\usepackage{rotating}

\usepackage{babel}

\shorttitle{Large-Scale Galactic Filaments}
\shortauthors{Catherine Zucker}

\begin{document}


\title{Physical Properties of Large-Scale Galactic Filaments}


\author{Catherine Zucker}
\affil{Harvard-Smithsonian Center for Astrophysics, Cambridge, MA 02138}
\author{Cara Battersby}
\affil{Harvard-Smithsonian Center for Astrophysics, Cambridge, MA 02138}
\affil{Department of Physics, University of Connecticut, Storrs,
CT 06269, USA}
\author{Alyssa Goodman}
\affil{Harvard-Smithsonian Center for Astrophysics, Cambridge, MA 02138}



\begin{abstract}
The characterization of our Galaxy's longest filamentary gas features has been the subject of several studies in recent years, producing not only a sizeable sample of large-scale filaments, but also confusion as to whether all these features (e.g. ``Bones", ``Giant Molecular Filaments") are the same. They are not. We undertake the first standardized analysis of the physical properties ($\rm{H_2}\;$column densities, dust temperatures, morphologies, radial column density profiles) and kinematics of large-scale filaments in the literature. We expand and improve upon prior analyses by using the same data sets, techniques, and spiral arm models to disentangle the filaments' inherent properties from selection criteria and methodology. Our results suggest that the myriad filament finding techniques are uncovering different physical structures, with length ($11-269\,\rm\,pc$), width ($1-40\,\rm\,pc$), mass ($\rm3\times10^3\; M_\odot-1.1\times10^{6}\;M_\odot$), aspect ratio (3:1-117:1), and high column density fraction (0.2-100\%) varying by over an order of magnitude across the sample of 45 filaments. We develop a radial profile fitting code, \texttt{RadFil}, which is publicly available. We also perform a \textit{position-position-velocity} (\textit{p-p-v}) analysis on a subsample and find that while 60-70\% lie spatially in the plane of the Galaxy, only 30-45\% concurrently exhibit spatial \textit{and} kinematic proximity to spiral arms. In a parameter space defined by aspect ratio, dust temperature, and column density, we broadly distinguish three filament categories, which could indicate different formation mechanisms or histories. Highly elongated ``Bone-like" filaments show the most potential for tracing gross spiral structure (e.g. arms, spurs), while other categories could be large concentrations of molecular gas (GMCs, core complexes).
\end{abstract}

\section{Introduction} \label{intro}
Since 2014, the discovery of the Nessie cloud's potential association with the ``spine" of the Scutum-Centaurus arm has served as a tantalizing case in point for how some of our Galaxy's longest, high density filamentary features might be shaped by the structural dynamics of the Milky Way. First identified in \citet{Jackson_2010} as an 80 pc $\times$ 0.5 pc velocity-contiguous filamentary infrared dark cloud (IRDC), Nessie hosts at least a dozen dense molecular cores which \citet{Jackson_2010} argue are likely birthplaces of high-mass stars. \citet{Goodman_2014} extend the analysis of  \citet{Jackson_2010} and find that not only is Nessie 2-to-5 times longer than originally claimed but also hypothesize that it might be the first in a class of filaments whose formation and persistence is likely governed by the gravity of the Galaxy. Not only does Nessie have an aspect ratio of at least 300:1, it also lies within a few parsecs of the physical Galactic midplane (at a distance of 3.1 kpc) and seems to trace out the spine of the Scutum-Centaurus arm in \textit{position-position-velocity (p-p-v)} space. This cumulative evidence led the authors to suggest that Nessie is a ``Bone" of the Milky Way---an ultradense, highly elongated filamentary molecular cloud whose formation and evolution could be intimately linked to spiral structure on a grander scale. 

Nessie has since become the archetype of large-scale ($\gtrsim 10 \rm \; pc$) Galactic filaments and has triggered a plethora of searches aimed at compiling a statistical sample of massive filamentary molecular clouds. The subsequent catalogs \citep[see \S \ref{sample_selection}, ][]{Ragan_2014, Wang_2015, Zucker_2015, Abreu_Vicente_2016, Wang_2016} are as diverse as the methodology used to obtain them. The various selection criteria employed for recent large-scale filament catalogs to be included in this study (see \S \ref{sample_selection}) are summarized in Table \ref{tab:selection_criteria}. The filaments are initially identified over a range of wavelengths (radio, sub-millimeter, far-infrared, mid-infrared, and near-infrared) and contain both ``by-eye" and automated searches. The filament kinematics are mapped via both low- ($^{13} \rm CO$) and high- ($\rm N_2H^+$, $\rm HCO^+$, and $\rm NH_3$) density gas tracers, and requirements for velocity contiguity vary from study to study. Some searches are blind, while others specifically target spiral-tracing filaments. However, each uses different spiral arm models and criteria to establish spiral arm association---a concern given that fits to the same arm in \textit{longitude-velocity} space can vary by at least $\rm 10 \; km \; s^{-1}$ \citep[see Figure 2 in][]{Zucker_2015}. 

While recent studies yield several dozen Galactic filaments, the disparity in methodologies makes it difficult to draw meaningful conclusions or reliably compare filaments across catalogs. To date, it is unclear how the physical properties of the filaments vary amongst catalogs, and how much variation can be attributed to inherent differences in the properties versus a lack of standardization in calculating them. Several key questions remain. Are specific ``types" of filaments more prone to lie in spiral arms, spurs, or interarm regions, and can these different populations be reliably distinguished observationally? Are the cataloged filaments physically different structures with different formation mechanisms? How are large-scale filaments shaped by or shaping the star formation process? Are some types of filaments related, but at different stages of evolution? To what extent are these catalogs of filaments hierarchical, and/or are the filaments themselves hierarchical, with small filaments inside larger ones? Which filaments are caused or maintained by the global gravitational potential of the Galaxy, and thus able to trace out the gravitational midplane and/or the Galaxy's structure?

Two components are required to answer these questions. The first component is the development of numerical simulations that can dynamically resolve both highly elongated filaments and the environments in which they form. The second is a systematic analysis of the physical properties of existing large-scale filaments catalogs for comparison with these simulations. 

Only recently have simulations of large-scale molecular filaments appeared in the literature. Simulations prior to 2014 have shown organized `spurs' trailing off spiral arms, which often appear to extend into long inter-arm features \citep[e.g.][]{Dobbs_2013, Kim_2002, Renaud_2013, Shetty_2006}. However, the first simulations capable of resolving Nessie-like features are those from \citet{Smith_2014}, based on the AREPO moving mesh code, capable of providing $\rm \approx 0.3 \; pc$ resolution in regions where the gas density $n > 10^3 \; \rm cm^{-3}$. \citet{Smith_2014} find that high density filaments with column densities and aspect ratios similar to Nessie tend to form in spiral arms, and in close proximity to the Galactic midplane. While the \citet{Smith_2014} simulations account for the chemical evolution of the gas, they only impose a four arm spiral potential and do not include self-gravity, stellar feedback, or magnetic fields. 

\citet{Duarte_Cabral_2016} and \citet{Duarte_Cabral_2017} use a smooth particle hydrodynamics simulation to explicitly study elongated molecular clouds in a two-armed spiral galaxy at lower resolution (regridded cell sizes of 1 and 5 pc) and lower density than the \citet{Smith_2014} simulations. \citet{Duarte_Cabral_2016} find that shear from Galactic rotation plays a critical role in filament formation, as the shearing motion tends to stretch out and align gas with spiral arms. Unlike \citet{Smith_2014}, \citet{Duarte_Cabral_2016} find that elongated gaseous features tend to form in the interarm regions, but that they reach their fullest extent at the deepest point in the spiral potential well just prior to arm entry. Like \citet{Smith_2014}, the \citet{Duarte_Cabral_2016, Duarte_Cabral_2017} simulations do not include magnetic fields, but they do include stellar feedback and self-gravity. Building upon existing work, upcoming early simulations---now starting to include MHD, self-gravity, supernovae feedback, and OB feedback---will provide even richer datasets for comparison with observational studies. 

This paper seeks to satisfy the second component necessary for successfully contextualizing the wide diversity of large-scale filaments---a comprehensive, standardized analysis of the physical properties of large-scale filaments in the literature. Towards the goal of synthesisizing discordant observational studies and meaningfully interpreting them in light of numerical simulations, we use the same data sets, techniques, and spiral arm models to analyze \textit{all} the filaments in the \textit{same framework}. Furthermore, we significantly expand and improve upon previous analyses, by systematically quantifying the filaments' column densities, dust temperatures, morphologies, and spatial and kinematic displacements from spiral arms. 

In section \S \ref{sample_selection} we discuss the major large-scale filament catalogs currently found in the literature and the subsample we select for inclusion in this study. In \S \ref{methodology} we define quantitative boundaries for the filaments, detail our methodology for calculating their physical properties and kinematics, and compare and contrast these results for the different catalogs in our sample. In \S \ref{discussion} we discuss the physical properties of the full sample in the context of Galactic structure and recent numerical simulations. We summarize our findings and conclusions in \S \ref{conclusion}.

\begin{table}
    \makebox[\textwidth]{\includegraphics[angle=90, width=10cm]{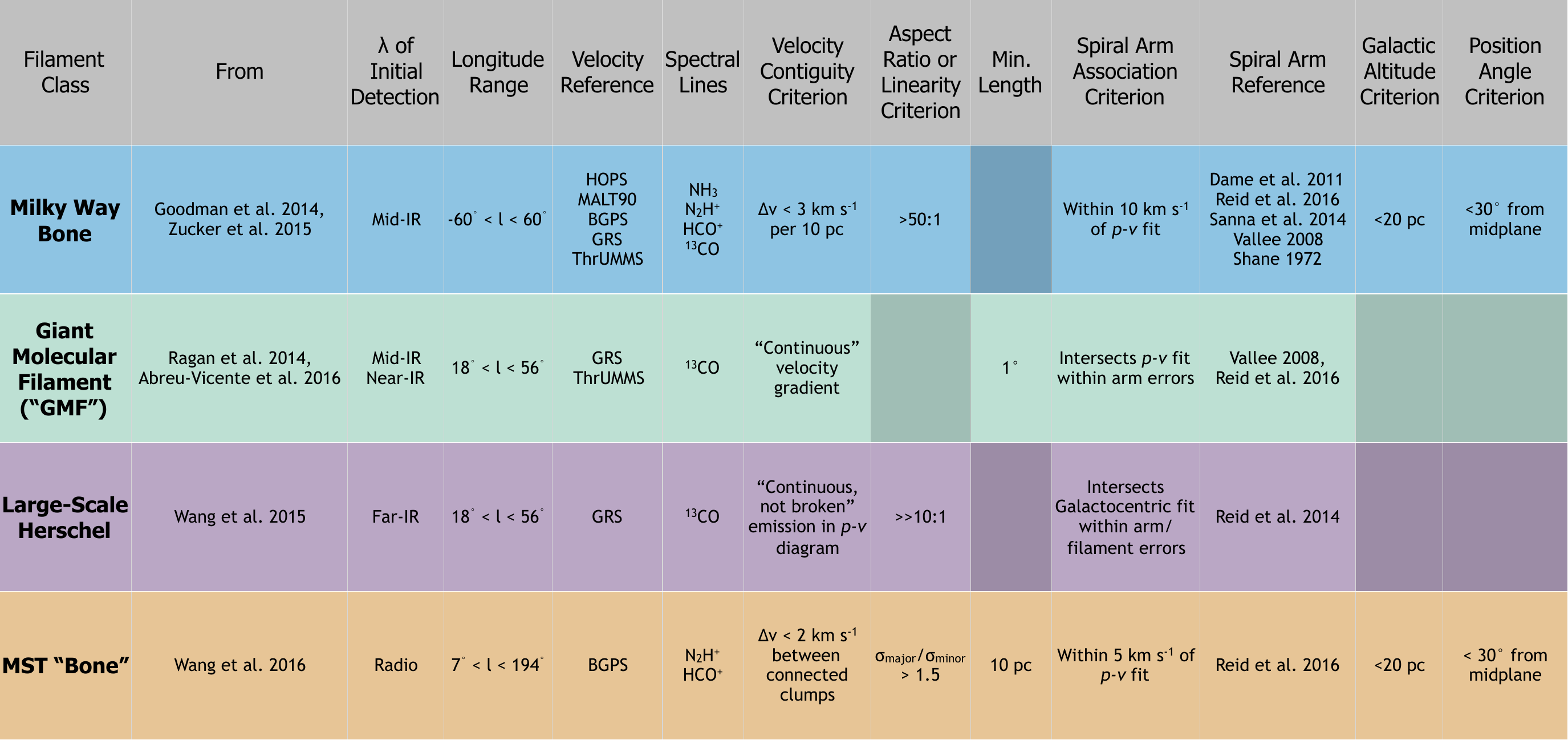}}
    \caption{  \label{tab:selection_criteria} The various selection criteria employed for major large-scale filament catalogs included in this study, taken from their original publications \citep{Ragan_2014, Abreu_Vicente_2016, Zucker_2015, Wang_2015, Wang_2016}. }
\end{table}

\section{Sample Selection} \label{sample_selection}
We draw our selection of large-scale filaments from several existing catalogs. To be included in this analysis, a filament must be at least 10 pc long, stated to be velocity contiguous in the original paper presenting it, and be part of a catalog whose members are purported to lie near or be associated with some aspect of spiral structure. Filaments that satisfy these criteria include the Giant Molecular Filaments from \citet{Ragan_2014} and \citet{Abreu_Vicente_2016}, the Large-Scale Herschel filaments from \citet{Wang_2015}, the Milky Way ``Bone" filaments from \citet{Zucker_2015} and the MST Filaments from \citet{Wang_2016}. Here we briefly summarize the key components of each large-scale filament catalog. For a more detailed overview of the initial selection of these filaments in their original publications, as well as the subsample selected for inclusion in this study, see \S \ref{gmf_appendix}, \ref{bone_appendix}, \ref{herschel_appendix}, and \ref{mst_appendix} in the Appendix. There is an additional sample of five large-scale molecular filaments towards the outer Galaxy presented in \citet{Du_2017}; however, these filaments are outside the longitude range of major inner Galactic plane surveys used to uniformly calculate physical properties throughout this study \citep[e.g. Hi-GAL, GRS; ][]{Molinari_2016, Jackson_2006}. Thus the \citet{Du_2017} filaments are excluded from this analysis. 

The Giant Molecular Filaments (``GMFs") from \citet{Ragan_2014} and \citet{Abreu_Vicente_2016} are first identified as near and mid-infrared extinction features via a visual search of the inner Galactic plane \citep[UKIDSS Galactic Plane Survey, GLIMPSE Survey, MIPSGAL Survey; ][]{Lucas_2008, Churchwell_2009, Carey_2009}. The GMFs' spatial extent ($\approx 100$ pc) and physical properties (e.g. $ \rm M \approx 10^{5} \; M_\odot$, dense gas fractions 1\%-40\%) are then derived via the morphology of the low density tracing $^{13} \rm CO$ emission enveloping these dense filamentary complexes. \citet{Ragan_2014} and \citet{Abreu_Vicente_2016} confirm velocity contiguity by collapsing the $^{13} \rm CO$ cubes in latitude bounds around the filament, and ensuring that the \textit{position-velocity} diagram exhibits unbroken emission. Together \citet{Ragan_2014} and \citet{Abreu_Vicente_2016} identify 16 GMFs (9 claimed to be associated with spiral arms), all of which are included in this study. 

\citet{Wang_2015} undertake a search for the coldest and densest filaments (``Large-Scale Herschel Filaments") by visually inspecting Herschel Hi-GAL images of the inner galaxy \citep{Molinari_2010} and confirming velocity contiguity via a custom $\rm ^{13} CO$ \textit{position-velocity} slice taken along the filament.  \citet{Wang_2015} uncover 9 filaments with typical lengths on the order of 40-100 pc, column densities $\approx 1 \times 10^{21} - 2\times 10^{22} \; \rm cm^{-2}$ and dust temperatures 17-21 K. Seven of these are claimed to be associated with spiral arms, and all nine filaments listed in Table 2 from \citet{Wang_2015} are included in this study. 

Next, \citet{Zucker_2015} carry out a by-eye search of the mid-IR GLIMPSE and MIPSGAL surveys \citep{Carey_2009, Churchwell_2009} specifically for Nessie analogues---the ``Bones of the Milky Way." \citet{Zucker_2015} catalog highly elongated extinction features (typical lengths 20-60 pc) that lie parallel and in close proximity to the plane-of-the-sky projections of known spiral arms. \citet{Zucker_2015} identify 10 candidates and develop a set of Bone criteria (with prescriptions for aspect ratio, spatial and kinematic proximity to spiral arms, and velocity contiguity) intended to differentiate features most likely to form due to the Milky Way's global spiral potential. To confirm velocity contiguity, \citet{Zucker_2015} take a custom $\rm ^{13} CO$ \textit{position-velocity} slice along the filament and ensure that dense gas catalog sources tracing the filament exhibit a gradient of $< 3 \rm \; km \; s^{-1}$ per 10 pc. Six of ten of these filaments meet all Bone criteria (with the other four failing the aspect ratio criterion\footnote{This paper determines the aspect ratio of the \citet{Zucker_2015} filaments more rigorously, using a custom-developed radial profile fitting analysis, and additional candidates that did not meet the extinction-based aspect ratio criterion in \citet{Zucker_2015} meet it in this paper, as discussed in \S \ref{lengths}}), and we include all ten listed in Table 2 from \citet{Zucker_2015}, plus Nessie \citep{Goodman_2014}, in this study (hereafter the ``Milky Way Bone" catalog). 

Finally, \citet{Wang_2016} produce a population of Minimum Spanning Tree (MST) Filaments by applying a Minimum Spanning Tree algorithm to dense molecular clumps from the Bolocam Galactic Plane Survey spectroscopic catalog \citep{Shirley_2013} in \textit{position-position-velocity} space. By connecting clumps with spatial ($< 0.1^\circ$) and kinematic ($<2 \; \rm km \; s^{-1}$) contiguity, \citet{Wang_2016} produce a sample of 54 dense Galactic filaments (lengths 10-300 pc). Thirteen of these 54 filaments also show some association with spiral arms, so \citet{Wang_2016} term them the MST ``Bone" filaments following the criteria outlined in \citet{Zucker_2015}. \citet{Wang_2016} argue that the physical properties of the MST ``Bones" are not differentiable from the larger MST filament sample. To prevent the analysis from becoming unwieldy, we only consider the 13 filaments in the MST ``Bone" sample, which puts the number of MST filaments on par with the other three catalogs. Of these thirteen MST ``Bone" filaments, one is outside the boundaries of the Herschel Hi-GAL survey (a key dataset in the forthcoming analysis), leaving twelve MST ``Bone" filaments available for analysis in this study. 

\begin{figure}[h!]
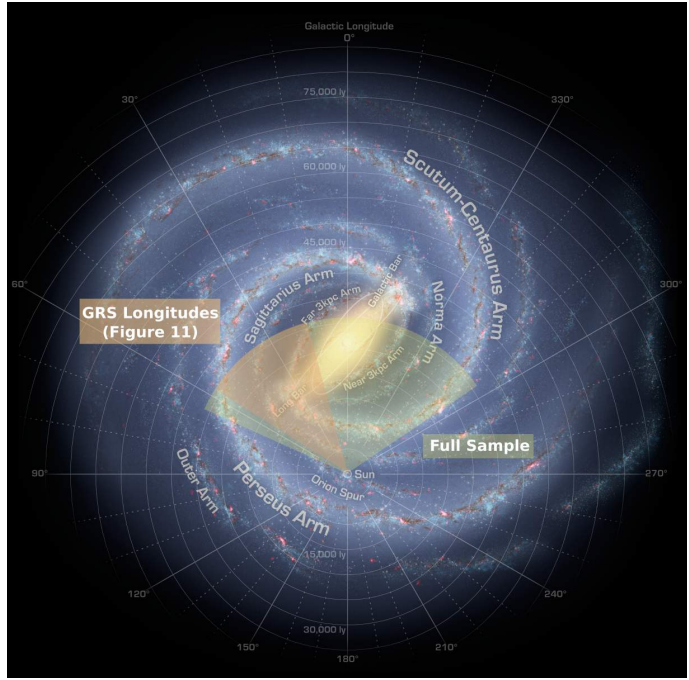

\begin{center}
\includegraphics[width=0.5\columnwidth]{{{top_down}}}
\caption{{\label{fig:top_down} Top down artist's conception view of the Milky Way Galaxy (Courtesy Robert Hurt; NASA). The longitude range of the full sample is shown as a yellow arc (with the radius of the arc set by the farthest filament distance, $\approx 10$ kpc (see Table \ref{tab:tab1}). Every filament we consider lies in this area. In orange we show the range of longitudes covered by the GRS survey of $\rm ^{13} CO$ in the first quadrant \citep{Jackson_2006}, which we use to perform a kinematic analysis on a subset of the filaments as discussed in \S \ref{ppv_summary} and Figure \ref{fig:pv_summary}. }}
\end{center}
\end{figure}

To complement these large-scale catalogs, there is an abundance of smaller-scale filament catalogs found in the literature, and a subset contains filaments longer than 10 pc in length \citep[see][]{Koch_2015,Schisano_2014,Li_2016}. However, the majority of these filaments are $\rm <10 \; pc$ \citep[see, for example, Figure 14 in][]{Li_2016}, making large-scale filaments the exception, rather than the norm. While a comparison of the filament properties within this hierarchy is worthy of further study, it is beyond the scope of this paper. To avoid selectively choosing filaments from a continuous length spectrum at some arbitrary ``large-scale" filament boundary, we exclude catalogs mainly composed of small-scale filaments. In total, there are 45 unique large-scale filaments included for analysis in this study. The filaments in the sample are confined to a small swath of area covering approximately two-thirds of the inner Galaxy, which is shown as a yellow polygon in a top down illustration of the Galaxy in Figure \ref{fig:top_down}. No filaments in the sample lie in the outer Galaxy.\footnote{Interested readers can view the overlapping footprints of the surveys used to identify the filaments using the WorldWide Telescope API at \href{http://milkyway3d.org/}{MilkyWay3D.org}}  

Few filaments appear in more than one of the catalogs we consider, as illustrated in Table \ref{tab:venntable}. Of the 45 filaments in the sample, only two filaments are identical across more than one catalog (the ``Snake" and ``Nessie", see Table \ref{tab:venntable}). For these two filaments, we include them in all filament catalogs in which they are found in Figures \ref{fig:colcomp}, \ref{fig:tempcomp}, \ref{fig:dgfcomp}, \ref{fig:lwa_comp}, and \ref{fig:armprops}. Four additional filaments share some spatial and kinematic overlap (i.e they are smaller filaments nested inside larger filaments), but this is uncommon. The lack of commonality between catalogs is not because each study examines a different spatial region. Prior studies have searched an identical swath of the Galaxy and found different filaments. For example, several prior studies have cataloged filaments in part of the first Galactic quadrant \citep[$ 18^\circ < l < 56^\circ$, corresponding to the footprint of the high-angular 46" resolution $\rm ^{13}CO$ survey from][]{Jackson_2006} and of the 22 filaments in this region, only two cataloged filaments share any spatial and kinematic overlap (Fil5 nested in GMF20 and G26 nested in GMF26, see Table \ref{tab:venntable}), accounting for $<$ 10 \% of the sample. 

\begin{table}
\begin{center}
  \includegraphics[width=0.75\textwidth]{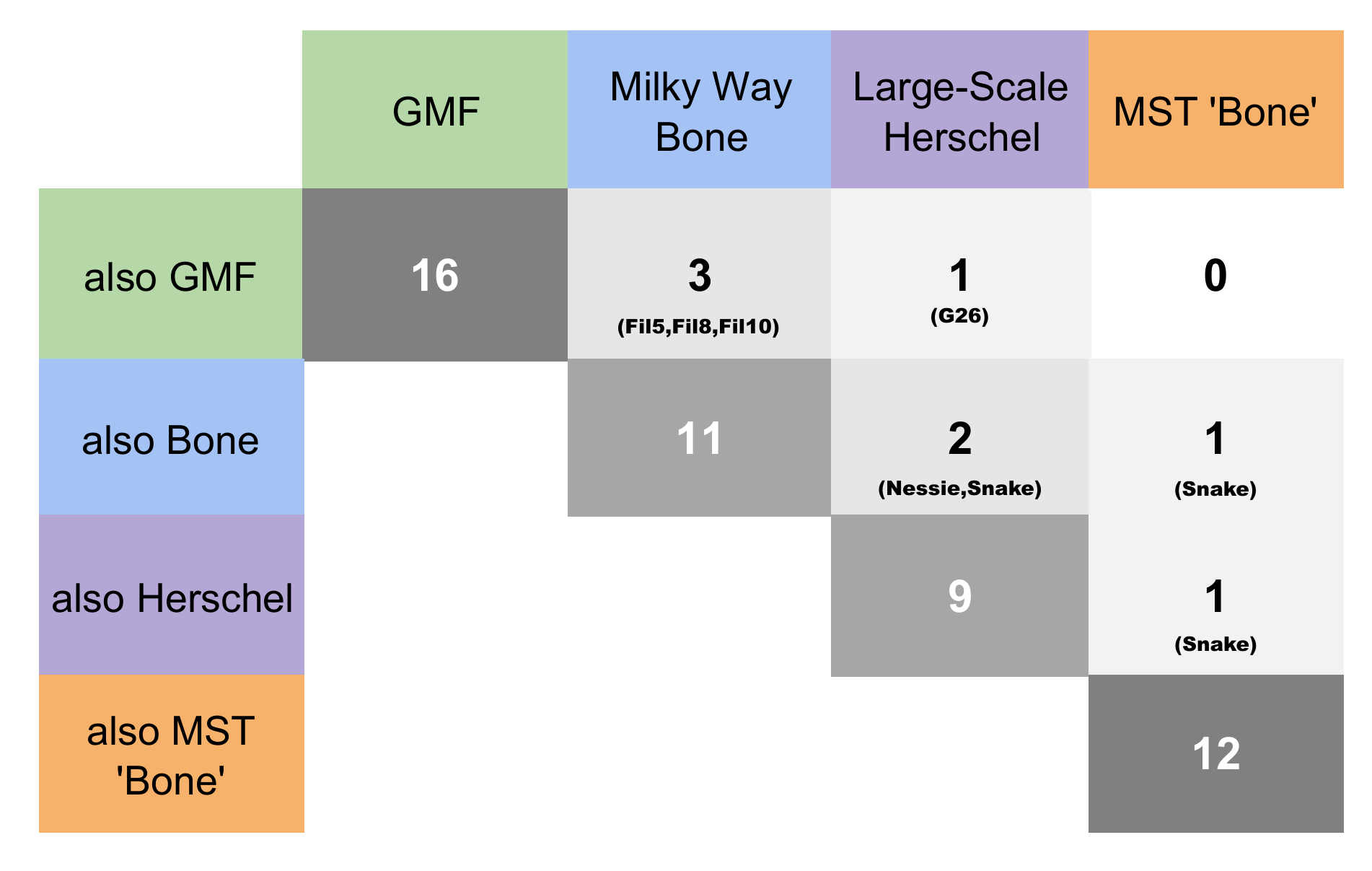}
    \caption{  \label{tab:venntable} Summary of the overlap between existing large-scale filament catalogs. Of the 45 filaments in our sample, only two filaments (Nessie, the ``Snake" filament) are identical across catalogs. An additional four filaments share some spatial and kinematic overlap with another filament in the sample (i.e. they are nested inside a larger filament), but this is uncommon. }
    \end{center}
\end{table}

\section{Methodology} \label{methodology}
As outlined in \S \ref{intro}, our goal is to produce a systematic reanalysis of the physical properties of major large-scale filament catalogs in the literature. This includes a measurement of the filaments' column densities and dust temperatures as well as their morphologies---all calculated using the same datasets and pipeline. We characterize our approach as follows:

To quantify the distribution of cold and dense gas, we perform a standardized analysis of the column densities and dust temperatures of the large-scale filaments in the sample, using the Herschel Hi-GAL survey of the Galactic plane \citep{Molinari_2016}, which provides coverage in five passbands at 70, 160, 250, 350, and 500 $\rm \mu m$. We choose the Hi-GAL survey due to its large dynamic range and its wavelength coverage near the peak of the spectral energy distribution for dust temperatures typical of large-scale filaments ($\rm 8 < T < 50 \;K$). It provides coverage of the entire inner Galactic plane ($\rm 68^\circ > l > -70^\circ, \;  |b| < 1^\circ$) with high resolution at its longest wavelength (35" nominal, diffraction-limited beam for the $500 \; \rm \mu m$ band). 

To gauge the morphology of the filaments (e.g. lengths, widths, aspect ratios) and to facilitate a uniform analysis of other properties, we redefine the boundaries of every filament in our sample quantitatively. \textit{While all the filaments are analyzed in the same framework, the fundamental differences in their selection criteria necessitated that we delineate the area of each filament class in different, yet standardized, ways}. While ideally we would apply the same column density threshold to the entire sample, doing so produces structures which are not filamentary and unrecognizable in comparison to their original studies in the majority of cases. Whenever possible, we use the Hi-GAL survey to delineate filament boundaries. If filaments cannot be delineated by-eye as a semi-continuous high-column-density feature in the Hi-GAL column density maps, we define their boundaries with the same dataset used to characterize them in their original publications. 

Because they exhibit high column density morphology along their entire length, we delineate the boundaries of the \citet{Zucker_2015} Milky Way Bones and the \citet{Wang_2015} Large-Scale Herschel Filaments by applying a column density threshold $1-2\sigma$ above the mean background column density using the custom Hi-GAL-based column density maps derived in this study (see \ref{bone_boundaries} for more details). Because the \citet{Ragan_2014} and \citet{Abreu_Vicente_2016} GMFs can only be visually traced via their low density gas, we apply an integrated intensity threshold to $\rm ^{13}CO$ zeroeth moment maps integrated over the velocity range of the filaments to delineate their boundaries. 

While the MST ``Bone" filaments are characterized by dense clumps, many of these clumps cannot be visually connected in Hi-GAL, so we offer two distinct methods of defining boundaries for the MST ``Bones" presented in \citet{Wang_2016}, both using the BGPS 1.1 mm data. In our `catalog-based' method, we define a polygon encompassing the outline of the ``tree" connecting all the dense molecular clumps \citep[from the spectroscopic catalog from][]{Shirley_2013} defining each MST Filament (see green outline in Figure \ref{fig:mst}). In our `continuum-based' method we apply a closed contour to the underlying BGPS 1.1 mm continuum emission enveloping the set of dense clumps from the catalog-based method (see yellow outline in Figure \ref{fig:mst}). While the catalog-based method assumes a cylindrical structure for each filament, the continuum-based method does not assume any geometry a priori. More detailed information on the application of the filament boundaries to all four classes can be found in \S \ref{gmf_boundaries}, \S \ref{bone_boundaries}, \S \ref{herschel_boundaries}, and \S \ref{mst_boundaries} in the Appendix. 

The final filament boundaries used in this analysis are shown via green contours in the Appendix (green and yellow contours for MST ``Bone" filaments---catalog- and continuum-based methods), and are available online at the \href{https://dataverse.harvard.edu/dataverse/Galactic-Filaments}{Large-Scale Galactic Filaments Dataverse}. Also on the Dataverse, we include a text file summarizing the contour levels used to define the boundaries around filaments in the sample. 

These quantitative boundaries, in combination with our Hi-GAL column density and dust temperature maps, are the basis for calculating the physical properties derived in this study, including:
(1) Distances (2) Nearest spiral arms (3) Column densities (4) Dust temperatures (5) Masses (6) High column density fractions (7) Cold \& high column density fractions (8) Lengths (9) Linear masses (1) Widths (11) Aspect ratios (12) Position angles (13) Galactic altitudes and (14) Kinematic displacements from spiral arms (for a subset of the sample with high angular resolution 46" CO data)

Details on how we calculate each property are outlined in subsections below and are summarized in Tables \ref{tab:tab1}, \ref{tab:tab2}, and \ref{tab:tab3}. A machine readable version of the combined table of properties (including all the data in Tables \ref{tab:tab1}, \ref{tab:tab2}, and \ref{tab:tab3}) is available for public download at the \href{https://dataverse.harvard.edu/dataverse/Galactic-Filaments}{Large-Scale Galactic Filaments Dataverse}, along with all the filament masks, the column density maps, and the dust temperature maps. 

\subsection{Distances and Nearest Spiral Arms} \label{distances}
Determining distances to gaseous features in the Galaxy is notoriously difficult. The simplest approach is to use an object's line-of-sight velocity, in combination with an assumed Galactic rotation curve, to place it at one or two possible ``kinematic" distances.\footnote{In the case of the inner Galaxy, there is a ``distance ambiguity" where the same line-of-sight velocity occurs at two points, where the (unmeasurable) transverse velocities would have opposite signs. For the outer Galaxy, there is a unique kinematic distance for any line-of-sight velocity.}  

Additional information, beyond kinematics, can also be used to constrain distance. For example, if we assume the Milky Way has a spiral pattern, then distance options can be informed by purported arms. If features lie near a Giant Molecular Cloud with a trigonometric parallax measurement, that parallax distance can be used to constrain the location of the source. Morphological features can also narrow down distance options.  For example, when a cloud (like many of the filaments discussed here) is seen as a dark silhouette against a bright background, one can assume, in trying to eliminate the kinematic distance ambiguity, that it is likely not at the ``far" distance option.

\begin{figure}[h!]
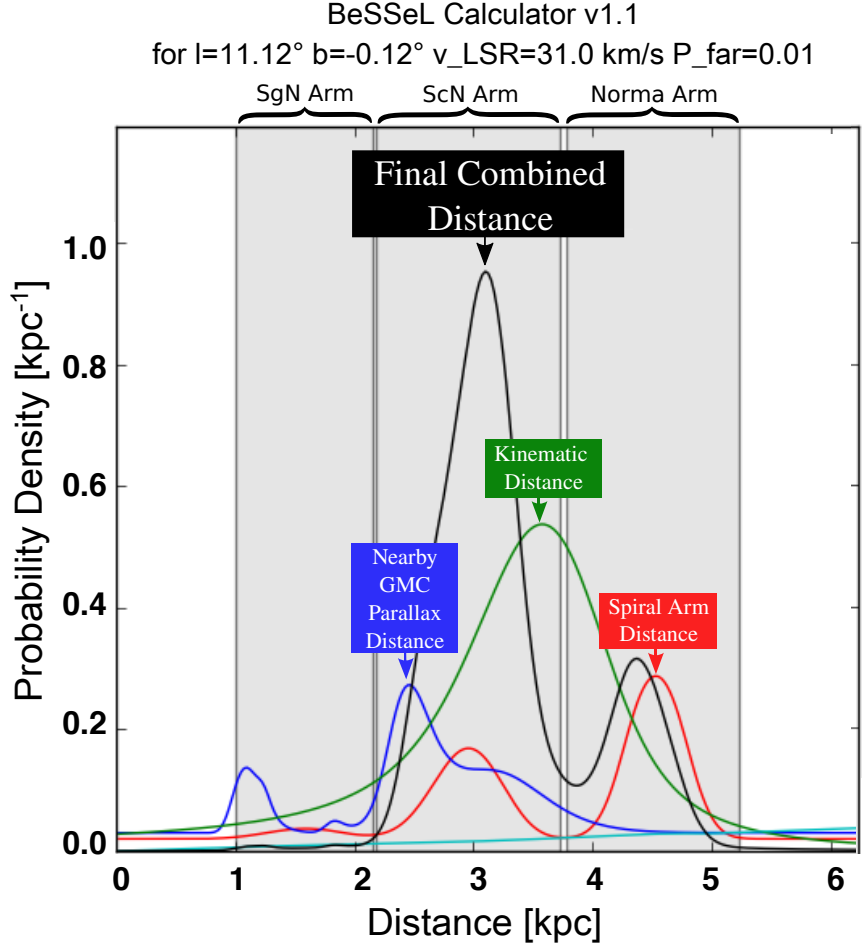

\begin{center}
\includegraphics[width=0.7\columnwidth]{{{reid2016_demo}}}
\caption{{\label{fig:reid_demo} The application of the Bayesian distance calculator from \citet{Reid_2016} to the ``Snake" filament (``Fil6" in Table \ref{tab:tab1}). Given a source's ($l,b,v$) values, along with the prior probability that it lies at the far distance ($P_{far}$), the calculator returns a combined distance probability density function (pdf) which is shown in black. The combined distance pdf is the product of individual distance pdfs based on information from kinematic distances (green curve), spiral arms (red curve), Galactic latitude (cyan curve), and proximity to a GMC with a parallax measurement (blue curve). The peak of the combined distance pdf indicates that the filament is most likely at a distance of $\rm 3.0 \pm 0.3 \; kpc$ and lies closest to the near Scutum-Centaurus arm in distance (``ScN", middle gray bar). }}
\end{center}
\end{figure}

Several constraints beyond kinematics are considered in the ``Bayesian Distance Calculator" presented in \citet{Reid_2016}, which we use to derive distances to the sample. The \citet{Reid_2016} calculator outputs a \textit{combined} distance probability density function (pdf) based on individual distance pdfs from current spiral arm models, kinematic distances, Galactic latitude, and proximity to a GMC with a parallax measurement.\footnote{The Bayesian Distance Calculator only takes into account GMCs with a parallax measurement, rather than any GMC along the line of sight. The rationale is that if one can associate a filament and a parallax source with the same GMC---based on linear separation in longitude, latitude, and velocity---then an additional constraint can be placed on the distance to the target filament. The entire ensemble of parallax sources is considered, and those with the smallest $(l,b,v)$ linear separation contribute the most weight. See \S 2.4 in \citet{Reid_2016} for more details.} In addition to a source's ($l,b,v$) values (we use the values shown in Columns 2-4 of Table \ref{tab:tab1}), the calculator requires the prior probability that the source lies at the far distance ($\rm P_{far}$), which is used to weight the importance of the near and far kinematic distances.\footnote{\citet{Ellsworth_Bowers_2013} are the first to use a Bayesian approach to distance determination by combining knowledge about a source's kinematic distance with prior external information from mid-IR extinction features and the Galactic distribution of molecular gas. \vadjust{\vskip\maxdimen}\citet{Ellsworth_Bowers_2013} find that only a few percent of dense molecular clumps with a high mid-IR contrast indicative of IRDC features lie at the far distance. The fraction at the far distance should be even lower in our mid-IR dark sample, as the \citet{Ellsworth_Bowers_2013} sources are more compact, and highly elongated extinction features are even less likely to lie at the far distance.  Adopting this $P_{far}$ criteria, all the filaments placed at the near distance in their original publications are also placed at the near distance in this analysis; the same is also true for the far distance.} Given the findings of \citet{Ellsworth_Bowers_2013}, we adopt $\rm P_{far}=0.01$ for all filaments that contain extinction features; otherwise we adopt $\rm P_{far}=0.5$. The application of the Bayesian distance calculator to a single filament in the sample (``Fil 6" i.e. the ``Snake" with $\{l,b,v, \rm P_{far}\}=\{11.1^\circ, -0.1^\circ, 31 \; km \; s^{-1}, 0.01\}$) is shown in Figure \ref{fig:reid_demo}. Combining all distance information, the calculator finds the most probable distance to be 3.0 kpc (black curve), which is closer than the most probable spiral arm ($\approx$ 4.5 kpc) and kinematic ($\approx$ 3.5 kpc) distances but farther than if the calculator only considers proximity to a GMC with a parallax measurement ($\approx$ 2.5 kpc). In general, the offset between the Bayesian distances and the pure kinematic distances in the sample is typically $ < 10-20\%$, though (rarely) it can be significantly higher if the distance is influenced by strong ($l,b,v$) proximity to a spiral arm trace or parallax measurement. 

In Figure \ref{fig:reid_demo}, the Bayesian distance calculator also returns the distance ranges of known spiral arms towards the filament (vertical gray rectangles in Figure \ref{fig:reid_demo}, corresponding to the near Sagittarius, Scutum, and Norma arms), which are used to inform the combined distance pdf.\footnote{The Bayesian distance calculator requires the formal probability of being in an arm be $>$ 1\% above the flat background probability. In Figure \ref{fig:reid_demo}, assuming that each spiral arm is centered at a distance $\mu$, the distance range over which the height of each Gaussian is $>$ 1\% above the background probability is $\mu \pm 2\sigma$. According to Figure 2 in \citet{Reid_2016}, the $1\sigma$ width of the arms along this sightline is approximately 300 pc. So the total width of each arm (gray bars in Figure \ref{fig:reid_demo}) would be four times the $1\sigma$ arm width, or $\approx$1200 pc, which is the approximate width of the spiral arms shown here.} Using the most probable \textit{combined} distance returned by the calculator (3.0 kpc for Fil 6), we determine whichever arm bounds the filament lies closest to in distance (e.g. Scutum-Centaurus for Fil 6). We then use the ($l,b,v$) trace of each filament's closest arm to calculate properties like spatial displacement from the arm at the ``combined" distance to the filament, the projected 2D angle between filament and arm, and the kinematic separation in \textit{longitude-velocity} space in \S \ref{posangle_altitude} and \S \ref{arm_proximity}. This method only allows us to calculate the arm each filament lies closest to, and not whether the filament actually lies \textit{in} that arm, versus an interarm region. 

Following the procedure shown in Figure \ref{fig:reid_demo} for the full sample, over half the filaments lie closest to the Scutum-Centaurus arm, at a distance of $\approx3-5$ kpc. The other half lie closest to the Norma arm and the Sagittarius arm, with a few outliers situated near the Local Spur and the Aquila Spur. The distance and the nearest spiral arm calculations are summarized in Table \ref{tab:tab1}. In the Appendix (\S \ref{gmf_gallery}, \ref{bone_gallery}, \ref{herschel_gallery}, \ref{mst_gallery}), we also overlay the $(l,b)$ tracks for each filament's nearest arm in lavender (if no arm is shown, this indicates it is outside the boundaries of the figure). 

\begin{deluxetable}{ccccccc}
\setlength{\tabcolsep}{12pt}
\tabletypesize{\scriptsize}
\renewcommand{\arraystretch}{0.75}
\colnumbers
\tablehead{\colhead{Name} & \colhead{Longitude} & \colhead{Latitude} & \colhead{Velocity} & \colhead{Type} & \colhead{Nearest Arm} & \colhead{Distance}  \\ \colhead{ } & \colhead{$\circ$} & \colhead{$\circ$} & \colhead{$\rm km \;  s^{-1}$} & \colhead{ } & \colhead{ } & \colhead{kpc}}
\startdata
Fil1 & 26.94 & -0.30 & 68 & Bone & ScN & 4.1 \\
Fil2 & 25.24 & -0.45 & 57 & Bone & ScN & 3.7 \\
Fil3 & 24.95 & -0.17 & 47 & Bone & ScN & 3.5 \\
Fil4 & 21.25 & -0.15 & 66 & Bone & Nor & 4.5 \\
Fil5 & 18.88 & -0.09 & 46 & Bone & ScN & 3.4 \\
Fil6* & 11.13 & -0.12 & 31 & Bone & ScN & 3.0 \\
Fil7 & 4.14 & -0.02 & 8 & Bone & ScN & 2.9 \\
Fil8 & 357.62 & -0.33 & 4 & Bone & CtN & 2.8 \\
Fil9 & 335.31 & -0.29 & -42 & Bone & CtN & 2.9 \\
Fil10 & 332.21 & -0.04 & -49 & Bone & CtN & 3.2 \\
Nessie** & 338.47 & -0.43 & -38 & Bone & CtN & 2.8 \\
F2 & 8.53 & -0.32 & 36 & MST & Nor & 4.4 \\
F3 & 8.76 & -0.37 & 38 & MST & Nor & 4.5 \\
F10 & 12.87 & -0.21 & 35 & MST & ScN & 3.0 \\
F13 & 14.07 & -0.49 & 21 & MST & SgN & 1.9 \\
F14 & 14.72 & -0.18 & 39 & MST & ScN & 3.1 \\
F15 & 14.20 & -0.19 & 40 & MST & ScN & 3.2 \\
F18 & 15.05 & -0.66 & 20 & MST & SgN & 1.9 \\
F28 & 25.30 & -0.22 & 63 & MST & ScN & 3.7 \\
F29 & 25.76 & -0.16 & 93 & MST & Nor & 5.6 \\
F37 & 37.39 & -0.07 & 57 & MST & SgF & 9.9 \\
F38 & 41.18 & -0.21 & 59 & MST & SgF & 9.0 \\
G24 & 24.00 & 0.48 & 96 & Herschel & Nor & 5.8 \\
G26 & 26.38 & 0.79 & 48 & Herschel & ScN & 3.0 \\
G28 & 28.68 & -0.28 & 88 & Herschel & ScN & 4.7 \\
G29 & 29.18 & -0.34 & 94 & Herschel & ScN & 5.0 \\
G47 & 47.06 & 0.26 & 58 & Herschel & SgF & 6.6 \\
G49 & 49.21 & -0.34 & 68 & Herschel & SgF & 5.7 \\
G64 & 64.27 & -0.42 & 22 & Herschel & LoS & 3.0 \\
GMF18 & 17.30 & 0.60 & 23 & GMF & SgN & 1.9 \\
GMF20 & 18.95 & 0.00 & 47 & GMF & ScN & 3.4 \\
GMF26 & 25.80 & 0.70 & 46 & GMF & ScN & 3.0 \\
GMF38a & 35.30 & 0.25 & 55 & GMF & AqS & 3.4 \\
GMF38b & 35.10 & -0.42 & 44 & GMF & SgN & 2.2 \\
GMF41 & 41.10 & -0.05 & 36 & GMF & SgN & 2.5 \\
GMF54 & 53.40 & 0.30 & 23 & GMF & LoS & 4.0 \\
GMF307 & 305.80 & 0.15 & -35 & GMF & CrN & 3.2 \\
GMF309 & 309.20 & -0.10 & -43 & GMF & CtN & 3.6 \\
GMF319 & 318.10 & -0.20 & -40 & GMF & CrN & 2.6 \\
GMF324 & 323.50 & -0.45 & -32 & GMF & CrN & 2.1 \\
GMF335a & 333.40 & -0.15 & -50 & GMF & CtN & 3.3 \\
GMF335b & 332.00 & -0.10 & -50 & GMF & CtN & 3.2 \\
GMF341 & 341.00 & -0.30 & -44 & GMF & CtN & 3.4 \\
GMF343 & 342.20 & 0.25 & -41 & GMF & CtN & 3.4 \\
GMF358 & 357.55 & -0.20 & 7 & GMF & CtN & 2.8 \\
\enddata
\caption{ \label{tab:tab1} Summary of large-scale filament properties computed in this study. The physical properties are as follows -- (1) Name of the filament  (2) Central longitude of the filament  (3) Central latitude of the filament (4) Central velocity of the filament (5) Filament type, sorted by original publication, with \citet{Zucker_2015} (Bone), \citet{Wang_2016} (MST), \citet{Wang_2015} (Herschel) and \citet{Ragan_2014, Abreu_Vicente_2016} (GMF) (6) Nearest spiral arm to the filament, determined using the Bayesian distance calculator from \citet{Reid_2016} (7) Distance in kpc derived using the Bayesian distance calculator from \citet{Reid_2016}}
\tablenotetext{*}{Fil6, colloquially known as the ``Snake" is also in the \citet{Wang_2015} Large-Scale Herschel filament sample as ``G11" and the the \citet{Wang_2016} MST Bone sample as ``F7"; it has been included in all three samples in Figures \ref{fig:colcomp}, \ref{fig:tempcomp}, \ref{fig:dgfcomp}, \ref{fig:lwa_comp}, and \ref{fig:armprops}}
\tablenotetext{**}{Nessie is also in the \citet{Wang_2015} Large-Scale Herschel filament sample as ``G339"; it has been included in both samples in Figures \ref{fig:colcomp}, \ref{fig:tempcomp}, \ref{fig:dgfcomp}, \ref{fig:lwa_comp}, and \ref{fig:armprops}. Due to the challenges of applying a semi-continuous closed contour to a 160+ pc long filament (see \S \ref{lengths}), we only consider the version of Nessie as originally defined in \citet{Jackson_2010}, even though Nessie is 2-5 times longer than originally claimed \citep{Goodman_2014}}
\end{deluxetable}

\subsection{Column Densities, Dust Temperatures, and Masses} \label{densities_temps}
To produce $\rm H_2$ column density and dust temperature maps, we perform pixel-by-pixel modified blackbody fits to the Hi-GAL 160, 250, 350, and 500 $\rm \mu m$ bands (the $70 \;  \micron$ band is excluded due to its high optical depth in very dense regions). We first convolve and regrid the 160, 250, and 350 $\rm \mu m$ bands to the 500 $\rm \mu m$ band grid and resolution. Though its nominal, diffraction-limited beam is 35", we convolve to the measured beam, found in Table 2 of \citet{Traficante_2011}. Since the measured beam has an $\approx 10\%$ ellipticity, we approximate the measured beam as circular and adopt the major axis beam size of 43" as the 500 $\rm \mu m$ band resolution.\footnote{Ideally, we would convolve each image to the same elliptical beam. However, according to \citet{Traficante_2011}, any individual scan has the long axis of its elliptical beam randomly oriented with respect to the scan position angle. Since the beam orientation of individual scans is unknown, we adopt the same symmetric beam assumption as \citet{Traficante_2011}.} The lower wavelength bands are convolved to a 43" resolution, and all column densities and dust temperatures are beam-averaged on this scale. Once convolved, we follow the convention of \citet{Battersby_2011} and adopt a mean molecular weight $\rm \mu_{H_2}$=2.8 \citep{Kauffmann_2008} and a gas-to-dust ratio of 100:1 \citep{Bohlin_1978}. We fix $\beta=1.75$, leaving only dust temperature, T, and column density $\rm N_{H_2}$ as the free parameters in the fit. To perform the fitting we use the \href{https://github.com/keflavich/HiGal_SEDfitter}{\textit{higal-sedfitter}} code \citep{Wang_2015}.

We conduct the SED fitting on both the original Hi-GAL fluxes and on fluxes that have been background subtracted. Due to the range of morphologies and size scales covered by our filament sample, picking an appropriate background subtraction method poses a challenge. We tested four different methods of background subtraction, before adopting the flat background subtraction method implemented in \citet{Juvela_2012}, which uses a circular reference area in a low emission region near the source to approximate the cirrus background. As \citet{Juvela_2012} argue, if the signal in the reference region is small, the resulting reference area provides a rough estimate of the total emission along the line of sight. While this does not account for structural variation in the background/foreground emission, it can be applied to filaments of all size scales. For more information on the three background subtraction methods tested but not implemented, see \S \ref{bg_subtract} in the Appendix. 

\begin{figure}[h!]
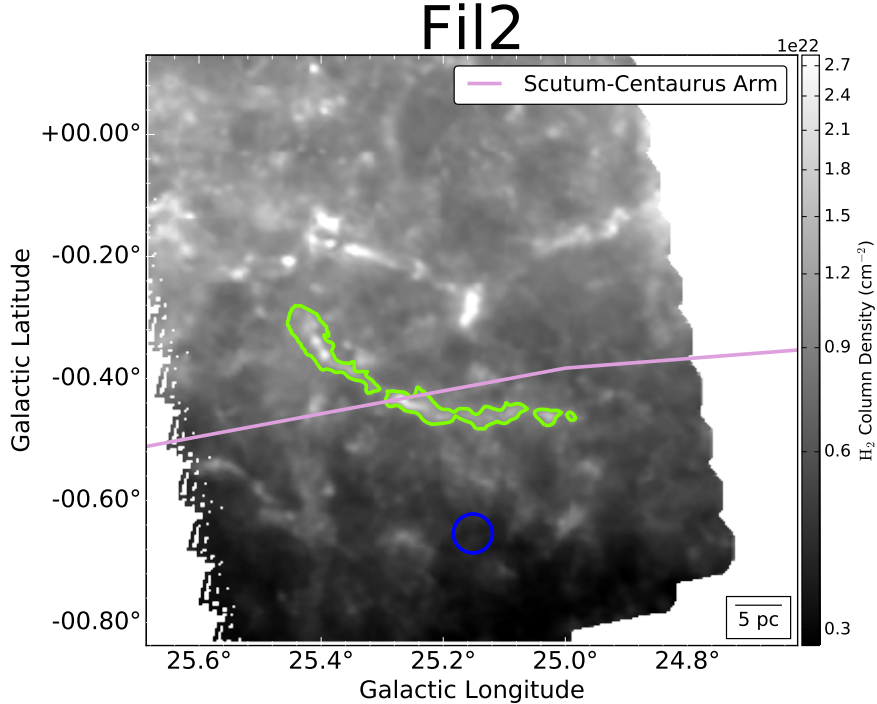

\begin{center}
\includegraphics[width=0.65\columnwidth]{{{Fil2_N_contours}}}
\caption{{\label{fig:colmap_ex} The unsubtracted Herschel column density map for a single filament in the sample (``Fil2"). The boundaries for the filament---used to calculate its physical properties throughout this study---are shown via the green contours, and are derived by applying a column density threshold $\approx 1\sigma$ above the mean background column density (see \S \ref{bone_boundaries} for details). The background column density is estimated within the blue circle, and this region is also used to determine a flux-subtracted, background-corrected column density for the filament, as discussed in \S \ref{densities_temps}. The $(l,b)$ track of the spiral arm the filament lies closest to in distance (see \S \ref{distances}) is overlaid in lavender \citep{Reid_2016}.}}
\end{center}
\end{figure}

For each source, we create a circular reference region with a radius of ten pixels ($\approx 0.03 ^\circ $), which is overlaid onto a low emission region outside but in close proximity to the boundaries of each filament. The flux within this aperture is measured in each band and subtracted, before running the pixel fitting code a second time on the flux-subtracted images.\footnote{In general, the column densities after background subtraction are $\approx60\%$ of their pre-subtracted values. Dense filaments without active star formation typically get $1 \rm \;  K$ cooler after background subtraction, though filaments which do contain a major star forming region typically get warmer by about $1 \rm \; K$.} The reference regions used for background subtraction are shown via the blue circles overlaid on the Herschel column density maps in the Appendix (\S \ref{gmf_gallery}, \ref{bone_gallery}, \ref{herschel_gallery}, \ref{mst_gallery}). There, we also overlay the quantitative boundaries of the filament (defined in \S \ref{sample_selection}) in green (green and yellow for the catalog- and continuum-based MST ``Bone" definitions). An example column density map for a single filament (``Fil2", see Table \ref{tab:tab1}) is shown in Figure \ref{fig:colmap_ex}.

Previously, only the \citet{Wang_2015} study has used the Hi-GAL survey to derive the column densities, dust temperatures, and masses of their sources. However, our methodology differs from the \citet{Wang_2015} study in a few ways. First, we use the 160, 250, 350, and 500 $\rm \mu m$ bands while \citet{Wang_2015} use the 70, 160, 250, and 350 $\rm \mu m$ bands. While excluding the longest wavelength band allows \citet{Wang_2015} to convolve to a higher resolution, the $70 \; \rm \mu m$ emission is generally assumed to be optically thick in the densest regions (i.e. much of our sample), with $\tau_{70} = 1$ at $N(H_2) \rm= 1.2 \times 10^{23} \; cm^{-2}$ according to \citet{Battersby_2011}. Its high optical depth is why the ``Snake" nebula, part of both the \citet{Wang_2015} and \citet{Zucker_2015} catalogs, is seen in absorption at $70 \; \rm \mu m$. When the emission becomes optically thick, fitting an optically thin greybody is no longer valid, so excluding the shortest wavelength Hi-GAL band is typical \citep[see][]{Battersby_2011,Peretto_2010,Zahorecz_2016,Longmore_2012}. Moreover, \citet{Battersby_2011} and others \citep{Finkbeiner_1999,Schnee_2008} argue that a large fraction of the dust emission at $70\; \micron$ is due to a separate population of very small dust grains that do not emit at longer wavelengths and whose dust temperatures are inconsistent with the colder equilibrium dust temperatures of large grains. 

Finally, to derive the filament masses we take the integral of the column densities across the filament masks, given by $\rm M_{tot}=\mu_{H_{2}}m_H\int N_{H_{2}} dA$, assuming $\mu_{H_{2}}=2.8$ \citep{Kauffmann_2008}. We approximate the integral by taking the sum over the column density in each pixel times its physical area. As seen in Table \ref{tab:tab2}, the masses of the large-scale filaments span almost three orders of magnitude: the lower mass bound for the large-scale filaments lies around $\rm 3 \times 10^{3} \; M_{\odot}$. The most massive filament in the sample is GMF335a at $\rm 1.1 \times 10^{6} \; M_{\odot}$. The GMFs are also the only class to contain any filaments greater than $\rm 1 \times 10^{5} \;  M_{\odot}$, which increases their average filament mass an order of magnitude higher than the other samples---all three of which have median masses on the order of $\rm 1 \times 10^{4} \;  M_{\odot}$.

To compare the average column density and dust temperature distributions of the large-scale filament samples, we collect every pixel within the quantitative filament boundaries for every filament in the sample. We compile these pixels into their different classes and create normalized column density and dust temperature PDFs that contain every pixel within every filament of that class. We plot the results of the column density analysis (in a log-log scale) in Figure \ref{fig:colcomp} and the dust temperature analysis in Figure \ref{fig:tempcomp}. 

As seen in Figure \ref{fig:colcomp}, the \citet{Wang_2016} MST ``Bone" filaments and the \citet{Zucker_2015} Milky Way Bones have the highest median column density ($\rm N_{H_2} = 1.0\times10^{22} \; cm^{-2}$), followed by the Large-Scale Herschel filaments ($\rm N_{H_2}  = 7.7\times10^{21} \; cm^{-2}$), and the Giant Molecular Filaments ($\rm N_{H_2}=4.8\times10^{21} \; cm^{-2}$). However, some of these distributions are clearly biased by selection criteria. For instance, the MST ``Bones" are identified by connecting dense clumps of molecular gas in $p-p-v$ space, which increases their relative fraction of very high column density gas in comparison to the other classes. 

In Figure \ref{fig:colcomp}, we also show in gray the distribution of a typical star forming region (L1689) in the nearby Ophiucus molecular cloud from \citet{Chen_2017} (see their Figures 2 \& 6). While the comparison is not 1:1 (we plot an entire class of filaments with multiple column density thresholds) the PDF of the GMFs in the high-density regime is almost identical to L1689 in Ophiucus. Compared to L1689, the Milky Way Bones, Large-Scale Herschel, and MST ``Bone" filaments all have larger fractional areas of their gas at higher column densities. The \citet{Zucker_2015} Bones also appear to have the steepest ``power law tail" towards higher column densities, which has been interpreted by some as evidence for the dominance of gravity over turbulence in the molecular cloud environment \citep[e.g.][]{Ballesteros_Paredes_2011}.

Unsurprisingly, Figure \ref{fig:tempcomp} shows that most of the gas in these filaments is cool, as suggested by the tight spread in median dust temperatures among all four classes, averaging $\approx 18-21 \rm \; K$.\footnote{While we do not measure the gas temperature, previous studies \citep[e.g.][]{Battersby_2014} have found that the gas and dust temperatures typically agree to within a few Kelvin in cold IRDCs.} The \citet{Zucker_2015} Milky Way Bones and the \citet{Wang_2015} Large-Scale Herschel filaments have the same median dust temperature ($\approx$ 18-19 K), while the MST ``Bones" and Giant Molecular Filaments have on average 1-2 K higher dust temperatures. 

\subsection{High Column Density Fraction and Potential Star Formation Activity} \label{dgf}
\begin{figure}[h!]
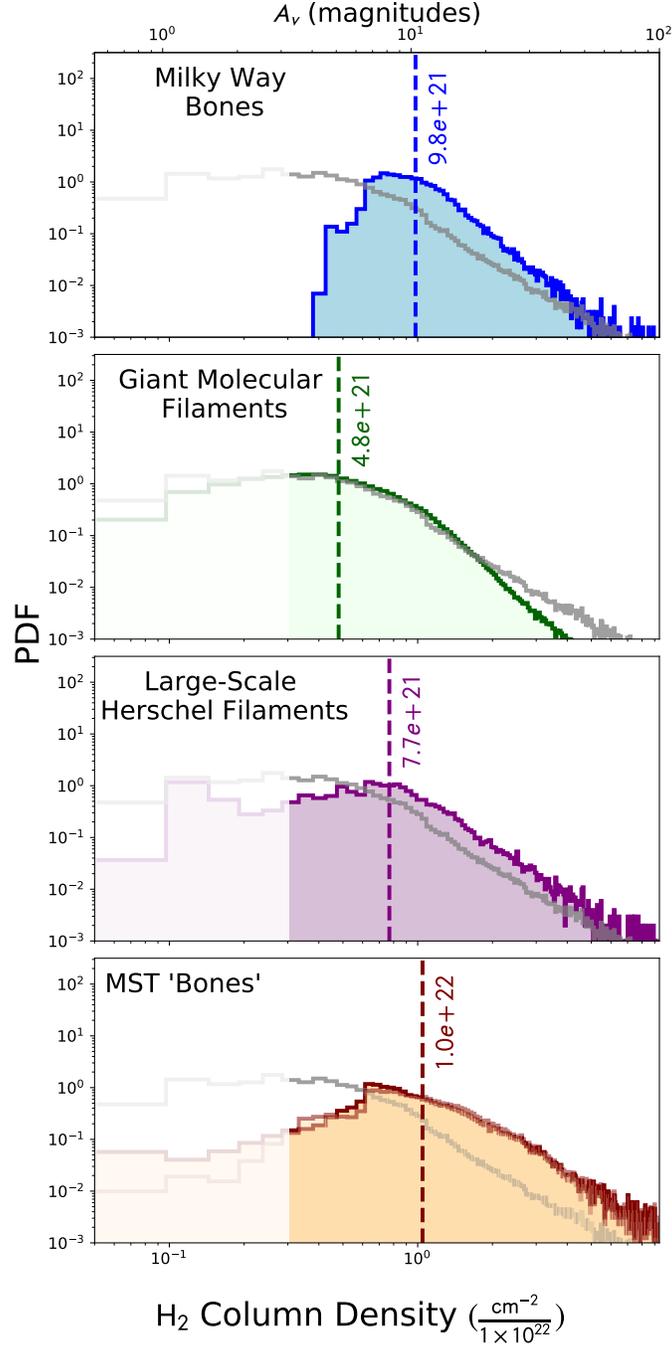

\begin{center}
\includegraphics[width=0.5\columnwidth]{{{Stacked_Col_Comp}}}
\caption{{\label{fig:colcomp} Herschel-derived $\rm H_2$ column density PDFs showing the column density distribution for each filament class as a whole. The histograms are normalized such that their total area is equal to one. We mark the median column density of each distribution via the vertical dashed lines. In each panel we additionally plot in gray the column density distribution of a typical star forming region in a nearby molecular cloud \citep[L1689 in Ophiuchus from][]{Chen_2017}. In this log-log view we mask out the low column density end ($N_{H_2} < 3 \times 10^{21} \; \rm cm^{-2}$) in white for visualization purposes. The corresponding $A_v$ values shown in the top axis are computed using the conversion $\frac{N_{H_2}}{A_v}=9.4 \times 10^{20} \rm \; cm^{-2}$ \citep{Bohlin_1978}. The filament classes span over a factor of two in median column density, with the MST ``Bones" and Milky Way Bones having the highest column density, followed by the Large-Scale Herschel filaments, and the GMFs. In the bottom panel, we show two distributions for the MST ``Bones", based on two definitions for the filaments outlined in \S \ref{mst_boundaries}. While the differences are minimal, the MST ``Bone" distribution based on the BGPS catalog data is plotted in dark orange while the BGPS continuum-based distribution is plotted in light orange.}}
\end{center}
\end{figure}

\begin{figure}[h!]
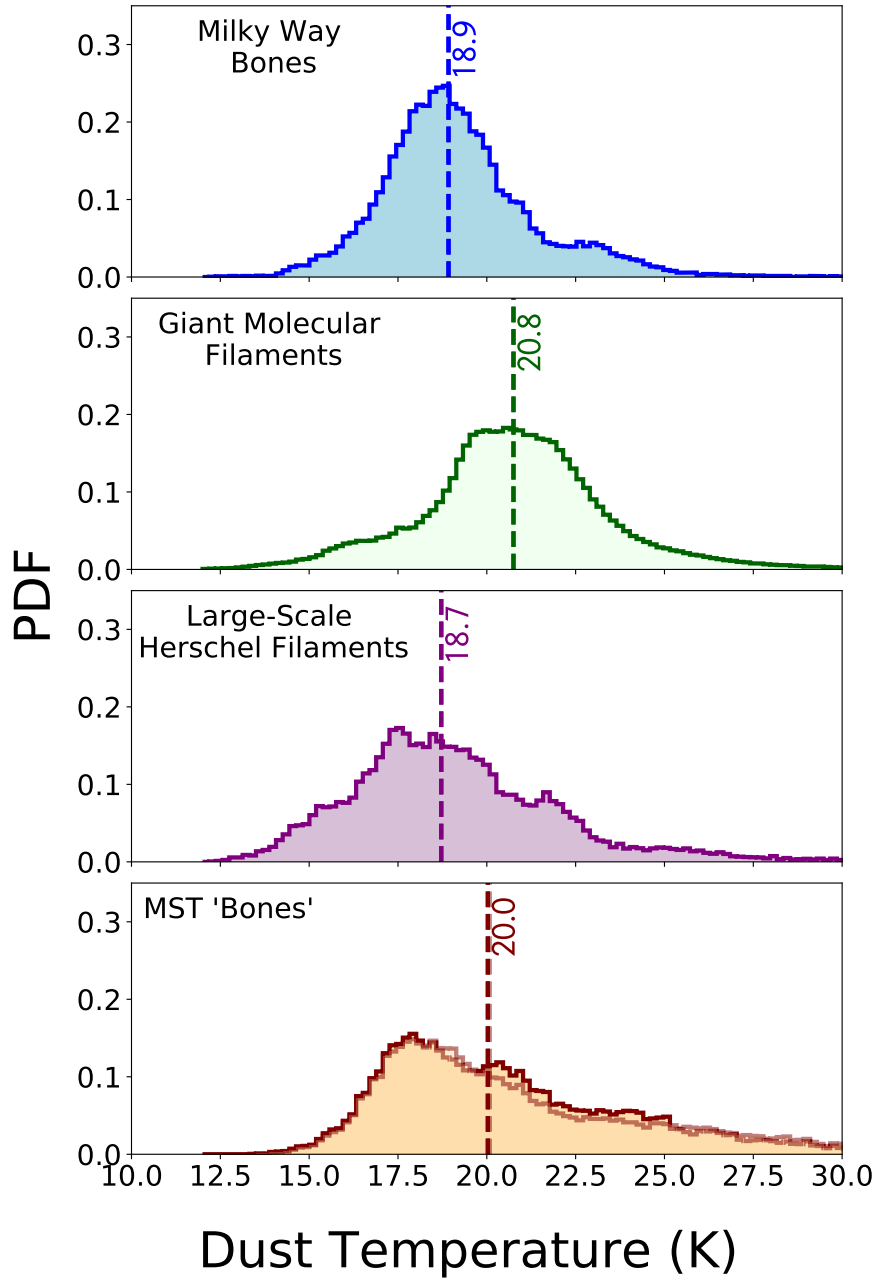

\begin{center}
\includegraphics[width=0.65\columnwidth]{{{Stacked_Temp_Comp}}}
\caption{{\label{fig:tempcomp} Herschel-derived dust temperature PDFs showing the dust temperatures distribution for each filament class as a whole. The histograms are normalized such that their total area is equal to one. We mark the median temperature of each distribution via the vertical dashed lines. The Milky Way Bones and the Large-Scale Herschel filaments tend to be on average about $1-2 \rm K$ cooler than the other classes. In the bottom panel, we show two distributions for the MST ``Bones", based on two definitions for the filaments outlined in \S \ref{mst_boundaries}. While the differences are minimal, the MST ``Bone" distribution based on the BGPS catalog data is plotted in dark orange while the BGPS continuum-based distribution is plotted in light orange. }}. 
\end{center}
\end{figure}

\begin{figure}[h!]
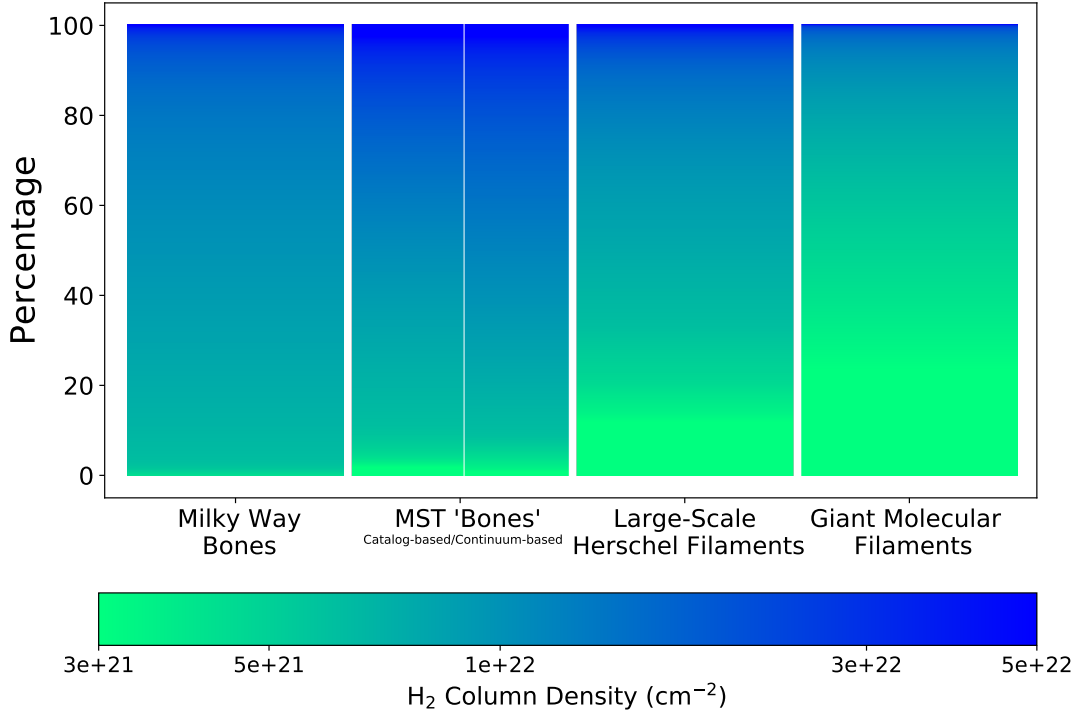

\begin{center}
\includegraphics[width=0.8\columnwidth]{{{DGF_Continuous}}}
\caption{{\label{fig:dgfcomp} High column density fraction analysis for the four filament classes included in this study, showing the relative fraction of a class' gas that falls within a set of continuous column density thresholds distributed between $N_{H_2} = 3.0 \times 10^{21} \; \rm cm^{-2}$ and $N_{H_2}=5.0 \times 10^{22} \; \rm cm^{-2}$. This is directly proportional to the relative area in each class occupied by gas of different column densities. One can determine the percentage of a class' gas lying below any column density by matching the color of the class's bar at that percentage to a corresponding column density in the colorbar. The Giant Molecular Filaments are dominated by low column density gas, with over 75\% of their gas falling below an $\rm H_2$ column density of $\rm 7.5\times10^{21} \;cm^{-2}$. At around 45\%, the MST ``Bone" filaments and the Milky Way Bone filaments have the largest fraction of gas above a column density of $\rm 1.0\times10^{22} \;cm^{-2}$. Gas with column densities capable of forming high-mass stars, $N_{H_2} \gtrsim \rm 2.5 \times 10^{22} \; \rm cm^{-2}$, \citep[see][]{Battersby_2017} is rare in all four classes, typically occupying only a few percent. For the MST ``Bones", we show high column density fractions based both on the BGPS catalog data (left bar) and the BGPS continuum data (right bar) definitions of this filament class, summarized in \S \ref{mst_boundaries}.  }}
\end{center}
\end{figure}

The variation in the column density distribution between classes (Figure \ref{fig:colcomp}) is better illustrated via a high column density fraction analysis, which we show in Figure \ref{fig:dgfcomp}.  Recall that we create catalog-specific column density distributions in Figure \ref{fig:colcomp} by binning the column density values found in every pixel within every filament of each catalog. We use this analysis to determine the fraction of gas in each filament class that falls within a continuous set of column density thresholds distributed between $N_{H_2} \rm = 3.0 \times 10^{21} \; cm^{-2}$ and $N_{H_2} =\rm 5.0 \times 10^{22} \; cm^{-2}$. This is equivalent to the fractional area occupied by gas of different column densities on the plane of the sky. Because the thresholds are continuous, one can interpret Figure \ref{fig:dgfcomp} by looking at the color corresponding to the percentage for each class. That color will correspond to a specific column density in the color bar, and it equates to the fraction of a class' area that falls below that column density value. While we cannot say for sure, the high column density fractions we report likely align well with dense gas fractions computed in three dimensions, due to the filamentary geometry of our sample. 

The Giant Molecular Filaments are dominated by gas at lower column densities, with over 75\% of their gas falling at a column density $N_{H_2} \rm < 7.5 \times 10^{21} \; cm^{-2}$, compared to $\approx$ 20-25\% for the Milky Way Bone and MST ``Bone" filaments and $\approx$ 50\% for the Large-Scale Herschel filaments. Adopting $N_{H_2} > 1.0 \times 10^{22} \rm \; cm^{-2}$ as a threshold for high column density gas, typical high column density fractions are 45\% - 50\% for the Milky Way Bone and MST ``Bone" filaments, 30\% for the Large-Scale Herschel filaments, and only around 10\% for the Giant Molecular Filaments. 

Our definition of ``star-forming" gas, with beam-averaged column densities sufficient for high-mass star formation \citep[$N_{H_2}  \gtrsim 2.5 \times 10^{22} \; \rm cm^{-2}$, see][]{Battersby_2017} is rare in all four classes, comprising about 3\% of the Milky Way Bones and Large-Scale Herschel filaments, compared to half a percent for the Giant Molecular Filaments. Unsurprising given their selection critieria (i.e. spatially and kinematically correlated massive clumps) the MST ``Bone" filaments have the highest fraction of ``star-forming" gas, but this still accounts for only $\approx$ 10\% of their total area. This might seem low, but we reiterate that our column densities are highly beam-diluted and represent the column densities averaged over our beamwidth of $43 \arcsec$ (the resolution of our longest wavelength Hi-GAL band), corresponding to $\approx 0.7$ pc at the median distance to our filament sample (3.3 kpc). As a result, our beam-diluted column densities will underestimate the true, peak column densities in dense regions typical of most of our sample \citep[see][]{Battersby_2011}

Finally, assuming that dust temperature above 25 K is most plausibly associated with star formation activity, we can estimate the \textit{fractional} area of each class over which star formation \textit{might} be actively taking place. This analysis follows from the work of \citet{Battersby_2017} which uses a $25 \rm \; K$  dust temperature threshold to differentiate between starry and starless clumps in dense molecular regions ($n \approx 10^{4}-10^{7} \; \rm cm^{-3}$). Adopting this definition, the MST ``Bone" filaments have the highest proportion of pixels with dust temperatures $>25 \;  \rm K$, at either $\approx$ 15\% or 22\% (for the catalog- and continuum-based method, where the former definition is based on the BGPS catalog data and the latter is based on the BGPS continuum-data, respectively; see \S \ref{mst_boundaries} for details on each method), followed by GMFs ($\approx$ 6\%), Large-Scale Herschel filaments ($\approx$ 5\%) and the Milky Way Bones ($\approx$ 2\%). For the Milky Way Bone filaments, this could be predicted observationally,  as all are originally identified as continuous mid-IR extinction features, and high star formation activity would disrupt the morphology of the filament. As these filaments are expected to produce high-mass stars \citep{Jackson_2010, Wang_2014} over the course of their lifetime, this would suggest that most of these filaments are at early stages in their evolution (less than a few million years), as otherwise, we would observe higher dust temperatures. 

\begin{deluxetable}{cccccc}
\setlength{\tabcolsep}{12pt}
\renewcommand{\arraystretch}{0.75}
\tabletypesize{\scriptsize}
\colnumbers
\tablehead{\colhead{Name} & \colhead{Median $\rm N_{H_2}$} & \colhead{Median Dust} & \colhead{High Column} & \colhead{Cold \& High Column} & \colhead{Mass} \\ \colhead{} &  \colhead{} & \colhead{Temperature} & \colhead{Density Fraction} & \colhead{Density Fraction} & \colhead{} \\ \colhead{ } &  \colhead{$\rm cm^{-2}$} & \colhead{K} & \colhead{ } & \colhead{ } & \colhead{$\rm M_\odot$}}
\startdata
Fil1 & 5.2e+21 & 22.3 & 0.03 & 0.03 & 3.8e+03 \\
Fil2 & 7.4e+21 & 20.6 & 0.18 & 0.08 & 1.0e+04 \\
Fil3 & 1.0e+22 & 17.5 & 0.53 & 0.53 & 3.0e+03 \\
Fil4 & 8.3e+21 & 19.4 & 0.23 & 0.17 & 7.8e+03 \\
Fil5 & 1.1e+22 & 20.6 & 0.71 & 0.30 & 3.0e+04 \\
Fil6* & 9.7e+21 & 16.8 & 0.49 & 0.48 & 1.5e+04 \\
Fil7 & 8.7e+21 & 19.2 & 0.30 & 0.27 & 2.5e+04 \\
Fil8 & 8.8e+21 & 18.9 & 0.36 & 0.34 & 9.9e+03 \\
Fil9 & 1.1e+22 & 18.1 & 0.66 & 0.60 & 2.3e+04 \\
Fil10 & 1.3e+22 & 18.8 & 0.93 & 0.76 & 3.8e+04 \\
Nessie** & 8.4e+21 & 18.9 & 0.32 & 0.24 & 4.2e+04 \\
F2 & 1.1e+22/1.3e+22 & 21.4/21.2 & 0.58/0.78 & 0.17/0.25 & 4.7e+04/4.7e+04 \\
F3 & 9.6e+21/1.5e+22 & 18.0/17.1 & 0.48/0.75 & 0.47/0.68 & 3.2e+04/3.2e+04 \\
F10 & 1.9e+22/1.9e+22 & 19.0/19.3 & 0.85/0.85 & 0.58/0.56 & 6.9e+04/6.9e+04 \\
F13 & 1.4e+22/1.4e+22 & 18.5/18.4 & 0.83/0.88 & 0.76/0.80 & 1.0e+04/1.0e+04 \\
F14 & 1.2e+22/1.5e+22 & 18.1/17.1 & 0.62/0.94 & 0.60/0.92 & 1.2e+04/1.2e+04 \\
F15 & 1.2e+22/1.2e+22 & 17.9/18.0 & 0.59/0.61 & 0.59/0.61 & 9.2e+03/9.2e+03 \\
F18 & 1.2e+22/9.8e+21 & 27.6/30.1 & 0.60/0.49 & 0.00/0.00 & 1.6e+04/1.6e+04 \\
F28 & 9.2e+21/9.8e+21 & 22.2/24.2 & 0.42/0.48 & 0.18/0.12 & 1.4e+04/1.4e+04 \\
F29 & 8.3e+21/1.1e+22 & 22.2/21.5 & 0.36/0.64 & 0.04/0.08 & 3.4e+04/3.4e+04 \\
F37 & 3.3e+21/4.2e+21 & 24.3/23.4 & 0.03/0.03 & 0.00/0.00 & 2.1e+04/2.1e+04 \\
F38 & 6.8e+21/6.5e+21 & 20.9/21.2 & 0.19/0.18 & 0.12/0.10 & 3.9e+04/3.9e+04 \\
G24 & 1.0e+22 & 19.4 & 0.50 & 0.42 & 4.5e+04 \\
G26 & 5.1e+21 & 18.1 & 0.10 & 0.10 & 7.3e+03 \\
G28 & 1.6e+22 & 19.0 & 1.00 & 0.78 & 2.6e+04 \\
G29 & 5.3e+21 & 21.7 & 0.06 & 0.06 & 1.9e+04 \\
G47 & 4.2e+21 & 17.6 & 0.07 & 0.07 & 2.8e+04 \\
G49 & 1.4e+22 & 24.3 & 0.78 & 0.02 & 5.7e+04 \\
G64 & 1.6e+21 & 16.0 & 0.00 & 0.00 & 4.0e+03 \\
GMF18 & 3.7e+21 & 22.7 & 0.03 & 0.02 & 4.7e+04 \\
GMF20 & 7.8e+21 & 21.5 & 0.22 & 0.09 & 8.4e+04 \\
GMF26 & 3.1e+21 & 21.0 & 0.01 & 0.01 & 1.3e+05 \\
GMF38a & 6.1e+21 & 20.2 & 0.11 & 0.06 & 6.9e+05 \\
GMF38b & 4.3e+21 & 20.3 & 0.06 & 0.05 & 5.0e+04 \\
GMF41 & 3.8e+21 & 20.0 & 0.01 & 0.01 & 3.9e+04 \\
GMF54 & 3.1e+21 & 18.6 & 0.03 & 0.02 & 4.5e+05 \\
GMF307 & 3.4e+21 & 22.1 & 0.02 & 0.00 & 5.5e+05 \\
GMF309 & 8.4e+21 & 16.2 & 0.33 & 0.33 & 7.3e+05 \\
GMF319 & 4.5e+21 & 19.6 & 0.04 & 0.04 & 2.6e+05 \\
GMF324 & 2.9e+21 & 19.6 & 0.00 & 0.00 & 6.4e+04 \\
GMF335a & 8.3e+21 & 22.0 & 0.35 & 0.04 & 1.1e+06 \\
GMF335b & 7.3e+21 & 21.4 & 0.22 & 0.13 & 1.8e+05 \\
GMF341 & 5.6e+21 & 21.5 & 0.11 & 0.04 & 6.3e+05 \\
GMF343 & 8.9e+21 & 19.4 & 0.38 & 0.32 & 1.0e+05 \\
GMF358 & 5.7e+21 & 21.8 & 0.06 & 0.04 & 1.7e+05 \\
\enddata
\caption{ \label{tab:tab2} Summary of large-scale filament properties computed in this study. For the filaments of type ``MST", properties in columns (2)-(6) are computed using two different boundary definitions for the filament (``catalog-based/continuum-based"); see \S \ref{mst_boundaries} for how MST boundaries are applied. The physical properties are as follows -- (1) Name of the filament from original publication (2) Median $\rm H_2$ column density inside the filament mask (3) Median dust temperature inside the filament mask (4) High column density fraction, defined as fraction of pixels in each filament's mask above a column density of $\rm 1\times 10^{22} \; cm^{-2}$ (5) Cold \& high column density fraction, defined as the fraction of pixels in each filament's mask above a column density of $\rm 1\times 10^{22} \; cm^{-2}$ and below a dust temperature of 20 K (6) Total mass derived from dust emission}
\tablenotetext{*}{Fil6, colloquially known as the ``Snake" is also in the \citet{Wang_2015} Large-Scale Herschel filament sample as ``G11" and the the \citet{Wang_2016} MST Bone sample as ``F7"; it has been included in all three samples in Figures \ref{fig:colcomp}, \ref{fig:tempcomp}, \ref{fig:dgfcomp}, \ref{fig:lwa_comp}, and \ref{fig:armprops}}
\tablenotetext{**}{Nessie is also in the \citet{Wang_2015} Large-Scale Herschel filament sample as G339; it has been included in both samples in Figures \ref{fig:colcomp}, \ref{fig:tempcomp}, \ref{fig:dgfcomp}, \ref{fig:lwa_comp}, and \ref{fig:armprops}. Due to the challenges of applying a semi-continuous closed contour to a 160+ pc long filament (c.f. \S \ref{lengths}), we only consider the version of Nessie as originally defined in \citet{Jackson_2010}, even though Nessie is 2-5 times longer than originally claimed \citep{Goodman_2014}}
\end{deluxetable}

\subsection{Lengths,Widths, Aspect Ratios} \label{lengths}
\begin{figure}[h!]
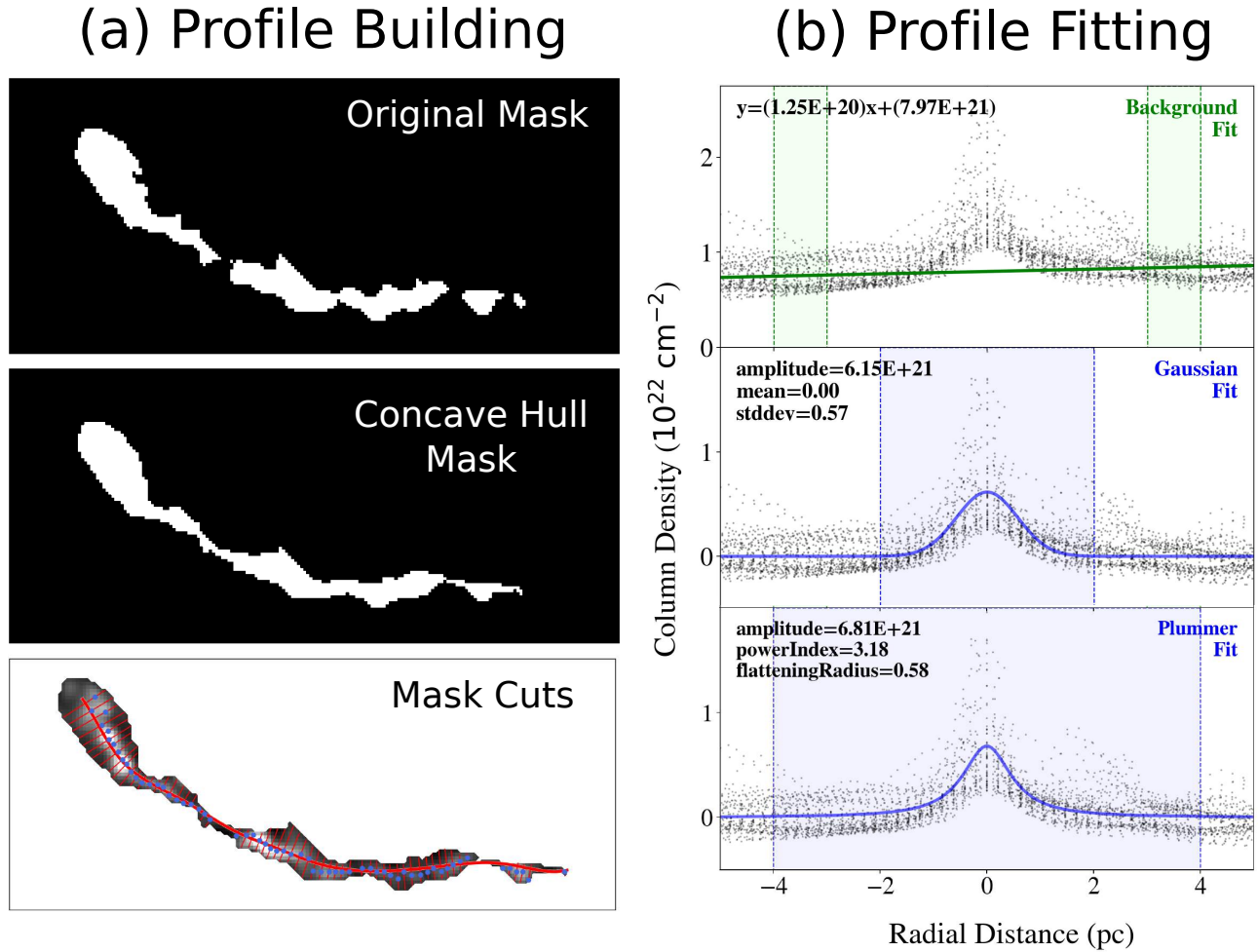

\begin{center}
\includegraphics[width=1.0\columnwidth]{{{New_RadFil_Figure}}}
\caption{{\label{fig:radfil} \textbf{a)} The process of building and fitting a radial column density profile for ``Fil2" (see Table \ref{tab:tab1} and Figure \ref{fig:colmap_ex}). First a filament mask is computed as outlined in \S \ref{bone_boundaries}. If the mask (Figure a., top) is not a single continuous closed contour, we apply an alpha-concave hull algorithm to the original mask to produce a connected feature (Figure a., middle). We create and smooth the filament spine (thick red line in bottom panel of Figure a.), make perpendicular cuts across it (thin red lines in bottom panel of Figure a.), and shift the profiles to the pixel with the peak column density across each cut (blue scatter points in bottom panel of Figure a.). \textbf{b)} We fit and subtract off a non-constant linear background (Figure b., top), before fitting Gaussian (Figure b., middle) and Plummer-like functions (Figure b., bottom) to the profiles. In each fit, the green and blue vertical dashed lines indicate the range in parsecs over which the background and Plummer/Gaussian functions are fit, respectively. }}
\end{center}
\end{figure}

We calculate the lengths of our filaments using the \href{https://github.com/e-koch/FilFinder}{FilFinder package} described in \citet{Koch_2015}. The package uses medial axis skeletonization to find filament spines by reducing filament masks to a one-pixel wide representation of the mask topology. We input masks derived from our filament boundaries outlined in \S \ref{sample_selection}. For filaments defined via a single, continuous closed contour, the FilFinder mask we use is identical to those shown in the Appendix. However, sometimes fluctuations in the background column density or integrated CO emission necessitate that we define the filament boundaries via a set of closely separated closed contours rather than a single continuous one. Because we calculate total length by finding the longest path through a set of connected skeletons, we transform the set of closed contours to a single continuous contour using an alpha-concave hull algorithm.\footnote{To compute the hulls we specifically use the alpha shape function originally found in this \href{https://github.com/dwyerk/boundaries/blob/master/concave_hulls.ipynb}{Github source code}. The alpha-concave hull is a polygon that envelopes the set of points defining the set of closed contours, such that the area of the polygon is minimized and any angle between border points is allowed. Unlike a convex hull, the alpha-concave hull representation of a set of points is not unique because the rigidity of the enveloping border depends on an alpha parameter; the larger the alpha parameter, the more closely the border follows the original boundaries of each polygon. We adopt the largest alpha parameter that can enclose our set of points.} Once each filament is represented by a continuous closed contour, we calculate the overall length of each filament by employing FilFinder's builtin ``longest path" option. An example of the concave hull mask for Filament 2 is shown in the middle panel of Figure \ref{fig:radfil}a. 

We determine the widths of the filaments using two different methods: in the first way, we compute a mask-based width for all the filaments using their boundaries. In the second way, we assume a cylindrical geometry for the filaments and fit Gaussians to the radial column density profiles of a subset of the sample displaying continuous, high column density filamentary structure. A third estimate of the width can be obtained by fitting a Plummer profile, as discussed in \S \ref{plummer}. The code required to perform these width analyses---including a jupyter notebook that users can run themselves---is publicly available online as part of the \href{https://github.com/catherinezucker/radfil}{\texttt{RadFil} package}. The \texttt{RadFil} package and its functionality will be described further in a forthcoming paper (Zucker \& Chen 2018, in prep) and more information is provided in \S \ref{radfil_appendix} in the Appendix. 

Our first estimate of the width involves taking perpendicular cuts across the spine, determining where the cut touches the edge of the mask on either side of the spine, and taking the median length of these cuts as the mask-based filament width (see the thin red cuts across the spine in the bottom panel of Figure \ref{fig:radfil}a). This mask-based width is determined for every filament in the sample and is given in Column (6) of Table \ref{tab:tab2}. For our second estimate of the width, we fit a Gaussian function to the radial column density profiles of a subset of filaments whose quantitative boundaries are defined via the Hi-GAL column density maps (the \citet{Zucker_2015} Bones and \citet{Wang_2015} Large-Scale Herschel Filaments). The Gaussian-based width is not determined for the GMF and MST ``Bone" samples. Specifically, we determine the pixel with the maximum column density value along each of the perpendicular cuts and build a radial column density profile along each cut with respect to this pixel. We then fit a Gaussian to the entire ensemble of radial column density profiles given by these set of cuts. Our Gaussian model has two free parameters (amplitude, standard deviation) and is given by $N(r)=a \exp{(\frac{-r^2}{2\sigma^2})}$, where a is the amplitude, and $\sigma$ is the standard deviation. We first subtract a background by fitting a first-order polynomial between a radial distance of 3 to 4 pc. All Gaussians are fitted out to a radius of 2 pc. We use the Gaussians to compute a deconvolved FWHM for the filaments using the best-fit $\sigma$ values, which are reported in Column (7) of Table \ref{tab:tab3}.\footnote{The deconvolved FWHM is computed using the formula from \citet{Konyves_2015}: $FWHM_{deconv}=\sqrt{(FWHM^2 - HPBW^2)}$ where FWHM is the FWHM we determine from the Gaussian fit and HPBW is the half-power beamwidth of our Hi-GAL column density maps ($43\arcsec$, converted to parsecs at the distance of each filament).} 

Several studies \citep{Smith_2014, Juvela_2012} have shown that the best-fit radial column density profile values are intimately dependent on background subtraction and fitting radius. As such, we provide alternative Gaussian-based widths for our filaments using different background subtraction and fitting radii in Table \ref{tab:profilealt} in the Appendix. The impact of fitting parameter variation is discussed at length in \S \ref{radfil_appendix} in the Appendix, but the average FWHM typically changes by a few tenths of a parsec for a reasonable range of fitting parameters. The process of building and fitting a radial column density profile for Filament 2 is shown in Figure \ref{fig:radfil}, and outlined in more detail in \S \ref{radfil_appendix} in the Appendix. 

\begin{figure}[h!]
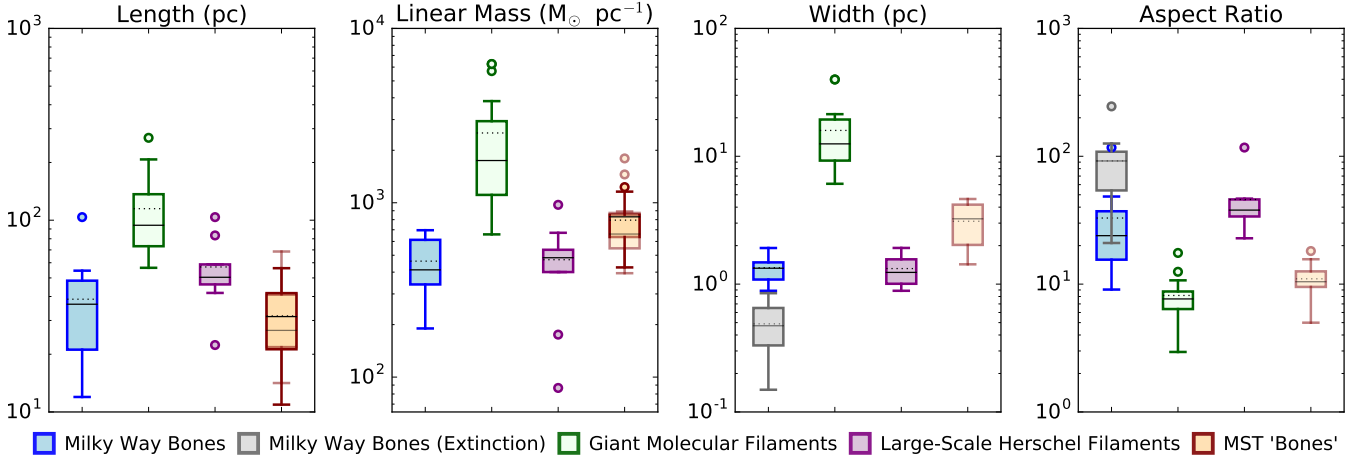

\begin{center}
\includegraphics[width=1.0\columnwidth]{{{length_linemass_width_aspect_newradfil}}}
\caption{{\label{fig:lwa_comp} Length, linear mass, width, and aspect ratio box-and-whisker plots for the four different filament classes in our sample. All four are plotted in log-scale. In each boxplot, the solid line indicates the median of the distribution while the dashed indicates the mean. The bottom and top edge of each box denotes the end of the first and third quartiles of the distribution, while the end of the ``whiskers" enclose all the data points that are not considered outliers, which are plotted as individual dots. We find significant variation in large-scale filament properties both among and between classes, with length, width, linear mass, and aspect ratio varying by one to two orders of magnitude from one end of the filament spectrum to the other. In panels one and two, we show two distributions for the MST ``Bone" lengths and linear masses---one from our catalog-based MST ``Bone" definition (dark orange) and the other from the continuum-based MST ``Bone" definition (light orange). In panels three and four, only the continuum-based MST ``Bone" widths and aspect ratios are available. In panels three and four, we additionally show in gray the extinction-derived widths and aspect ratios of the Milky Way Bone catalog, which are all originally identified as IRDCs. }}
\end{center}
\end{figure}

Box-and-whisker plots for the length, linear mass, width, and aspect ratio distributions are shown as side-by-side panels in Figure \ref{fig:lwa_comp}. As seen in the left-most panel of Figure \ref{fig:lwa_comp}, the median lengths of the four different filament classes span approximately a factor of three, with the GMFs having the highest median length (94 pc), followed by the Large-Scale Herschel filaments (50 pc), the Milky Way Bones (36 pc) and the MST ``Bone" filaments (31 pc for the catalog-based method, 27 pc for the continuum-based method). 

In the middle left panel of Figure \ref{fig:lwa_comp}, we see that the median linear masses for the four classes vary by a factor of four: the GMFs have the highest linear mass ($\approx \rm 1500 \; M\odot \; pc^{-1}$), with the average MST ``Bone" linear mass about half that value ($\approx 700-800 \; \rm M_\odot \; pc^{-1}$). The Milky Way Bones and the Large-Scale Herschel filaments have the lowest linear masses, with both distributions centered around $\approx 400-500 \; \rm M_\odot \; pc^{-1}$. This is about two orders of magnitude higher than the typical linear mass observed for smaller-scale filaments in the Herschel Gould Belt Survey of local molecular clouds \citep{Koch_2015}, with values of $\approx 5 \; \rm M_\odot \; pc^{-1}$---though some filaments do show linear masses as high as $100-200 \; \rm  M_\odot \; pc^{-1}$ \citep{Arzoumanian_2011}.

In the middle right panel of Figure \ref{fig:lwa_comp}, we show the distribution of filament widths. For the Milky Way Bone and Large-Scale Herschel filaments, we plot the full-width half max (FWHM) of the filaments (deconvolved with the beam) derived from the Gaussian fits to the radial column density profiles; alternative mask-based widths are shown in Table \ref{tab:tab3} and typically agree within a factor of two. For the Giant Molecular Filaments and the MST ``Bone" filaments, we only show a mask-based width, as no radial column density profile analysis is possible for these filaments due to a lack of continuous high column density structure in their column density maps. Regardless of width type, the median filament width across catalogs spans approximately one order of magnitude, with both the \citet{Wang_2015} and \citet{Zucker_2015} filaments possessing a median width of $\rm \approx 1.3 \; pc$, the MST ``Bones" $\approx$ 3 pc, and the GMFs $\approx$ 12 pc.  

All the \citet{Zucker_2015} Milky Way Bones are originally identified as mid-IR extinction features. In addition to showing the Hi-GAL emission-based widths (dark blue box and whiskers), we also show in Figure \ref{fig:lwa_comp} the dust extinction-based widths obtained via Gaussian fitting of column density maps derived from the GLIMPSE $\rm 8 \; \mu m$ images (gray box and whiskers). The IRDC features are skeletonized in the same fashion as the emission-based masks, and cuts are taken at the same sampling interval; more details on this procedure can be found in \S \ref{irdc_profiles} in the Appendix. Comparing the two distributions, typical extinction widths are over a factor of two smaller than the emission widths, though the discrepancy can be as high as a factor of six (e.g. for Filament 2).\footnote{Two compounding factors likely contribute to the smaller extinction widths. The first factor is the significantly higher resolution of the Spitzer-GLIMPSE survey: one pixel element in the GLIMPSE $8 \; \rm \micron$ maps is $1.2"$, which is over an order of magnitude smaller than the pixel element used in our Hi-GAL column density map analysis. Second, we are likely sensitive to different parts of the cloud, with the extinction widths tracing only the very densest and coldest elements. The dispersion in the Milky Way Bone extinction widths is noticeably higher than the emission widths, likely due to how dust emission maps to dust extinction at $8 \rm \; \mu m$ and how the visibility of the feature depends on a constant illuminating background.}

Finally, in the right-most panel of Figure \ref{fig:lwa_comp} we plot the distribution of aspect ratios, which unsurprisingly shows the same bimodality evident in panel two. The \citet{Wang_2015} Large-Scale Herschel filaments have the highest median aspect ratio (38:1) followed by the Milky Way Bones (24:1), the MST ``Bone" filaments (11:1) and the GMFs (8:1). If one uses the extinction widths to calculate the aspect ratios for the Milky Way Bones (gray box-and-whiskers in panel three), \textit{their median aspect ratio rises to $\approx$ 80:1}. In \citet{Zucker_2015}, the minimum extinction-based aspect ratio required for Bone classification was 50:1, and all the \citet{Zucker_2015} filaments which satisfied this criterion originally also satisfy it here, plus an additional two filaments. Given the more rigorous determinations of filament length and width described above, in total 8/10 filaments in the \citet{Zucker_2015} meet all the Bone criteria, with prescriptions for aspect ratio, velocity contiguity, position angle, and Galactic plane proximity. 

\subsubsection{Plummer Profiles} \label{plummer}
In addition to Gaussians, we also fit Plummer-like profiles to the same subset of filaments displaying continuous high column density structure in their Hi-GAL column density maps (the \citet{Zucker_2015} Milky Way Bones and \citet{Wang_2015} Large-Scale Herschel Filaments). The Plummer-like cylindrical model is the same as that found in \citet{Cox_2016}: 

\begin{equation}
N(r) = \frac{N_0}{[{1+{(\frac{r}{R_{flat}})^2}]}^{\; \frac{p-1}{2}}}
\end{equation}

where $N_0$ is the maximum peak column density along the spine, $p$ is the index of the density profile, and $R_{flat}$ is the inner flattening radius. We fit the profile out to a radius of 4 pc and adopt the same background subtraction radii as the Gaussian fits (3 to 4 pc). 

In Figure \ref{fig:plummer_comp} we plot histograms of the Plummer ``$p$" index obtained from the Plummer-like fits to the radial column density profiles for large-scale filaments (blue) and small-scale filaments (red). The large-scale sample includes all filaments in the Milky Way Bone and Large-Scale Herschel catalogs, while the small-scale sample is taken from the \citet{Arzoumanian_2011} study of 27 filaments in the IC5146 molecular cloud. One of the key results from the \citet{Arzoumanian_2011} study is that the density profiles of small-scale filaments fall off like $r^{-1.5}-r^{-2.5}$ at large radii. When $r>>R_{flat}$, $N(r) = \frac{N_0}{[{1+({\frac{r}{R_{flat}})^2}]}^{\; \frac{p-1}{2}}} \propto r^{-p}$, so the index of the density profile reduces to the Plummer ``$p$" index of these filaments ($\approx 1.5-2.5$). This range of Plummer ``$p$" indices appears constant across small-scale filament studies, and is also seen in the \citet{Juvela_2012} of filaments embedded in cold clouds identified by Planck, and in the \citet{Palmeirim_2013} study of filaments in the B211/L1495 region of Taurus. The shallow power law profiles have also been observed in simulations of small-scale filaments without the need for magnetic support, and could simply be a consequence of their formation in a turbulent molecular cloud environment \citep{Smith_2014a}. While pervasive, the shallow power law index is inconsistent with the theoretical predictions for an isothermal, self-gravitating, axisymmetric cylinder in hydrostatic equilibrium, whose profile should fall off as $r^{-4}$ (Plummer ``$p$" of 4.0) at large radii \citep{Ostriker_1964}. As shown in Figure \ref{fig:plummer_comp}, the large-scale filament profiles are more consistent with the \citet{Ostriker_1964} model, with a median Plummer ``$p$" index $\approx 3.15$, higher than seen in small-scale samples. 

One relevant variable could potentially be how overpressured the filaments are with respect to their environment, where ``overpressure" is defined as the ratio of the filament's central pressure to the external pressure \citep{Fischera_2012}. Modeling filaments as isothermal, self-gravitating infinite cylinders, \citet{Fischera_2012} find that filaments which are highly overpressured (by a factor of 20) show $r^{-4}$ profiles at large radii (Plummer ``$p$" of 4.0), consistent with the classical \citet{Ostriker_1964} result, while filaments which are only mildly overpressured (factor of 6-12) display the shallow $r^{-2}$ profile. Under the assumption that our large-scale filaments can be modeled as isothermal cylinders, they may have higher relative central pressures than small-scale filaments, consistent with a factor of 16 for our median $r^{-3.15}$ profile at large radii \citep{Fischera_2012}. However, without results from numerical simulations, it is difficult to speculate on the exact origin of the steeper density profiles we observe. 

Another possible explanation could simply be difference in background type. For small-scale filaments in clouds like IC5146 \citep{Arzoumanian_2011} the region used for background subtraction is still deeply embedded in the larger GMC environment. This is not the case for large-scale filaments---we are fitting a larger fraction of the cloud, so the background contribution from the GMC itself is diminished. 

Unfortunately, like the Gaussian best-fit parameters, the Plummer-like best fit parameters are also \textit{highly} dependent on the choice of fitting and background subtraction radii. Given the general uncertainty surrounding the Plummer-like fits, and the fact that this uncertainty has also been voiced in other studies \citep{Smith_2014a}, we are reticent to attribute the large-scale filaments' steeper profiles solely to the physical conditions in the cloud as opposed to systematics in the fitting process. While all combinations of background subtraction radii and cutoff radii indicate steeper profiles than seen in small-scale filaments, it does have a fundamental effect on the physical significance one can attribute to the average Plummer ``$p$" value of the distribution. Variations in the Plummer ``$p$" value given different background subtraction and fitting radii are summarized in Table \ref{tab:profilealt} and discussed at length in \S \ref{radfil_appendix} in the Appendix. In short, the average ``$p$" value of the distribution can vary by at least a half, though it typically falls between 2.75 and 3.25 when adopting a reasonable range of fitting parameters. 

\begin{deluxetable}{ccccccccccc}
\renewcommand{\arraystretch}{0.75}
\setlength{\tabcolsep}{2pt}
\tabletypesize{\scriptsize}
\setlength{\tabcolsep}{6pt}
\colnumbers
\tablehead{\colhead{Name} & \colhead{Position} & \colhead{Galactic} & \colhead{Length} & \colhead{Linear} & \colhead{Mask} & \colhead{FWHM} & \colhead{FWHM} & \colhead{Aspect Ratio} & \colhead{Aspect Ratio} & \colhead{$\rm \Delta v_{arm}$ \vspace{-0.5em}} \\ \colhead{} & \colhead{Angle} & \colhead{Altitude} & \colhead{} & \colhead{Mass} & \colhead{Width} & \colhead{$\rm _{(emission)}$} & \colhead{$\rm _{(extinction)}$} & \colhead{} & \colhead{$\rm _{(extinction)}$} & \colhead{}\\ \colhead{ } & \colhead{$\circ$} & \colhead{pc} & \colhead{pc} & \colhead{$\rm M_\odot \; pc^{-1}$} & \colhead{pc} & \colhead{pc} & \colhead{ } & \colhead{ } & \colhead{ } & \colhead{$\rm km \; s^{-1}$}}
\startdata
Fil1 & 1 & 18 & 20 & 190 & 1.3 & 1.5 & 0.2 & 13 & 92 & 0.5 \\
Fil2 & 33 & 4 & 37 & 280 & 1.2 & 1.1 & 0.1 & 33 & 246 & 4.2 \\
Fil3 & 14 & 13 & 12 & 254 & 0.9 & 1.3 & 0.6 & 9 & 21 & 15.4 \\
Fil4 & 10 & 2 & 20 & 399 & 1.9 & 1.9 & 0.3 & 10 & 78 & 8.2 \\
Fil5 & 15 & 16 & 52 & 585 & 3.0 & 1.1 & 0.4 & 48 & 126 & 1.2 \\
Fil6* & 7 & 13 & 22 & 672 & 2.2 & 1.0 & 0.7 & 23 & 30 & -- \\
Fil7 & 11 & 12 & 45 & 551 & 3.0 & 1.9 & 0.5 & 24 & 95 & -- \\
Fil8 & 15 & 4 & 24 & 412 & 1.6 & 1.4 & 0.4 & 18 & 58 & -- \\
Fil9 & 20 & 10 & 36 & 640 & 1.9 & 1.5 & 0.7 & 25 & 50 & -- \\
Fil10 & 6 & 4 & 55 & 695 & 2.5 & 1.3 & 0.6 & 41 & 94 & -- \\
Nessie** & 5 & 7 & 104 & 404 & 1.7 & 0.9 & 0.8 & 117 & 122 & -- \\
F2 & 2/5 & 15/14 & 40/26 & 1158/1796 & --/4.4 & -- & -- & --/5 & -- & -- \\
F3 & 17/14 & 17/18 & 37/37 & 851/845 & --/3.0 & -- & -- & --/12 & -- & -- \\
F10 & 14/7 & 5/6 & 56/47 & 1229/1453 & --/4.6 & -- & -- & --/10 & -- & -- \\
F13 & 15/14 & 3/3 & 24/25 & 425/394 & --/1.4 & -- & -- & --/18 & -- & -- \\
F14 & 34/17 & 4/2 & 15/20 & 841/611 & --/1.5 & -- & -- & --/12 & -- & -- \\
F15 & 2/3 & 3/2 & 11/14 & 843/650 & --/1.5 & -- & -- & --/9 & -- & -- \\
F18 & 12/8 & 7/6 & 18/17 & 865/891 & --/3.5 & -- & -- & --/4 & -- & -- \\
F28 & 25/30 & 13/14 & 26/27 & 527/499 & --/2.9 & -- & -- & --/9 & -- & 1.1 \\
F29 & 26/21 & 74/73 & 42/38 & 799/868 & --/3.6 & -- & -- & --/10 & -- & 75.3 \\
F37 & 8/6 & 22/21 & 42/48 & 503/428 & --/4.1 & -- & -- & --/11 & -- & 1.2 \\
F38 & 6/6 & 26/23 & 47/68 & 819/563 & --/4.4 & -- & -- & --/15 & -- & 1.3 \\
G24 & 30 & 32 & 83 & 537 & 1.9 & 1.9 & -- & 43 & -- & 25.6 \\
G26 & 24 & 70 & 42 & 175 & 1.7 & 1.2 & -- & 34 & -- & 19.1 \\
G28 & 10 & 4 & 50 & 507 & 1.8 & 1.1 & -- & 47 & -- & 17.3 \\
G29 & 8 & 16 & 48 & 400 & 5.8 & 1.6 & -- & 31 & -- & 17.7 \\
G47 & 32 & 49 & 59 & 483 & 4.6 & 1.7 & -- & 35 & -- & 1.4 \\
G49 & 14 & 9 & 59 & 972 & 2.8 & 1.5 & -- & 38 & -- & 9.0 \\
G64 & 37 & 30 & 46 & 86 & 1.9 & 1.0 & -- & 46 & -- & -- \\
GMF18 & 9 & 4 & 56 & 838 & 8.2 & -- & -- & 7 & -- & 4.0 \\
GMF20 & 18 & 17 & 62 & 1353 & 6.7 & -- & -- & 9 & -- & 0.9 \\
GMF26 & 16 & 64 & 104 & 1199 & 12.1 & -- & -- & 9 & -- & 18.7 \\
GMF38a & 2 & 11 & 269 & 2566 & 15.3 & -- & -- & 18 & -- & 22.8 \\
GMF38b & 28 & 10 & 76 & 658 & 6.1 & -- & -- & 12 & -- & 14.5 \\
GMF41 & 76 & 24 & 59 & 665 & 8.0 & -- & -- & 7 & -- & 2.3 \\
GMF54 & 3 & 11 & 80 & 5690 & 9.6 & -- & -- & 8 & -- & 21.3 \\
GMF307 & 4 & 11 & 207 & 2640 & 39.9 & -- & -- & 5 & -- & -- \\
GMF309 & 4 & 23 & 117 & 6243 & 39.9 & -- & -- & 3 & -- & -- \\
GMF319 & 12 & 9 & 127 & 2064 & 19.3 & -- & -- & 7 & -- & -- \\
GMF324 & 2 & 19 & 84 & 767 & 11.5 & -- & -- & 7 & -- & -- \\
GMF335a & 1 & 12 & 169 & 6251 & 21.3 & -- & -- & 8 & -- & -- \\
GMF335b & 14 & 5 & 70 & 2607 & 13.4 & -- & -- & 5 & -- & -- \\
GMF341 & 3 & 9 & 165 & 3828 & 19.7 & -- & -- & 8 & -- & -- \\
GMF343 & 18 & 34 & 74 & 1415 & 12.9 & -- & -- & 6 & -- & -- \\
GMF358 & 16 & 10 & 117 & 1428 & 10.9 & -- & -- & 11 & -- & -- \\
\enddata
\caption{\label{tab:tab3} Summary of large-scale filament properties computed in this study. For the filaments of type ``MST", properties in columns (2), (3), (4), \& (5) are computed using two different boundary definitions for the filament (``catalog-based/continuum-based"); see \S \ref{mst_boundaries} for how MST boundaries are applied. The physical properties are as follows -- (1) Name of the filament from original publication (2) Absolute 2D projected position angle between the filament and the midplane of the arm it lies closest to in distance (3) 2D projected separation between the filament and the midplane of the spiral arm it lies closest to in distance (4) Total length of the filament (5) Linear mass of the filament (6) Median width of the mask defining the boundary of each filament (7) Herschel, dust-emission-based FWHM derived from Gaussian fits to radial column density profiles (8) GLIMPSE-Spitzer, dust-extinction-based FWHM derived from Gaussian fits to radial column density profiles (9) Aspect ratio, derived from Col 18. for Bones/Herschel type filaments and Col 17. for MST/GMF type filaments (10) Aspect ratio derived using extinction width from Col. 19 (11) Minimum displacement from the arm in Col 6. of Table \ref{tab:tab1} in \textit{longitude-velocity} space}
\tablenotetext{*}{Fil6, colloquially known as the ``Snake" is also in the \citet{Wang_2015} Large-Scale Herschel filament sample as ``G11" and the the \citet{Wang_2016} MST Bone sample as ``F7"; it has been included in all three samples in Figures \ref{fig:colcomp}, \ref{fig:tempcomp}, \ref{fig:dgfcomp}, \ref{fig:lwa_comp}, and \ref{fig:armprops}}
\tablenotetext{**}{Nessie is also in the \citet{Wang_2015} Large-Scale Herschel filament sample as ``G339"; it has been included in both samples in Figures \ref{fig:colcomp}, \ref{fig:tempcomp}, \ref{fig:dgfcomp}, \ref{fig:lwa_comp}, and \ref{fig:armprops}. Due to the challenges of applying a semi-continuous closed contour to a 160+ pc long filament (c.f. \S \ref{lengths}), we only consider the version of Nessie as originally defined in \citet{Jackson_2010}, even though Nessie is 2-5 times longer than originally claimed \citep{Goodman_2014}}
\end{deluxetable}

\begin{figure}[h!]
\begin{center}
\includegraphics[width=0.7\columnwidth]{{{Density_Profile_Comp}}}
\caption{{\label{fig:plummer_comp} A comparison of the power lax index ``$p$" derived from Plummer-like fits to the radial column density profiles of large-scale filaments \citep{Zucker_2015,Wang_2015} (blue) and small-scale filaments \citep{Arzoumanian_2011} (red). The vertical line indicates the median power law index for each sample. The large-scale filaments show significantly steeper profiles than small-scale filaments (with widths $\approx$ 0.1-0.3 pc) and are more consistent with the theoretical profile of an isothermal, self-gravitating cylinder \citep{Ostriker_1964}, though this could be due to background subtraction effects.}}
\end{center}
\end{figure}

\subsection{Position Angle and Galactic Plane Separation} \label{posangle_altitude}
We calculate position angle (2D projected angle of the filament with respect to a spiral arm) and Galactic plane separation (2D projected separation between the filament and the plane of a spiral arm) for the entire sample. Recall in \S \ref{posangle_altitude} that we ``associate" each filament with a spiral arm by using the output of the Bayesian distance calculator to determine which \citet{Reid_2016} spiral feature the filament lies closest to in distance (listed in Column 6 of Table \ref{tab:tab1}). In this analysis, the ($l$,$b$) trace of the nearest spiral arm is used to calculate the position angle and Galactic plane separation of the filament regardless of whether the filament actually lies \textit{in} the arm. 

To calculate plane-of-the-sky position angle we first fit a major axis to each filament. We treat every pixel within the filament mask as a point in a scatter plot and fit a line to the set of points. We repeat the same process for the set of ($l$,$b$) points defining the \citet{Reid_2016} spiral arm within the longitude range of the filament.\footnote{In most cases, the fit to the ($l$,$b$) arm points in the longitude range of the filament will be identical to the original spiral arm trace. However, \citet{Reid_2016} fits are not log-spirals, but rather connect the GMCs that trace the arms. When the filament is coincident with a GMC ``connection" point, the local spiral arm trace will be a combination of two lines with two different slopes, in which case the spiral arm trace will be the best fit to these two lines. In cases where the filament's central longitude is beyond the published longitude range of the spiral arm we extrapolate the trace to the filament's longitude.} The position angle we report is the difference between the filament position angle ($\theta_{fil}$) and the arm position angle ($\theta_{arm}$), where $\theta_{fil}$ and $\theta_{arm}$ are parameterized by the best-fit filament slope ($m_{fil}$) and the best-fit arm slope ($m_{arm}$), respectively. Specifically, $\theta_{fil}=\arctan{(m_{fil})}$ and $\theta_{arm}=\arctan{(m_{arm})}$, so the overall difference in position angle is given by the difference between $\arctan{(m_{fil})}$ and $\arctan{(m_{arm}})$.

To calculate Galactic plane separation, we take every point along the ``spine" of the filament (determined via medial axis skeletonization of the filament masks, see \S \ref{lengths}) and find the minimum angular distance between that spine point and the filament's assigned arm. Specifically, we calculate the angular distance between the spine point and every point along the arm, and take the minimum of these values as the minimum displacement for that point. We convert minimum arm distance for each spine point to a physical separation in parsecs using the distance returned from the Bayesian distance calculator. Finally we adopt the mean of all the minimum spine-to-arm distances as the final plane separation for the filament. The latitudes of the \citet{Reid_2016} spiral arm traces along the same line-of-sight do not always increase monotonically with distance---as would be expected if the IAU defined mid-plane is tilted at some angle with respect to a flat physical Galactic midplane \citep[see Figure 3 from][]{Goodman_2014}---so assuming such a schematic would produce different separations. 

\begin{figure}[h!]
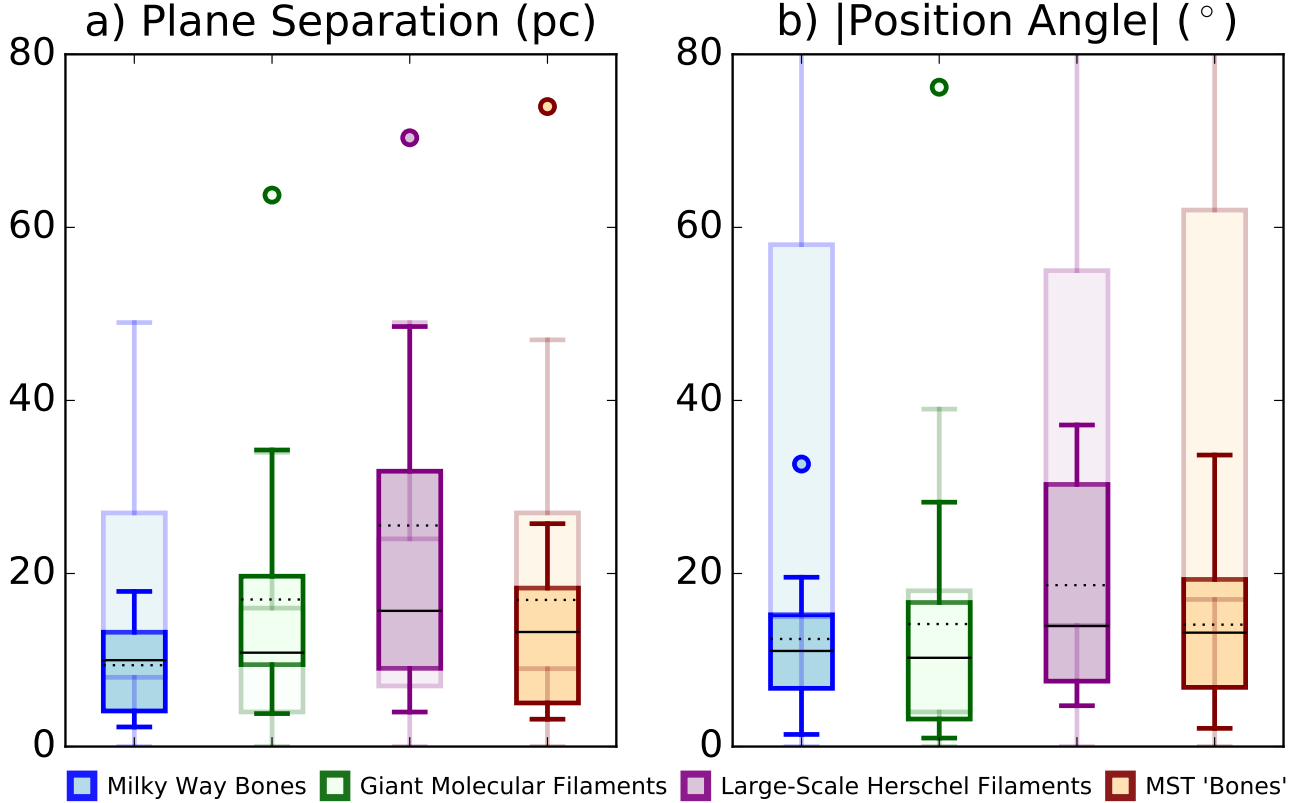

\begin{center}
\includegraphics[width=1.0\columnwidth]{{{PosAngle_Altitude}}}
\caption{{\label{fig:armprops} Galactic plane separation (left panel) and absolute position angle (right panel) box-and-whisker plots for the four different filament classes in our sample; only the distribution for the MST ``Bones" based on the catalog data are shown, but the tracks for the MST ``Bones" based on the continuum data are functionally equivalent (see Table \ref{tab:tab3}). In each boxplot, the solid line indicates the median of the distribution while the dotted line indicates the mean. The bottom and top edges of each box denote the end of the first and third quartiles of the distribution, while the end of the ``whiskers" enclose all the data points that are not considered outliers, which are plotted as individual dots. In the background we also overlay the expected plane separation and position angles of a random control sample based on a Monte Carlo analysis (light colored box and whiskers). This is akin to what the distributions would look like if filaments with lengths typical of each class are dropped randomly in a box with the stipulation that each filament must be completely ``observed" given the boundaries of current Galactic plane surveys.
}}
\end{center}
\end{figure}

We show box and whisker plots of Galactic plane separation (left panel) and position angle (right panel) in Figure \ref{fig:armprops}.  As seen in the left panel, only $\approx$ 15\% of the filaments lie more than 30 pc above the physical Galactic midplane of their associated arm. For reference, at Galactocentric radii typical of our filaments (5-6 kpc), \citet{Nakanishi_2006} find the height of the molecular $\rm H_2$ disk, as traced by $^{12}$CO (1-0) observations, to be 51 pc.\footnote{\citet{Nakanishi_2006} define the ``height" of the molecular $\rm H_2$ disk to be the height at which the disk has half the number density of the midplane, where the midplane is defined as $b=0^\circ$. See \S 3.3 in \citet{Nakanishi_2006} for details.} The \citet{Zucker_2015} Bones are closest to the plane (median scale height of 10 pc) and all classes have a median displacement $\leq 16$ pc. However, each of the GMF, Herschel and MST classes has at least one filament above an altitude of 60 pc. In terms of position angle, $\approx 85\%$ of the sample is aligned with the Galactic plane, to within $30^{\circ}$, and all but one filament within $45^{\circ}$. The notable exception is GMF41, which is almost completely perpendicular to the plane. The median position angle for all four classes lies within $15^\circ$ of zero. 

In addition to showing the plane separation and position angles for the four filament classes, in Figure \ref{fig:armprops} we also overlay the expected plane separation and position angles of a random control sample based on a Monte Carlo analysis (light-colored box and whiskers). This is akin to what the distributions would look like if filaments with lengths and widths typical of each class are dropped randomly in a box with the stipulation that each filament must be completely``observed" given the boundaries of current Galactic plane surveys. This gives a sense of how far each distribution is from random. More details on the methodology used to derive the random control samples for each class is discussed in \S \ref{control_discussion} in the Appendix and a histogram of these same distributions is shown in Figure \ref{fig:control} in the Appendix. The typical plane separations for the GMFs and the Large-Scale Herschel Filaments are consistent with their random control samples, in that these are the altitudes one would expect randomly given the observational biases of current Galactic plane surveys. In contrast the observed altitudes of the Milky Way Bone distribution is closer to the Galactic plane than one would expect randomly. The same is true for the MST ``Bones," but recall that we subselected 25\% of the sample which \citet{Wang_2016} determined to be close to the Galactic plane, so including the entire ``MST" sample would skew the altitudes closer to the random control sample. With the exception of the Giant Molecular Filaments, the rest of the filament classes tend to be be more parallel to the plane than expected randomly. 

Both Galactic altitude and position angle are 2D projected quantities, and do not represent 3D orientation with respect to spiral arms. It can only eliminate filaments that are significantly misaligned (e.g. GMF41), which have position angles $>>30 ^\circ$. Nevertheless, we find that our results are in qualitative agreement with numerical simulations.  Figure 10 from \citet{Goodman_2014} shows an edge on view of the total column density of structures forming in the mid-plane of a spiral galaxy from \citet{Smith_2014}. In this simulation, no structures seen in projection with total edge on column density $\rm >10^{22}\; cm^{-2}$ (reminiscent of the Milky Way Bones or the Large-Scale Herschel filaments) are forming between 20 pc and 100 pc (the upper bound of the simulation). No filaments are rotated more than $\approx 30^{\circ}$ from the Galactic mid-plane, though this is difficult to constrain given saturation near the midplane \citep{Goodman_2014}. The position angles of lower column density filaments (e.g. the GMFs) are also consistent with those found for simulated Giant Molecular Clouds with aspect ratios $> 8$ from Figure 12 in \citet{Duarte_Cabral_2016}. Both distributions have a median position angle of $\approx 10^{\circ}$, and both include a few outliers inclined $>30^{\circ}$.

\subsection{Position-Position-Velocity Summary of Large-Scale Filament and Spiral Arm Models} \label{ppv_summary}
When attempting to associate filaments with spiral arm models, the velocity of the gas is the most relevant criterion, as it is free of the uncertainties that riddle conversion to distance. The most common method to trace spiral arms is log-spiral  fitting to the arm locus in \textit{longitude-velocity} space. Spiral models can differ by more than $10 \; \rm km \; s^{-1}$, depending on both the spiral tracer used (e.g. CO, HI) and the tangent points deduced from these tracers \citep[see Figure 2 from][]{Zucker_2015}. 

Given the deviations from model to model, it is critical to compare all the large-scale filaments within the same spiral arm framework, and we do so by adopting the \textit{longitude-velocity} fits from \citet{Reid_2016}. The \citet{Reid_2016} fits depart from convention, in that they are not log-spirals, but rather connect the major emission features defining each spiral arm segment in \textit{longitude-velocity} space, mainly using the \citet{Dame_2001} CO survey. The \citet{Reid_2016} model alone traces the ($l$,$b$,$v$) structure of the majority of large-scale CO features (including far spiral arms and local features), so it provides a uniform comparison to CO signatures observed throughout the Galaxy. 

To compare the velocity structure of our filaments to the \citet{Reid_2016} spiral arms, we use a Gaussian fitting code to determine the central velocities of our filaments at evenly spaced longitude intervals along the filament masks. Because we need consistent CO velocity information for all filaments in our kinematic analysis, we only consider the 22 filaments lying within the GRS longitude range \citep{Jackson_2006}, which is shown as an orange arc in a top down view of the Galaxy in Figure \ref{fig:top_down}. We convolve and regrid each of the filament masks to the angular resolution and grid of the GRS survey. We divide the pixels in the new GRS masks into $0.1^\circ$ longitude bins. In each bin, we collapse over all the pixels that fall within the mask, and perform a single or multi-component (up to $\approx$ 5) Gaussian fit to each bin's spectrum. 

The Gaussian fitting is performed interactively using the python package $pyspeckit$ \citep{Ginsburg_2011}. As most of the $\rm ^{13}CO$ spectra do contain multiple components, we differentiate between the different \textit{longitude-velocity} ``tracks" by plotting the filament's $^{13}\rm CO$ $l-v$ diagrams, along with all Gaussian components and supplementary dense gas measurements \citep[from the HOPS, BGPS, ATLASGAL catalog sources inside the mask, see][]{Urquhart_2014,Shirley_2013,Purcell_2012a,Jackson_2013}. We choose the $\rm ^{13}CO$ component that is contiguous along the filament and associated with the dense gas velocities (where available) and the velocity of the filament from its original publication (where reported).\footnote{Given complications due to self-absorption and feature blending this Gaussian fitting should only be used to determine approximate central velocities (within a few $\rm km\; s^{-1}$) at even intervals along the filament. We plan to explore filament kinematics, including detailed velocity gradients and linewidths in a future paper, using the SCOUSE multi-component spectral fitting engine from \citet{Henshaw_2016}. For now, we simply wish to determine how \textit{longitude-velocity} traces of our filaments compare to \textit{longitude-velocity} traces of major spiral segments.}

The results of this analysis are shown in the \textit{longitude-velocity} diagram in the top panel of Figure \ref{fig:pv_summary}. In the bottom panel, we also show a plane-of-the-sky map over an identical longitude range (a fourth quadrant version of this map is shown in Figure \ref{fig:lb_fourthquad} in the Appendix). Overlaid in black are major spiral arm segments from \citet{Reid_2016} (see caption for details). In the top panel, alongside the spiral arms, we show central velocities (colored-coded by large-scale filament catalog) obtained by fitting Gaussians to the filaments' emission spectra in $0.1^\circ$ longitude bins, as described above. 

The filaments display a significant range of \textit{l-v} orientations with respect to spiral features. Some filaments (e.g. Filaments 1, 2, 5) have \textit{l-v} tracks aligned with spiral arms (consistent with being ``spiral arm" tracing filaments). Other filaments have \textit{l-v} tracks slightly inclined to spiral arm fits (e.g. GMF18) and could potentially trace a spur or feather-like feature trailing off arms. We also observe a large fraction ($\approx \frac{1}{3}$) of filaments with velocity gradients almost directly perpendicular to spiral arm traces (e.g. G24), possibly constituting an interarm filament. This is not to say that all of these filaments are associated with spiral features, just that some fraction of the filaments have velocity signatures consistent with large-scale CO emission features corresponding to spiral arms, spurs, and interarm regions \textit{within the \citet{Reid_2016} spiral arm model}. An alternate scenario is that a fraction of these filaments bear little relationship to spiral structure. While it has been shown that spiral arms host a higher density of molecular clouds \citep[e.g.][]{Moore_2012}, whether the longest molecular filaments are forming due to the presence of the arm itself is an entirely different question and will impact whether the filaments themselves can be used to pinpoint the location of spiral features (see \S \ref{bone_discussion}). We also caution that velocity gradients across the filament could result in different filament orientations in \textit{p-p-p} space at different distances (i.e. velocity gradients could map to different physical gradients if the filament lies on the near versus far side of the Galaxy). We know very little about the velocity structure of spiral arms, let alone that of spurs and interarm structure---an issue complicated by the fact that spiral arm traces are very model dependent \citep[see, for instance, Figure 2 in][]{Zucker_2015}. However, the development of improved numerical simulations of the dense gas in spiral galaxies \citep{Smith_2014} could better constrain our models (i.e. through the creation of synthetic CO cubes of large kpc-scale swaths of spiral galaxies, which could then be compared with the kinematic structure of dense filaments in the same simulation). 

Besides a large range in \textit{l-v} orientations with respect to spiral arms, we also observe an uncanny \textit{l-v} association with the ``far" Sagittarius (SgF) arm for several of the filaments (e.g. GMF26, G26, GMF38a, Fil3) despite ample evidence that these filaments lie on the near side of the Galaxy. The support for their lying at the near distance is as follows. GMF38a contains five or six prominent filamentary extinction features and placing this GMF at the far distance would also place the extinction features at $\approx 10 \rm \; kpc$, behind the Galactic center. Given that elongated mid-IR extinction features very commonly need to lie at the near distance (in front of the mid-IR bright background of the Galactic center) to be visible \citep{Ellsworth_Bowers_2013}, it is highly unlikely that all these extinction features are at the far distance. There is also a trigonemtric parallax source G035.02+00.34 \citep{Wu_2014} located within the boundaries of GMF38a (both kinematically and spatially) that would place it at the near distance, around 2.3 kpc. The W44 supernovae remnant also encompasses the right half of GMF38a, so it is possible that the two sides of GMF38a are at two different distances, with CO emission coincident with GMF38b (just below GMF38a on the \textit{l-v} diagram) pushed to higher velocity by the expanding shell. While GMF38a contains prominent extinction features, Fil3 \textit{is} a prominent filamentary mid-IR extinction feature, containing dense clumps with peak optical depths at $8\; \micron$ $>$ 5 \citep{Peretto_2009}. 

 A similar scenario occurs for GMF26 and G26. While aligned with the SgF arm, GMF26 and G26 are located near $b=0.8^\circ$, so while they are kinematically consistent with the SgF arm, they are a degree above the physical Galactic midplane of that arm on the plane of the sky. The majority of the distance estimates from \citet{Ellsworth_Bowers_2015} for BGPS sources kinematically coincident with all four of these filaments also prefer the near distance. This analysis places all four filaments at the near distance, though this is artificial as we assumed a very low prior probability of them being at the far distance given the prevalence of their extinction features (see \S \ref{distances}). The more plausible scenario is that these filaments are interarm or loosely associated with the Scutum-Centaurus near arm. 
 
The probable lack of association of these filaments with the SgF arm, despite their alignment with the $l-v$ trace of that arm, might raise concerns about the confidence with which we can associate filaments with particular spiral features. There are two compounding factors that make association with the SgF arm unreliable near $l=30^\circ$. Associating filaments with spiral arm fits should be considered less reliable in regions with anomalous velocity signatures, including regions in close proximity to supernovae remnants or to the non-circular motions of the Galactic bar. \citet{Reid_2016} compare their Bayesian distances to 62 HII regions with the distances to the same regions derived via the kinematic distance method in \citet{Anderson_2012}, where the distance ambiguity has been resolved using HI self-absorption. \citet{Reid_2016} find agreement on the resolution of the near/far distance ambiguity in over 90\% of cases, with only 6/62 sources being discrepant. Four of those sources are likely interacting with the edge of the bar or the W44 supernovae remnant, and mistakenly placed at the far distance. A majority of these anomalous sources also lie at the same longitudes and velocities as the filaments which show uncanny proximity to the SgF $l-v$ trace, suggesting that these filaments are either interacting with W44 and the bar and/or the SgF trace needs to be revised over these longitudes. The second compounding factor is that there are zero parallax measurements constraining the $(l,b,v)$ location of the SgF arm below $l \lesssim 40^\circ$, so SgF trace is being extrapolated to the longitudes of our filaments based on pitch angles fit to parallax data at higher longitudes. Since pitch angle is known to vary as a function of azimuthal angle, the fit to this arm at low longitudes is poorly constrained, and should be considered with caution \citep{Reid_2016}. In contrast, we should have significantly more confidence in $(l,b,v)$ fits to the near Scutum-Centaurus and Norma arms, which are constrained by dozens of parallax measurements, so the spiral association of filaments like Filament 5 \citep{Zucker_2015} can be made with higher certainty. 

\subsection{Spatial and Kinematic Proximity to Spiral Arms} \label{arm_proximity}
\begin{sidewaysfigure}
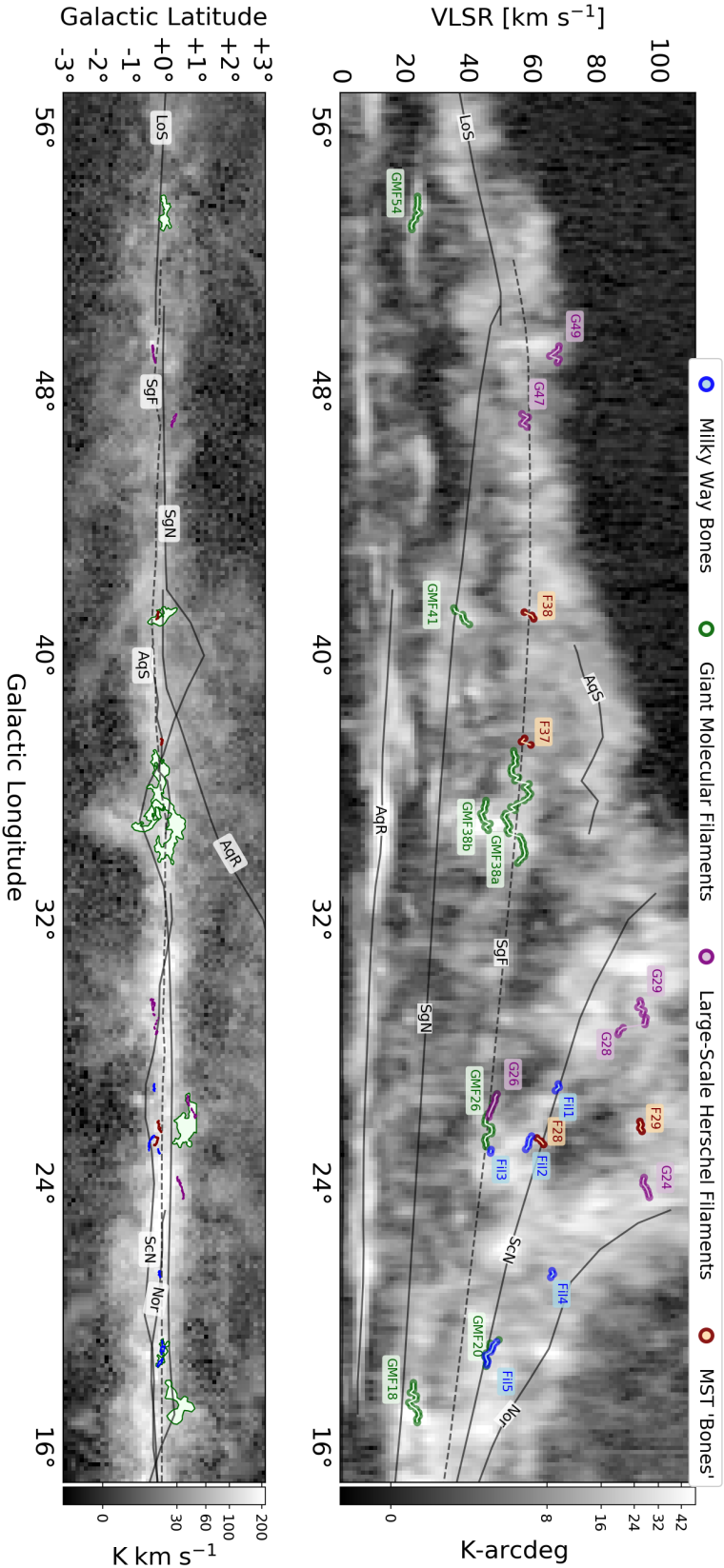

\begin{center}
\includegraphics[width=22cm]{{{Summary_Combined_CO}}}
\caption{{\label{fig:pv_summary} \textbf{Top:} \textit{Longitude-velocity} diagram of spiral arm models and large-scale filaments in a section of the first Galactic quadrant. The MST ``Bones" based on the catalog data are shown, but the tracks for the MST ``Bones" based on the continuum data are functionally equivalent. Background grayscale shows $^{12} \rm CO$ emission from the \citet{Dame_2001} survey, integrated between $-1^\circ < b < 1^\circ$. Spiral arm traces are taken from \citet{Reid_2016} and include the Local Spur (LoS), the Aquila Rift (AqR), the Aquila Spur (AqS), the Sagittarius Near Arm (SgN), the Sagittarius Far Arm (SgF), the Scutum Near Arm (ScN), and the Norma arm (Nor). The colored dots correspond to the four different large-scale filament catalogs included in this study, derived through Gaussian fitting of the filament's CO emission in $0.1^\circ$ longitude intervals (see \S \ref{ppv_summary}). \textbf{Bottom:} Plane-of-the-sky map of the same section of the first Galactic quadrant. Background grayscale also shows $^{12} \rm CO$ emission from the \citet{Dame_2001} survey, now integrated between $\rm 0-110 \; km \; s^{-1}$. The filament masks are overlaid to scale and color-coded according to filament catalog. The same spiral arm models are shown as in the top panel. 
}}
\end{center}
\end{sidewaysfigure}

\begin{figure}[h!]
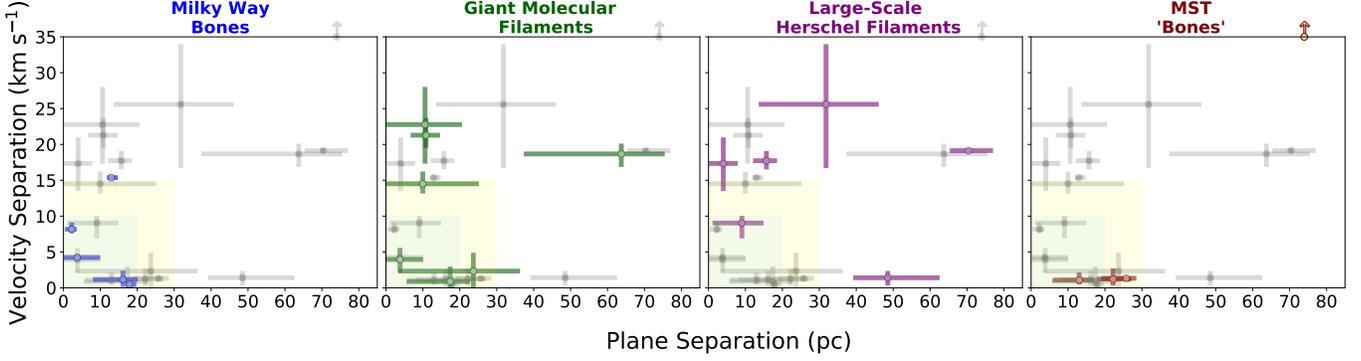

\begin{center}
\includegraphics[width=1.0\columnwidth]{{{lbv_pub_final}}}
\caption{{\label{fig:lbv} Side-by-side comparison of filaments' proximity to the \citet{Reid_2016} (``R16") spiral arms in both \textit{position-position} and \textit{position-velocity} space. Only filaments with high resolution CO data ($16^\circ < l < 56^\circ$) are shown (22/45 filaments in the full sample). A different filament class is highlighted in each panel, with the underlying distribution of all filament classes shown in grayscale in all panels.  In panel three, only the distribution for the MST ``Bones" based on the catalog data are shown but the displacement values for the MST ``Bones" based on the continuum data are functionally equivalent. The center of each cross marks the mean spatial and kinematic displacement between the filament and the most probable arm it is associated with, while the cross indicates the range of spatial and kinematic displacements observed over the entire length of the filament. In green shading (lower left hand corners) we highlight the region of parameter space indicative of strong spiral arm proximity, where the filament lies $\rm < 10 \; km \; s^{-1}$ and $\rm < 20 \; pc$ from a spiral arm trace. Using these criteria, only 8/22 (35\%) of filaments lie very close to spiral arms in ($l,b,v$) space. Weaker spiral arm proximity criteria, $\rm < 15 \; km \; s^{-1}$ and $\rm < 30 \; pc$ (light yellow shading), yields 12/22 filaments (55\%) in this region of parameter space.}}
\end{center}
\end{figure}

We can synthesize the information shown in Figure \ref{fig:pv_summary} by combining knowledge about a filament's proximity to a spiral arm in both \textit{position-position} space (bottom panel of Figure \ref{fig:pv_summary}) and in \textit{position-velocity} space (top panel of Figure \ref{fig:pv_summary}). A filament which is truly coincident with an arm in \textit{p-p-v} space will lie close to the arm in both diagrams, in addition to being aligned with the arm on the plane of the sky (to within $\approx 30^\circ$, as most long filaments are; see Figure \ref{fig:armprops}). The result of this analysis is illustrated in Figure \ref{fig:lbv}, where we show each filament's kinematic displacement from a spiral arm as a function of its spatial displacement (i.e. its Galactic plane separation). 

We establish the Galactic plane separation using the analysis presented in Figure \ref{fig:armprops} and \S \ref{posangle_altitude}. The kinematic displacement is calculated almost identically to spatial displacement, except now in \textit{(l,v)} space.\footnote{In more detail, we use the \textit{l-v} tracks for each filament shown in the top panel of Figure \ref{fig:pv_summary} to compute the filament's vertical displacement from the arm in units of $\rm km \; s^{-1}$. If a filament is outside the longitude range of the published arm fits from \citet{Reid_2016} (e.g. F29 associated with the Norma arm) we linearly extrapolate the \textit{l-v} arm fit to the longitude of the filament. Then, at every \textit{l,v} point along the filament sampled in $0.1 ^\circ$ longitude bins (and derived from Gaussian fitting to each bin's spectrum) we take the absolute difference between the filament velocity and the arm velocity at the same longitude to derive the kinematic displacement} Because the spatial and kinematic displacement vary across the filament, we mark the mean spatial and kinematic displacement with a scatter point in Figure \ref{fig:lbv}, and use the ``errorbars" to indicate the range of spatial and kinematic displacements observed across the filament's entire length. 

We calculate proximity between the filament and the arm with which the Bayesian distance estimator associates it (see \S \ref{distances}), which is not necessarily the arm it lies closest to in either \textit{position-position} or \textit{position-velocity} space. This is relevant in cases like GMF38a, which shows near perfect kinematic and spatial alignment with the far Sagittarius arm (at 10 kpc), but has ancillary information (e.g. parallax measurements, mid-IR extinction features) indicating that it lies at the near distance (at a few kpc). While this has a trivial effect on our conclusions, we also caution that filaments which lie just beyond the tangent points of spiral arms (e.g. G24) will perform poorly in the kinematic analysis---because the arm exhibits a steep velocity gradient just prior to the filament's latitude, extrapolating the arm to the longitude of the filament results in a large kinematic displacement.

To quantify ``spiral feature proximity," we demarcate two zones in Figure \ref{fig:lbv}: filaments exhibiting ``strong spiral arm proximity" lie within $\rm 10 \; km \; s^{-1}$ and within 20 pc of their nearest spiral arm in 2D projected space (green boxes in lower left hand corners of Figure \ref{fig:lbv}), while filaments exhibiting weak proximity lie within $\rm 15 \; km \; s^{-1}$ and within 30 pc (light yellow boxes in lower left corners of Figure \ref{fig:lbv}). While trends differ by catalog, our results in Figure \ref{fig:lbv} indicate that most long molecular filaments do not exhibit strong proximity to spiral arms, and this is particularly true for filaments uncovered via a blind visual search of the Galactic plane. We strongly emphasize that we only compare to one spiral arm model \citep[hereafter R16]{Reid_2016} and while it is the most comprehensive to date, the R16 (\textit{l,b,v}) traces are not necessarily correct or complete. Our results only hold within the context of the R16 spiral arm model, and could change with the adoption of different spiral arm models.

We consider the class-by-class R16 spiral arm proximity statistics for the 22/45 large-scale filaments in the sample with $\rm ^{13}CO$ data. Due in large part to selection criteria (these filaments are sub-selected to lie near spiral arms), 4 of 5 of the \citet{Zucker_2015} Milky Way Bones exhibit strong proximity to their R16 arm; the fifth one, Fil3, lies just over $\rm 15 \; km \; s^{-1}$ from the Scutum-Centaurus near arm, which is also noted in \citet{Zucker_2015}. We find that 1 of 4 of the MST ``Bones" exhibit strong proximity to their R16 arm, while 3 of 4 exhibit weaker proximity. The fourth one (F29) lies over 70 pc and over $\rm 70 \; km \; s^{-1}$.\footnote{F29 should arguably never have been classified as a ``Bone" in the original \citet{Wang_2016} study. While F29 lies close to the Sagittarius far arm in \textit{l-v} space, ancillary information \citep[mid-IR extinction features, maximum likelihood estimation performed on BGPS clumps from][]{Ellsworth_Bowers_2015}, indicates that it lies at the near distance, and indeed \citet{Wang_2016} places it there, while still allowing filaments to qualify as ``spiral arm" filaments if it lies near any arm on $either$ side of the Galaxy.}

Unsurprisingly, filaments uncovered through blind searches of the Galactic plane \citep{Wang_2015,Abreu_Vicente_2016,Ragan_2014} show less proximity to R16 spiral features. 2 of 7 GMFs from \citet{Ragan_2014} exhibit strong proximity to R16 spiral features, while 4 of 7 meet our looser definition. The \citet{Wang_2015} Large-scale Herschel filaments have the lowest number of filaments in close proximity to R16 spiral arms; only 1 of 6 Large-Scale Herschel filaments in this longitude range lies near a spiral feature---G49 at 9 pc and $\rm 9 \; km \; s^{-1}$ away. In this same longitude range, \citet{Wang_2015} find that 5/6 of the same filaments are ``spiral arm" filaments, but they use a ``top-down" Galactocentric view of the Galaxy, and classify filaments as spiral arm filaments if they overlap an arm within the distance error of the arm or filament. Using ($l,b,v$) associations could be more robust, given compounding distance uncertainties on both arm and filament in a Galactocentric-based view. 

While our statistics are skewed by including some filaments self-selected to lie near spiral arms, we determine that 35\% (8/22) of filaments in this subsample exhibit strong proximity to R16 spiral features, while 12/22 (55\%) exhibit weaker proximity. Therefore, strong alignment ($< \rm 10 \; km \; s^{-1}$, $<$ 20 pc) to R16 spiral arms is not favored among major large-scale filament catalogs. \textit{Furthermore, this should be considered an upper limit, as once again, we pre-select some filaments likely to lie near spiral arms based on previous analyses}. In their original study \citet{Wang_2016} find that only about 20\% of their sample lie near spiral arms (the MST ``Bones" included in this study), given somewhat stricter criteria of $\rm 5 \; km \; s^{-1}$ and 20 pc. This is much less of an issue in \citet{Zucker_2015}---they find that if a filament is long, skinny, dense, and parallel to the Galactic plane, 9/10 times it lies within $\rm 10 \; km \; s^{-1}$ and 20 pc of its most probable spiral arm. 

Finally, we analyze what fraction lies ``in the plane of the Galaxy" versus ``in an arm of the Galaxy." Of the 22 filaments in our subsample, between 64\% and 77\% of the filaments are $<$ 20-30 pc from the physical Galactic midplane, and all but one of those filaments (60\%-73\%) are also aligned with the Galactic plane, possessing position angles $<$ $30^\circ - 45^\circ$.  However, if we institute the additional criterion that the filaments must also be located within $\rm 10-15 \; km \; s^{-1}$ from that same R16 arm, those statistics drop by a factor of $1.5-2\times$. \textit{Within the context of the R16 spiral arm model}, this suggests that while filaments lying ``in the plane" of the Galaxy (possessing low position angles and Galactic altitudes) is common, lying ``in an arm" of the Galaxy (being in the plane and additionally being kinematically consistent with a purported spiral arm) is more rare. 

\newpage

\begin{figure}[h!]
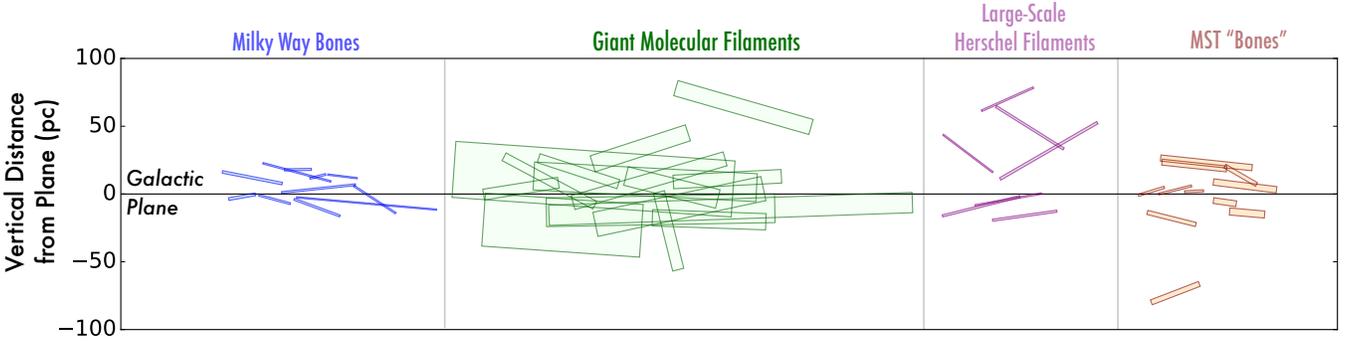

\begin{center}
\includegraphics[width=1.0\columnwidth]{{{revised_cartoon}}}
\caption{{\label{fig:cartoon} Idealized comparison of the different large-scale filament catalogs. Each filament is shown as a rectangle, with the same length and width as summarized in Table \ref{tab:tab2} and shown in Figure \ref{fig:lwa_comp}. The height above or below the Galactic plane, as well as the position angle, are the same as given in Table \ref{tab:tab3}, except the absolute value is not taken.}}
\end{center}
\end{figure}

\section{Discussion} \label{discussion}
All of the filaments included in this study are considered ``large-scale" filaments in the literature. Yet our results indicate that there is significant variation in large-scale filament properties, especially between catalogs, but also within catalogs. This may indicate that there are different types of filaments, possibly with unique formation mechanisms or evolutionary histories, being picked up by the myriad ``filament finding" techniques. Figure \ref{fig:cartoon} summarizes some of the many differences among classes, using the length, width, aspect ratio, and position angles of the filaments derived in this paper. Each filament is shown as a rectangle with the same length and width as those in Figure \ref{fig:lwa_comp}. The Galactic plane separation and position angle of the filaments are the same shown in Figure \ref{fig:armprops}, except the absolute value is not taken. 

Given the fledgling state of this research area, agreeing on uniform nomenclature for the variety of large-scale filaments we see poses a challenge. However, we can broadly differentiate between the various large-scale filament types in a region of parameter space defined by the filaments' cold, high column density fraction and aspect ratio, as illustrated in Figure \ref{fig:filpops}. On the vertical axis we show the filaments' cold \& high column density fraction, defined as the fraction of pixels inside each filament mask with $\rm H_2$ column densities greater than $\rm 1 \times 10^{22} \; cm^{-2}$ and dust temperatures lower than $\rm 20 \; K$. On the horizontal axis we show the aspect ratio. 

Our choice of parameter space is motivated in part by simulations. In the \citet{Smith_2014} simulations, we see that spine-like filaments potentially forming primarily along spiral features in the Galaxy should be highly elongated and have very high column densities, and we use aspect ratio and high column density fraction as a proxy for what we observe qualitatively. We require that the gas have both high column densities \textit{and} low dust temperatures because these Bone-like features should be in the earliest stages of their evolution, as otherwise, they might be disrupted by the feedback from high-mass stars or entrance into the arm itself \citep{Duarte_Cabral_2016}, making it more difficult to ascertain what structure (both morphologically and kinematically) the filament had prior to these scenarios occurring. 

In Figure \ref{fig:filpops} the filament catalogs \textit{roughly} lie in three zones of parameter space. The first zone is characterized by filaments with low aspect ratios (5:1-10:1) and low cold \& high column density fractions (1\%-10\%). The Giant Molecular Filaments (``GMFs") from \citet{Ragan_2014} and \citet{Abreu_Vicente_2016} occupy this region of parameter space and are consistent with being the elongated tail of the GMC aspect ratio distribution. The second zone harbors filaments possessing high aspect ratios ($>$20:1) and large cold \& high column density fractions (10\%-50\%). The \citet{Zucker_2015} Bones and \citet{Wang_2015} Large-Scale Herschel filaments dominate this portion of the diagram; we denote them potential ``Bone candidates," as they are all long and skinny, high column density features, but do not exclusively contain filaments from the \citet{Zucker_2015} Milky Way Bone catalog. The third category occupies a smaller zone of parameter space and possesses filaments with low aspect ratios (8:1-15:1) but very large cold \& high column density fractions (10\%-75\%). Most of the MST ``Bone" filaments lie in this region, and their morphology is consistent with being elongated sets of dense core complexes, which trace out the densest velocity contiguous portions of molecular clouds. Whether the filaments in some regions produce or evolve into filaments in other categories is difficult to quantify without more advanced numerical simulations. 

Assuming that the three populations are well-differentiated along catalog boundaries (which is a very rough approximation given the broad overlap between regions) we characterize typical cold \& high column density fractions and aspect ratios for each category. This is shown in Table \ref{tab:filpops_tab} which summarizes the 16th, 50th, and 84th percentiles of the cold \& high column density fraction and aspect ratio distributions for each class.\footnote{The 50th percentile represents the median value of the distribution, while the 84th and 16th percentiles correspond to $\pm 1 \sigma$ from the median value assuming a Gaussian distribution} Approximately 70\% of each population should be expected to fall within the lower and upper bounds of each property listed in the table. 

There is a large degree of dispersion along both axes (aspect ratio and cold \& high column density fraction), which could have a number of underlying causes. The dispersion in aspect ratio could be due in large part to projection effects. The best we have at this time is simple projected lengths, from which aspect ratios are calculated.\footnote{Based on analysis performed in \citet{Zucker_2015} and in \S \ref{ppv_summary} it is probable that Nessie and Fil1 lie along the Scutum-Centaurus arm. Under this hypothesis, we can speculate that filaments which lie perpendicular to our line-of-sight \citep[e.g. Nessie;][]{Goodman_2014} will have much larger projected lengths than filaments which lie near the tangent points of spiral arms. \citep[e.g. Fil1, lying near the tangent point of the Scutum-Centaurus arm at $l=30^\circ$;][]{Zucker_2015}, where our line of sight is likely parallel to the long axis of the filament} It is possible that filaments in the ``elongated dense core complex" and the ``elongated molecular filament" categories could have both higher aspect ratios and lower column densities if projection effects were removed. However, based on the results from \citet{Myers_1991}, it is unlikely that the true aspect ratios of these filaments would be significantly higher than their projected aspect ratios, unless every filament was highly elongated along the line of sight. \citet{Myers_1991} determine the true aspect ratios for an ensemble of identical prolate spheroids whose long axes are oriented randomly with respect to an observer. The median aspect ratios for the ``elongated dense core complex" and the ``elongated molecular filament" samples are 11:1 and 8:1, respectively. Assuming that these categories can be modeled as prolate spheroids, \citet{Myers_1991} find that the true mean aspect ratios for these populations would only be slightly higher than their projected aspect ratios, at 16:1 and 11:1 for the ``elongated dense core complex" and the ``elongated molecular filament" categories, respectively \citep[see Figure 3 in][]{Myers_1991}. In order to have a significant effect on the categories shown in Figure \ref{fig:filpops}, a majority of the filaments in the low aspect ratio categories would need to be highly elongated along the line of sight, which would make them extreme outliers based on the projected median aspect ratio for these populations shown in Table \ref{tab:filpops_tab}. While beyond the scope of this paper, we plan to explore the impact of projection effects on the filament properties we infer from the observations using numerical simulations in a future work, by calculating how column densities, lengths, temperatures, and aspect ratios change based on the inclination angle of the observer. 

In addition to projection effects, the second effect is a possible correlation between gas density and Galactocentric radius. \citet{Miville_Deschenes_2016} find, for instance, that the $\rm H_2$ density of molecular clouds in their Milky Way-wide catalog is dependent upon their location in the Galaxy, with the Scutum-Centaurus arm in the inner Galaxy hosting an overabundance of high $\rm H_2$ density clouds compared to interarm regions or more minor arms \citep[e.g. near the Carina-Sagittarius arm, in Figure 27;][]{Miville_Deschenes_2016}. Our filaments span $120^\circ$ in longitude and lie at distances between $1-10 \; \rm kpc$, so this diversity of environments could have a non-trivial effect on the spread of properties we see both in and between classes. We could in theory test the effect of Galactocentric environment by comparing the densities of the filaments at different distances. However, we also suffer from small number statistics.

We elaborate upon the key characteristics and possible formation mechanisms of each filament population in more detail below. 

\begin{figure}[h!]
\begin{center}
\includegraphics[width=0.75\columnwidth]{{{filpops_pub_newradfil}}}
\caption{{\label{fig:filpops} \textbf{Top:} The distribution of cold \& high column density fraction versus aspect ratio for all four catalogs included in this study. The cold \& high column density fraction is defined as the fraction of a filament's gas above an $\rm H_2$ column density of $1\times10^{22} \; \rm cm^{-2}$ and below a dust temperature of 20 K. For the MST ``Bone" filaments, we only show the distribution based on the continuum data as width analysis was not available for the catalog based definition (see \S \ref{mst_boundaries}).  Furthermore, two aspect ratio distributions are shown for the Milky Way Bones and Large-Scale Herschel filaments: the darker blue and purple markers show the aspect ratio computed using a width derived from radial column density profile fitting, while the lighter purple and blue markers show the aspect ratio computed using a mask-based width (see \S \ref{lengths}}). For the Giant Molecular Filaments and the MST ``Bone" filaments, we are only able to compute a mask-based width. \textbf{Bottom:} We observe a differentiation of filament catalogs in this region of parameter space; rough outlines for each ``population" are shown by applying a convex hull algorithm to the bulk of the points constituting each class.} 
\end{center}
\end{figure}

\begin{table}[h!]
  \begin{center}
  \includegraphics[width=1.0\linewidth]{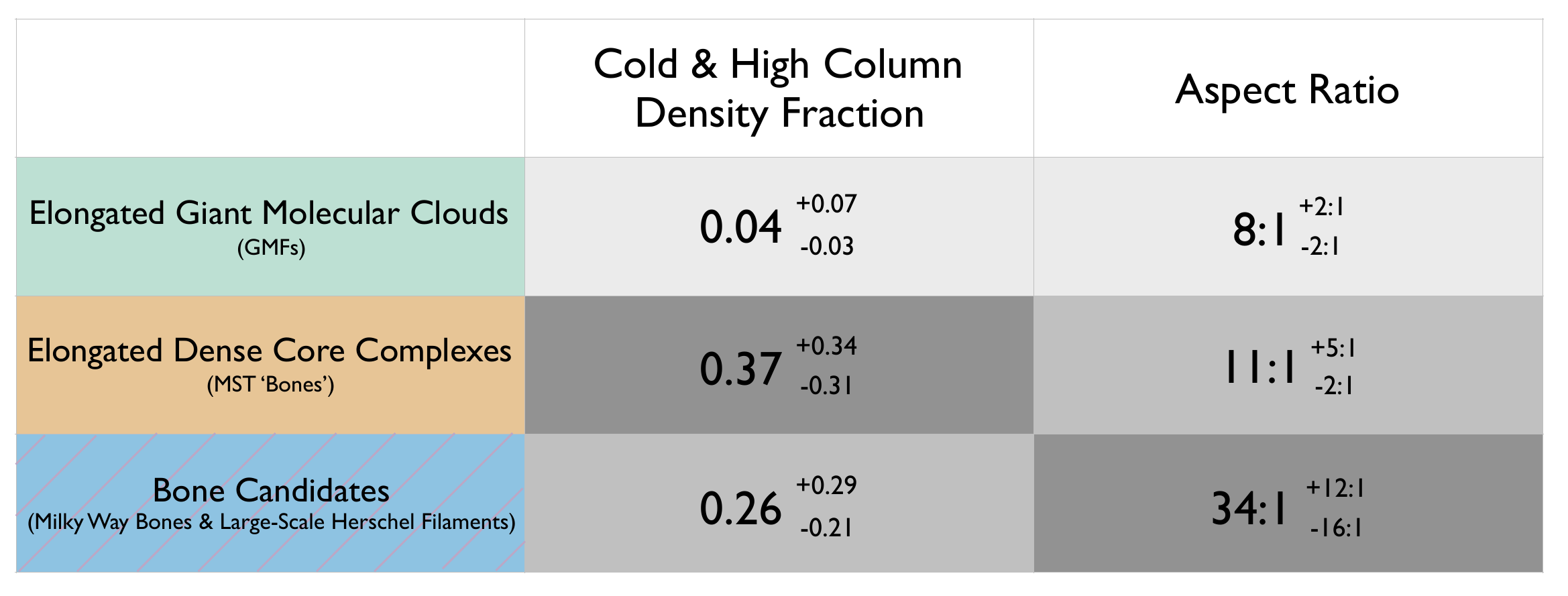}
    \caption{  \label{tab:filpops_tab} We summarize the characteristic aspect ratios and cold \& high column density fractions for the multi-modal filament population proposed in Figure \ref{fig:filpops}. We assume the categories are differentiated along catalog lines, as indicated in Figure \ref{fig:filpops}, and take the 16th, 50th, and 84th percentiles of each category. The average value is the 50th percentile, while the upper bound is the difference between the 50th and 84th percentiles, and the lower bound is the difference between the 18th and 50th percentiles. Approximately 70\% of the population should fall within the lower and upper bounds for aspect ratio and cold \& high column density fraction. }
  \end{center}
\end{table}

\subsection{Giant Molecular Filaments: Elongated Giant Molecular Clouds?}  \label{gmf_discussion}
Giant Molecular Filaments (GMFs) occupy the bottom left half of Figure \ref{fig:filpops} and are characterized by lower aspect ratios and lower cold \& high column density fractions. However, the average GMF aspect ratio (8:1) is still higher than observed in the typical Giant Molecular Cloud (GMC). As shown to be the case in recent numerical and observational studies of GMCs, the average GMC traced by CO emission has an aspect ratio $\lesssim$ 2:1. For instance, \citet{Miville_Deschenes_2016} find a typical molecular cloud aspect ratio of 1.5:1 for their Galaxy-wide catalog of GMCs, produced by decomposing each sightline in the \citet{Dame_2001} CO survey into a set of Gaussians and applying a hierarchical clustering algorithm to identify coherent structures. \citet{Duarte_Cabral_2016} simulate a region of the \citet{Dobbs_2013} spiral galaxy at higher resolution and apply the SCIMES spectral clustering algorithm \citep{Colombo_2015} to the synthetic CO (1-0) emission, deriving a population of Giant Molecular clouds with a median aspect ratio of 1.9:1.\footnote{Many of these analyses use a dendrogram-based technique \citep{Rosolowsky_2008} which is limited in its ability to find very elongated structures. Thus, the typical observed aspect ratios of molecular clouds might be underestimated.}

One possible scenario, also presented in \citet{Duarte_Cabral_2016}, is that GMFs are an elongated, skinny subset of GMCs. The longest (lengths $\approx$ 100 pc) and most massive molecular clouds ($\rm M >  10^4 \; M_\odot$) at the very tail end of the distribution of aspect ratios from the \citet{Duarte_Cabral_2016} study are generally consistent with the GMF properties derived in this study when accounting for the lower spatial and density resolution of the simulation. For clouds delineated via CO (1-0) emission, \citet{Duarte_Cabral_2016} find a peak aspect ratio of $\approx$ 5:1. Given that the widths of GMF-like structures are unresolved at the \citet{Duarte_Cabral_2016} 10 pc linear resolution, this peak aspect ratio should be even higher, and agrees well with the typical range of aspect ratios we calculate for the GMFs ($\approx$ 5:1-10:1). 

Next, we consider the density structure of GMFs. As evident in \S \ref{gmf_gallery} and Figure \ref{fig:dgfcomp}, these filaments contain dense gas in the form of cores and clumps, filamentary complexes, and in some cases high density, highly-elongated ``Bone" like features, but the latter are the exception, not the norm. Only four of sixteen of the GMFs contain filaments found in the dense large-scale filament catalogs of \citet{Zucker_2015} or \citet{Wang_2015}, though this can be attributed in part to selection effects (e.g. some of the filaments inside the GMFs are very highly inclined to the plane of the Galaxy, and those types of filaments are explicitly excluded from the \citet{Zucker_2015} search). There is clearly some correlation between large-scale ($>1^\circ$) molecular CO features and large-scale filamentary dense gas features, but it is currently unclear whether the former requires the latter's formation. We see several filaments (e.g. G24, Fil2, etc) with lengths between 20-80 pc and no evidence of being embedded in a larger GMF-like environment, so either the GMF has dispersed over time, was never there to begin with, or has not yet formed. 

It is also possible that the properties of GMFs, including aspect ratio and density, are strongly dependent on environment. One formation mechanism for GMFs, as proposed in \citet{Duarte_Cabral_2016}, is that these filaments form solely in the interarm region via Galactic shear and reach their highest densities and aspect ratios at the deepest point in the gravitational potential well, just prior to arm entry. The creation of elongated molecular clouds via large-scale gas motions is also consistent with the work of \citet{Koda_2006}, which finds that molecular clouds are systematically elongated along the Galactic plane---likely due to the driving energy produced by Galactic rotation---and that some reach aspect ratios in excess of 5:1. However, many of the synthetic GMFs in the interarm regions start off as $\approx$ 500 pc long mostly atomic structures \citep[][$\frac{M_{H_2}}{M_H} \approx 0.05$, see their Figure 6]{Duarte_Cabral_2017} and do not begin to resemble anything we see observationally until they enter spiral arms, gain more molecular mass, and become more visible in the CO (1-0) line. The current \citet{Duarte_Cabral_2017} simulations only trace very low column density gas at low spatial resolution---with a regridded cell size of 1-5 pc and a maximum characteristic density  $\approx 1\times 10^{21}$ \citep[see Figure 8 \& A1][]{Duarte_Cabral_2016}. Future high resolution simulations should hopefully shed additional light on the evolutionary histories of GMFs and what implications this has---if any---for higher density, Bone-like structures potentially nested within them. 
\subsection{MST Filaments: Elongated Dense Core Complexes?} \label{mst_discussion}
The analysis put forth in this paper indicates that MST ``Bone" filaments occupy a unique zone of parameter space, characterized by large cold \& high column density fractions ($\approx 10-75\%$) and much lower aspect ratios ($ \approx $ 10:1). Our aspect ratio analysis is in strong disagreement with the results from \citet{Wang_2016}. Recall that the MST filaments in \citet{Wang_2016} are identified by grouping neighboring dense BGPS clumps into complexes in an automated way, using a Minimum Spanning Tree algorithm.  A filament is determined to be linear if the standard deviation of the BGPS clumps along the major axis is 1.5 times the standard deviation of the clumps along the minor axis. However, we find the \citet{Wang_2016} linearity criterion is, in most cases, not restrictive enough to identify features that \textit{look} highly elongated on the plane of the sky. If one assumes no geometry a priori for the morphology of MST ``Bone" filaments (as in our ``continuum-based" definition, see \S \ref{mst_boundaries} and Figure \ref{fig:mst}) the aspect ratio for the same filament can be as much as $36\times$ lower than in \citet{Wang_2016}.\footnote{This is well illustrated in the case of F18 (Figure \ref{fig:mst} in the Appendix). According to \citet{Wang_2016} the aspect ratio of this filament is 108:1, calculated by summing all the edges in the tree (thin green line in Figure \ref{fig:mst}) and dividing by the average width of the velocity contiguous BGPS clumps (0.34 pc, colored points in Figure \ref{fig:mst}). By assuming no geometry a priori (our ``continuum-based" definition; yellow contours in Figure \ref{fig:mst}) we calculate a new length (17 pc) and new width (4 pc) as outlined in \S \ref{lengths} in this study, determining an aspect ratio of 3:1 for this filament, which is over a factor of 30 lower than the original \citet{Wang_2016} value of 108:1.}

In light of this, it is necessary to discuss the potential formation mechanisms for the MST ``Bone" filaments. One scenario \citep{Wang_2016} is that MST filaments can be approximated as cylinders, within which clumps form at evenly spaced intervals due to sausage instability \citep{Ostriker_1964}. However, without very strict requirements on linearity, there exists a degeneracy between massive clumps that form in an isothermal cylinder via sausage instability \citep{Jackson_2010} and massive clumps that form any other way, and can simultaneously be traced out by a continuous line in \textit{position-position-velocity} space. One alternative scenario is that massive clumps form inside the filamentary networks inside GMCs, becoming massive by preferentially accreting gas from the greater gravo-turbulent molecular cloud environment \citep[see][]{Smith_2011}. Given the low aspect ratios and lack of high column density filamentary morphology, we conclude that the clumps within the MST filaments could have a variety of formation mechanisms---not necessarily limited to the sausage instability mode of cylindrical fragmentation evident in filaments like Nessie \citep[e.g.][]{Jackson_2010}. However, in the high linearity limit ($>>10$), the MST ``Bone" filaments and the Milky Way Bone filaments would be the same. Only in such a limit would it be reasonable to assume that the set of cores formed via cylindrical fragmentation (as likely occurs in Nessie) versus any other mechanism causing cores to form in the same molecular cloud and show contiguity in \textit{position-position-velocity} space.

\subsection{Milky Way Bones and Large-Scale Herschel Filaments: Bone Candidates (The ``Spines" of Spiral Features?)} \label{bone_discussion}
The third potential category encompasses a set of highly-elongated ($\approx$ 1 pc wide, $>$ 25 pc long), high column density filaments, which show contiguous, filamentary morphology. Most of the \citet{Zucker_2015} Bone and \citet{Wang_2015} Large-Scale Herschel filaments occupy this category. Since the \citet{Zucker_2015} Bone criteria are based on IRDCs, and many of the Herschel filaments are only observed in dust emission, we refer to this class as ``Bone candidates" pending the results of numerical simulations and a larger statistical sample. Given their very large cold \& high column density fractions and aspect ratios, it is plausible that these filaments are formed via a Galactic-scale mechanism, rather than a scaled-up version of small-scale filament formation. 

In one popular small-scale filament formation scenario, as highlighted in recent numerical simulations \citep[see][]{Smith_2014a}, filaments we observe in regions like Taurus are filamentary clumps or ``fibers" \citep{Hacar_2013} swept together into a continuous structure by a combination of turbulence and the large-scale non-spherical gravitational collapse of a molecular cloud. However, placing a very generous upper limit on the maximum aspect ratio of a typical Giant Molecular cloud ($\approx$ 15:1 according to \citet{Duarte_Cabral_2016}), one may not be able form a cloud-scale, contiguous filament exceeding this aspect ratio by self-gravity alone, particularly given that the collapse itself is non-uniform due to large-scale density perturbations within the molecular cloud. As outlined in this paper, filaments which constitute the bulk of the ``Bone candidate" sample typically have aspect ratios of 20:1 and above. Two additional diagnostics (linear mass density and power law index of the Plummer-like density profile) also show, unsurprisingly, that these Bone-like features lie in a different regime than small-scale ones, with both higher linear mass densities (see Figure \ref{fig:lwa_comp}) and steeper Plummer profiles (see Figure \ref{fig:plummer_comp}). 

Current numerical simulations are insufficient to probe the formation mechanisms of these dense ``Bone-like" Galactic filaments. While gravity certainly plays a role, we argue that instead of forming via local gravitational instabilities, these ``Bone-like" filaments may require a large-scale gravitational force (e.g. like the one associated with the midplane of a rotating disk galaxy) to form and maintain the highly linearized morphology we see over tens of parsec scales. If this is the case, these filaments could have important implications for tracing two things. First, they may be able to trace the midplane vertically, which only requires that they form via the instabilities associated with a flattened, rotating disk (without the requirement of spiral arms). And second, they may be able to trace spiral structure in that plane (e.g. spiral arms, spurs, and feathers), as their formation could be due in part to the large-scale gas dynamics of spiral galaxies (either due to a spiral potential well or due to Galactic shear followed by gas compression as filaments enter arms). This is consistent with recent numerical simulations from \citet{Smith_2014} which qualitatively show filamentary features with widths, densities, and aspect ratios compatible with the majority of our Bone candidate sample forming in the mid-planes of spiral galaxies. The evidence presented in this work suggests that these filaments could pin down the gravitational midplane and potentially the spiral structure in that plane, regardless of whether they reach alignment in or just prior to arm entry, at the deepest point in the gravitational potential well.

We also consider which subset of the ``Bone candidate" sample shows potential for tracing Galactic structure. While there are a few outliers that easily fail any reasonable criterion for potential spiral arm association, $\approx$ 75\% of the filaments in the ``Bone candidate" sample have lengths on the order of $\approx 20-80 \; \rm pc$, aspect ratios exceeding 20:1 in either extinction or emission, Galactic altitudes $\lesssim 20-30 \; \rm pc$, and position angles $\lesssim 30^\circ$, so they possess properties consistent with very nascent results from numerical simulations \citep{Smith_2014} when neglecting kinematic information (which is not yet available in these simulations). 

The next step is to develop the next generation of these simulations and extract a sample of synthetic filaments worthy of comparison with observations. Such a correlative study could allow better refinement of the \citet{Zucker_2015} Bone criteria, and could calibrate the confidence with which we could associate particular types of filaments with larger scale spiral features.

Pending the results of numerical simulations---and a targeted, expanded search for more Bone-like features---it may be possible to use these filaments as anchors of Galactic structure, pinning down the location of the physical Galactic midplane or specific spiral features with high resolution at particular regions in \textit{l,b,v} space. Combined with pre-existing, gross knowledge of Galactic structure from masers \citep{Reid_2014}, 3D dust extinction mapping \citep{Green_2015}, stellar densities (e.g. LSST, Gaia) and molecular cloud catalogs \citep{Rice_2016, Miville_Deschenes_2016} large-scale Bone-like filaments could help us iteratively refine our current understanding of Milky Way structure over the course of the next decade. 

\section{Conclusion} \label{conclusion}
We perform a comprehensive, standardized analysis of the physical properties of 45 large-scale Galactic filaments derived from four different catalogs in the literature, from \citet{Zucker_2015} [Milky Way Bones], \citet{Ragan_2014} and \citet{Abreu_Vicente_2016} [Giant Molecular Filaments], \citet{Wang_2015} [Large-Scale Herschel Filaments], and \citet{Wang_2016} [MST ``Bone" filaments]. Each catalog uses fundamentally different selection criteria and methodology, so we have reanalyzed and significantly expanded upon previous work to disentangle inherent properties of the filaments from the diversity of methods employed to calculate them. Our conclusions are as follows:

\begin{enumerate} 
\item{We find significant variation in the physical properties of large-scale filaments not only between catalogs, but also within catalogs, suggesting that different filament finding techniques are uncovering a wide range of physical structures. The length (11-269 pc), width (1-40 pc), mass ($\rm 3\times10^{3} - 1.1\times10^{6} \; M_\odot$), aspect ratio (3:1-117:1), linear mass ($80-6300 \; \rm M_\odot \;  pc ^{-1})$, and median column density ($\rm 1\times10^{21} - 2 \times 10^{22} \; cm^{-2}$) vary by one to two orders of magnitude across the large-scale filament spectrum. Only a few properties remain consistent across the full sample. Most notably, 85\% of the filaments in our full sample lie mostly parallel to ($ < 45 ^\circ$) and in close proximity to ($ < 30 \; \rm pc$) the physical Galactic midplane on the plane-of-the-sky, though in several cases this is what would be expected at random given the observational biases of current Galactic plane surveys.}

\item{We can very broadly distinguish between different large-scale filament catalogs in a zone of parameter space defined by the filaments' cold \& high column density fraction and aspect ratio. Based on their physical properties, we identify three categories. 

\begin{enumerate}

\item We first identify a category of large-scale filaments (Giant Molecular Filaments or ``GMFs") that are consistent with being the high-aspect-ratio tail of the GMC distribution (aspect ratio $\approx$ 8:1, cold \& high column density fraction $<$ 10 \%) 

\item A second category could be elongated dense core complexes tracing out networks of dense compact sources embedded in GMCs (aspect ratio $\approx$ 10:1, 10\% $<$  cold \& high column density fraction $<$ 75\%) 

\item A third category consists of highly elongated, high column density filaments (aspect ratio $\gtrsim$ 20:1, 10\% $<$  cold \& high column density fraction $<$ 50\%) some fraction of which could constitute the ``Bones" of the Galaxy. 

\end{enumerate}
 
At the very least, we determine that a single umbrella term for all large-scale filaments (e.g. ``Bones", ``GMFs") is clearly insufficient to convey the differences in the filament physical properties we observe, so more nuanced nomenclature, like that used here, needs to be adopted in the field.}

\item{For the subset of our full-sample showing continuous, high density structure in Herschel column density maps (18/45 filaments), we fit Plummer-like and Gaussian functions to radial column density profiles. We determine that our large-scale filaments exhibit much steeper density profiles than smaller-scale filaments embedded in nearby molecular clouds \citep{Arzoumanian_2011, Cox_2016, Palmeirim_2013}. Large-scale filaments have a typical Plummer ``$p$" index of the density profile between $\approx 2.75-3.25$ and a typical width on the order of $1.3 \; \rm pc$, while small-scales filaments have $p \approx 1.5-2.5$ and widths on the order of 0.1-0.3 pc. Both the width and the $p$ index are dependent upon the fitting distance and background subtraction method used in measuring the profiles. The new profile fitting code, \href{https://github.com/catherinezucker/radfil}{\texttt{RadFil}}, is publicly available. }

\item{We undertake a detailed \textit{position-position-velocity (p-p-v)} analysis of a subset of the sample with high resolution CO emission data (22/45 filaments in the range $18^\circ < l < 56^\circ$). We compare the \textit{p-p-v} track of each filament with the \textit{p-p-v} track of the nearest spiral feature from \citet{Reid_2016} (R16) by quantifying displacement from the spiral feature in both \textit{p-p} and \textit{p-v} space. We find that only 35\% of large-scale filaments exhibit strong proximity to R16 spiral arms, both spatially and kinematically ($\rm < 20 \; pc,\; < 10 \; km \; s^{-1}$), and this should be considered an upper limit given that some classes were pre-selected to lie near spiral arms. We further find that filaments lying ``in the plane" of the Galaxy (aligned and at close separation from the midplane in 2D projected space) is more common than filaments lying both ``in the plane" \textit{and} potentially at a distance of an arm. That is, of the filaments lying in the disk (z $< 20$ pc and $\theta <30^\circ$), only half of those are also kinematically consistent with known spiral arms ($\rm \Delta v_{arm} < 10 \; km \; s^{-1}$). However, if a filament is long, skinny, high column density, and parallel to the Galactic plane, it is more likely to kinematically lie near spiral arms than those that are not.}

\item{The very longest and densest ``Bone-like" filaments might be formed and maintained by the external gravity of the Galaxy (as opposed to local gravitational instabilities), and thus have some utility in delineating spiral structure (e.g. spiral arms, spurs, feathers, interarm regions) or the gravitational midplane of the Galaxy.}

\end{enumerate}

\acknowledgments
We would like to thank Amy Cohn for her contributions to the kinematic analysis presented in \S \ref{arm_proximity}. Likewise, we would like to thank Tom Dame and Mark Reid for valuable discussion regarding the CO spiral arm traces and the Galactic structure of the Milky Way as constrained by maser parallax observations. Tom Robitaille's work on the \href{http://www.glueviz.org}{\texttt{glue}} visualization software made many of the conclusions presented in this paper possible. We would like to thank Rowan Smith for informing our current and future work on the comparison between observational and synthetic large-scale Galactic filaments. Finally, we would like to thank our anonymous referee, who provided a very constructive and thorough review of our manuscript, and whose feedback improved the quality of this work considerably. 

\bibliographystyle{apj}
\bibliography{full_article_revision}

\begin{thebibliography}{}
\expandafter\ifx\csname natexlab\endcsname\relax\def\natexlab#1{#1}\fi

\bibitem[{Abreu-Vicente {et~al.}(2016)Abreu-Vicente, Ragan, Kainulainen,
  Henning, Beuther, \& Johnston}]{Abreu_Vicente_2016}
Abreu-Vicente, J., Ragan, S., Kainulainen, J., {et~al.} 2016, Astronomy {\&}
  Astrophysics, 590, A131

\bibitem[{{Anderson} {et~al.}(2012){Anderson}, {Bania}, {Balser}, \&
  {Rood}}]{Anderson_2012}
{Anderson}, L.~D., {Bania}, T.~M., {Balser}, D.~S., \& {Rood}, R.~T. 2012,
  \apj, 754, 62

\bibitem[{Arzoumanian {et~al.}(2011)Arzoumanian, Andr{\'{e}}, Didelon,
  Könyves, Schneider, Men'shchikov, Sousbie, Zavagno, Bontemps, Francesco,
  Griffin, Hennemann, Hill, Kirk, Martin, Minier, Molinari, Motte, Peretto,
  Pezzuto, Spinoglio, Ward-Thompson, White, \& Wilson}]{Arzoumanian_2011}
Arzoumanian, D., Andr{\'{e}}, P., Didelon, P., {et~al.} 2011, Astronomy {\&}
  Astrophysics, 529, L6

\bibitem[{Ballesteros-Paredes {et~al.}(2011)Ballesteros-Paredes,
  V{\'{a}}zquez-Semadeni, Gazol, Hartmann, Heitsch, \&
  Col{\'{\i}}n}]{Ballesteros_Paredes_2011}
Ballesteros-Paredes, J., V{\'{a}}zquez-Semadeni, E., Gazol, A., {et~al.} 2011,
  Monthly Notices of the Royal Astronomical Society, 416, 1436

\bibitem[{Barnes {et~al.}(2015)Barnes, Muller, Indermuehle, O'Dougherty, Lowe,
  Cunningham, Hernandez, \& Fuller}]{Barnes_2015}
Barnes, P.~J., Muller, E., Indermuehle, B., {et~al.} 2015, {ApJ}, 812, 6

\bibitem[{{Battersby} {et~al.}(2014){Battersby}, {Bally}, {Dunham}, {Ginsburg},
  {Longmore}, \& {Darling}}]{Battersby_2014}
{Battersby}, C., {Bally}, J., {Dunham}, M., {et~al.} 2014, \apj, 786, 116

\bibitem[{Battersby {et~al.}(2010)Battersby, Bally, Jackson, Ginsburg, Shirley,
  Schlingman, \& Glenn}]{Battersby_2010}
Battersby, C., Bally, J., Jackson, J.~M., {et~al.} 2010, The Astrophysical
  Journal, 721, 222

\bibitem[{Battersby {et~al.}(2017)Battersby, Bally, \&
  Svoboda}]{Battersby_2017}
Battersby, C., Bally, J., \& Svoboda, B. 2017, The Astrophysical Journal, 835,
  263

\bibitem[{Battersby {et~al.}(2011)Battersby, Bally, Ginsburg, Bernard, Brunt,
  Fuller, Martin, Molinari, Mottram, Peretto, Testi, \&
  Thompson}]{Battersby_2011}
Battersby, C., Bally, J., Ginsburg, A., {et~al.} 2011, Astronomy {\&}
  Astrophysics, 535, A128

\bibitem[{Bohlin {et~al.}(1978)Bohlin, Savage, \& Drake}]{Bohlin_1978}
Bohlin, R.~C., Savage, B.~D., \& Drake, J.~F. 1978, The Astrophysical Journal,
  224, 132

\bibitem[{Carey {et~al.}(2009)Carey, Noriega-Crespo, Mizuno, Shenoy, Paladini,
  Kraemer, Price, Flagey, Ryan, Ingalls, Kuchar, Gon{\c{c}}alves, Indebetouw,
  Billot, Marleau, Padgett, Rebull, Bressert, Ali, Molinari, Martin, Berriman,
  Boulanger, Latter, Miville-Deschenes, Shipman, \& Testi}]{Carey_2009}
Carey, S.~J., Noriega-Crespo, A., Mizuno, D.~R., {et~al.} 2009, {PUBL} {ASTRON}
  {SOC} {PAC}, 121, 76

\bibitem[{{Chen} {et~al.}(2017){Chen}, {Burkhart}, {Goodman}, \&
  {Collins}}]{Chen_2017}
{Chen}, H., {Burkhart}, B., {Goodman}, A.~A., \& {Collins}, D.~C. 2017, ArXiv
  e-prints, arXiv:1707.09356

\bibitem[{Churchwell {et~al.}(2009)Churchwell, Babler, Meade, Whitney,
  Benjamin, Indebetouw, Cyganowski, Robitaille, Povich, Watson, \&
  Bracker}]{Churchwell_2009}
Churchwell, E., Babler, B.~L., Meade, M.~R., {et~al.} 2009, {PUBL} {ASTRON}
  {SOC} {PAC}, 121, 213

\bibitem[{Colombo {et~al.}(2015)Colombo, Rosolowsky, Ginsburg, Duarte-Cabral,
  \& Hughes}]{Colombo_2015}
Colombo, D., Rosolowsky, E., Ginsburg, A., Duarte-Cabral, A., \& Hughes, A.
  2015, Monthly Notices of the Royal Astronomical Society, 454, 2067

\bibitem[{Cox {et~al.}(2016)Cox, Arzoumanian, Andr{\'{e}}, Rygl, Prusti,
  Men'shchikov, Royer, K{\'{o}}sp{\'{a}}l, Palmeirim, Ribas, Könyves, Bernard,
  Schneider, Bontemps, Merin, Vavrek, de~Oliveira, Didelon, Pilbratt, \&
  Waelkens}]{Cox_2016}
Cox, N. L.~J., Arzoumanian, D., Andr{\'{e}}, P., {et~al.} 2016, Astronomy {\&}
  Astrophysics, 590, A110

\bibitem[{Dame {et~al.}(2001)Dame, Hartmann, \& Thaddeus}]{Dame_2001}
Dame, T.~M., Hartmann, D., \& Thaddeus, P. 2001, The Astrophysical Journal,
  547, 792

\bibitem[{Dame \& Thaddeus(2011)}]{Dame_2011}
Dame, T.~M., \& Thaddeus, P. 2011, {ApJ}, 734, L24

\bibitem[{{Dobbs} \& {Pringle}(2013)}]{Dobbs_2013}
{Dobbs}, C.~L., \& {Pringle}, J.~E. 2013, \mnras, 432, 653

\bibitem[{{Du} {et~al.}(2017){Du}, {Xu}, {Yang}, \& {Sun}}]{Du_2017}
{Du}, X., {Xu}, Y., {Yang}, J., \& {Sun}, Y. 2017, \apjs, 229, 24

\bibitem[{Duarte-Cabral \& Dobbs(2016)}]{Duarte_Cabral_2016}
Duarte-Cabral, A., \& Dobbs, C.~L. 2016, Mon. Not. R. Astron. Soc., 458, 3667

\bibitem[{{Duarte-Cabral} \& {Dobbs}(2017)}]{Duarte_Cabral_2017}
{Duarte-Cabral}, A., \& {Dobbs}, C.~L. 2017, ArXiv e-prints, arXiv:1706.05421

\bibitem[{Ellsworth-Bowers {et~al.}(2015)Ellsworth-Bowers, Rosolowsky, Glenn,
  Ginsburg, II, Battersby, Shirley, \& Svoboda}]{Ellsworth_Bowers_2015}
Ellsworth-Bowers, T.~P., Rosolowsky, E., Glenn, J., {et~al.} 2015, The
  Astrophysical Journal, 799, 29

\bibitem[{Ellsworth-Bowers {et~al.}(2013)Ellsworth-Bowers, Glenn, Rosolowsky,
  Mairs, Evans, Battersby, Ginsburg, Shirley, \& Bally}]{Ellsworth_Bowers_2013}
Ellsworth-Bowers, T.~P., Glenn, J., Rosolowsky, E., {et~al.} 2013, The
  Astrophysical Journal, 770, 39

\bibitem[{Finkbeiner {et~al.}(1999)Finkbeiner, Davis, \&
  Schlegel}]{Finkbeiner_1999}
Finkbeiner, D.~P., Davis, M., \& Schlegel, D.~J. 1999, The Astrophysical
  Journal, 524, 867

\bibitem[{Fischera \& Martin(2012)}]{Fischera_2012}
Fischera, J., \& Martin, P.~G. 2012, Astronomy {\&} Astrophysics, 542, A77

\bibitem[{Ginsburg \& Mirocha(2011)}]{Ginsburg_2011}
Ginsburg, A., \& Mirocha, J. 2011, Astrophysics Source Code Library,
  doi:2011ascl.soft09001G

\bibitem[{Ginsburg {et~al.}(2013)Ginsburg, Glenn, Rosolowsky, Ellsworth-Bowers,
  Battersby, Dunham, Merello, Shirley, Bally, II, Stringfellow, \&
  Aguirre}]{Ginsburg_2013}
Ginsburg, A., Glenn, J., Rosolowsky, E., {et~al.} 2013, The Astrophysical
  Journal Supplement Series, 208, 14

\bibitem[{Goodman {et~al.}(2014)Goodman, Alves, Beaumont, Benjamin, Borkin,
  Burkert, Dame, Jackson, Kauffmann, Robitaille, \& Smith}]{Goodman_2014}
Goodman, A.~A., Alves, J., Beaumont, C.~N., {et~al.} 2014, {ApJ}, 797, 53

\bibitem[{Green {et~al.}(2015)Green, Schlafly, Finkbeiner, Rix, Martin,
  Burgett, Draper, Flewelling, Hodapp, Kaiser, Kudritzki, Magnier, Metcalfe,
  Price, Tonry, \& Wainscoat}]{Green_2015}
Green, G.~M., Schlafly, E.~F., Finkbeiner, D.~P., {et~al.} 2015, The
  Astrophysical Journal, 810, 25

\bibitem[{Hacar {et~al.}(2013)Hacar, Tafalla, Kauffmann, \&
  Kov{\'{a}}cs}]{Hacar_2013}
Hacar, A., Tafalla, M., Kauffmann, J., \& Kov{\'{a}}cs, A. 2013, Astronomy {\&}
  Astrophysics, 554, A55

\bibitem[{Henshaw {et~al.}(2016)Henshaw, Longmore, Kruijssen, Davies, Bally,
  Barnes, Battersby, Burton, Cunningham, Dale, Ginsburg, Immer, Jones, Kendrew,
  Mills, Molinari, Moore, Ott, Pillai, Rathborne, Schilke, Schmiedeke, Testi,
  Walker, Walsh, \& Zhang}]{Henshaw_2016}
Henshaw, J.~D., Longmore, S.~N., Kruijssen, J. M.~D., {et~al.} 2016, Monthly
  Notices of the Royal Astronomical Society, 457, 2675

\bibitem[{Jackson {et~al.}(2010)Jackson, Finn, Chambers, Rathborne, \&
  Simon}]{Jackson_2010}
Jackson, J.~M., Finn, S.~C., Chambers, E.~T., Rathborne, J.~M., \& Simon, R.
  2010, {ApJ}, 719, L185

\bibitem[{Jackson {et~al.}(2006)Jackson, Rathborne, Shah, Simon, Bania,
  Clemens, Chambers, Johnson, Dormody, Lavoie, \& Heyer}]{Jackson_2006}
Jackson, J.~M., Rathborne, J.~M., Shah, R.~Y., {et~al.} 2006, The Astrophysical
  Journal Supplement Series, 163, 145

\bibitem[{Jackson {et~al.}(2013)Jackson, Rathborne, Foster, Whitaker, Sanhueza,
  Claysmith, Mascoop, Wienen, Breen, Herpin, Duarte-Cabral, Csengeri, Longmore,
  Contreras, Indermuehle, Barnes, Walsh, Cunningham, Brooks, Britton, Voronkov,
  Urquhart, Alves, Jordan, Hill, Hoq, Finn, Bains, Bontemps, Bronfman, Caswell,
  Deharveng, Ellingsen, Fuller, Garay, Green, Hindson, Jones, Lenfestey, Lo,
  Lowe, Mardones, Menten, Minier, Morgan, Motte, Muller, Peretto, Purcell,
  Schilke, Bontemps, Schuller, Titmarsh, Wyrowski, \& Zavagno}]{Jackson_2013}
Jackson, J.~M., Rathborne, J.~M., Foster, J.~B., {et~al.} 2013, Publications of
  the Astronomical Society of Australia, 30, doi:10.1017/pasa.2013.37

\bibitem[{Juvela {et~al.}(2012{\natexlab{a}})Juvela, Malinen, \&
  Lunttila}]{Juvela_2012a}
Juvela, M., Malinen, J., \& Lunttila, T. 2012{\natexlab{a}}, Astronomy {\&}
  Astrophysics, 544, A141

\bibitem[{Juvela {et~al.}(2012{\natexlab{b}})Juvela, Ristorcelli, Pagani, Doi,
  Pelkonen, Marshall, Bernard, Falgarone, Malinen, Marton, McGehee, Montier,
  Motte, Paladini, T{\'{o}}th, Ysard, Zahorecz, \& Zavagno}]{Juvela_2012}
Juvela, M., Ristorcelli, I., Pagani, L., {et~al.} 2012{\natexlab{b}}, Astronomy
  {\&} Astrophysics, 541, A12

\bibitem[{Kauffmann {et~al.}(2008)Kauffmann, Bertoldi, Bourke, Evans, \&
  Lee}]{Kauffmann_2008}
Kauffmann, J., Bertoldi, F., Bourke, T.~L., Evans, N.~J., \& Lee, C.~W. 2008,
  Astronomy and Astrophysics, 487, 993

\bibitem[{{Kim} \& {Ostriker}(2002)}]{Kim_2002}
{Kim}, W.-T., \& {Ostriker}, E.~C. 2002, \apj, 570, 132

\bibitem[{Koch \& Rosolowsky(2015)}]{Koch_2015}
Koch, E.~W., \& Rosolowsky, E.~W. 2015, Mon. Not. R. Astron. Soc., 452, 3435

\bibitem[{{Koda} {et~al.}(2006){Koda}, {Sawada}, {Hasegawa}, \&
  {Scoville}}]{Koda_2006}
{Koda}, J., {Sawada}, T., {Hasegawa}, T., \& {Scoville}, N.~Z. 2006, \apj, 638,
  191

\bibitem[{{K{\"o}nyves} {et~al.}(2015){K{\"o}nyves}, {Andr{\'e}},
  {Men'shchikov}, {Palmeirim}, {Arzoumanian}, {Schneider}, {Roy}, {Didelon},
  {Maury}, {Shimajiri}, {Di Francesco}, {Bontemps}, {Peretto}, {Benedettini},
  {Bernard}, {Elia}, {Griffin}, {Hill}, {Kirk}, {Ladjelate}, {Marsh}, {Martin},
  {Motte}, {Nguy{\^e}n Luong}, {Pezzuto}, {Roussel}, {Rygl}, {Sadavoy},
  {Schisano}, {Spinoglio}, {Ward-Thompson}, \& {White}}]{Konyves_2015}
{K{\"o}nyves}, V., {Andr{\'e}}, P., {Men'shchikov}, A., {et~al.} 2015, \aap,
  584, A91

\bibitem[{Li {et~al.}(2016)Li, Urquhart, Leurini, Csengeri, Wyrowski, Menten,
  \& Schuller}]{Li_2016}
Li, G.-X., Urquhart, J.~S., Leurini, S., {et~al.} 2016, Astronomy {\&}
  Astrophysics, 591, A5

\bibitem[{Longmore {et~al.}(2012)Longmore, Rathborne, Bastian, Alves, Ascenso,
  Bally, Testi, Longmore, Battersby, Bressert, Purcell, Walsh, Jackson, Foster,
  Molinari, Meingast, Amorim, Lima, Marques, Moitinho, Pinhao, Rebordao, \&
  Santos}]{Longmore_2012}
Longmore, S.~N., Rathborne, J., Bastian, N., {et~al.} 2012, {ApJ}, 746, 117

\bibitem[{Lucas {et~al.}(2008)Lucas, Hoare, Longmore, Schröder, Davis,
  Adamson, Bandyopadhyay, de~Grijs, Smith, Gosling, Mitchison,
  G{\'{a}}sp{\'{a}}r, Coe, Tamura, Parker, Irwin, Hambly, Bryant, Collins,
  Cross, Evans, Gonzalez-Solares, Hodgkin, Lewis, Read, Riello, Sutorius,
  Lawrence, Drew, Dye, \& Thompson}]{Lucas_2008}
Lucas, P.~W., Hoare, M.~G., Longmore, A., {et~al.} 2008, Monthly Notices of the
  Royal Astronomical Society, 391, 136

\bibitem[{Miville-Desch{\^{e}}nes {et~al.}(2016)Miville-Desch{\^{e}}nes,
  Murray, \& Lee}]{Miville_Deschenes_2016}
Miville-Desch{\^{e}}nes, M.-A., Murray, N., \& Lee, E.~J. 2016, The
  Astrophysical Journal, 834, 57

\bibitem[{{Molinari} {et~al.}(2010){Molinari}, {Swinyard}, {Bally}, {Barlow},
  {Bernard}, {Martin}, {Moore}, {Noriega-Crespo}, {Plume}, {Testi}, {Zavagno},
  {Abergel}, {Ali}, {Andr{\'e}}, {Baluteau}, {Benedettini}, {Bern{\'e}},
  {Billot}, {Blommaert}, {Bontemps}, {Boulanger}, {Brand}, {Brunt}, {Burton},
  {Campeggio}, {Carey}, {Caselli}, {Cesaroni}, {Cernicharo}, {Chakrabarti},
  {Chrysostomou}, {Codella}, {Cohen}, {Compiegne}, {Davis}, {de Bernardis}, {de
  Gasperis}, {Di Francesco}, {di Giorgio}, {Elia}, {Faustini}, {Fischera},
  {Fukui}, {Fuller}, {Ganga}, {Garcia-Lario}, {Giard}, {Giardino}, {Glenn},
  {Goldsmith}, {Griffin}, {Hoare}, {Huang}, {Jiang}, {Joblin}, {Joncas},
  {Juvela}, {Kirk}, {Lagache}, {Li}, {Lim}, {Lord}, {Lucas}, {Maiolo},
  {Marengo}, {Marshall}, {Masi}, {Massi}, {Matsuura}, {Meny}, {Minier},
  {Miville-Desch{\^e}nes}, {Montier}, {Motte}, {M{\"u}ller}, {Natoli}, {Neves},
  {Olmi}, {Paladini}, {Paradis}, {Pestalozzi}, {Pezzuto}, {Piacentini},
  {Pomar{\`e}s}, {Popescu}, {Reach}, {Richer}, {Ristorcelli}, {Roy}, {Royer},
  {Russeil}, {Saraceno}, {Sauvage}, {Schilke}, {Schneider-Bontemps},
  {Schuller}, {Schultz}, {Shepherd}, {Sibthorpe}, {Smith}, {Smith},
  {Spinoglio}, {Stamatellos}, {Strafella}, {Stringfellow}, {Sturm}, {Taylor},
  {Thompson}, {Tuffs}, {Umana}, {Valenziano}, {Vavrek}, {Viti}, {Waelkens},
  {Ward-Thompson}, {White}, {Wyrowski}, {Yorke}, \& {Zhang}}]{Molinari_2010}
{Molinari}, S., {Swinyard}, B., {Bally}, J., {et~al.} 2010, \pasp, 122, 314

\bibitem[{Molinari {et~al.}(2016)Molinari, Schisano, Elia, Pestalozzi,
  Traficante, Pezzuto, Swinyard, Noriega-Crespo, Bally, Moore, Plume, Zavagno,
  di~Giorgio, Liu, Pilbratt, Mottram, Russeil, Piazzo, Veneziani, Benedettini,
  Calzoletti, Faustini, Natoli, Piacentini, Merello, Palmese, Grande,
  Polychroni, Rygl, Polenta, Barlow, Bernard, Martin, Testi, Ali, Andr{\'{e}},
  Beltr{\'{a}}n, Billot, Carey, Cesaroni, Compi{\`{e}}gne, Eden, Fukui,
  Garcia-Lario, Hoare, Huang, Joncas, Lim, Lord, Martinavarro-Armengol, Motte,
  Paladini, Paradis, Peretto, Robitaille, Schilke, Schneider, Schulz,
  Sibthorpe, Strafella, Thompson, Umana, Ward-Thompson, \&
  Wyrowski}]{Molinari_2016}
Molinari, S., Schisano, E., Elia, D., {et~al.} 2016, Astronomy {\&}
  Astrophysics, 591, A149

\bibitem[{{Moore} {et~al.}(2012){Moore}, {Urquhart}, {Morgan}, \&
  {Thompson}}]{Moore_2012}
{Moore}, T.~J.~T., {Urquhart}, J.~S., {Morgan}, L.~K., \& {Thompson}, M.~A.
  2012, \mnras, 426, 701

\bibitem[{{Myers} {et~al.}(1991){Myers}, {Fuller}, {Goodman}, \&
  {Benson}}]{Myers_1991}
{Myers}, P.~C., {Fuller}, G.~A., {Goodman}, A.~A., \& {Benson}, P.~J. 1991,
  \apj, 376, 561

\bibitem[{Nakanishi \& Sofue(2006)}]{Nakanishi_2006}
Nakanishi, H., \& Sofue, Y. 2006, Publications of the Astronomical Society of
  Japan, 58, 847

\bibitem[{Ostriker(1964)}]{Ostriker_1964}
Ostriker, J. 1964, The Astrophysical Journal, 140, 1056

\bibitem[{Palmeirim {et~al.}(2013)Palmeirim, Andr{\'{e}}, Kirk, Ward-Thompson,
  Arzoumanian, Könyves, Didelon, Schneider, Benedettini, Bontemps, Francesco,
  Elia, Griffin, Hennemann, Hill, Martin, Men'shchikov, Molinari, Motte, Luong,
  Nutter, Peretto, Pezzuto, Roy, Rygl, Spinoglio, \& White}]{Palmeirim_2013}
Palmeirim, P., Andr{\'{e}}, P., Kirk, J., {et~al.} 2013, Astronomy {\&}
  Astrophysics, 550, A38

\bibitem[{Peretto \& Fuller(2009)}]{Peretto_2009}
Peretto, N., \& Fuller, G.~A. 2009, Astronomy and Astrophysics, 505, 405

\bibitem[{Peretto {et~al.}(2016)Peretto, Lenfestey, Fuller, Traficante,
  Molinari, Thompson, \& Ward-Thompson}]{Peretto_2016}
Peretto, N., Lenfestey, C., Fuller, G.~A., {et~al.} 2016, Astronomy {\&}
  Astrophysics, 590, A72

\bibitem[{Peretto {et~al.}(2010)Peretto, Fuller, Plume, Anderson, Bally,
  Battersby, Beltran, Bernard, Calzoletti, DiGiorgio, Faustini, Kirk,
  Lenfestey, Marshall, Martin, Molinari, Montier, Motte, Ristorcelli,
  Rod{\'{o}}n, Smith, Traficante, Veneziani, Ward-Thompson, \&
  Wilcock}]{Peretto_2010}
Peretto, N., Fuller, G.~A., Plume, R., {et~al.} 2010, Astronomy and
  Astrophysics, 518, L98

\bibitem[{Purcell {et~al.}(2012{\natexlab{a}})Purcell, Longmore, Walsh,
  Whiting, Breen, Britton, Brooks, Burton, Cunningham, Green, Harvey-Smith,
  Hindson, Hoare, Indermuehle, Jones, Lo, Lowe, Phillips, Thompson, Urquhart,
  Voronkov, \& White}]{Purcell_2012}
Purcell, C.~R., Longmore, S.~N., Walsh, A.~J., {et~al.} 2012{\natexlab{a}},
  Monthly Notices of the Royal Astronomical Society, 426, 1972

\bibitem[{Purcell {et~al.}(2012{\natexlab{b}})Purcell, Longmore, Walsh,
  Whiting, Breen, Britton, Brooks, Burton, Cunningham, Green, Harvey-Smith,
  Hindson, Hoare, Indermuehle, Jones, Lo, Lowe, Phillips, Thompson, Urquhart,
  Voronkov, \& White}]{Purcell_2012a}
---. 2012{\natexlab{b}}, Monthly Notices of the Royal Astronomical Society,
  426, 1972

\bibitem[{Ragan {et~al.}(2014)Ragan, Henning, Tackenberg, Beuther, Johnston,
  Kainulainen, \& Linz}]{Ragan_2014}
Ragan, S.~E., Henning, T., Tackenberg, J., {et~al.} 2014, Astronomy {\&}
  Astrophysics, 568, A73

\bibitem[{Reid {et~al.}(2016)Reid, Dame, Menten, \& Brunthaler}]{Reid_2016}
Reid, M.~J., Dame, T.~M., Menten, K.~M., \& Brunthaler, A. 2016, {ApJ}, 823, 77

\bibitem[{Reid {et~al.}(2014)Reid, Menten, Brunthaler, Zheng, Dame, Xu, Wu,
  Zhang, Sanna, Sato, Hachisuka, Choi, Immer, Moscadelli, Rygl, \&
  Bartkiewicz}]{Reid_2014}
Reid, M.~J., Menten, K.~M., Brunthaler, A., {et~al.} 2014, {ApJ}, 783, 130

\bibitem[{{Renaud} {et~al.}(2013){Renaud}, {Bournaud}, {Emsellem}, {Elmegreen},
  {Teyssier}, {Alves}, {Chapon}, {Combes}, {Dekel}, {Gabor}, {Hennebelle}, \&
  {Kraljic}}]{Renaud_2013}
{Renaud}, F., {Bournaud}, F., {Emsellem}, E., {et~al.} 2013, \mnras, 436, 1836

\bibitem[{Rice {et~al.}(2016)Rice, Goodman, Bergin, Beaumont, \&
  Dame}]{Rice_2016}
Rice, T.~S., Goodman, A.~A., Bergin, E.~A., Beaumont, C., \& Dame, T.~M. 2016,
  The Astrophysical Journal, 822, 52

\bibitem[{Rosolowsky {et~al.}(2010)Rosolowsky, Dunham, Ginsburg, Bradley,
  Aguirre, Bally, Battersby, Cyganowski, Dowell, Drosback, Evans, Glenn,
  Harvey, Stringfellow, Walawender, \& Williams}]{Rosolowsky_2010}
Rosolowsky, E., Dunham, M.~K., Ginsburg, A., {et~al.} 2010, The Astrophysical
  Journal Supplement Series, 188, 123

\bibitem[{{Rosolowsky} {et~al.}(2008){Rosolowsky}, {Pineda}, {Kauffmann}, \&
  {Goodman}}]{Rosolowsky_2008}
{Rosolowsky}, E.~W., {Pineda}, J.~E., {Kauffmann}, J., \& {Goodman}, A.~A.
  2008, \apj, 679, 1338

\bibitem[{Sanna {et~al.}(2014)Sanna, Reid, Menten, Dame, Zhang, Sato,
  Brunthaler, Moscadelli, \& Immer}]{Sanna_2014}
Sanna, A., Reid, M.~J., Menten, K.~M., {et~al.} 2014, {ApJ}, 781, 108

\bibitem[{Schisano {et~al.}(2014)Schisano, Rygl, Molinari, Busquet, Elia,
  Pestalozzi, Polychroni, Billot, Carey, Paladini, Noriega-Crespo, Moore,
  Plume, Glover, \& V{\'{a}}zquez-Semadeni}]{Schisano_2014}
Schisano, E., Rygl, K. L.~J., Molinari, S., {et~al.} 2014, The Astrophysical
  Journal, 791, 27

\bibitem[{Schnee {et~al.}(2008)Schnee, Li, Goodman, \& Sargent}]{Schnee_2008}
Schnee, S., Li, J., Goodman, A.~A., \& Sargent, A.~I. 2008, The Astrophysical
  Journal, 684, 1228

\bibitem[{Shane(1972)}]{Shane_1972}
Shane, W. 1972, {A\&A}, 16, 118

\bibitem[{{Shetty} \& {Ostriker}(2006)}]{Shetty_2006}
{Shetty}, R., \& {Ostriker}, E.~C. 2006, \apj, 647, 997

\bibitem[{Shirley {et~al.}(2013)Shirley, Ellsworth-Bowers, Svoboda, Schlingman,
  Ginsburg, Rosolowsky, Gerner, Mairs, Battersby, Stringfellow, Dunham, Glenn,
  \& Bally}]{Shirley_2013}
Shirley, Y.~L., Ellsworth-Bowers, T.~P., Svoboda, B., {et~al.} 2013, The
  Astrophysical Journal Supplement Series, 209, 2

\bibitem[{Smith {et~al.}(2011)Smith, Slater, Fellhauer, Goodwin, \&
  Assmann}]{Smith_2011}
Smith, R., Slater, R., Fellhauer, M., Goodwin, S., \& Assmann, P. 2011, Monthly
  Notices of the Royal Astronomical Society, no

\bibitem[{Smith {et~al.}(2014{\natexlab{a}})Smith, Glover, Clark, Klessen, \&
  Springel}]{Smith_2014}
Smith, R.~J., Glover, S. C.~O., Clark, P.~C., Klessen, R.~S., \& Springel, V.
  2014{\natexlab{a}}, Monthly Notices of the Royal Astronomical Society, 441,
  1628

\bibitem[{Smith {et~al.}(2014{\natexlab{b}})Smith, Glover, \&
  Klessen}]{Smith_2014a}
Smith, R.~J., Glover, S. C.~O., \& Klessen, R.~S. 2014{\natexlab{b}}, Monthly
  Notices of the Royal Astronomical Society, 445, 2900

\bibitem[{Traficante {et~al.}(2011)Traficante, Calzoletti, Veneziani, Ali,
  de~Gasperis, Giorgio, Faustini, Ikhenaode, Molinari, Natoli, Pestalozzi,
  Pezzuto, Piacentini, Piazzo, Polenta, \& Schisano}]{Traficante_2011}
Traficante, A., Calzoletti, L., Veneziani, M., {et~al.} 2011, Monthly Notices
  of the Royal Astronomical Society, 416, 2932

\bibitem[{Urquhart {et~al.}(2014)Urquhart, Moore, Csengeri, Wyrowski, Schuller,
  Hoare, Lumsden, Mottram, Thompson, Menten, Walmsley, Bronfman, Pfalzner,
  Konig, \& Wienen}]{Urquhart_2014}
Urquhart, J.~S., Moore, T. J.~T., Csengeri, T., {et~al.} 2014, Monthly Notices
  of the Royal Astronomical Society, 443, 1555

\bibitem[{Vall\'{e}e(2008)}]{Vallee_2008}
Vall\'{e}e, J.~P. 2008, {AJ}, 135, 1301

\bibitem[{Wang {et~al.}(2016)Wang, Testi, Burkert, Walmsley, Beuther, \&
  Henning}]{Wang_2016}
Wang, K., Testi, L., Burkert, A., {et~al.} 2016, The Astrophysical Journal
  Supplement Series, 226, 9

\bibitem[{Wang {et~al.}(2015)Wang, Testi, Ginsburg, Walmsley, Molinari, \&
  Schisano}]{Wang_2015}
Wang, K., Testi, L., Ginsburg, A., {et~al.} 2015, Monthly Notices of the Royal
  Astronomical Society, 450, 4043

\bibitem[{Wang {et~al.}(2014)Wang, Zhang, Testi, v.~d. Tak, Wu, Zhang, Pillai,
  Wyrowski, Carey, Ragan, \& Henning}]{Wang_2014}
Wang, K., Zhang, Q., Testi, L., {et~al.} 2014, Monthly Notices of the Royal
  Astronomical Society, 439, 3275

\bibitem[{Wu {et~al.}(2014)Wu, Sato, Reid, Moscadelli, Zhang, Xu, Brunthaler,
  Menten, Dame, \& Zheng}]{Wu_2014}
Wu, Y.~W., Sato, M., Reid, M.~J., {et~al.} 2014, Astronomy {\&} Astrophysics,
  566, A17

\bibitem[{Zahorecz {et~al.}(2016)Zahorecz, Jimenez-Serra, Wang, Testi,
  T{\'{o}}th, \& Molinari}]{Zahorecz_2016}
Zahorecz, S., Jimenez-Serra, I., Wang, K., {et~al.} 2016, Astronomy {\&}
  Astrophysics, 591, A105

\bibitem[{Zucker {et~al.}(2015)Zucker, Battersby, \& Goodman}]{Zucker_2015}
Zucker, C., Battersby, C., \& Goodman, A. 2015, {ApJ}, 815, 23

\end{thebibliography}

\section{Appendix}

\subsection{Random Control Samples for Position Angle and Galactic Plane Separation Distributions} \label{control_discussion}
\begin{figure}[h!]
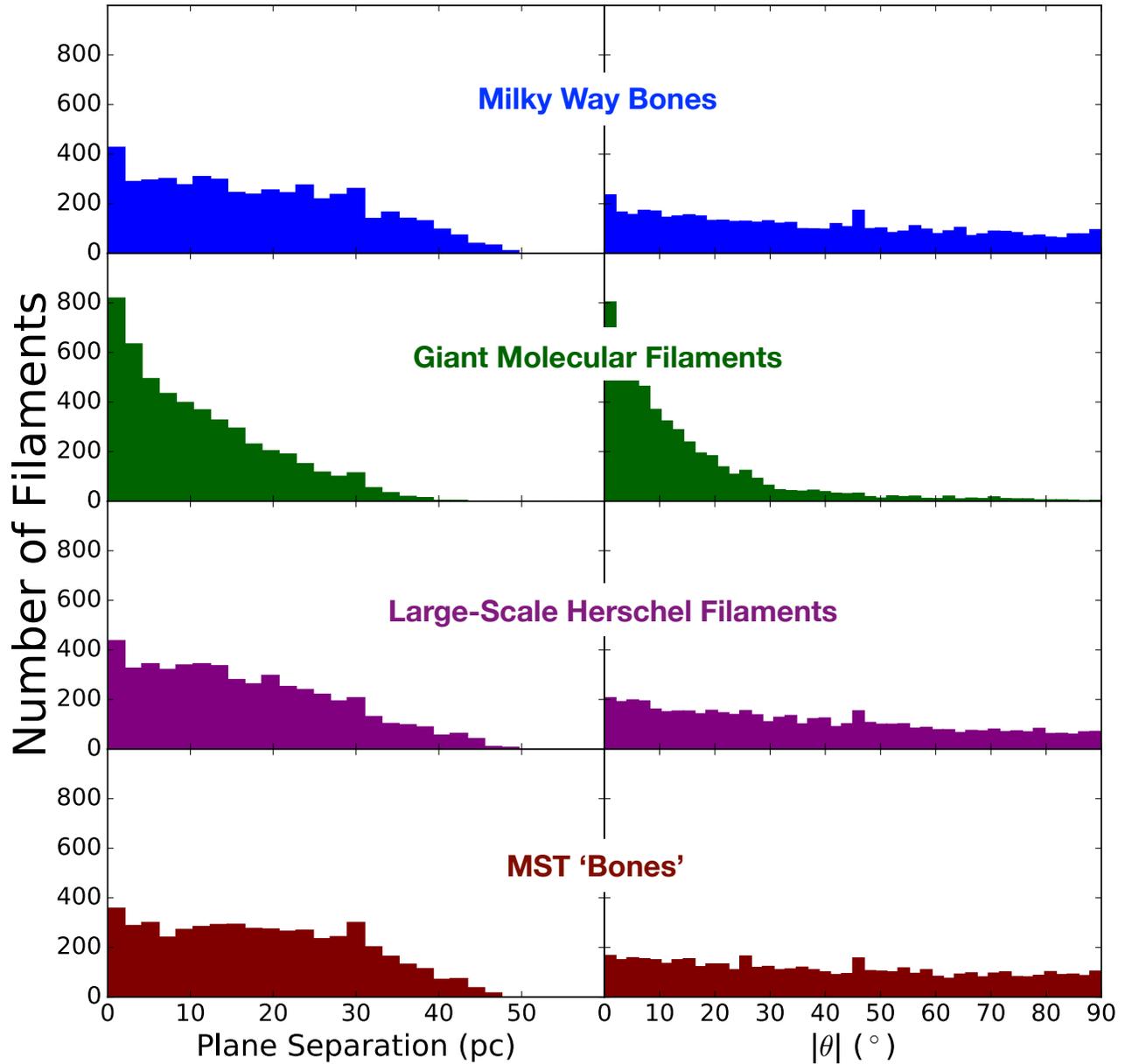

\begin{center}
\includegraphics[width=1.0\columnwidth]{{{posangle_altitude_control}}}
\caption{{\label{fig:control} Distribution of plane separation and position angles for randomly generated control samples corresponding to the four different filament classes.  The distributions are produced via a Monte Carlo analysis as described in \S \ref{control_discussion}. 
}}
\end{center}
\end{figure}

In Figure \ref{fig:armprops} we show the plane separation and position angle distributions of a control sample corresponding to each filament class (light box and whiskers), to get a sense of how each observed distribution differs from a random one. To do this, we perform a Monte Carlo analysis. First, we construct a box with a length of 5 kpc (the approximate size of the swath of the inner Galaxy at the typical distance to the near Scutum-Centaurus arm, where many of our filaments reside). At the median distance to our filaments (3.3 kpc), the GLIMPSE-Spitzer survey \citep{Churchwell_2009} and the Herschel Hi-GAL survey \citep{Molinari_2016} span $-1^\circ < b < 1 ^\circ$, which corresponds to 114 pc. We set the height or our box equal to 114 pc, with half the box above and half the box below the ``plane" of the Galaxy (0 pc). Then, for each filament class, we generate filaments with lengths and widths randomly sampled between the minimum and maximum lengths and widths observed within that class (see Table \ref{tab:tab3}). Then we place the center of each filament randomly inside the box (of size $5 {\rm \; kpc} \times 114 \; \rm{pc}$) with a random position angle drawn between $-90^\circ < \theta < 90^\circ$. If the filament falls entirely inside the box (consistent with all our filaments lying well inside the boundaries of GLIMPSE and Hi-GAL), it counts as ``observed" and enters the control sample shown in Figure \ref{fig:armprops}. We iteratively repeat this process until we obtain 5000 filaments in the control sample for each class. In addition to being shown as light-colored box and whiskers in Figure \ref{fig:armprops}, histograms of the Galactic plane separation and position angle for the control samples are shown in Figure \ref{fig:control}. 

The median plane separation of the random control samples for the Milky Way Bones, GMFs, Large-Scale Herschel, and MST `Bone' filaments is 17 pc, 9 pc, 15 pc, and 18 pc respectively (in comparison to the observed medians of 10 pc, 11 pc, 16 pc, and 13 pc). The median position angles of the random control samples for the Milky Way Bones, GMFs, Large-Scale Herschel, and MST `Bone' filaments is $34^\circ$, $9^\circ$, $32^\circ$, and $37^\circ$ respectively (in comparison to the observed medians of $11^\circ$, $10^\circ$, $14^\circ$, and $13^\circ$). Except for the GMFs, the position angles we observe tend to be more significant than the altitudes, in that the position angles tend to be less consistent with the random control sample. Of all four classes, the position angles and altitudes of the Milky Way Bones are the least consistent with being drawn at random. While this is part of the Milky Way Bone selection criteria \citep[see][]{Zucker_2015}, no long and skinny extinction features were found significantly inclined to or separated from the Galactic plane in their initial search, suggesting that very long and skinny high density extinction features are preferentially oriented parallel and in close proximity to the Galactic plane (likely due to how they form, see \S \ref{bone_discussion}).

\begin{sidewaysfigure}
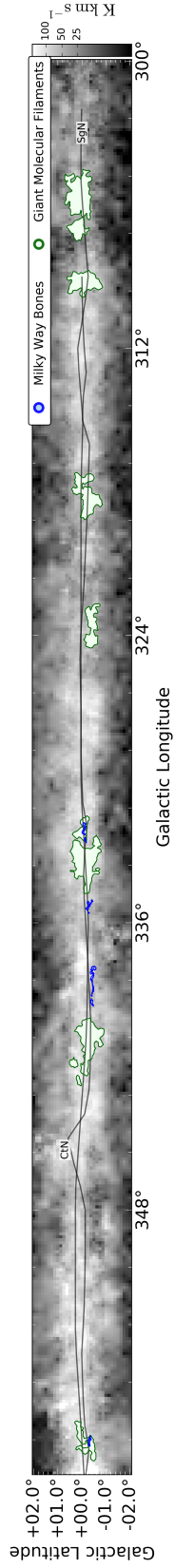

\begin{center}
\includegraphics[width=22cm]{{{lb_summary_fourth_quadrant}}}
\caption{{\label{fig:lb_fourthquad}  Plane-of-the-sky map of the fourth Galactic quadrant. Background grayscale shows $^{12} \rm CO$ emission from the \citet{Dame_2001} survey integrated between $\rm -100$ and $\rm 30 \; km \; s^{-1}$. The filament masks are overlaid to scale and color-coded according to filament catalog. The spiral arm models for two spiral arms---``CtN" (Scutum-Centaurus near) and ``SgN" (Sagittarius-near)--are overlaid as thin black lines. 
}}
\end{center}
\end{sidewaysfigure}

\subsection{Giant Molecular Filaments} \label{gmf_appendix}
\subsubsection{Original Selection} \label{orig_gmfs}
\citet{Ragan_2014} conduct a study in search of ``Giant Molecular Filaments" (GMFs)---velocity coherent, lower gas density $^{13} \rm CO$ filaments they identified using extinction \citep[UKIDSS Galactic Plane Survey]{Lucas_2008} or absorption \citep[GLIMPSE and MIPSGAL Surveys]{Churchwell_2009,Carey_2009} in a visual search of the first Galactic quadrant. \citet{Ragan_2014} search the GLIMPSE-Spitzer and UKIDSS-GPS surveys for filamentary extinction or absorption features that appear ``to extend $\approx1^\circ$ end to end and be identified by at least three group members." Gaps in extinction are allowed, as long as three authors agree that the structure continued further. Essentially, all co-authors carry out a by-eye inspection of the GLIMPSE and UKIDSS data, and if at least three agree on any individual identification, it is included in the catalog. Once an extinction or absorption feature is identified, \citet{Ragan_2014}'s procedure requires that the filament also appear velocity ``coherent," with a smooth velocity gradient, in the $^{13} \rm CO$ Galactic Ring Survey \citep{Jackson_2006}. Though dense gas is required, the GMFs are defined via the extent of their significant (above $ \rm 1 \; K\; km\; s^{-1}$) low-density tracing $^{13} \rm CO$ emission. Specifically, \citet{Ragan_2014} delineate the GMFs by creating a mask from the significant GRS $^{13}\rm CO$ emission ($ \rm >1 \; K\; km\; s^{-1}$) in the integrated intensity zeroeth moment maps, taken within the central velocity range of the filaments listed in their Table 2.

The \citet{Ragan_2014} study yields seven GMFs with lengths $\approx 100 \; \rm pc$ and masses between $10^4-10^5 \; \rm M_\sun$. Using the \citet{Vallee_2008} spiral arm models, six of the GMFs are declared to be interarm filaments, save GMF20.0-17.9 which is said to be a spur of the Scutum-Centaurus arm. \citet{Zucker_2015}, however, show that the truly velocity-contiguous portion of GMF20.0-17.9 is actually aligned with the Scutum-Centaurus arm, and likely traces it as well \citep[see detailed discussion in \S 3 of][]{Zucker_2015}.  

\citet{Abreu_Vicente_2016} repeat the same procedure as \citet{Ragan_2014} in the fourth quadrant, with lower resolution $^{13} \rm CO$ ThrUMMS data \citep{Barnes_2015} in lieu of GRS data, finding nine additional GMFs. Similar to \citet{Ragan_2014}, \citet{Abreu_Vicente_2016} approximate the GMFs via the extent of this $^{13} \rm CO$, except now they define regions of significant emission as those above $\rm >1.5 \; K\; km\; s^{-1}$ and they sometimes integrate over a different velocity range than the central velocity range of the filaments listed in their Table 2. 

Using the updated spiral arm models of \citet{Reid_2014}, \citet{Abreu_Vicente_2016} find six of the new GMFs to be associated with the Scutum-Centaurus arm while three are declared to be ``interarm" filaments. \citet{Abreu_Vicente_2016}  also reanalyze the filaments of \citet{Ragan_2014} using the updated \citet{Reid_2014} model and find three original GMFs (including GMF20.0-17.9) to now be associated with spiral structure (one with Scutum and two with Sagittarius), for a total of 9/16 GMFs showing spiral arm association. In both cases, a GMF is said to be ``associated" with a spiral arm if it intersects at any point a spiral \textit{longitude-velocity} fit within the arm's relative velocity error. 

\subsubsection{Sample Selected for Inclusion in this Study} \label{gmf_subsample}
In the present analysis, we include every Giant Molecular Filament (GMF) listed in Table 2 from \citet{Ragan_2014} (7 filaments) and \citet{Abreu_Vicente_2016} (9 filaments).

\subsubsection{Boundary Definition Employed in this Study} \label{gmf_boundaries}
To quantitatively measure the GMF properties, we standardize the definition of the GMFs by integrating over a consistent velocity range and applying a higher integrated intensity threshold to better isolate the GMFs from low-level background emission. Using the \href{https://spectral-cube.readthedocs.io/en/latest/}{spectral-cube} python package, we create $^{13}\rm CO$ GRS (1st quadrant) or ThrUMMS (4th quadrant) zeroeth moment maps integrated over the velocity range of the filaments listed in Table 2 of both \citet{Ragan_2014} and \citet{Abreu_Vicente_2016}. We then apply contours ranging between $ \rm 2-5 \; K\; km\; s^{-1}$. Adopting lower thresholds does not produce a closed contour and does not produce a filamentary structure, which is why \citet{Ragan_2014} overlay higher integrated intensity contours on top of their original masks to better highlight the filamentary morphology and the dense gas structure (see white contours in the \citet{Ragan_2014} Appendix). 

The two notable exception to the contour application procedure defined above are GMF324.5-321.4 from \citet{Abreu_Vicente_2016} and GMF20.0-17.9 from \citet{Ragan_2014}. According to Table 2 of \citet{Abreu_Vicente_2016}, GMF324.5-321.4 has no velocity gradient, so the velocity at both ends of the filament is equal to $\rm -32 \;km \;s^{-1}$. There is not enough significant emission at $\rm -32 \; km \; s^{-1}$ to define a closed contour, so we integrate between $\rm [-33,-31] \; km \; s^{-1}$ and apply a contour at $\rm 1.0\; km \; s^{-1}$. Additionally, GMF20.0-17.9 has been shown to exhibit two velocity breaks so we only integrate over the velocity contiguous region of emission between $\rm [44,49] \; km \; s^{-1}$. Finally, we apply a smoothness level between $\approx5-15$ to each contour, such that the contour is only evaluated at every 5-15 pixels (see \href{http://ds9.si.edu/doc/user/contour/index.html}{the SAO ds9 website} for more information on smoothing). The smoothing helps negate the effects of the low level $\approx 0.1 \; \rm K$ rms noise of the radio data, which makes the contours easier to skeletonize (a procedure used to determine the topology of the masks, and thus the filament lengths in \S \ref{lengths}). These $\rm ^{13} \rm CO$ contours become the new boundaries of the GMFs, and they are shown as green contours overlaid on the Herschel column density maps in \S \ref{gmf_gallery}. There is a minor background subtraction issue in the column density map for GMF309, which is due to a calibration error in the raw Hi-GAL flux images downloaded from the \href{https://tools.asdc.asi.it/HiGAL.jsp}{Hi-GAL server} for the $160 \micron$ band (i.e. neighboring tiles coincident with this filament exhibit a large discontinuous jump in mean flux in the $160\micron$ band). We expect this to have a trivial effect on our results. 

\subsubsection{Giant Molecular Filament Gallery} \label{gmf_gallery}


\begin{figure}[!htb]
\begin{center}
\includegraphics[width=0.6\linewidth]{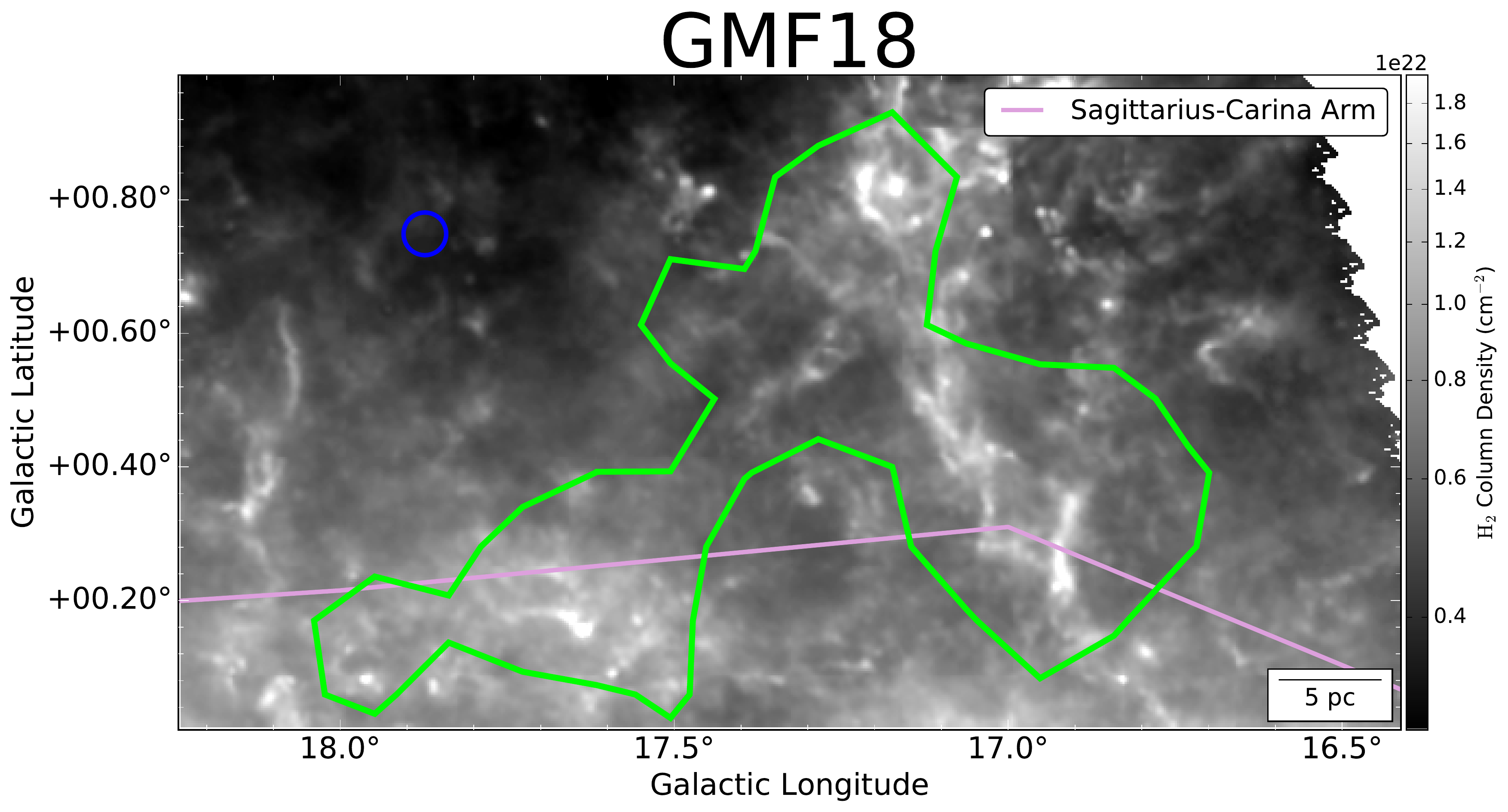}  \\  
\includegraphics[width=0.55\linewidth]{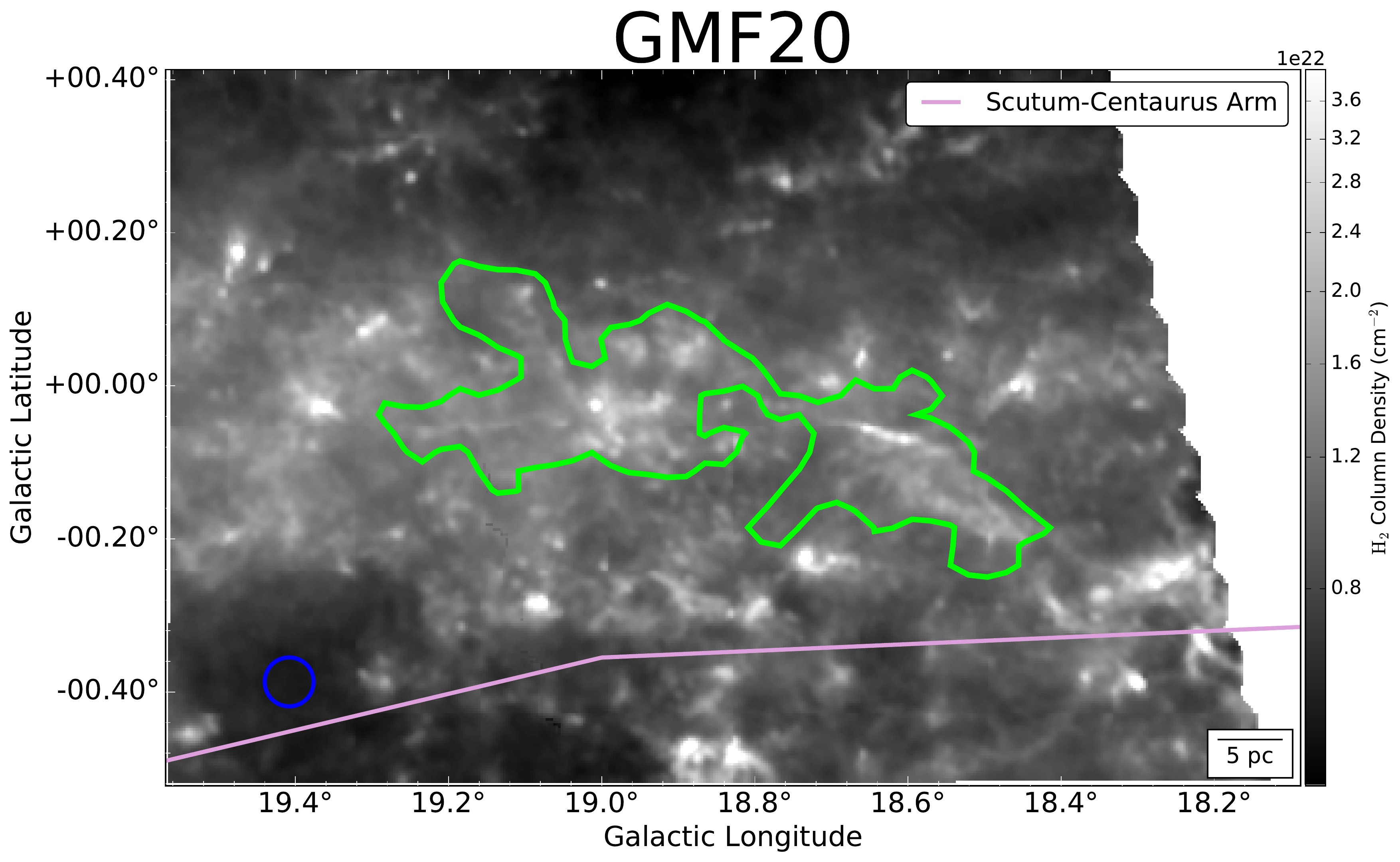} \\  
\includegraphics[width=0.7\linewidth]{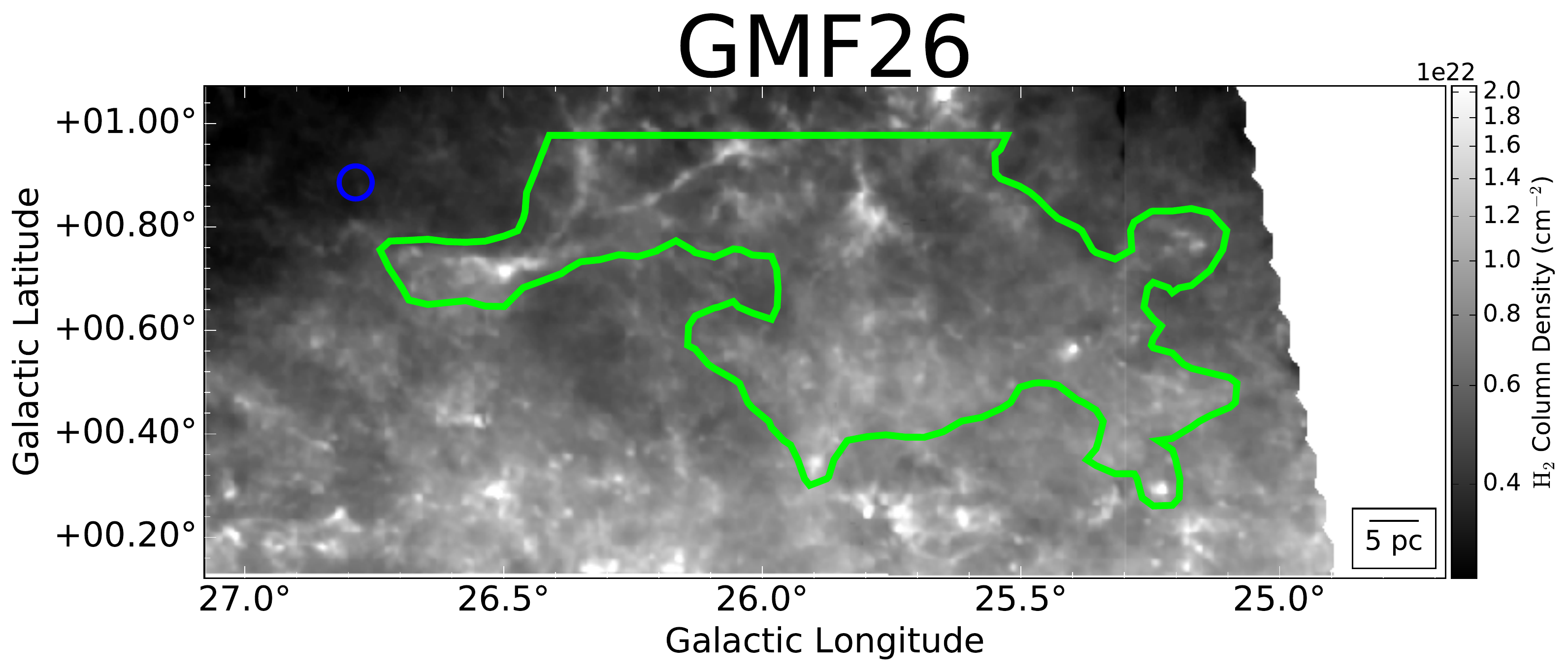}  \\ 
 \end{center}
\end{figure}

\begin{figure}[!htb]
\begin{center}
\includegraphics[width=0.7\linewidth]{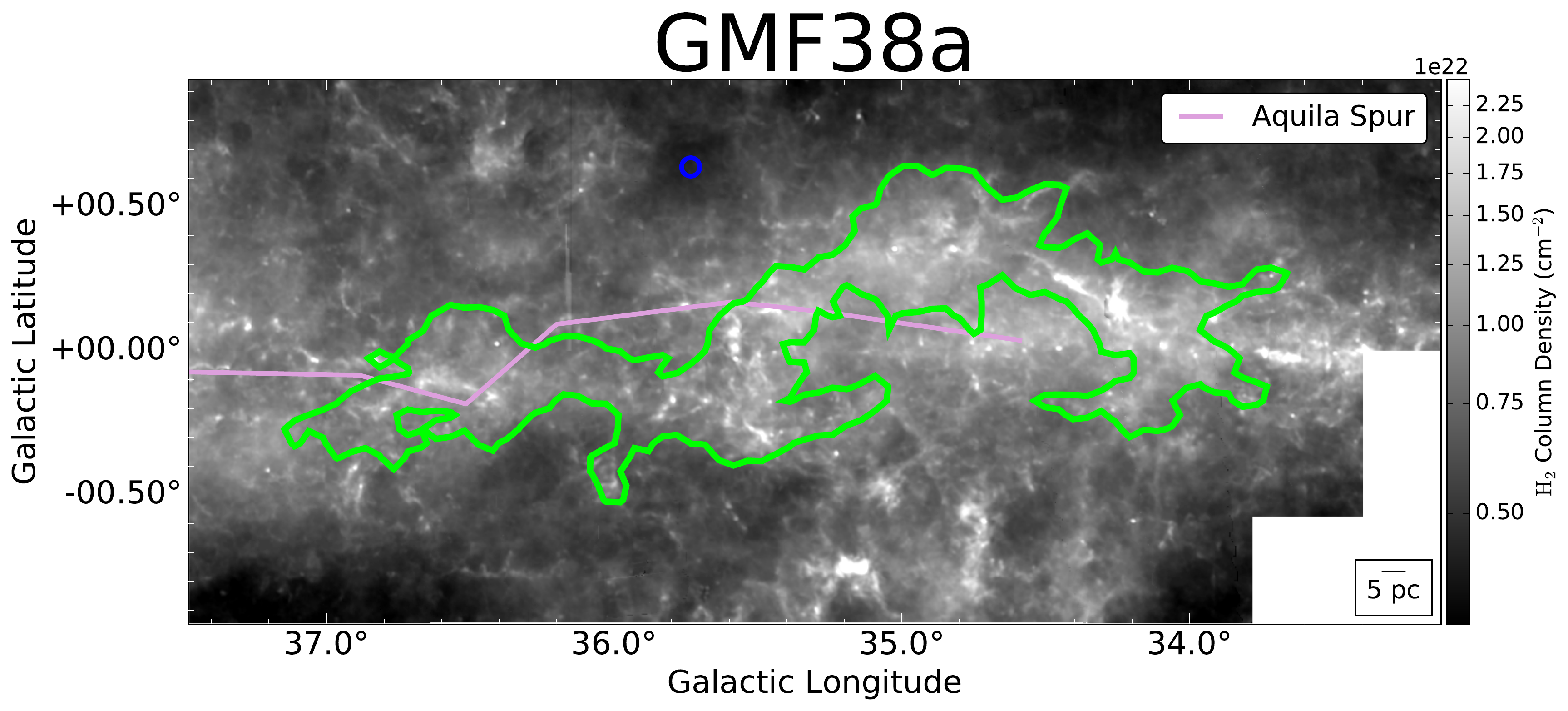} \\  
\includegraphics[width=0.5\linewidth]{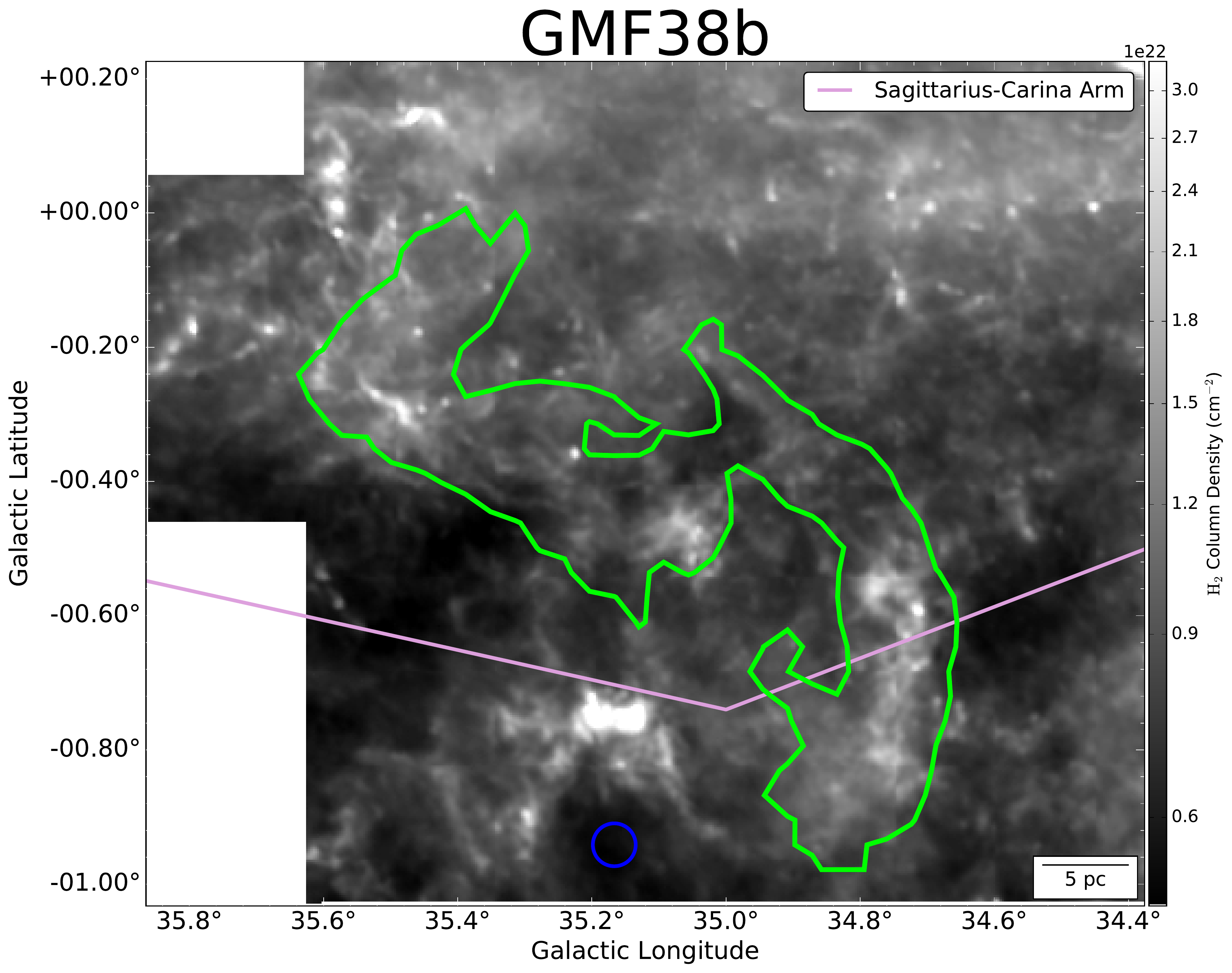} \\  
\includegraphics[width=0.5\linewidth]{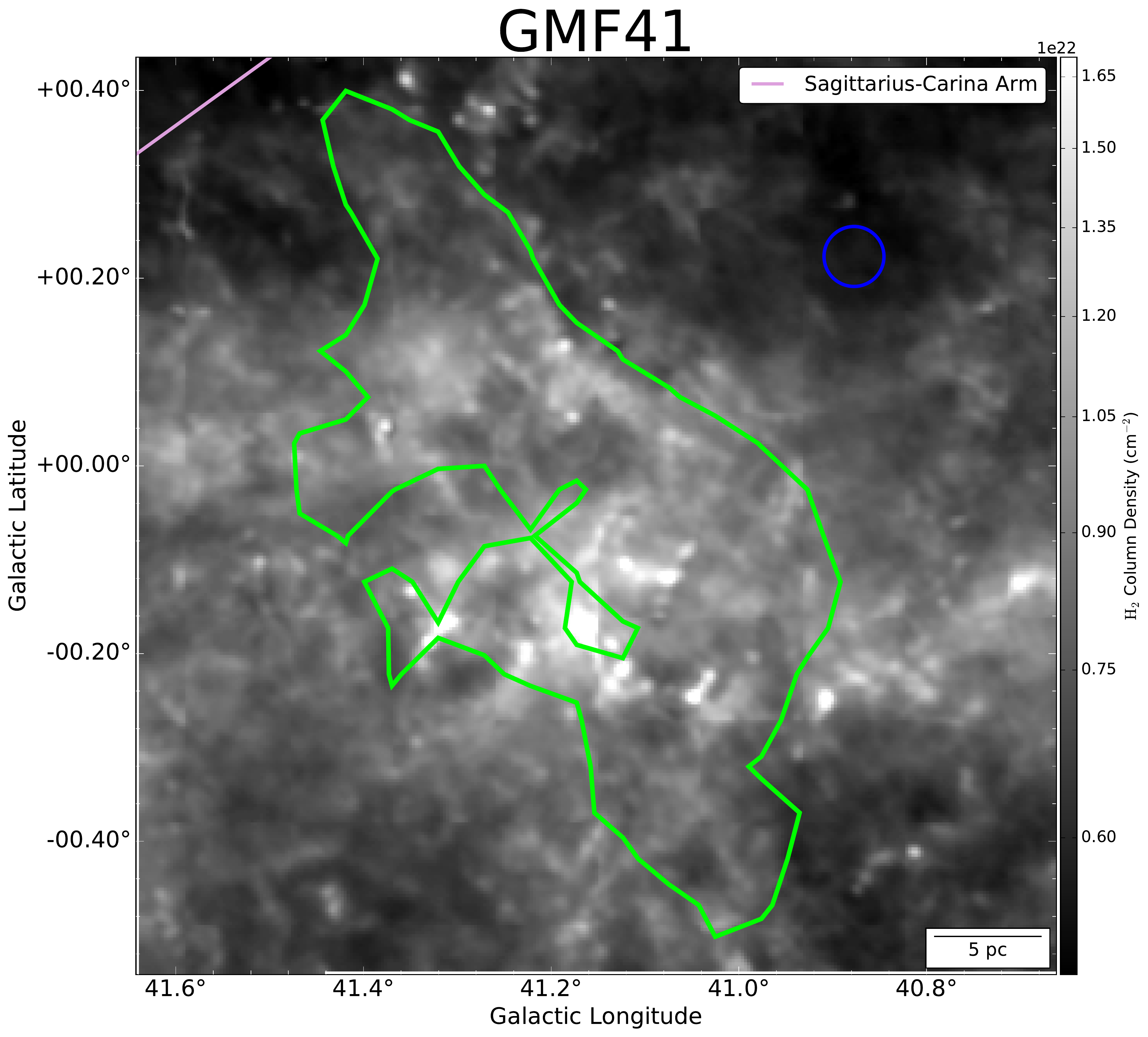} \\ 
 \end{center}
\end{figure}

\newpage

\begin{figure}[!htb]
\begin{center}
\includegraphics[width=0.55\linewidth]{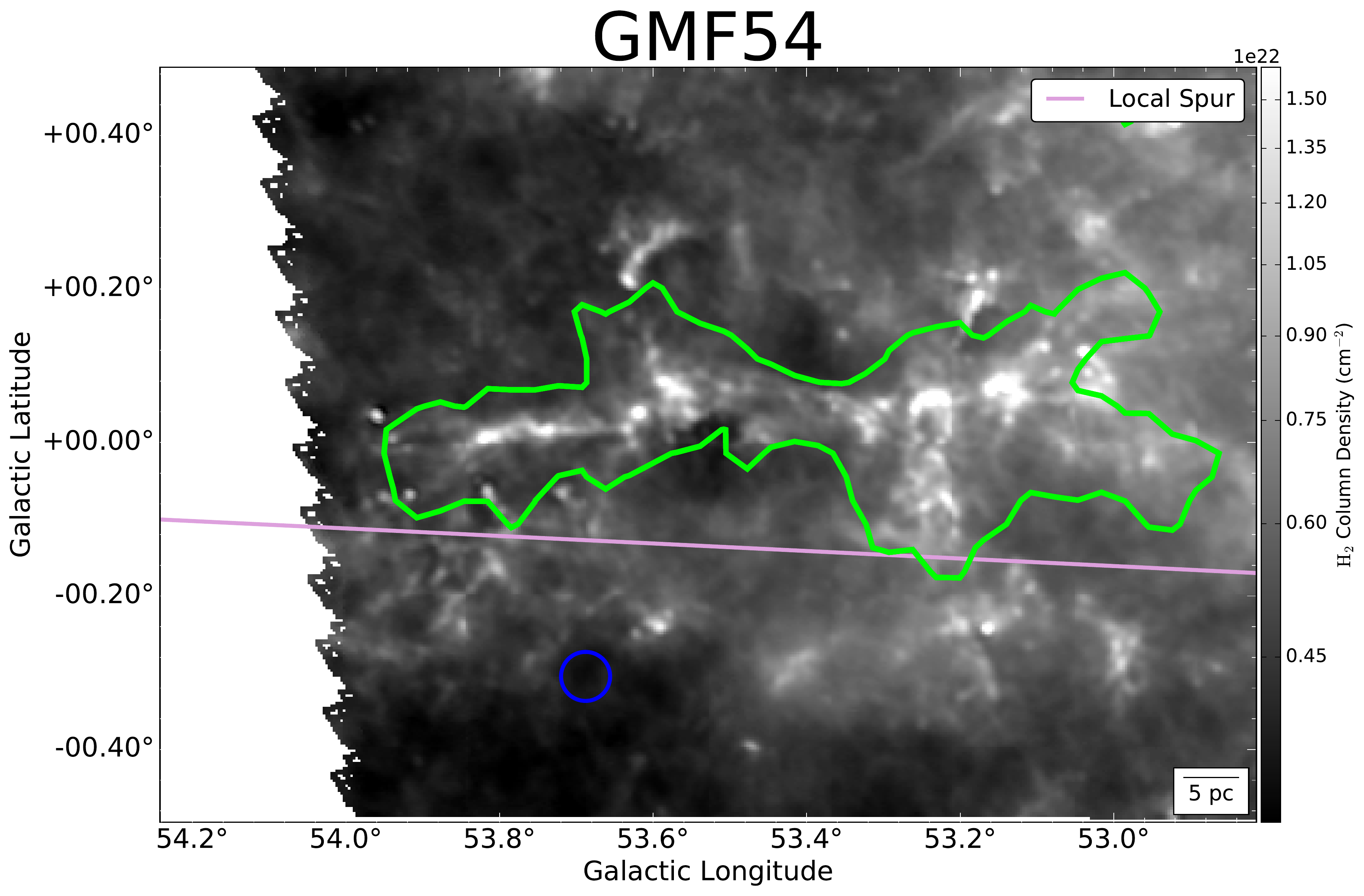} \\
\includegraphics[width=0.55\linewidth]{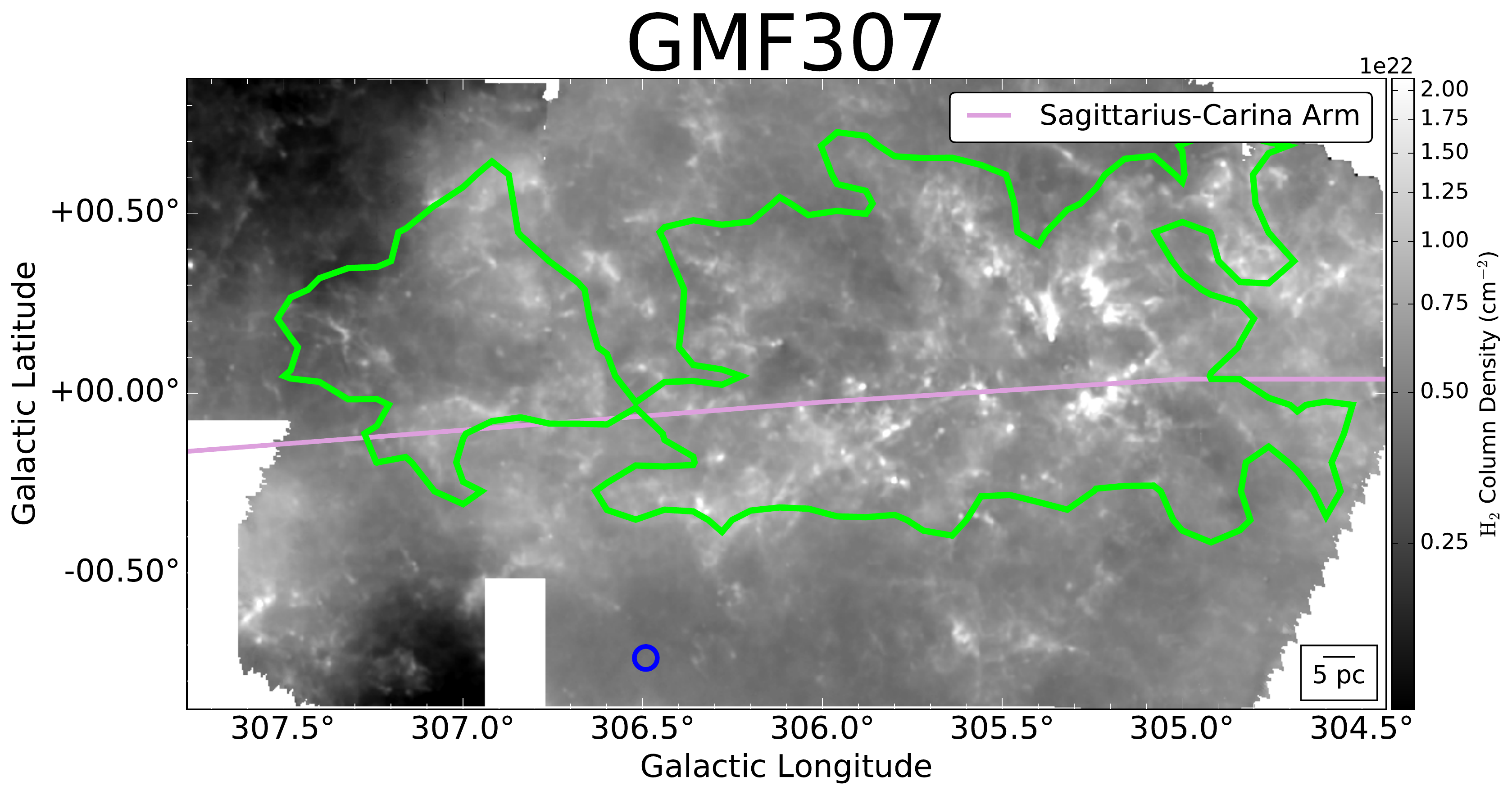} \\
\includegraphics[width=0.4\linewidth]{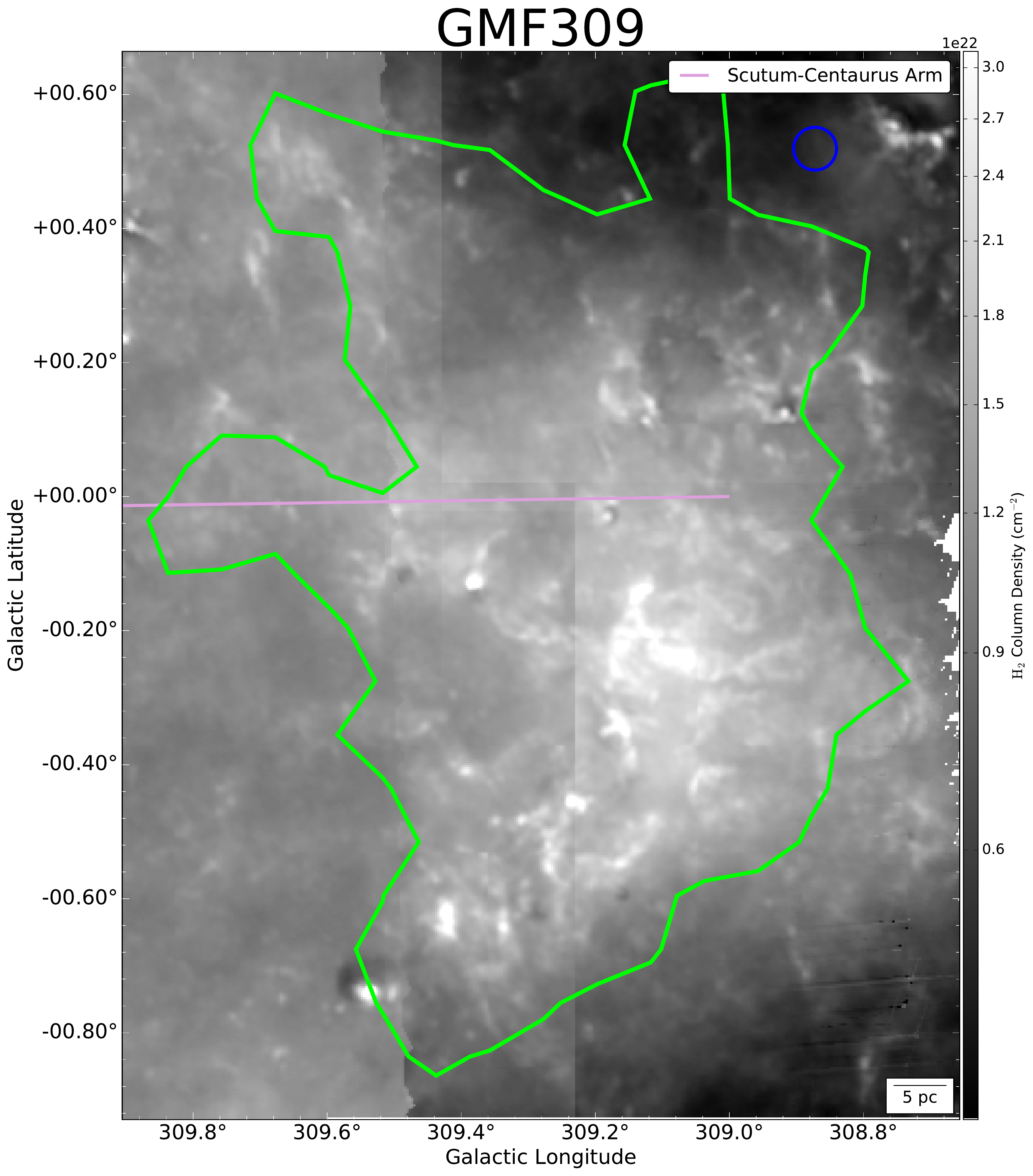} \\
 \end{center}
\end{figure}

\newpage

\begin{figure}[!htb]
\begin{center}
\includegraphics[width=0.65\linewidth]{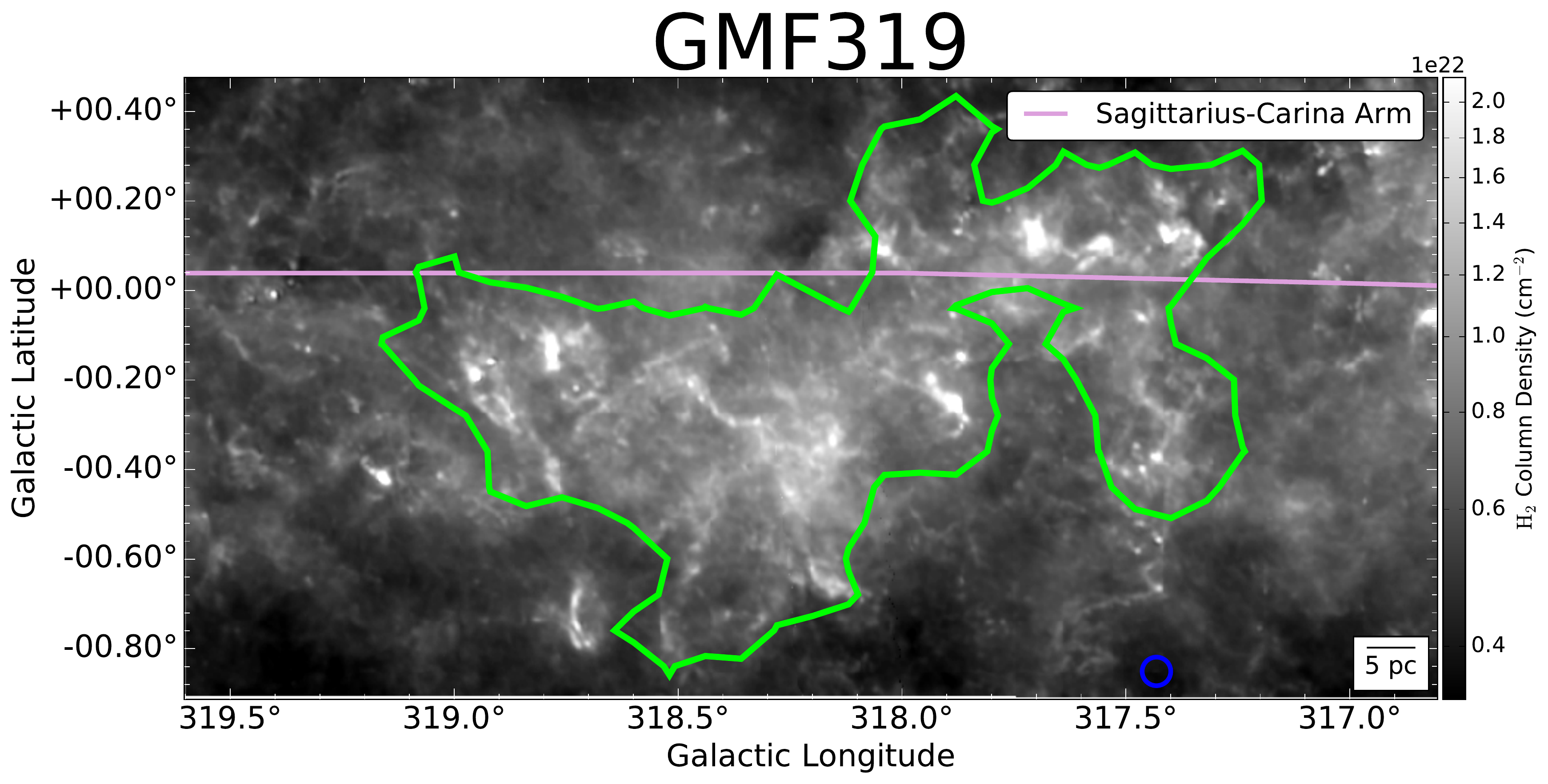} \\
\includegraphics[width=0.75\linewidth]{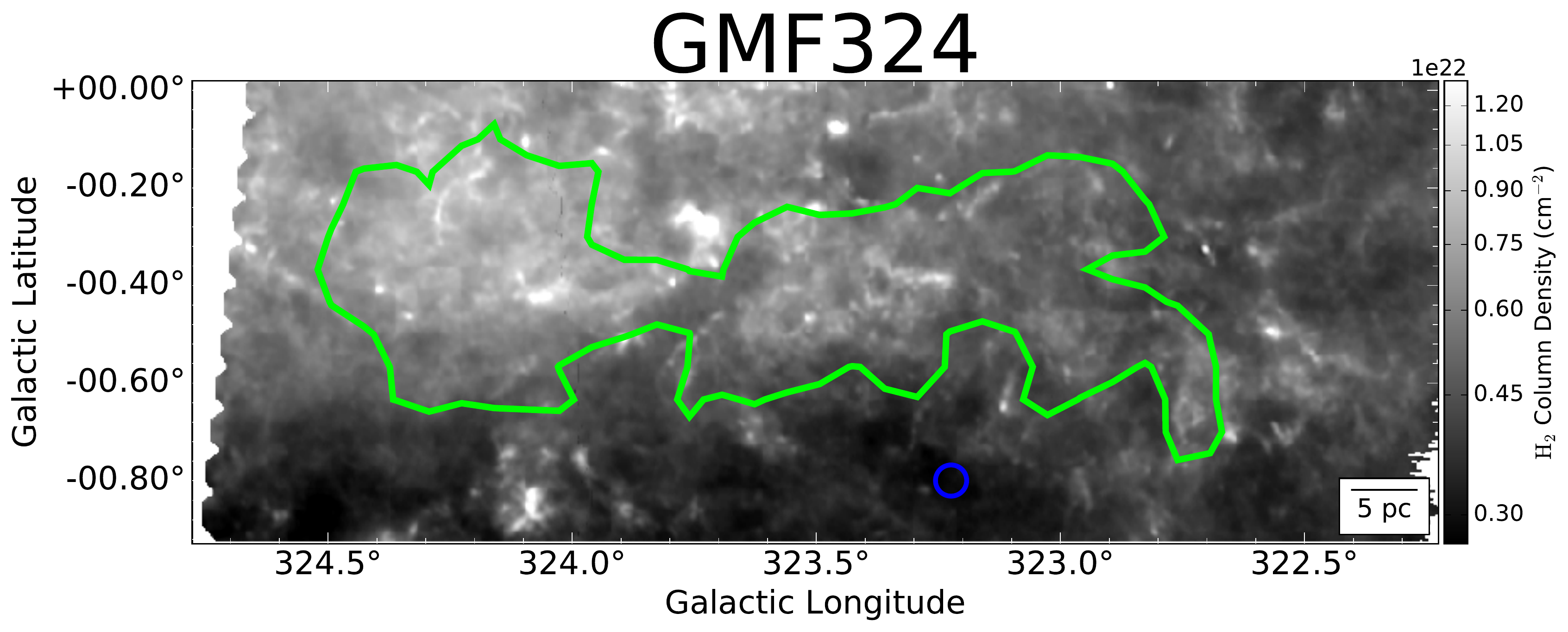} \\
\includegraphics[width=0.75\linewidth]{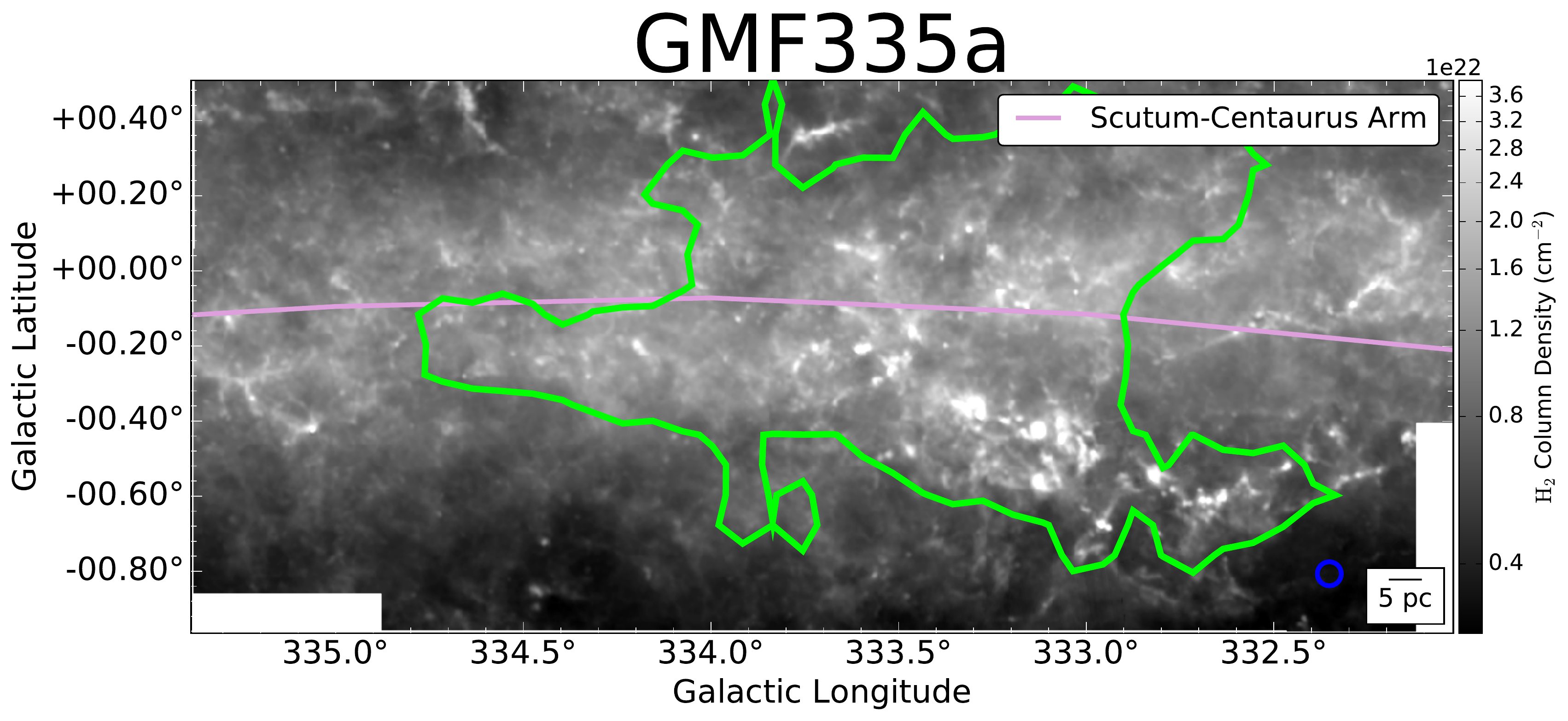} \\

 \end{center}
\end{figure}

\newpage

\begin{figure}[!htb]
\begin{center}
\includegraphics[width=0.6\linewidth]{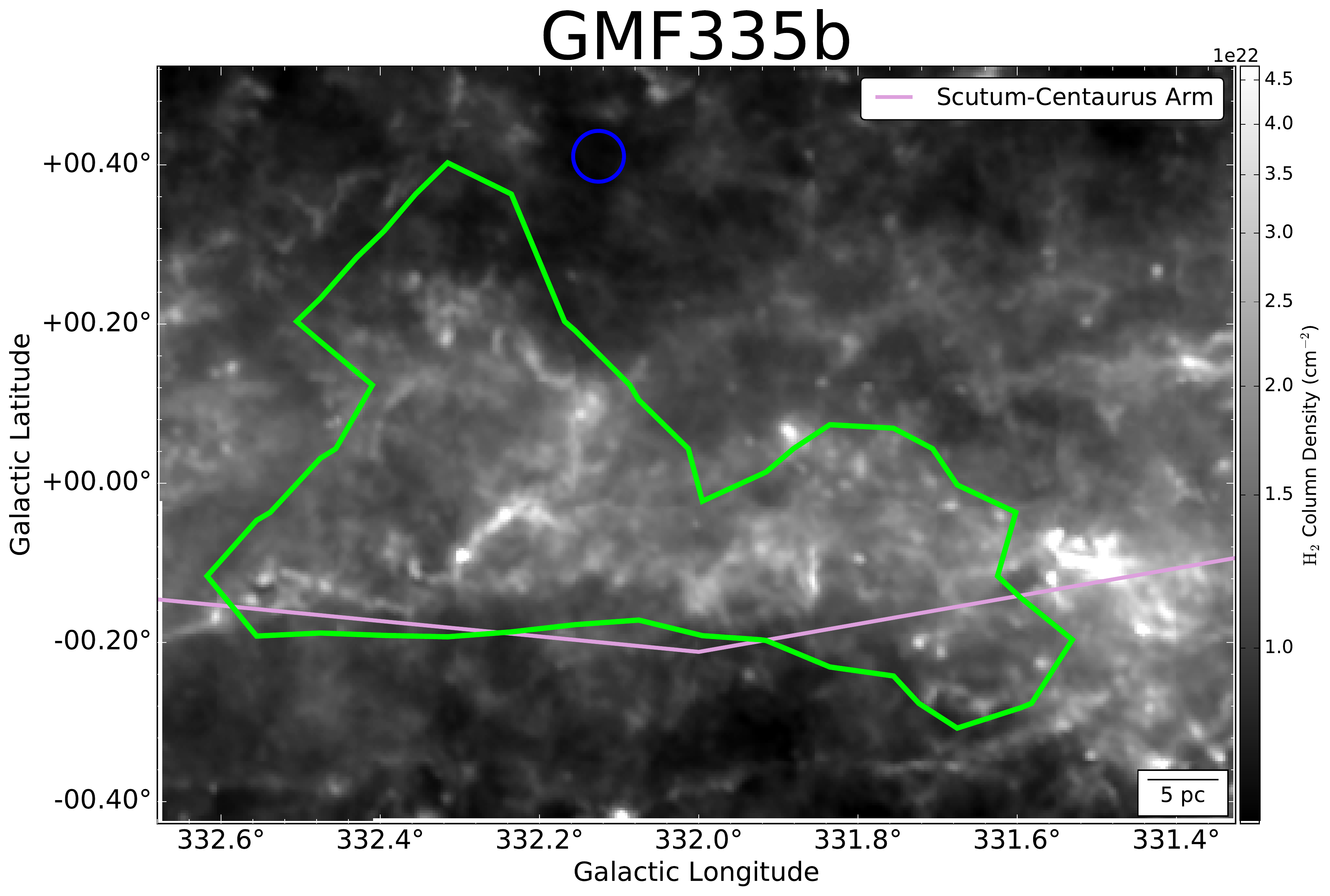} \\
\includegraphics[width=0.7\linewidth]{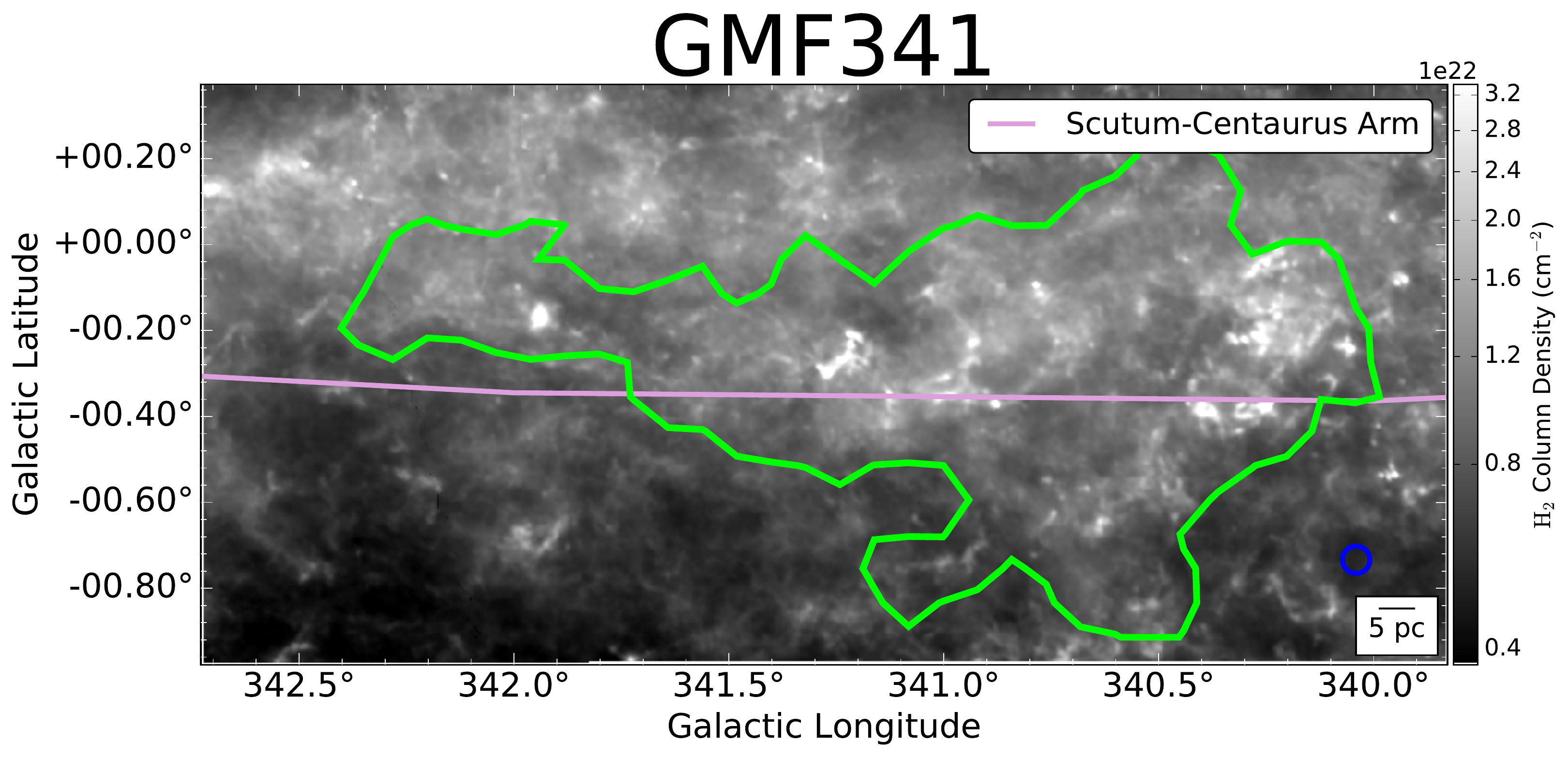} \\
\includegraphics[width=0.75\linewidth]{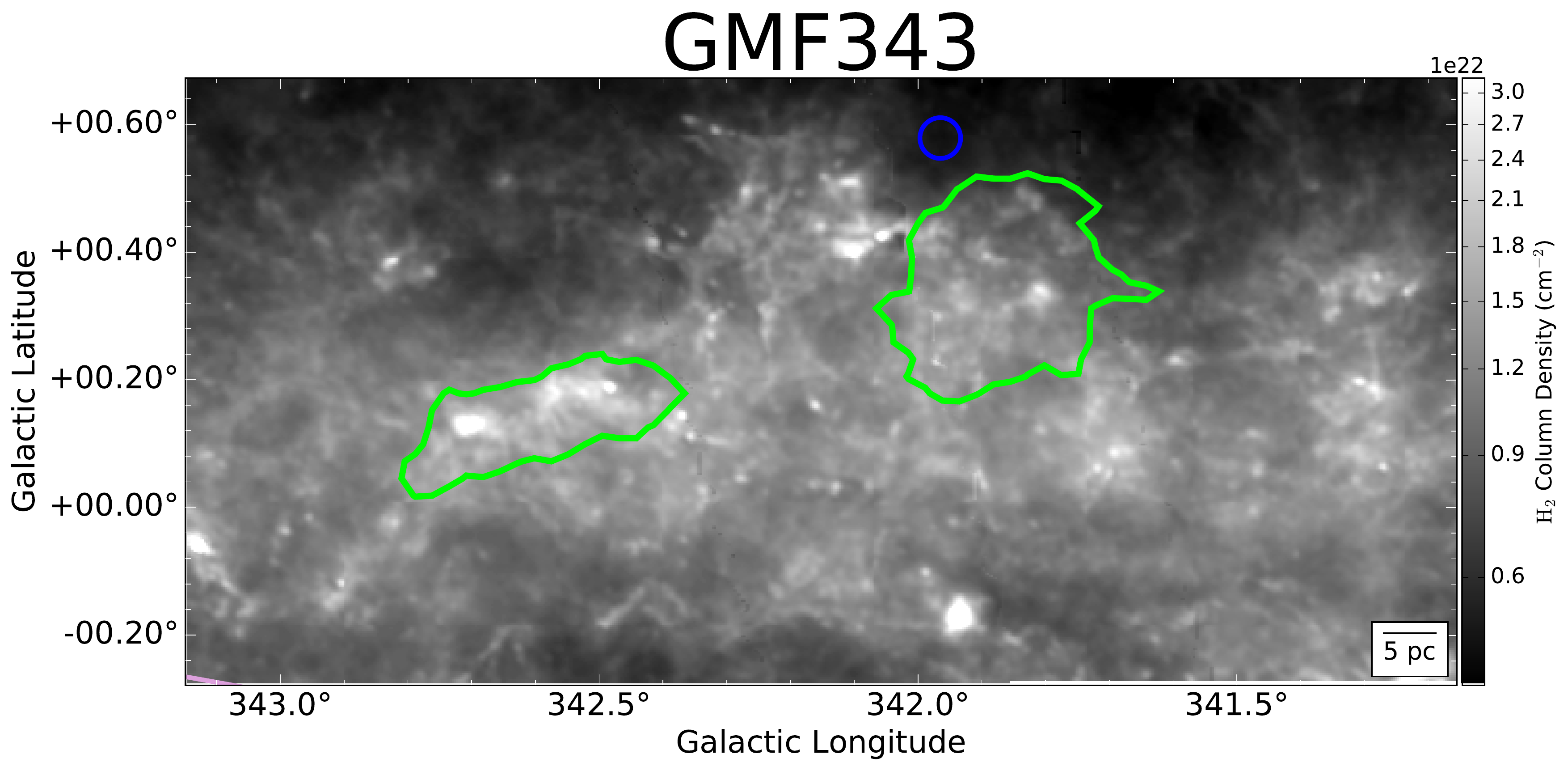} \\
 \end{center}
\end{figure}

\newpage

\begin{figure}[!htb]
\begin{center}
\includegraphics[width=0.75\linewidth]{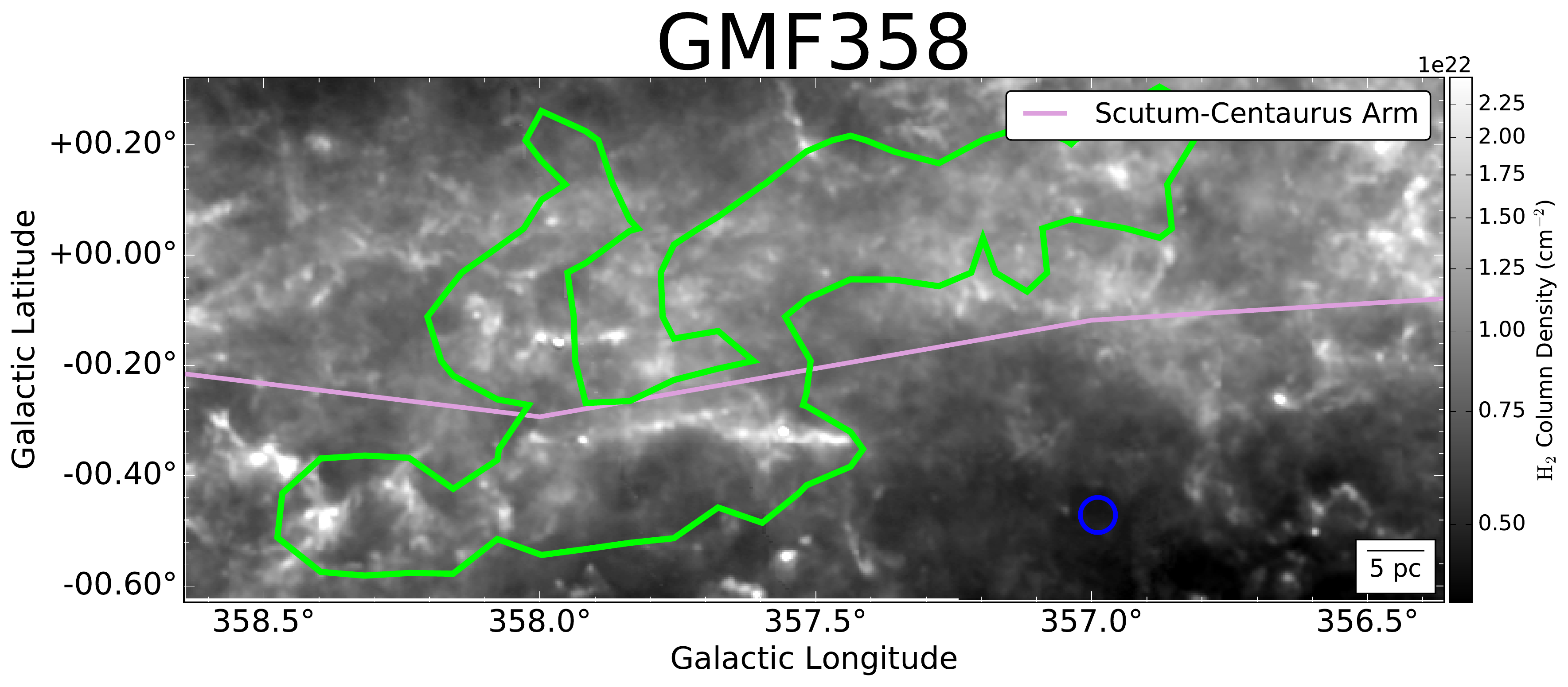} \\
 \end{center}
\end{figure}

\clearpage 
\subsection{Milky Way Bones} \label{bone_appendix}
\subsubsection{Original Selection} \label{orig_bones}
\citet{Zucker_2015} carry out a targeted follow-up search specifically for Nessie analogs---the ``Bones" of the Milky Way. Within the first and fourth Galactic quadrants, \citet{Zucker_2015} identify candidates for follow-up analysis by visually inspecting combined images from the GLIMPSE and MIPSGAL surveys \citep{Churchwell_2009,Carey_2009} for long, skinny filamentary infrared dark clouds parallel and in close proximity to the plane of the sky projections of known spiral arms. \citet{Zucker_2015} assess velocity contiguity in two ways: first, using existing dense molecular clump catalogs \citep[i.e. the BGPS, MALT90, and HOPS surveys;][]{Shirley_2013,Jackson_2013,Purcell_2012} and second, by taking a customized low-density tracing $\rm^{13}  CO$ \citep[i.e. the GRS survey, the ThrUMMS Survey;][]{Jackson_2006, Barnes_2015} \textit{position-velocity} slice tracing the dense spine of the mid-IR extinction feature. These velocities are compared to existing log-spiral \textit{longitude-velocity} fits taken from the literature \citep{Dame_2011,Sanna_2014,Shane_1972,Vallee_2008}. \citet{Zucker_2015} then develop a quantitative set of criteria intended to differentiate filaments likely to be associated with major spiral features. These include prescriptions for density (continuous mid-IR extinction feature), aspect ratio ($>50:1$), velocity contiguity ($< 3 \; \rm km \; s^{-1}$ per 10 pc), spiral arm association (within 10 $\rm km \; s^{-1}$ of any log-spiral fit to major arms), position angle ($<30^\circ$ from physical Galactic midplane), and Galactic scale height ($<20$ pc from physical Galactic midplane). Of the ten candidates identified visually, six of them meet all criteria and are classified as Galactic ``Bones," with the other four failing the aspect ratio criterion. Several filaments possess dense gas sources whose velocity gradients match the \citet{Dame_2011} fit to the Scutum-Centaurus arm, with the most prominent being the 50 pc long Filament 5 (``BC 18.88-0.09"), which is coincident with velocity contiguous part of the \citet{Ragan_2014} GMF20.0-17.9.

\subsubsection{Sample Selected for Inclusion in this Study} \label{bone_subsample}
We include all ten filaments from Table 2 in \citet{Zucker_2015}, plus Nessie \citep{Goodman_2014} in the present study (the ``Milky Way Bone" catalog).  

\subsubsection{Boundary Definition Employed in this Study} \label{bone_boundaries}
While the \citet{Zucker_2015} Bones are mainly delineated via their mid-IR extinction features, for the purposes of this study, we define them the same way we do the \citet{Wang_2015} Large-Scale Herschel filaments. We create new Herschel column density and dust temperature maps tailored to this study (see \S \ref{densities_temps} for more details) and apply closed contours at a level $\approx1-2\sigma$ above the mean background column density in each image. Specifically, we take the mean column density in a circular reference area in a low emission region near the source which we approximate as the background column density (this is also used to perform the actual background subtraction on the $160-500 \; \micron$ fluxes, see \S \ref{densities_temps}). These reference regions are shown as unfilled blue circles in \S \ref{bone_gallery}. We then take the standard deviation ($\sigma$) of the column density values for all the pixels in the image and adopt a contour level between one to two standard deviations ($1-2\sigma$) above the mean background column density. Note that these thresholds are applied to the column density maps without background subtraction, but we derive all physical properties from the background subtracted ones. We avoid defining filament boundaries using the dust temperature maps as they are more affected by star formation activity. In cases where there is a foreground or background structure at close projection but shown to be kinematically unassociated, we mask out these regions by hand in the data visualization software package \texttt{glue}\footnote{http://glueviz.org} before applying the contours. These closed contours become the boundaries of the \citet{Zucker_2015} Bones employed throughout this work. These contours are shown in green and overlaid on the Herschel Column Density maps in \S \ref{bone_gallery}.

\subsubsection{Milky Way Bone Filament Gallery} \label{bone_gallery}

\begin{figure}[!htb]
\begin{center}
\includegraphics[width=0.47\linewidth]{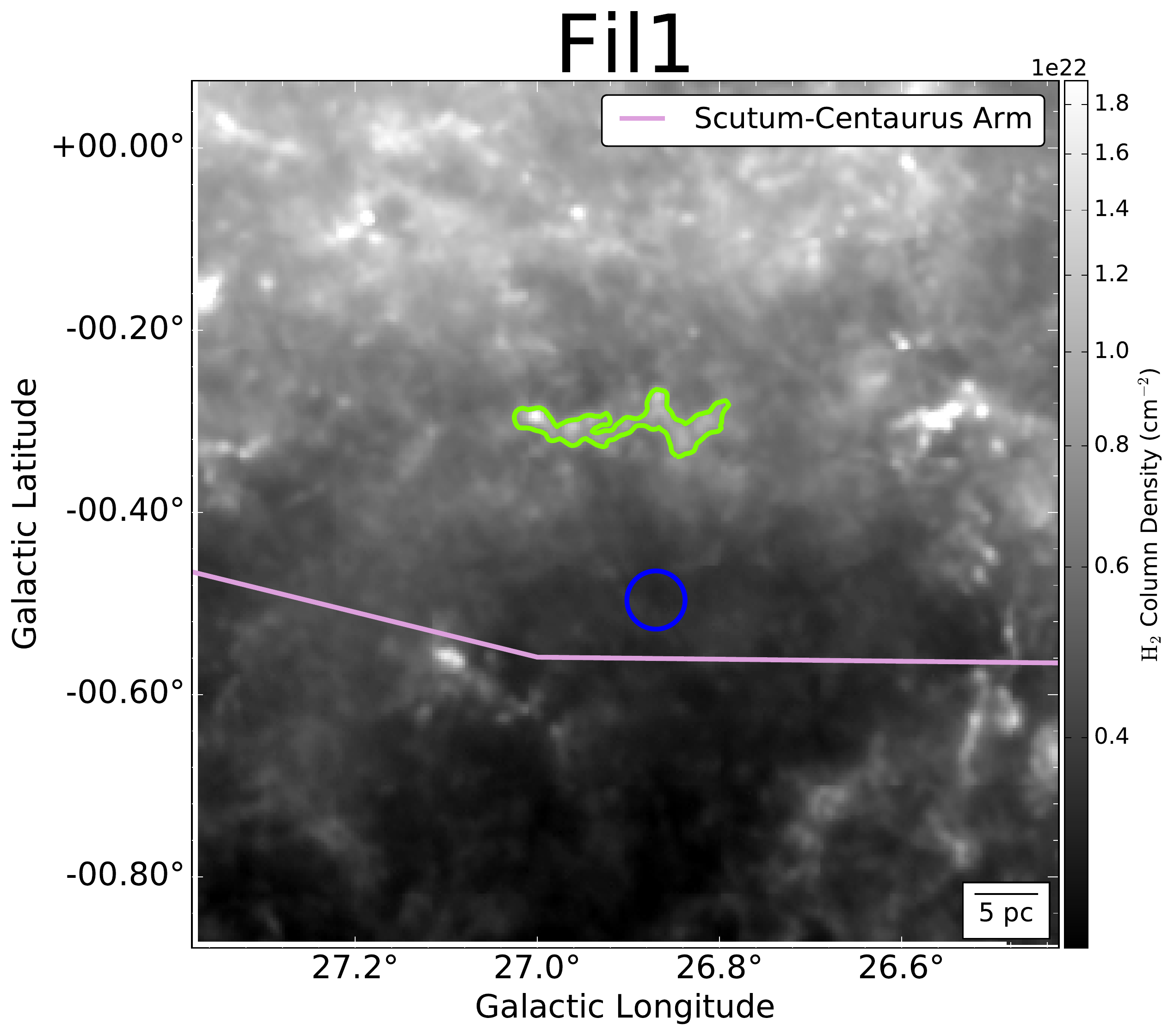} \hfill
\includegraphics[width=0.47\linewidth]{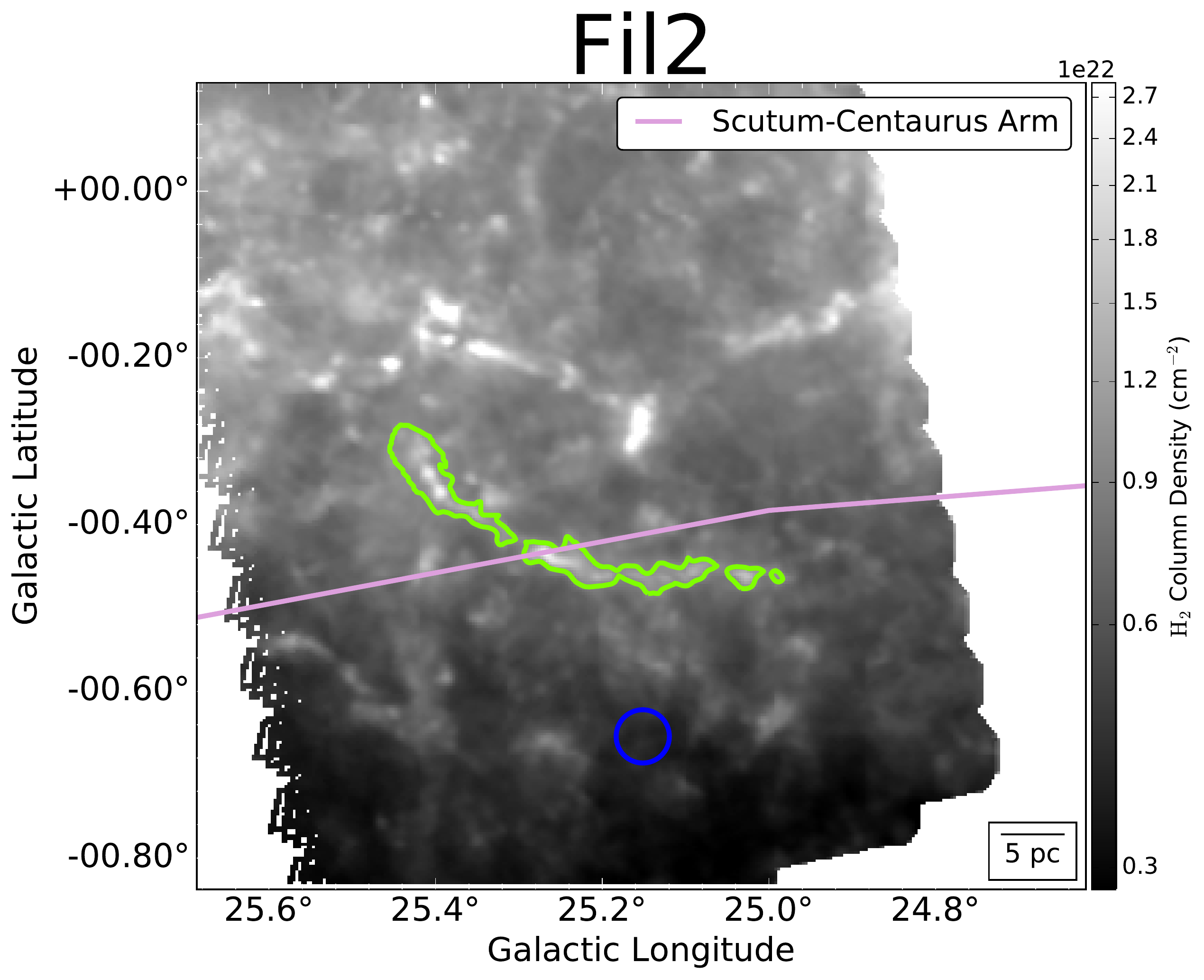} \\
\includegraphics[width=0.47\linewidth]{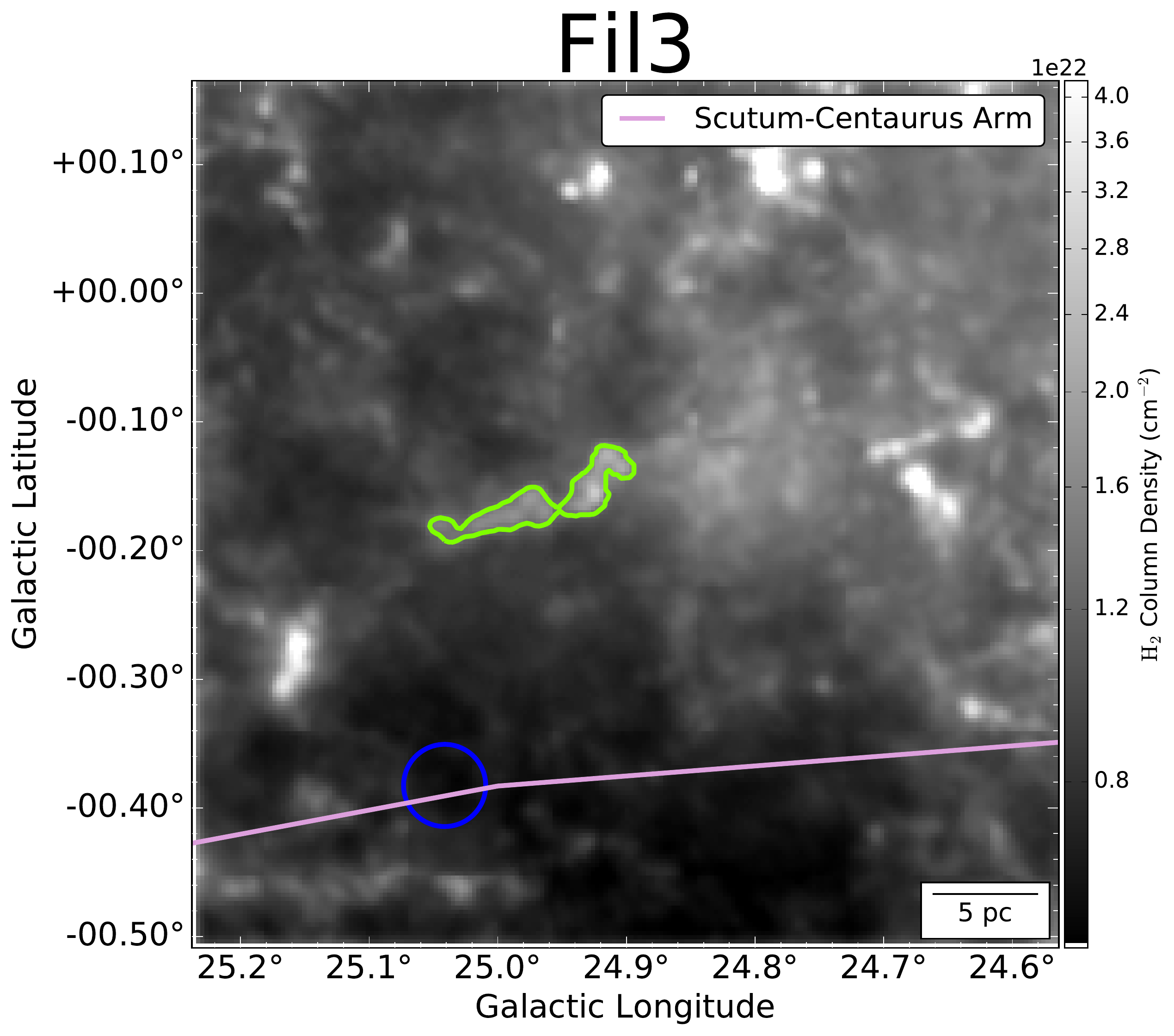} \hfill
\includegraphics[width=0.47\linewidth]{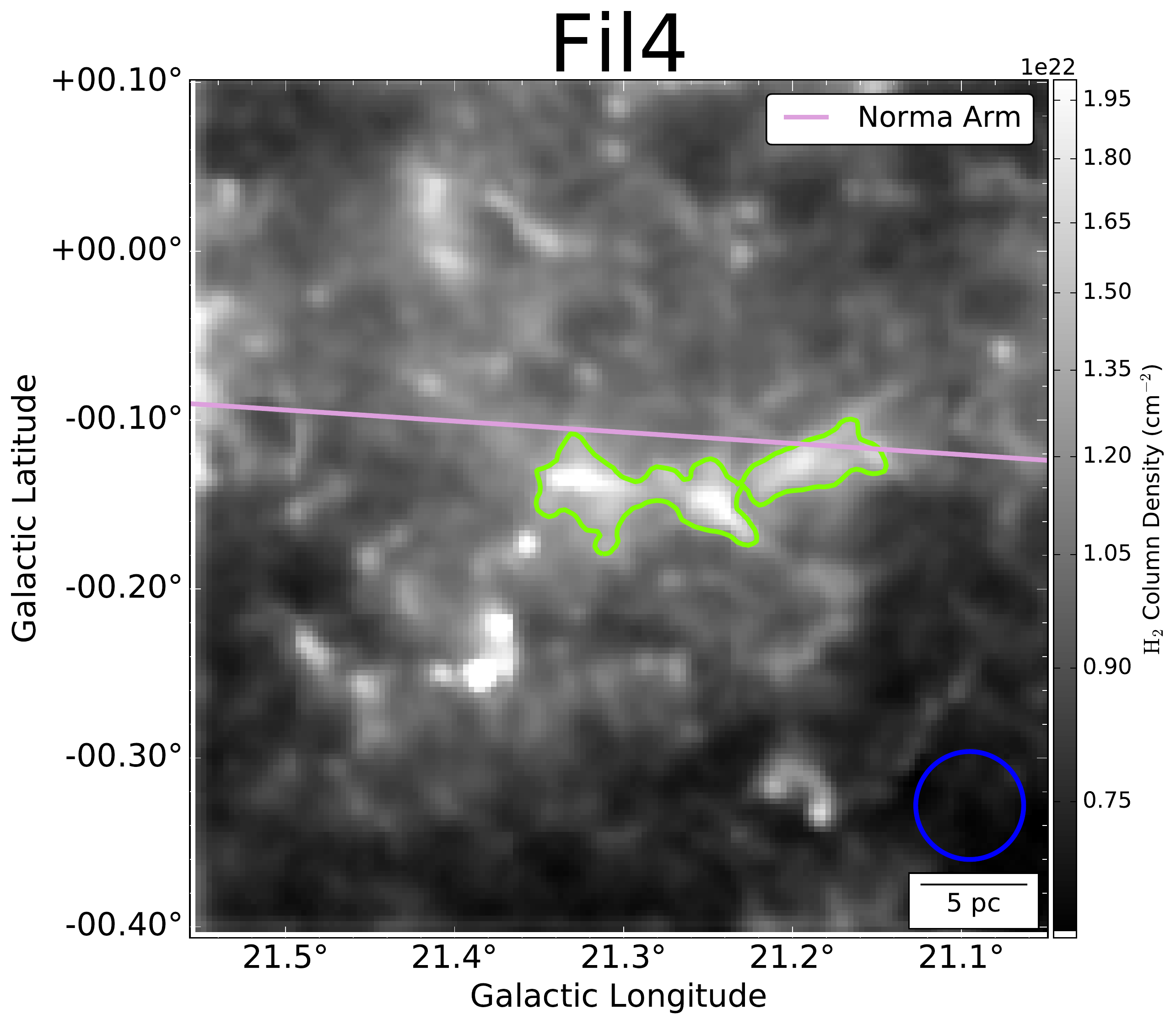} \\
\end{center}
\end{figure}

\newpage

\begin{figure}[!htb]
\begin{center}
\includegraphics[width=0.5\linewidth]{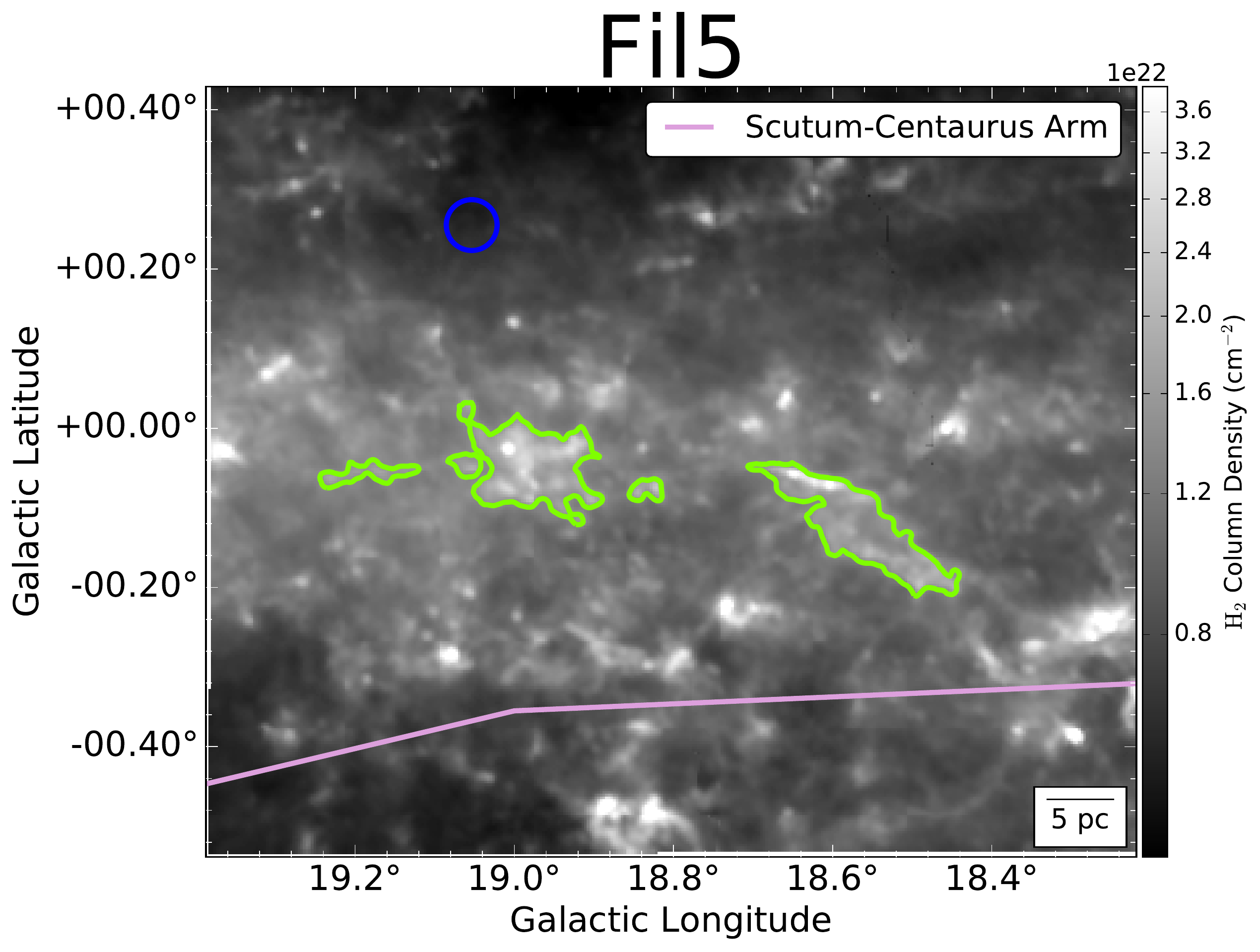} \hfill
\includegraphics[width=0.47\linewidth]{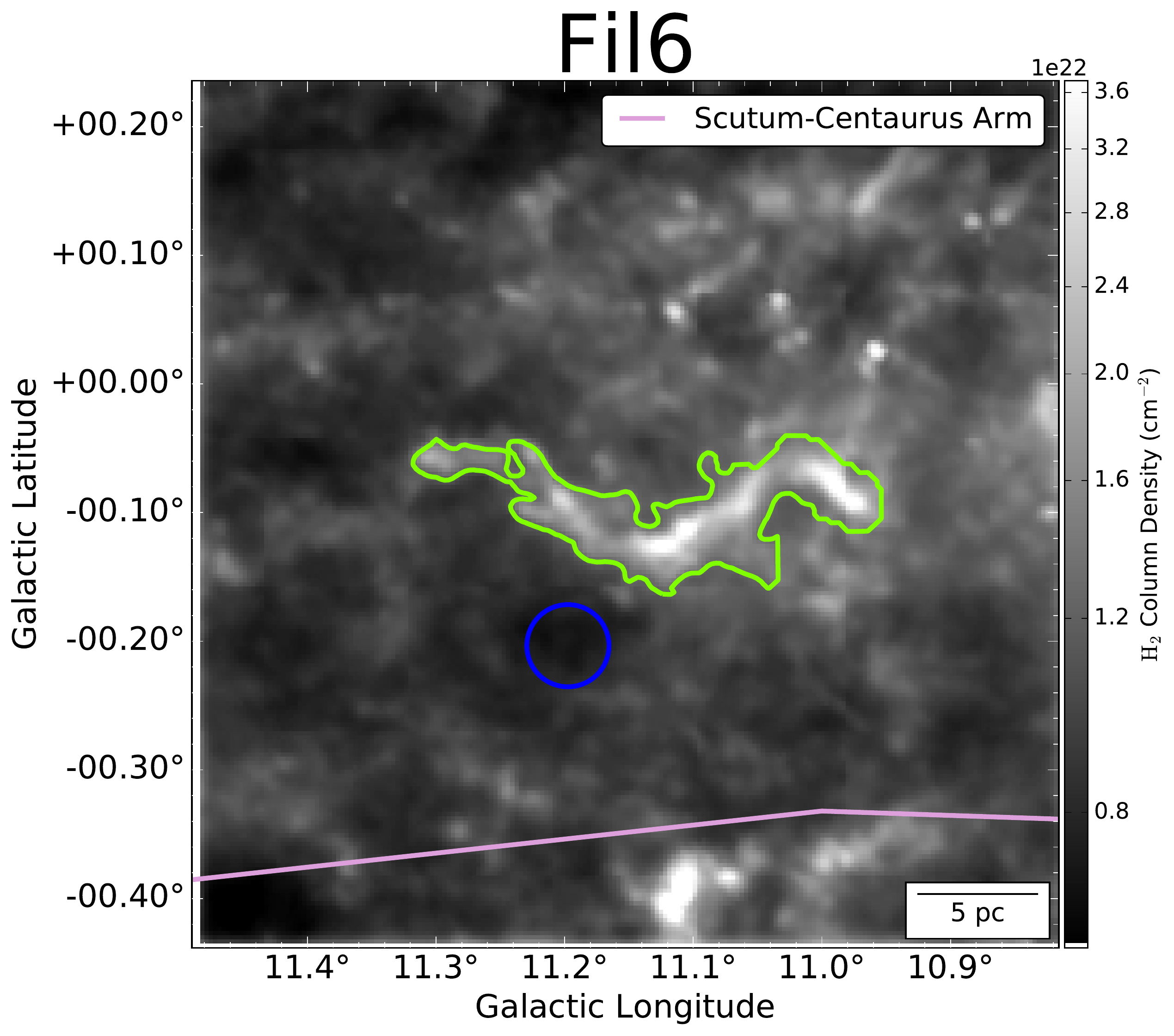} \\
\includegraphics[width=0.47\linewidth]{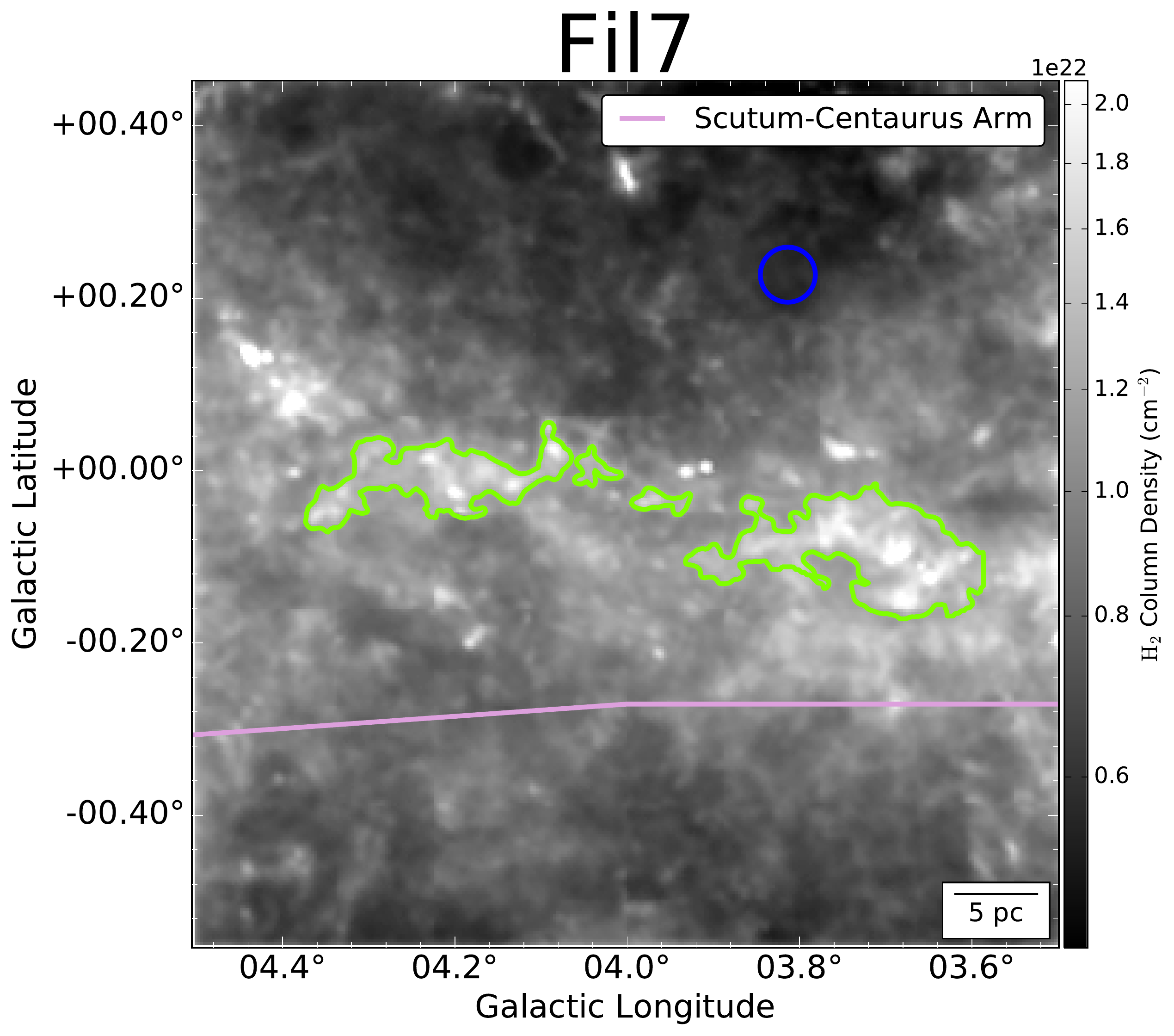} \hfill
\includegraphics[width=0.47\linewidth]{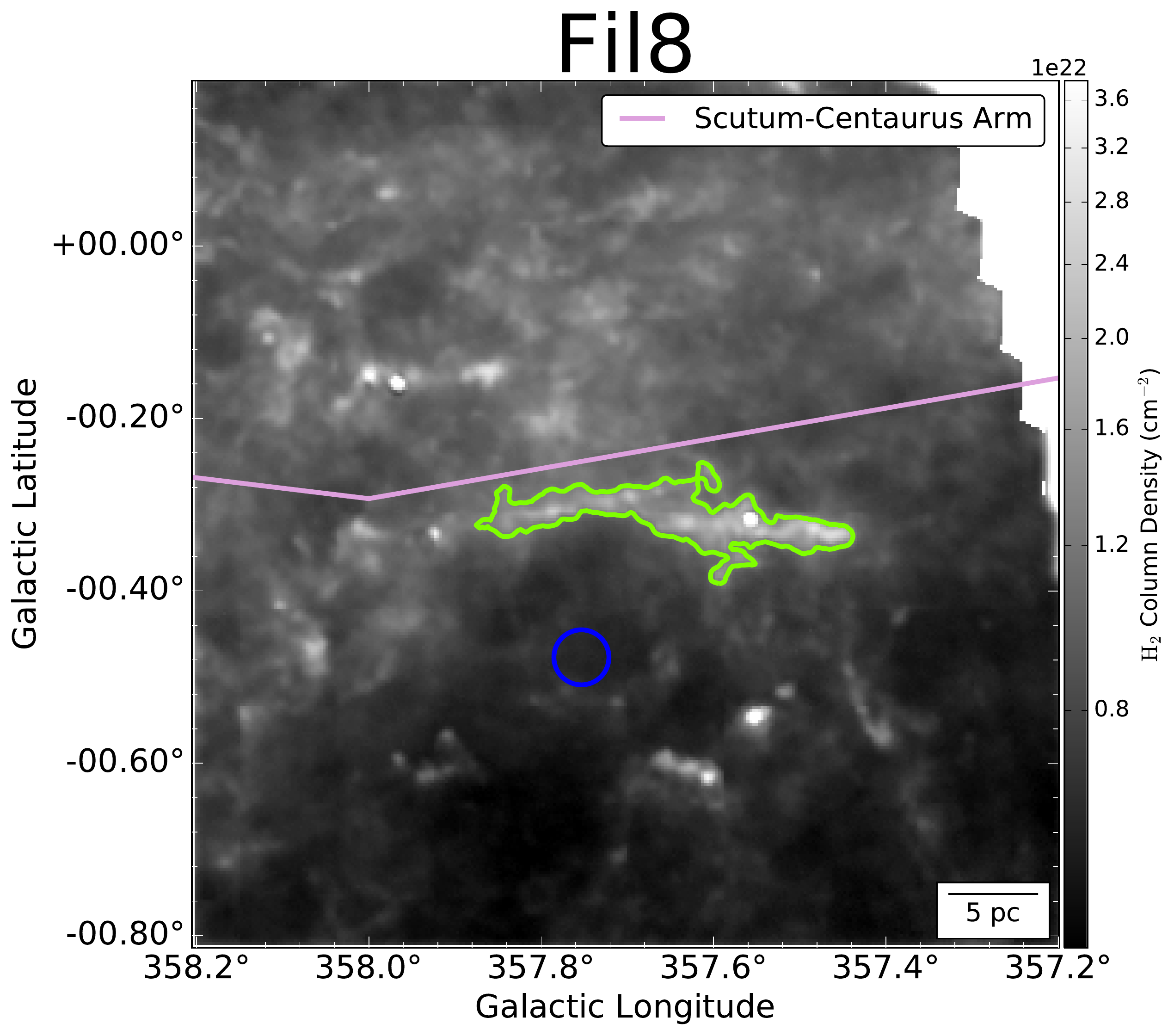} \\
\end{center}
\end{figure}

\newpage

\begin{figure}[!htb]
\begin{center}
\includegraphics[width=0.47\linewidth]{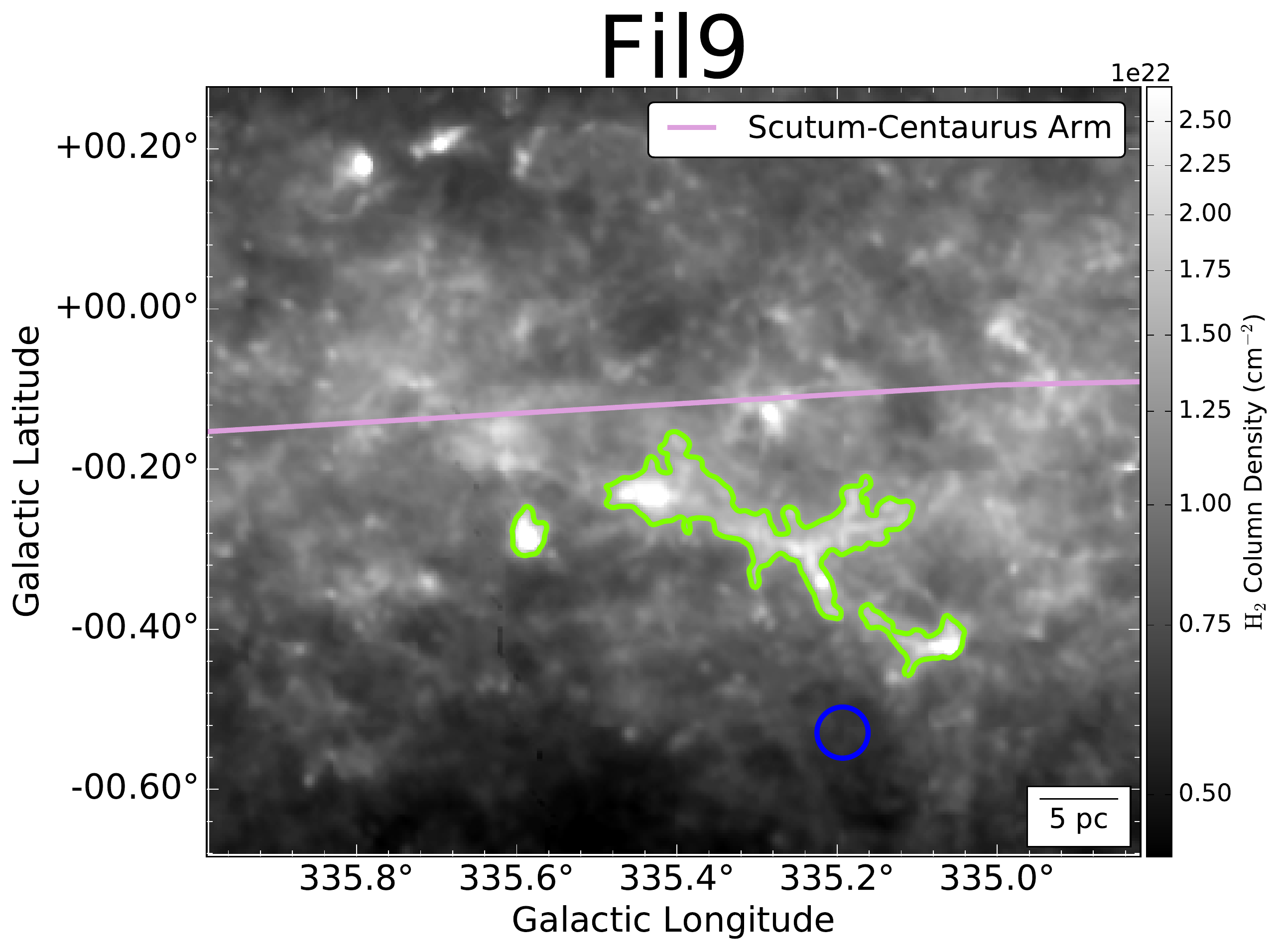} \hfill
\includegraphics[width=0.47\linewidth]{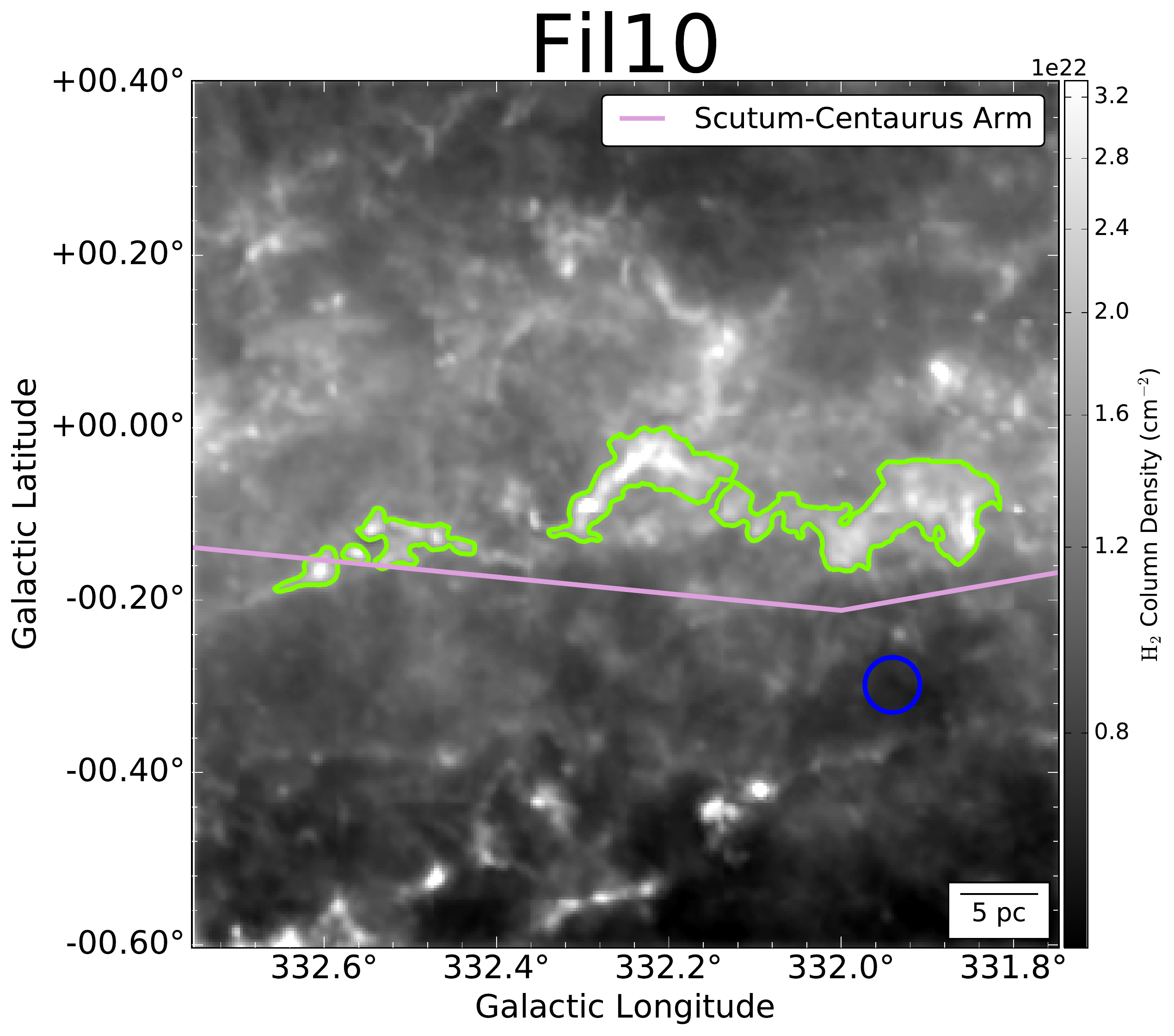} \\
\includegraphics[width=1.0\linewidth]{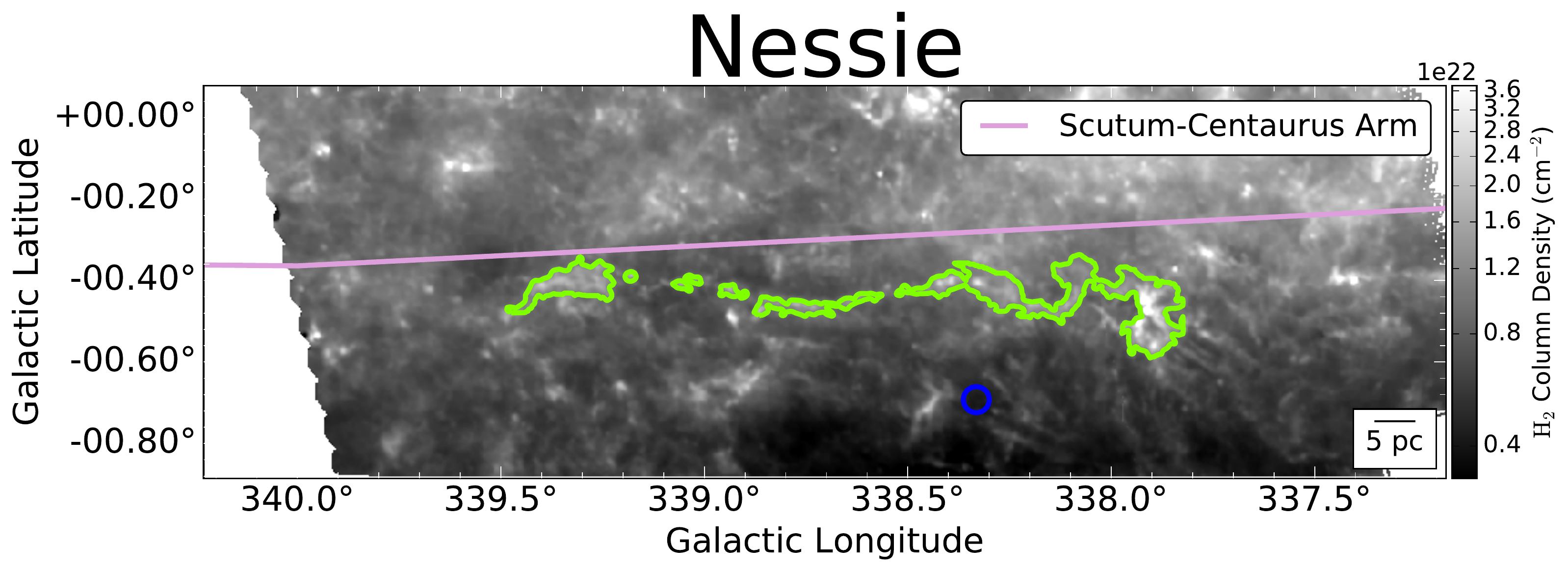} \\

\end{center}
\end{figure}

\clearpage

\subsection{Large-Scale Herschel Filaments} \label{herschel_appendix}
\subsubsection{Original Selection}
\citet{Wang_2015} identify Galactic filaments through their dust emission. \citet{Wang_2015} use the far-infrared Herschel Hi-GAL survey \citep{Molinari_2016} to select quiescent filaments (aspect ratio $>>10$) that exhibit high column density morphology and systematically lower dust temperatures with respect to their surroundings. Like \citet{Ragan_2014}, \citet{Wang_2015} confirm velocity contiguity using lower density $\rm ^{13}CO$ emission \citep[GRS Survey,][]{Jackson_2006} by extracting a \textit{position-velocity} diagram along the curvature of the filament. Once again, the velocity coherence criterion is qualitative, requiring that the \textit{position-velocity} slice exhibit ``continuous, not broken, emission." In total, \citet{Wang_2015} identify nine filaments with Galactic plane separation $z<60\; \rm pc$, seven of which are declared to be associated with a spiral arm. \citet{Wang_2015} use the \citet{Reid_2014} model in Galactocentric coordinates (a top down view of the Galaxy); if any part of the filament overlaps a spiral arm within the distance error of the arm or the filament, it is presumed to be a spiral-arm filament. 

\subsubsection{Sample Selected for Inclusion in this Study} \label{orig_herschel}
We include all nine Large-Scale Herschel filaments listed in Table 1 of \citet{Wang_2015} in this study (9 filaments). 

\subsubsection{Boundary Definition Employed in this Study} \label{herschel_boundaries}
\citet{Wang_2015} analyze the key properties of the Large-scale Herschel filaments by measuring their column densities and dust temperatures within a polygon encompassing the high column density regions of the filament in the Herschel column density maps/dust temperature maps. We still define the Herschel filaments using the high column density regions, but we take a more quantitative approach by delineating the filament boundaries via a closed contour rather than a free-form polygon. The procedure for defining these boundaries is the same as in \S \ref{bone_boundaries}, and the resulting contours are shown in green and overlaid on the Herschel Column Density maps in \S \ref{herschel_gallery}.

\subsubsection{Large-Scale Herschel Filament Gallery} \label{herschel_gallery}

\begin{figure}[!htb]
\begin{center}
\includegraphics[width=0.5\linewidth]{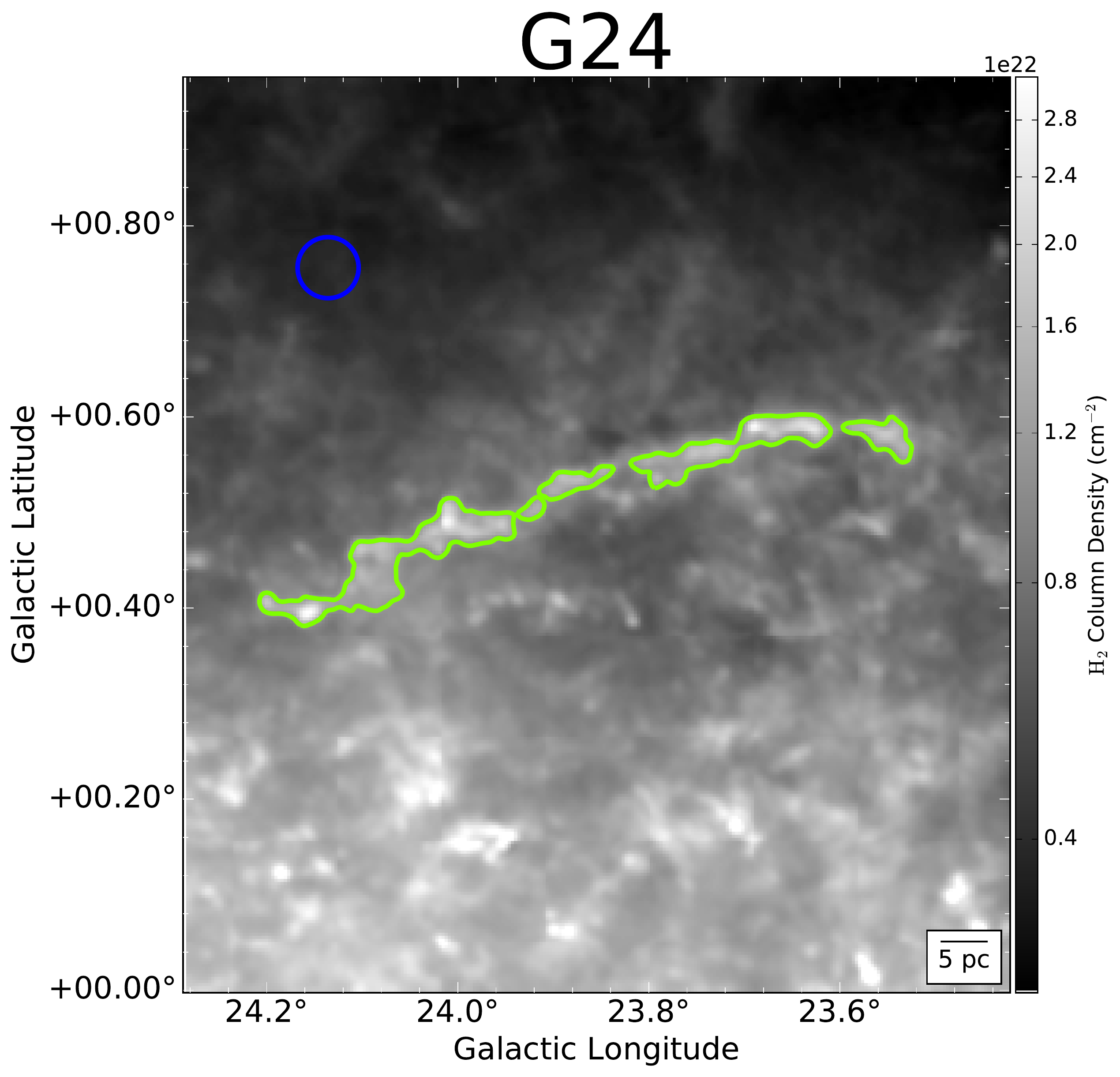} \hfill
\includegraphics[width=0.47\linewidth]{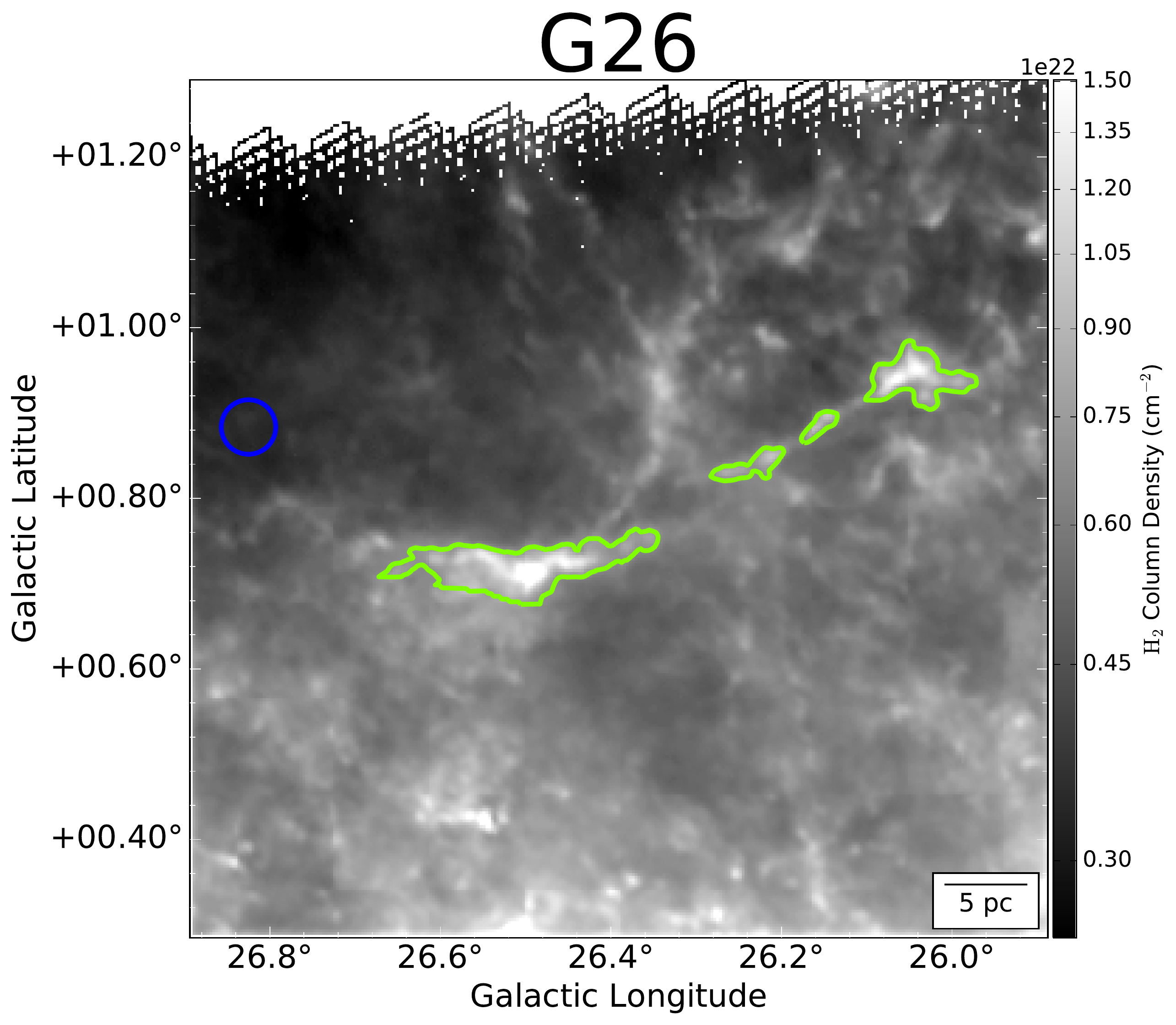} \\
\includegraphics[width=0.47\linewidth]{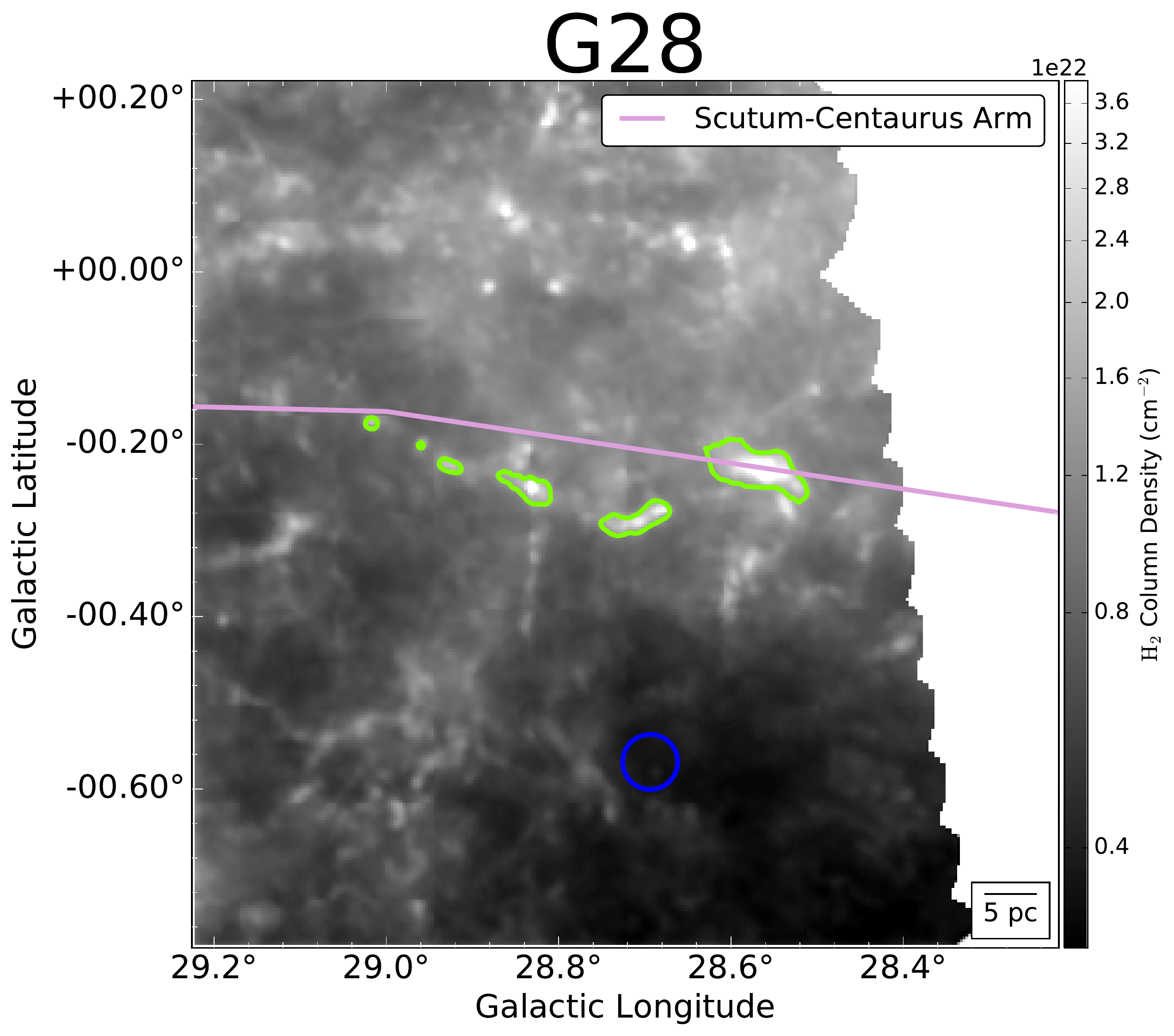} \hfill
\includegraphics[width=0.47\linewidth]{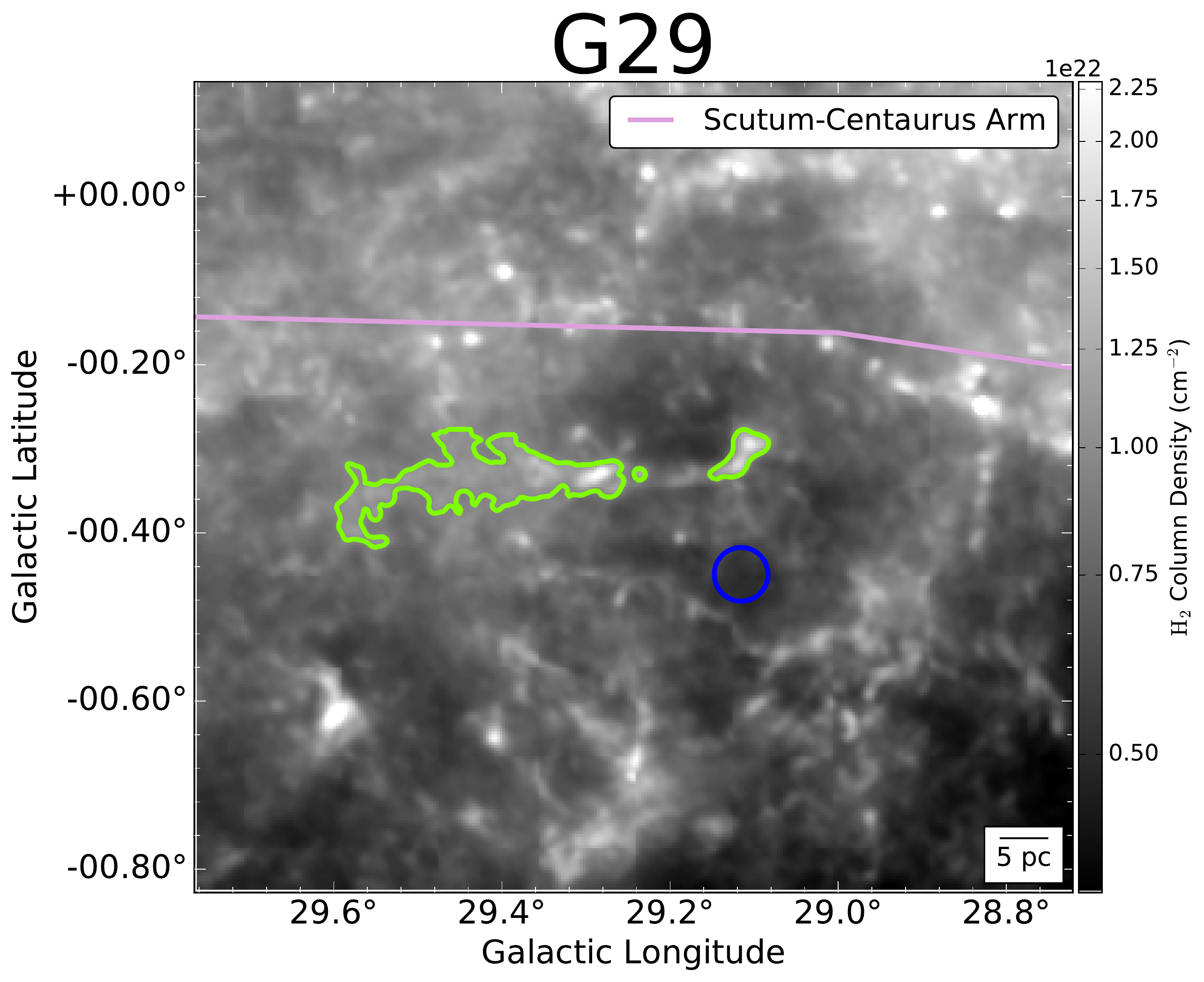} \\
\end{center}
\end{figure}

\newpage

\begin{figure}[!htb]
\begin{center}
\includegraphics[width=0.5\linewidth]{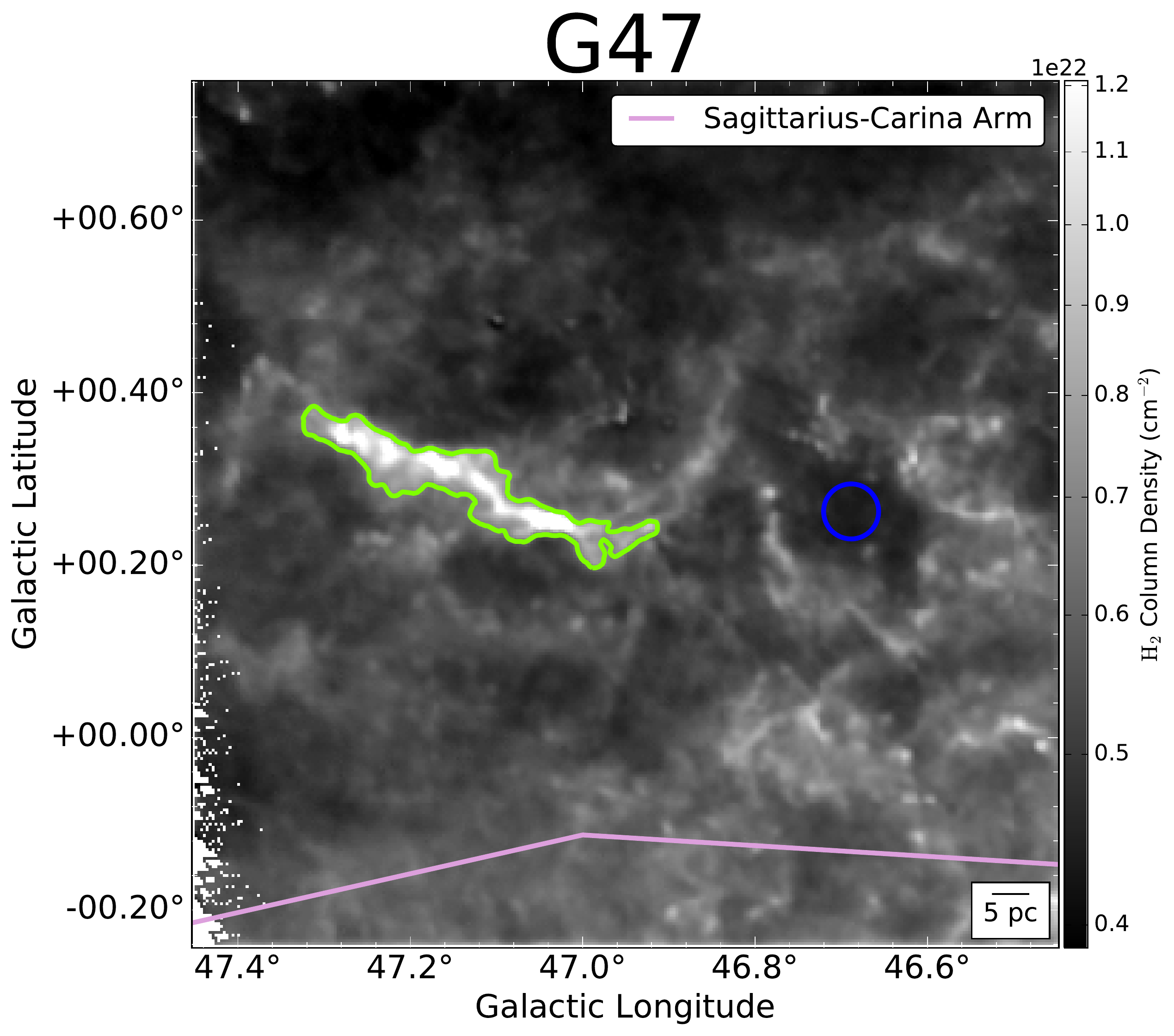} \hfill
\includegraphics[width=0.47\linewidth]{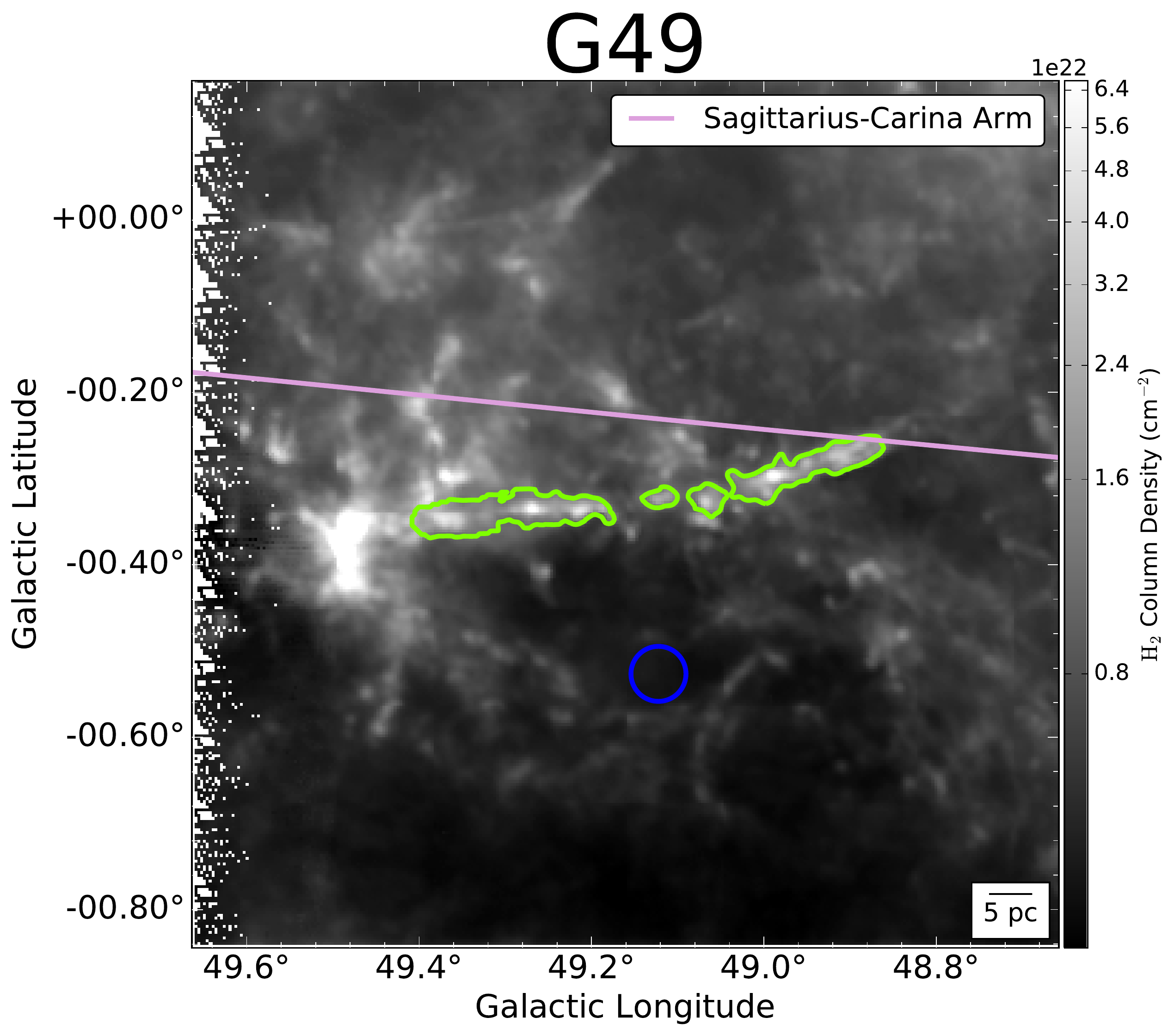} \\
\includegraphics[width=0.47\linewidth]{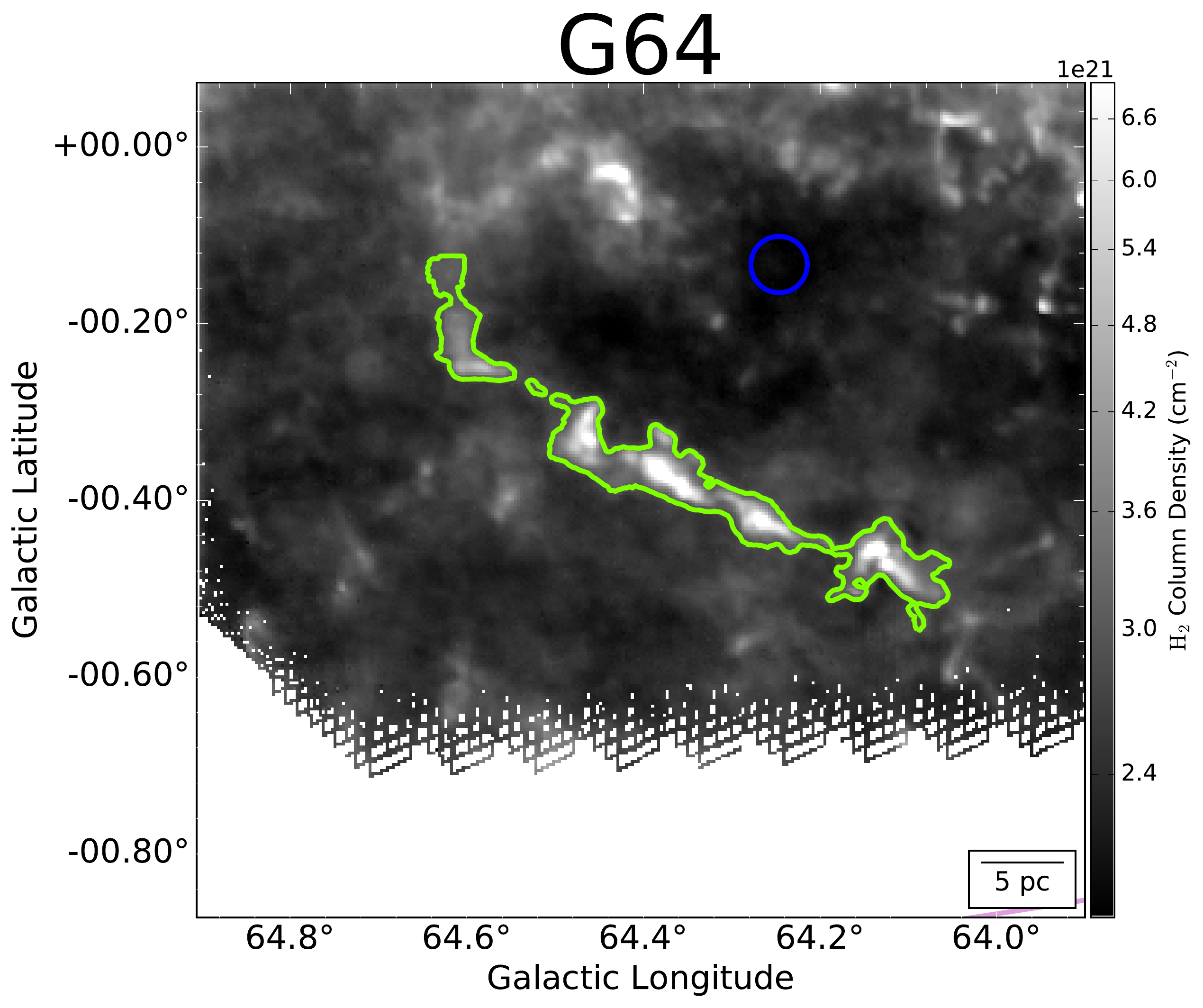} \hfill
\end{center}
\end{figure}

\clearpage

\subsection{Minimum Spanning Tree (MST) Filaments} \label{mst_appendix}
\subsubsection{Original Selection}
\citet{Wang_2016} use an automated minimum spanning tree (MST) approach to connect dense molecular BGPS \citep{Ginsburg_2013} clumps in \textit{position-position-velocity space}. The MST algorithm connects all the nodes (e.g. the BGPS clumps with spectroscopic data from the \citet{Shirley_2013} catalog) in a graph so as to minimize the sum of the edges between them, where an edge is the separation between a pair of clumps. \citet{Wang_2016} customize their MST algorithm such that all trees must contain five BGPS sources, with a maximum edge length of $0.1^\circ$ and a maximum velocity difference of 2 $\rm km\;s^{-1}$ between connected clumps. Furthermore, \citet{Wang_2016} institute a linearity criteria of 1.5, defined such that the standard deviation of the filament's clumps along the major axis must be 1.5x the standard deviation along the minor axis. Finally, \citet{Wang_2016} impose a minimum physical length criterion of 10 pc. The \citet{Wang_2016} study finds that 54 structures satisfy these five criteria and can reasonably be called large-scale filaments. Using the quantitative Bone criteria from \citet{Zucker_2015} as a guide, \citet{Wang_2016} also develop criteria that MST ``Bones" must satisfy on top of the large-scale filament criteria. The MST ``Bones" must also lie within 20 pc and of the physical Galactic midplane, run parallel to a spiral arm on the plane of the sky (to within $30^\circ$) and have a flux-weight LSR velocity within $\rm 5 \; km \; s^{-1}$ of a \citet{Reid_2014, Reid_2016} spiral arm. 13/54 large-scale filaments satisfy the additional MST ``Bone" criteria.

\subsubsection{Sample Selected for Inclusion in this Study} \label{mst_subsample}
Though \citet{Wang_2016} identify 54 large-scale filaments using their automated minimum spanning tree method, we only consider the 13 filaments they define as MST ``Bones", which must also show this additional association with spiral structure.  We exclude F48 because it is outside the boundaries of the Herschel Hi-GAL survey, which serves as a key dataset in this work. This leaves twelve MST ``Bones" available for inclusion in this study: F2, F3, F7, F10, F13, F14, F15, F18, F28, F29, F37, and F38. 

\begin{figure}[h!]
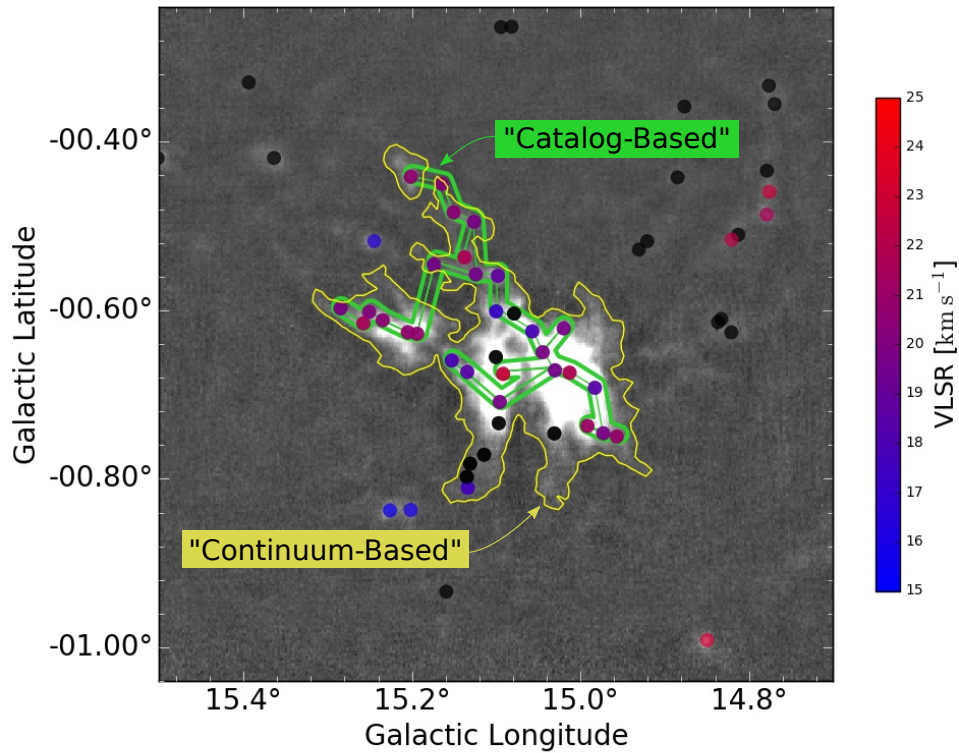

\begin{center}
\includegraphics[width=0.75\linewidth]{{{MST_Figure}}}
\caption{{\label{fig:mst} We show the two different boundary definitions for the MST ``Bone" filament F18---one based on the BGPS catalog data \citep{Shirley_2013}, as in the original \citet{Wang_2016} study (the ``catalog-based" definition; thick green outline), and one based on the underlying BGPS continuum emission data from \citet{Ginsburg_2013} (the ``continuum-based" definition; yellow outline). The background grayscale is the BGPS continuum emission data from \citet{Ginsburg_2013}. The dots indicate BGPS catalog sources derived from the continuum maps \citep{Rosolowsky_2010}, which also have follow-up spectroscopic data from \citet{Shirley_2013}. The dots are color-coded by their LSR Velocity, with black indicating that no velocity information is available. With the thin green line, we indicate the original edges of the Minimum Spanning Tree for F18 from \citet{Wang_2016}; note that the edges connect a set of spectroscopic clumps from the BGPS catalog, which act as the ``nodes" of the tree. We demarcate our catalog-based definition of F18 via the thick green line, created by finding all points within $1\arcmin$ of the edges and connecting them on the plane of the sky. We indicate the boundary of our continuum-based MST ``Bone" filaments via the yellow contours, which is a threshold applied to the original continuum map at a level of $\rm 0.04 \; Jy \; beam^{-1 }$. It contains all the clumps defining the original catalog-based definition, plus a few extra for which there is no spectroscopic data. For more information on the rationale behind the continuum-based definition see \S \ref{mst_boundaries}.}}
\end{center}
\end{figure}

\subsubsection{Boundary Definition Employed in this Study} \label{mst_boundaries}
We adopt two different definitions for the MST ``Bones," and define two sets of quantitative boundaries. For our catalog-based definition, we only consider the BGPS spectroscopic catalog data \citep{Shirley_2013}, as in \citet{Wang_2016}. For our continuum-based definition, we also consider the information in the BGPS continuum emission maps \citep{Ginsburg_2013}. \\

\textbf{Catalog-Based MST ``Bone" Filaments}:
\begin{enumerate}
\item{For our first definition, we define the boundaries of the MST ``Bone" filaments similarly to \citet{Wang_2016}---by using the edges of each filament's minimum spanning tree put forth in that study. These edges act as connection ``nodes" between the BGPS spectroscopic catalog clumps. We create a five pixel buffer ($\approx 1 \arcmin$) on either side of each edge in the Hi-GAL column density and dust temperature maps. This is akin to the \citet{Wang_2016} procedure of estimating the MST filament masses by measuring the 1.1 mm dust emission flux ``in a polygon encompassing the filament guided by the MST."  Specifically, we use the buffer function from the \href{http://toblerity.org/shapely/manual.html}{\texttt{Shapely}} python package, which ``returns an approximate representation of all points within a given distance" of some geometric object, which in our case is a set of connected lines defining the edges of each tree. The catalog-based MST ``Bone" Filament boundaries are shown via the green outlines overlaid on the Herschel column density maps in \S \ref{mst_gallery}.}
\end{enumerate}

\textbf{Continuum-Based MST ``Bone" Filaments}
\begin{enumerate}
\item{For our second definition, we define the boundaries of the MST ``Bone" filaments by isolating the larger emission structure in which the individual nodes (e.g. the BGPS clumps) defining each tree are embedded. We do this by overplotting the nodes of each tree onto the original continuum emission maps and then varying the emission threshold (between $\approx 0.03-0.06 \; \rm Jy \; beam^{-1}$) and adopting one which recovers the maximum number of BGPS cores contained in the original trees. We also apply a smoothness level of four using the ds9 contour smoothing option, such that each contour is only evaluated at every fourth pixel. The smoothing helps negate the effects of spatial filtering and the rms noise, which makes the contours easier to skeletonize (a procedure used to determine the topology of the masks and the filament lengths in \S \ref{lengths}). The continuum-based MST ``Bone" boundaries are shown via the yellow contours overlaid on the Herschel column density maps in \S \ref{mst_gallery}. 

One might question why this new approach is necessary. The original MST boundaries outlined in \citet{Wang_2016} provide no estimate on the width of the MST ``Bones". In their original publication, \citet{Wang_2016} assign a width to each MST filaments by taking the average of the major axes of all the individual BGPS catalog clumps in each tree. However, this assumes that all the clumps in any given tree formed in a cylindrical structure---otherwise, the trees might not be filaments, but simply a complex of BGPS clumps that are contiguous in velocity. Our new method for defining the morphology of the MST ``Bone" filaments (isolating the larger emission structure enveloping the clumps) does not assume any geometry a priori. Applying a closed contour is consistent with how we define the boundaries of the three other catalogs discussed herein. }

\end{enumerate}

\subsubsection{Comparison of the Catalog- and Continuum-Based Methods for Defining MST ``Bone" Filament Boundaries} \label{mst_boundaries_comparison}
The two different methods for defining the boundaries of a single MST ``Bone" filament (F18) are shown in Figure \ref{fig:mst}. The background shows the BGPS 1.1 mm continuum emission map from \citet{Ginsburg_2013}. The thin green line shows the edges of the initial minimum spanning tree for F18 from \citet{Wang_2016}. Note that this thin green line connects a set of BGPS clumps that are contiguous in velocity space, which are shown in Figure \ref{fig:mst} as circles colored according to their LSR velocity. The thick green outline framing the thin green line is our first set of quantitative boundaries (the ``catalog-based" definition), produced by finding all the points within $1\arcmin$ of the thin green line and connecting them on the plane of the sky. We plot our second set of boundaries for F18 (the ``continuum-based" definition) with yellow contours, which corresponds to an emission level of $0.04 \rm \; Jy/beam$. In doing so we include all the clumps belonging to the original tree, plus a few additional clumps for which there is no spectroscopic information (shown in black). The BGPS spectroscopic catalog \citep{Shirley_2013} only includes about 50\% of the clumps identified in the original continuum catalog \citep{Rosolowsky_2010}, so this is common. While all the original clumps from \citet{Wang_2016}  for F18 are located within our new boundary (shown in yellow) this is not always true, and is dependent on whether a clump on the borders of any individual tree can be sensibly associated with the bulk of the emission defining each tree. This is obviously not an exact science, but we reiterate that all of our boundaries are available for download at the \href{https://dataverse.harvard.edu/dataverse/Galactic-Filaments}{Large-Scale Galactic Filaments Dataverse}, so anyone can see exactly what we did. Whenever possible, we calculate the physical properties of the MST ``Bone" filaments using both sets of boundaries (``catalog-based" and ``continuum-based").
\subsubsection{MST ``Bone" Filament Gallery} \label{mst_gallery}

\begin{figure}[!htb]
\begin{center}
\includegraphics[width=0.5\linewidth]{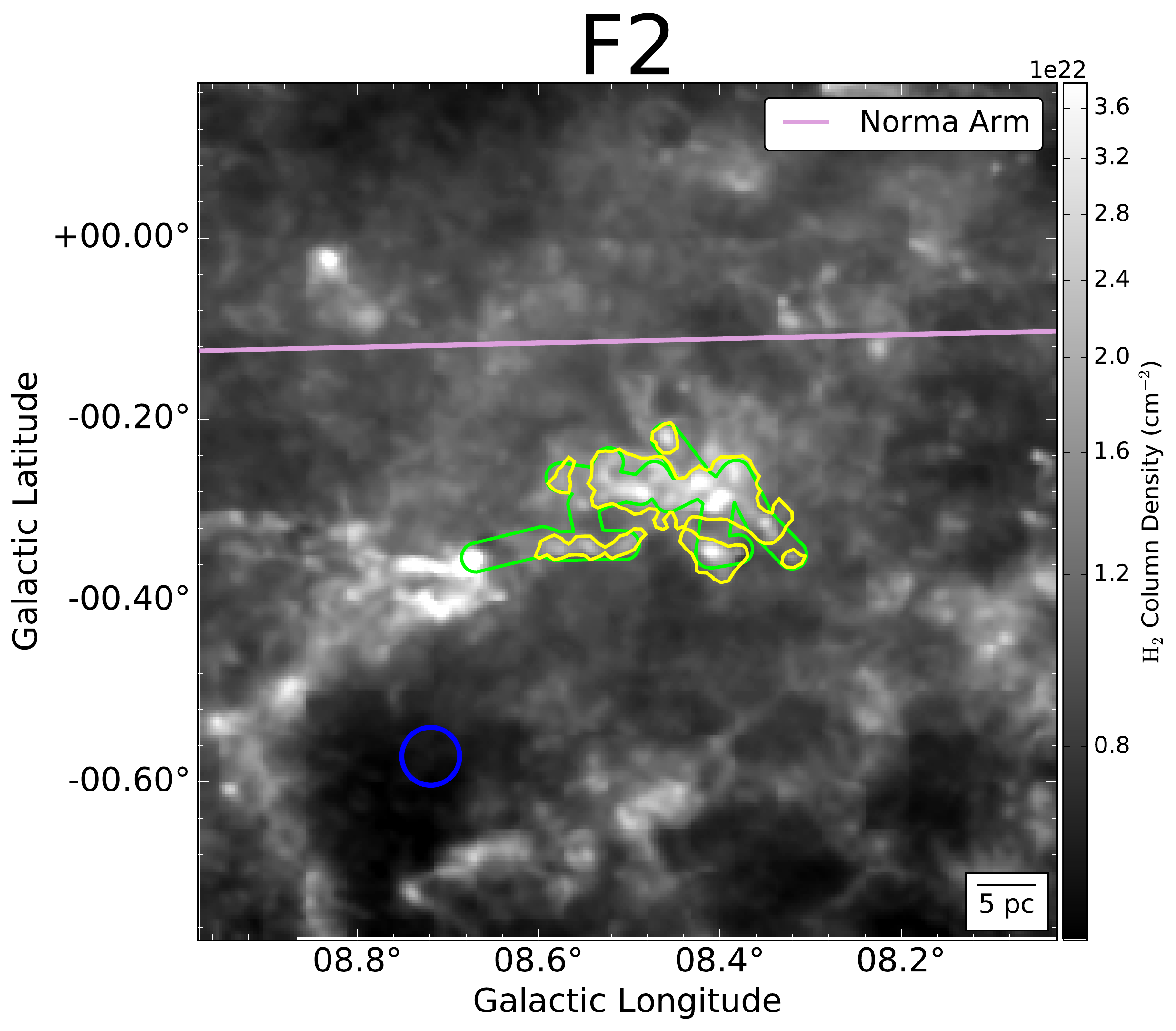} \hfill
\includegraphics[width=0.47\linewidth]{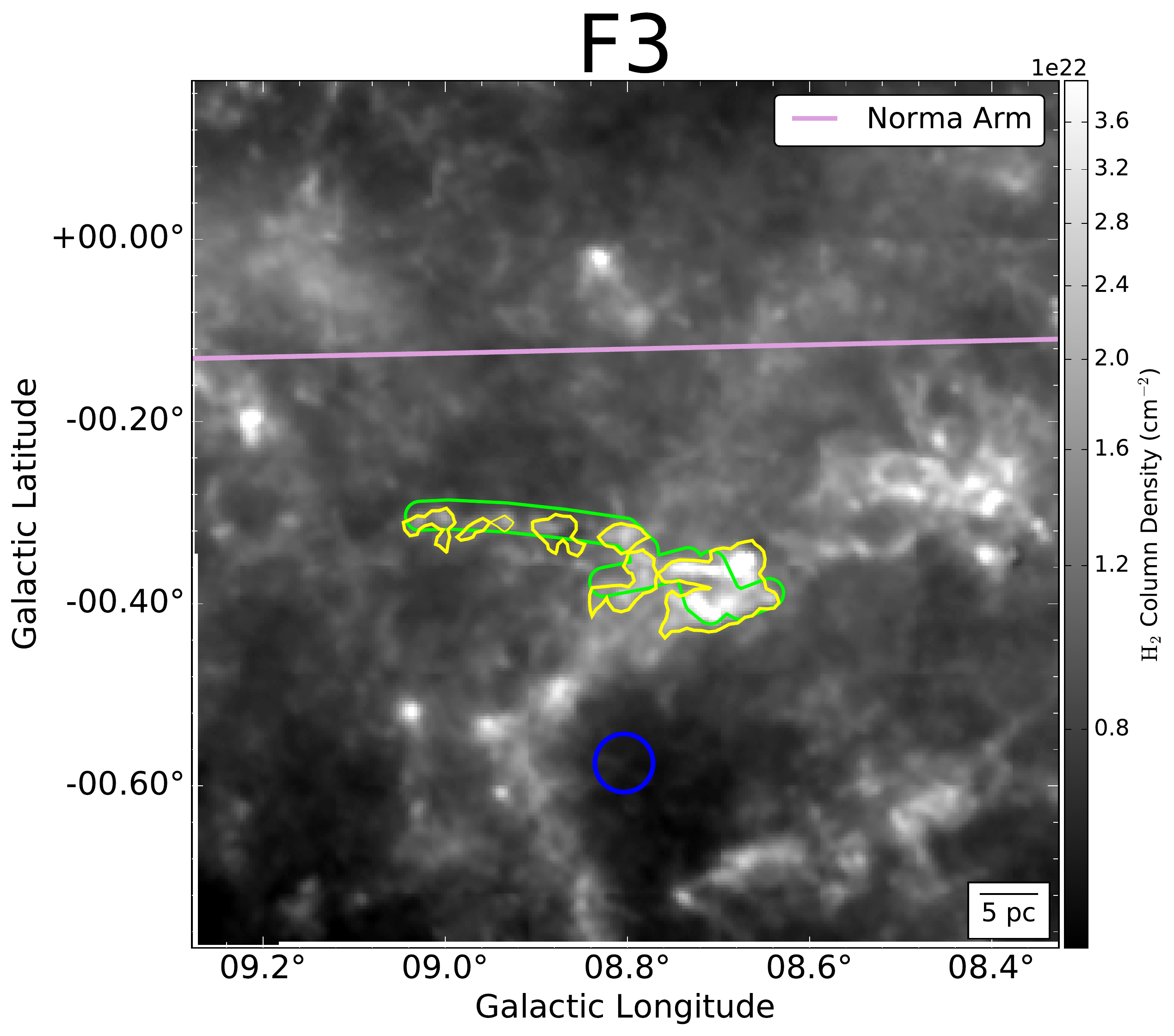} \\
\includegraphics[width=0.47\linewidth]{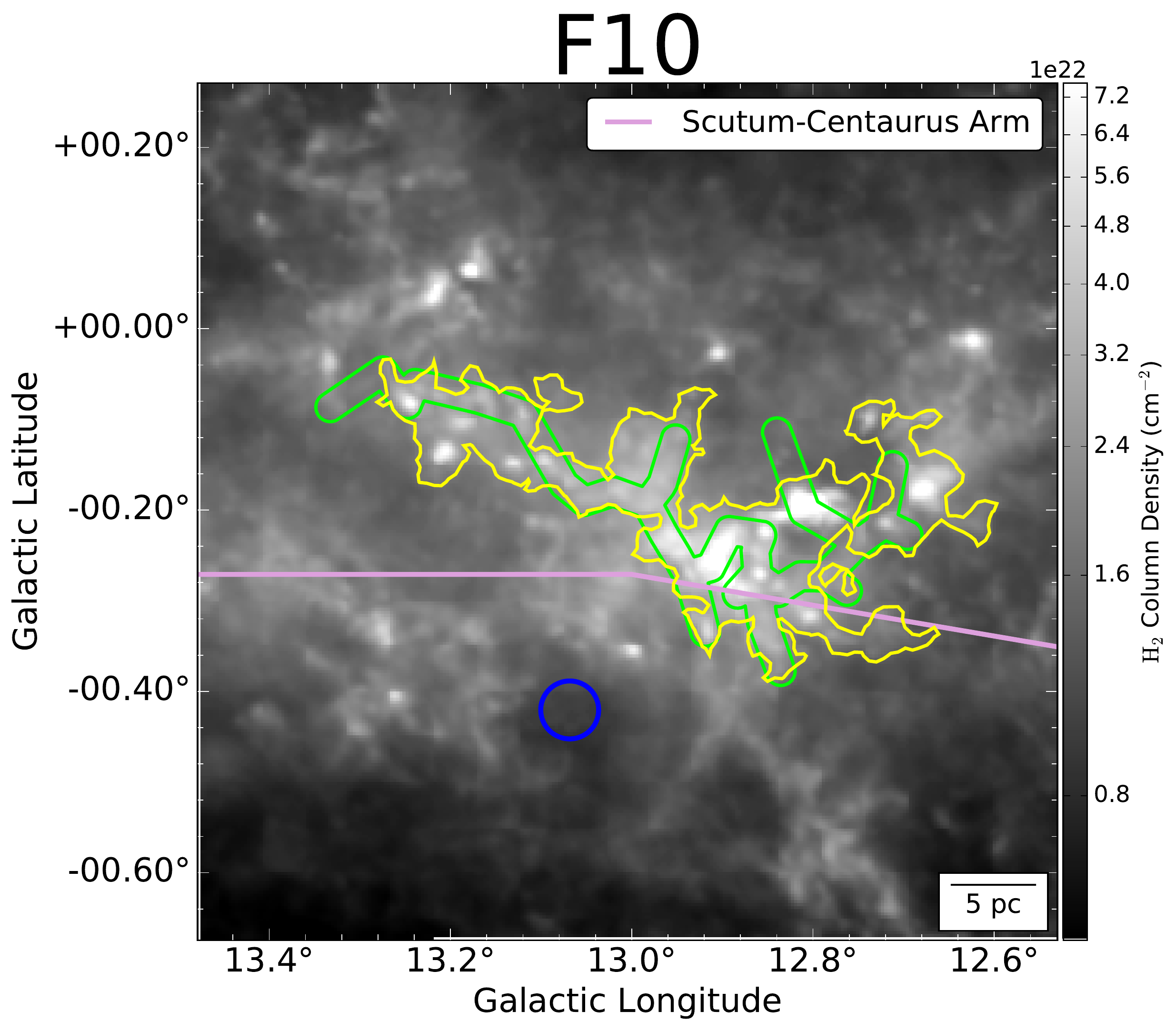} \hfill
\includegraphics[width=0.47\linewidth]{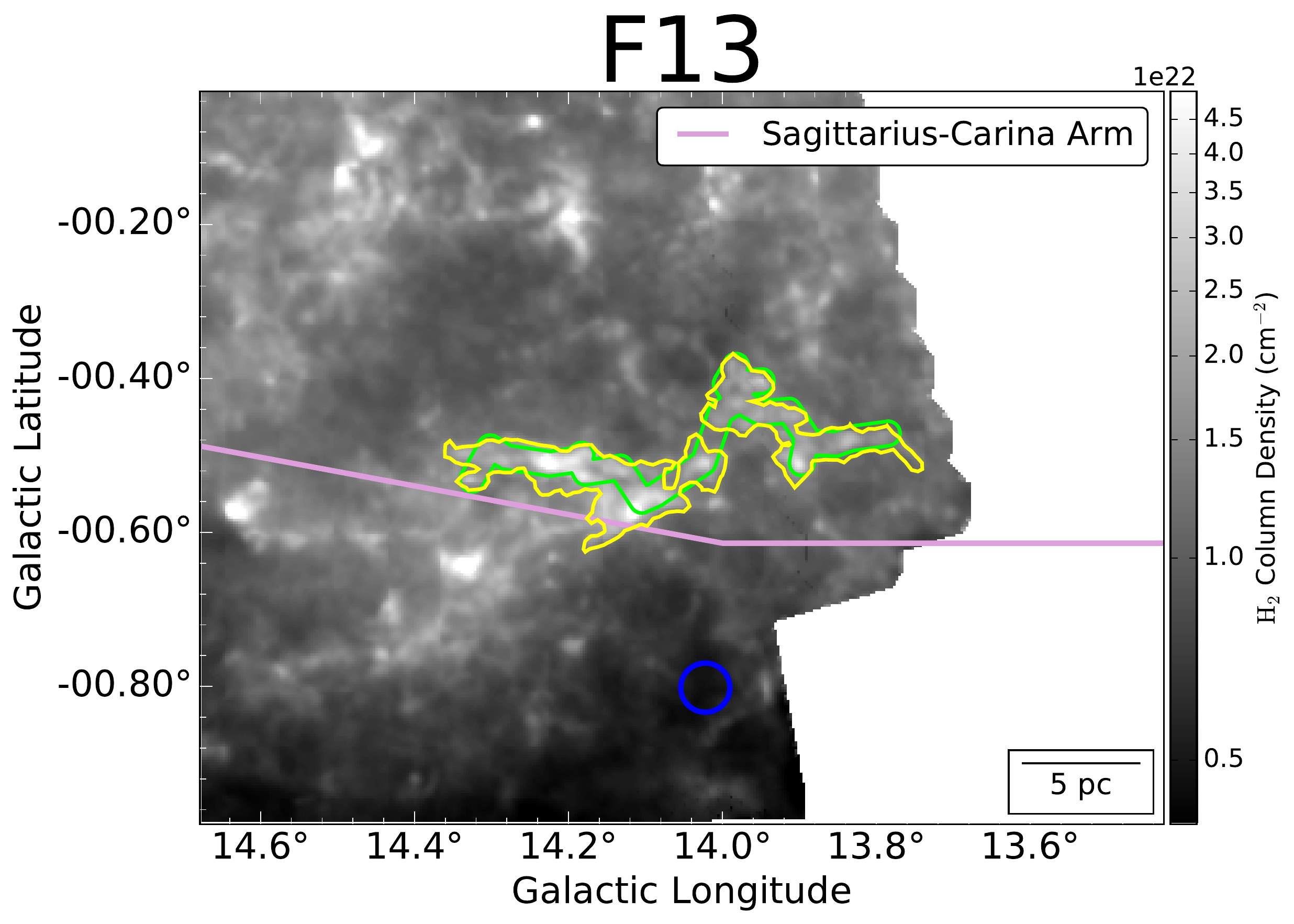}  \\
\end{center}
\end{figure}

\newpage

\begin{figure}[!htb]
\begin{center}
\includegraphics[width=0.5\linewidth]{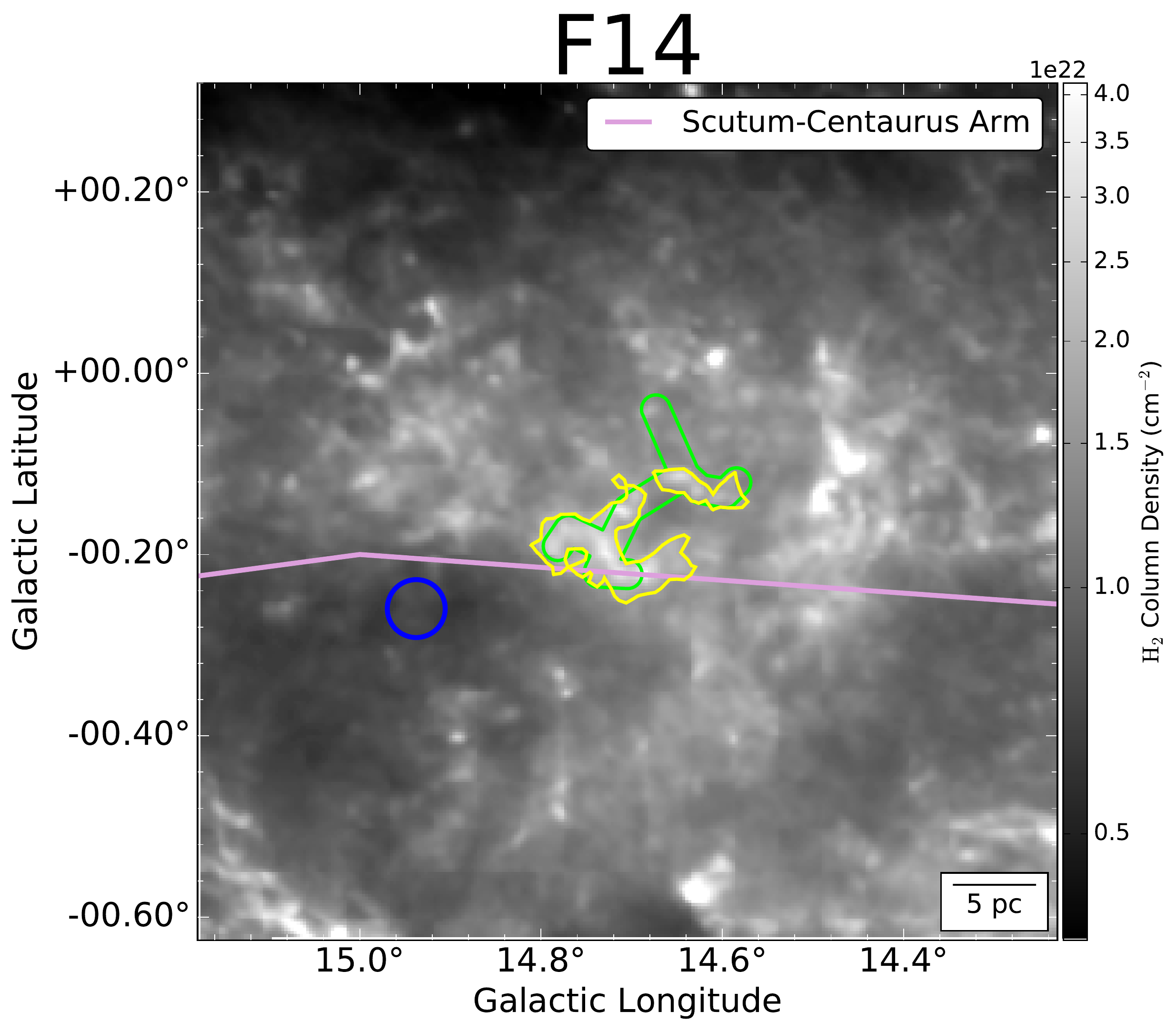} \hfill
\includegraphics[width=0.47\linewidth]{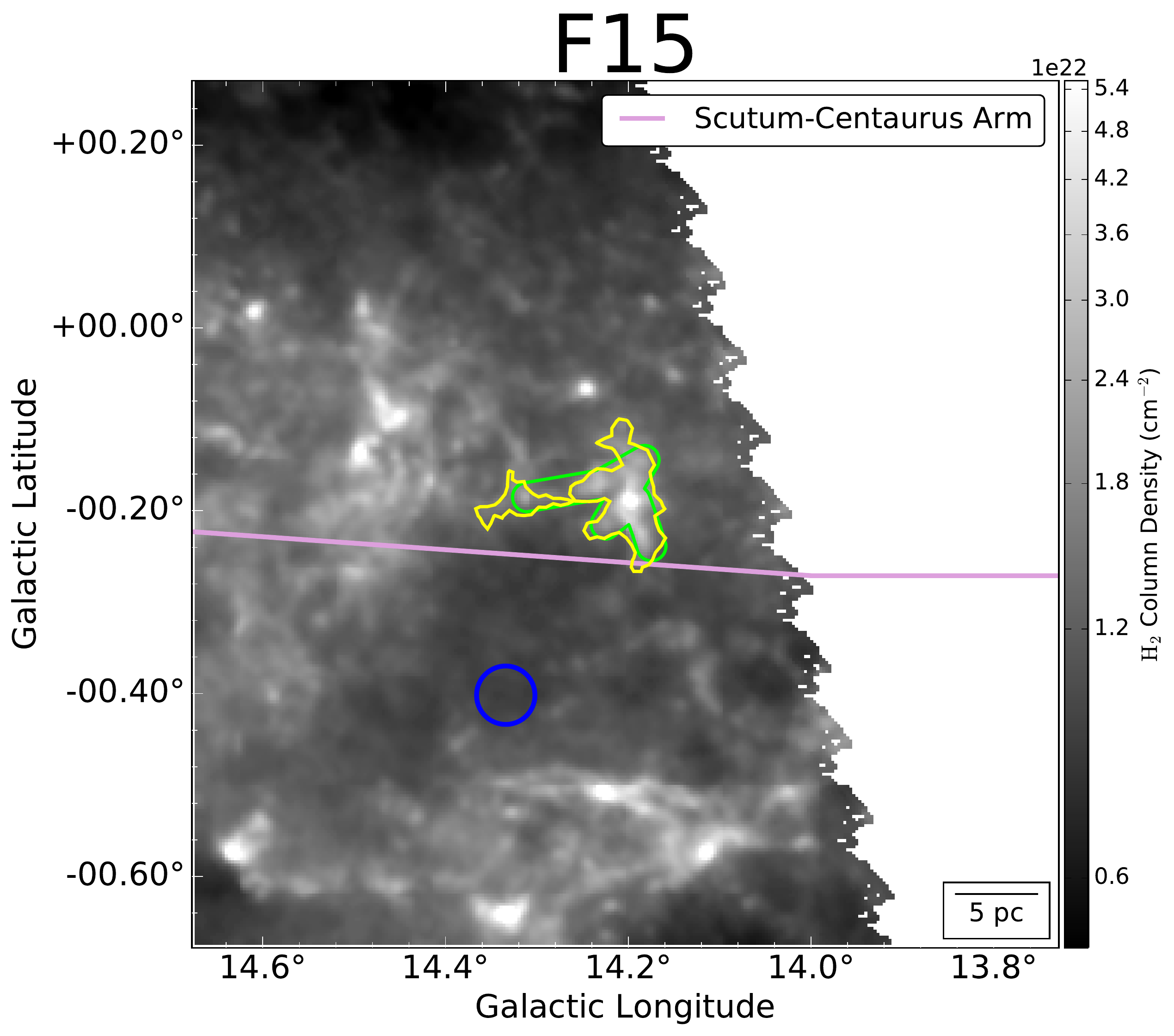} \\
\includegraphics[width=0.47\linewidth]{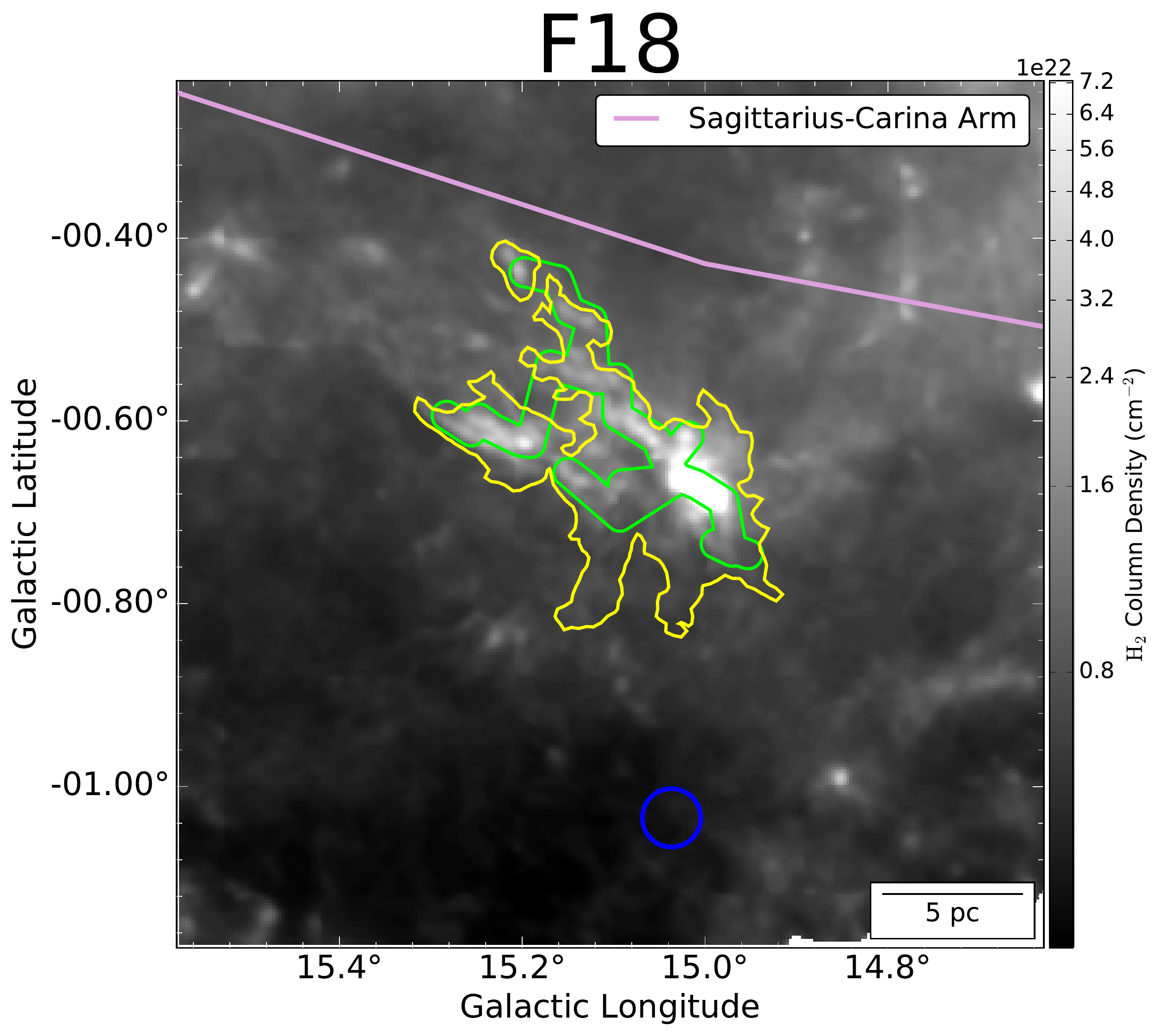} \hfill
\includegraphics[width=0.47\linewidth]{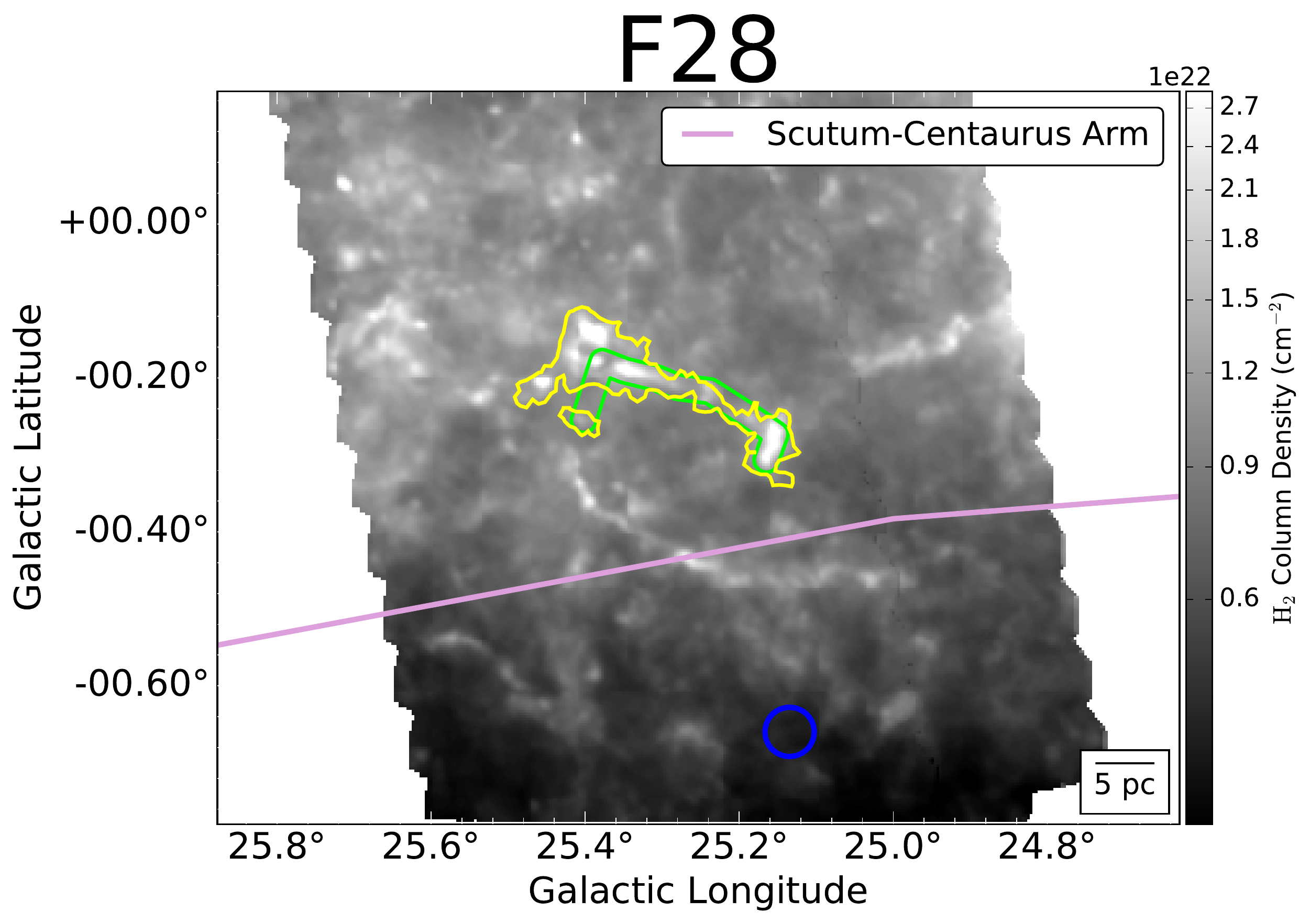}  \\
\end{center}
\end{figure}

\newpage

\begin{figure}[!htb]
\begin{center}
\includegraphics[width=0.5\linewidth]{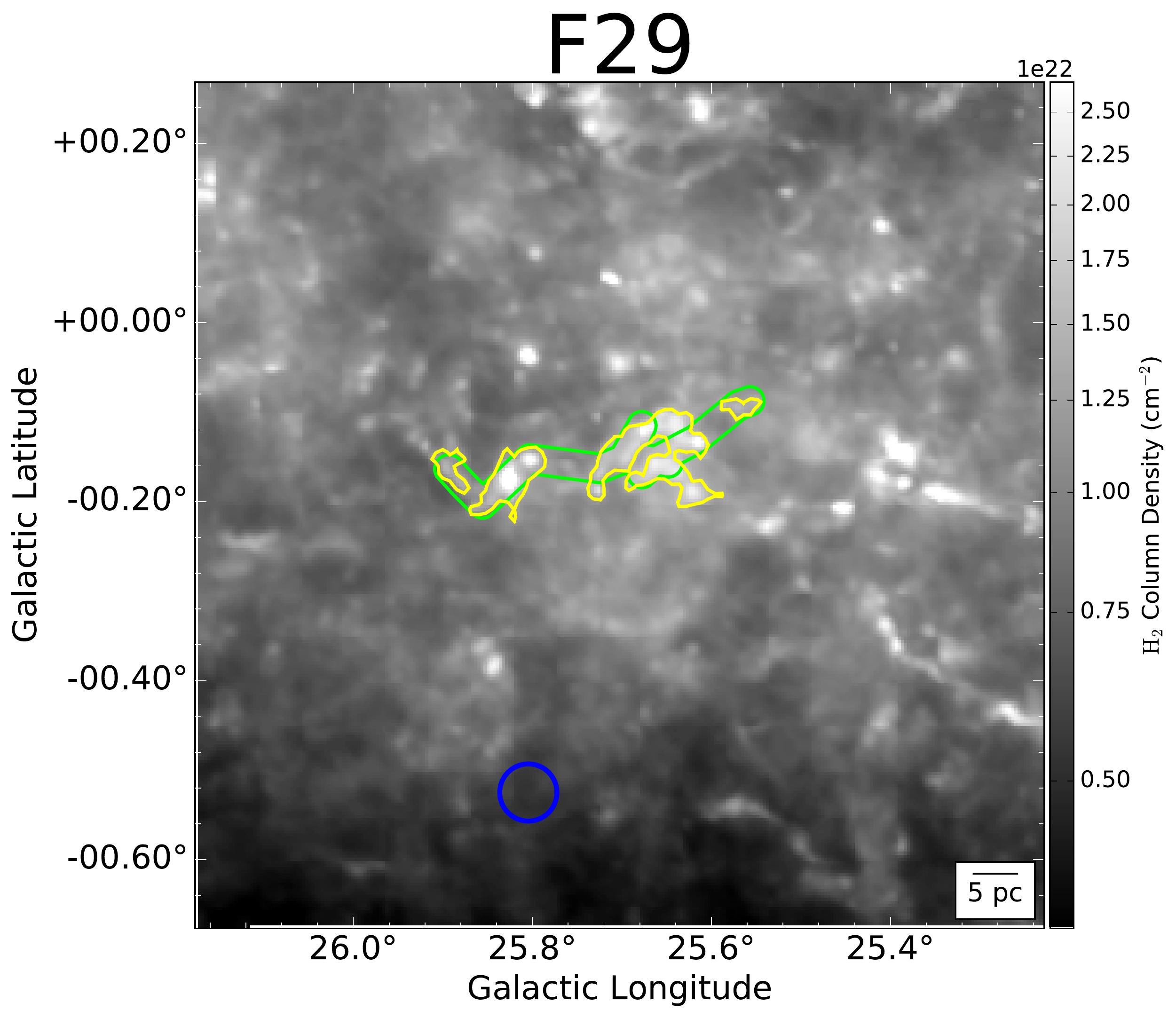} \hfill
\includegraphics[width=0.47\linewidth]{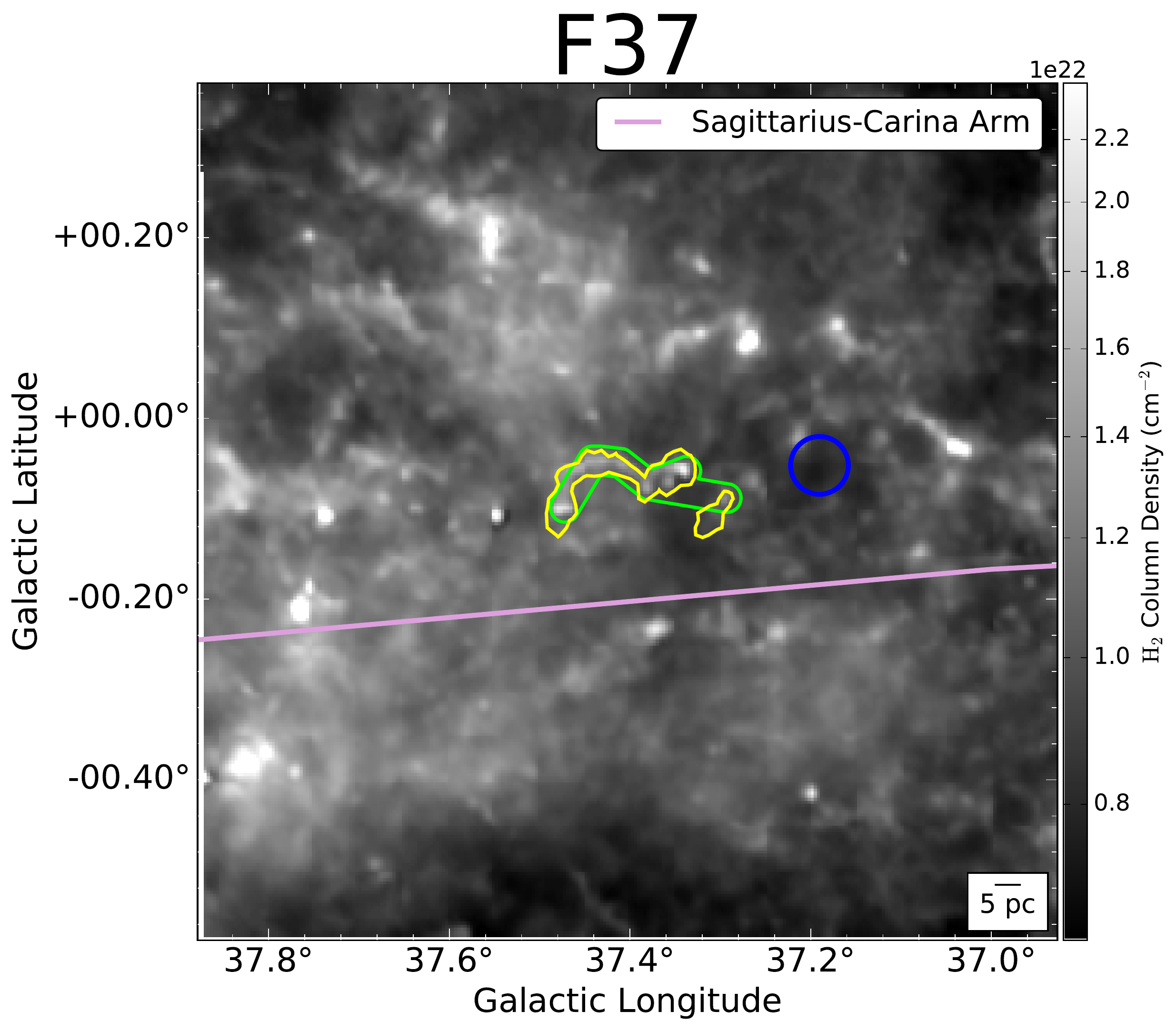} \\
\includegraphics[width=0.47\linewidth]{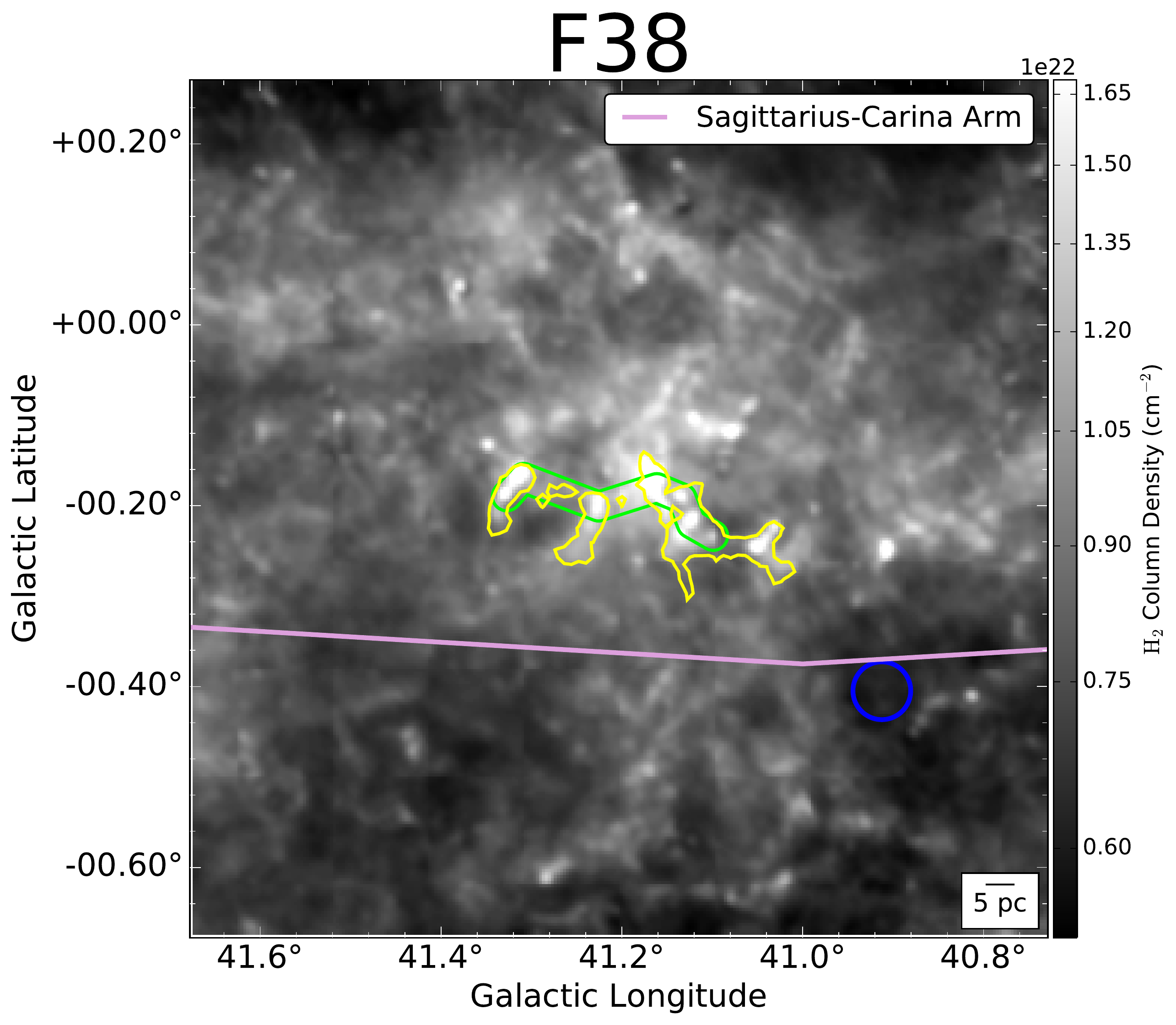} \hfill
\end{center}
\end{figure}

\clearpage

\subsection{Background Subtraction of the Hi-GAL Fluxes} \label{bg_subtract}
Prior to performing the column density analysis outlined in \S \ref{densities_temps}, we test four different background subtraction methods applied to the $160-500\; \micron$ fluxes, from \citet{Battersby_2011}, \citet{Juvela_2012}, \citet{Wang_2015}, and \citet{Peretto_2016}. We implement the flat background subtraction method adopted in \citet{Juvela_2012}. The \citet{Battersby_2011} approach fits Gaussians in latitude to each Galactic longitude, which are then subtracted from the original $500\; \rm \mu m$ fluxes. This process is iterated upon until the fit converges, and the resulting difference image forms the basis of the background mask. While this method was successfully applied to ``the Snake" (see Table \ref{tab:tab1}), it requires the user to adopt a sigma-threshold to delineate source from background, and a large range in thresholds would have to be adopted to successfully isolate each source in the sample. Next we implement the Fourier Transform method from \citet{Wang_2015}, which separates each flux image into low frequency (large-scale background/foreground structure) and high frequency (small-scale structure) components. This method is effective for the \citet{Wang_2015} filaments and most of the \citet{Zucker_2015} filaments, but poorly suited for the GMFs---because they cover tens of square arcminutes on the sky and exhibit significant variability in their fluxes they include both high and low spatial frequency components, which necessitates masking out large regions of these filaments. Generally, it also fails to recover structure at high Galactic latitudes.  Next we consider the background subtraction that \citet{Peretto_2016} use for the Herschel counterparts of their Spitzer infrared dark cloud catalog, containing over 11,000 dense clumps. Rather than performing background subtraction on the flux images, \citet{Peretto_2016} do so on the column density maps: they apply a 10' wide median filter to the column density maps and subtract this median component to create the final maps. While effective for a large number of compact sources, using a single median filter width produces artifacts in a subset of subtracted images, and we decide against adopting several different filter widths for the range of filament sizes found in our sample.

As noted in \S \ref{densities_temps}, we settle on the flat background subtraction method discussed in \citet{Juvela_2012}, which uses a circular reference area in a low emission region near the source to approximate the cirrus background. While it cannot account for structural variations in the background, the \citet{Juvela_2012} method can be applied to the wide diversity of filament size scales observed in this study. 

\subsection{Radial Column Density Profiles and Dependence of Best-Fit Values on Fitting Parameters} \label{radfil_appendix}
\subsubsection{Building and Fitting the Radial Column Density Profiles}
In Figure \ref{fig:radfil} we summarize the radial column density profile fitting process for a single filament in the sample (``Fil2"). All of the profile fitting is performed via our publicly available fitting code \href{https://github.com/catherinezucker/radfil}{\texttt{RadFil}}. First, we delineate the morphology of the filament via a closed contour, in this case by applying a threshold a few sigma above the mean background column density in the custom Hi-GAL-based column density maps derived in this study. This filament mask is shown in white in the top panel of Figure \ref{fig:radfil}a. When the filament cannot be delineated via a single continuous closed contour (e.g. for ``Fil2") we transform the set of discrete contours into a single connected feature using a concave hull algorithm. The concave hull mask for ``Fil2" is shown in the middle panel of Figure \ref{fig:radfil}a.  Next, a ``spine" of the filament is determined using medial axis skeletonization of the filament's concave hull mask. The result of this process---a one pixel wide representation of the mask's topology--is computed for the filament, and a smoothed version of that spine is shown via the thick red line in the bottom panel of Figure \ref{fig:radfil}a. Then, the derivative is computed along the filament mask, which is used to derive the tangent line at even intervals ($\rm \approx 0.5 \; pc$) across the spine. The line perpendicular to the tangent is then computed and overlaid on the filament mask, as shown via the thin red cuts across the spine in the bottom panel of Figure \ref{fig:radfil}a. The median length of these red lines is the ``mask" width for the filament shown in Column (6) of Table \ref{tab:tab3}. For the radial column density profile fitting, we find the pixel with the maximum column density along each of the red cuts but confined to the mask (shown as blue scatter points in the bottom panel of Figure \ref{fig:radfil}a) and shift the center of the profile to that value.  For those filaments represented by a concave hull, we discard cuts that are outside the original masks so as not to include regions with low signal. We also exclude active star forming regions (e.g. in the middle of Filament 5). We then build up a radial column density profile along each cut with respect to that pixel, and the profiles for each cut are overlaid together in each of the panels in Figure \ref{fig:radfil}b (transparent gray scatter points). We perform our fitting on the unbinned profiles, rather than taking some representative average profile, as we have found this to be more robust, particularly for the Plummer-like fits. 

To perform the fitting, we first subtract off a background (modeled as a first-order polynomial) by fitting a line between radii of 3 and 4 pc (vertical green dashed lines in the top panel of Figure \ref{fig:radfil}b). The line with the best-fit to the background is shown in solid green and this is the amplitude of the background that is subtracted off at each radius. The subtracted profile is shown in the two bottom panels of Figure \ref{fig:radfil}b. In the middle panel of Figure \ref{fig:radfil}b, we show in solid blue the best-fit Gaussian profile for ``Fil2", computed using all the data within $\rm \pm 2 \; pc$ of r=0 pc, as indicated via the vertical blue dashed lines. Since only the inner widths of the profiles are truly Gaussian, we fit out to the minimum radius where we start to see a flattening of the Gaussian profile, which is $\approx 2.0$ pc. The same procedure is followed with the Plummer-like fit in the bottom panel of Figure \ref{fig:radfil}b, except now all data within $\rm \pm 4 \; pc$ of r=0 pc is used, as indicated again via the vertical blue dashed lines.

\subsubsection{Uncertainties in the Best-Fit Values}
Several studies in the past have shown that both Gaussian fits and Plummer-like fits to radial column density profiles are intimately dependent upon the background subtraction method and the fitting distance. Most notably, \cite{Smith_2014a}, in their analysis of the synthetic profiles of small-scale filaments, find that their average best fit FWHM obtained with a fitting distance of 1 pc is 1.5 times larger than the FWHM obtained with a fitting distance of 0.35 pc. Similarly, \citet{Juvela_2012a}, who also characterize the synthetic profiles of small-scale filaments, find that the amount of noise added to their data can affect the Plummer parameters by tens of percent, with line-of-sight confusion also playing a role. Given these large uncertainties, we adopted a range of values for our fitting distances and background subtraction radii, chose a representative set for analysis in the main text, and report and discuss the values obtained with other background subtraction methods/fitting distances here and in Table \ref{tab:profilealt}. Recall that for our representative widths in Figure \ref{fig:lwa_comp}, we adopted a Gaussian fitting radius of 2 pc and background subtraction radii of 3 and 4 pc, which produces an average FWHM for the filaments of 1.3 pc. If we keep the background subtraction radii fixed but halve (1.0 pc) or quarter (0.5 pc) the fitting distance it decreases the median width of the filaments by 0.2 and 0.5 pc, respectively.

In addition to the Gaussian FWHM values, the index of the density profile ``$p$" of the Plummer-like fits are particularly sensitive to different background subtraction and fitting radii. In Figure \ref{fig:plummer_comp} we adopt representative background subtraction radii of 3 and 4 pc and a fitting radius of 4 pc. This produces a median ``$p$" index of 3.15. Varying the fitting radius between 3 and 5 pc, and estimating the background in different radii within the same range can change the median Plummer $p$ indices by at least a half, though it does typically fall between 2.9 to 3.3. While the median changes, the range of $p$-values seen across the sample typically stays constant, and almost always falls between 2.5 and 5.0. Finally, we further find that insufficient background subtraction (such that the profile does not plateau to zero at large radii) is a major factor in pushing the median Plummer ``$p$" index of our large-scale filament distribution down to lower values. Adopting a fitting radius of 4 pc and estimating the background between 3-4 pc (as adopted in this study) produces a median Plummer ``$p$" value of $\approx 3.15$, but adopting the same fitting radius of 4 pc and estimating the background between 4-5 pc produces a median Plummer ``$p$" value of 2.75. The less aggressive background subtraction means that the profiles do not always flatten at zero, which \textit{tends} to produce lower flattening radii ($\rm R_{flat}$), which is directly correlated with the index of the density profile ``$p$" \citep[see][]{Smith_2014a}, producing a lower ``$p$" value.

\subsection{Dust-Extinction Based Profiles} \label{irdc_profiles}
As the \citet{Zucker_2015} Bones are originally identified via their mid-IR extinction features, we also fit a radial column density profile to $\rm H_2$ column density maps derived from the GLIMPSE $8\; \rm \mu m$ images. The results of this analysis are shown in the third and fourth panels of Figure \ref{fig:lwa_comp}. To do this we create surface mass density maps from the $8 \; \rm \mu m$ maps following the procedure outlined in \citet{Battersby_2010}. We then convert to column density by dividing the surface mass densities by $\rm \mu_{H_{2}}m_H$, with the mean molecular weight $\rm \mu_{H_{2}}$ equal to 2.8 \citep{Kauffmann_2008}. As the visibility of extinction features is strongly dependent upon fluctuations in background emission, applying a semi-continuous column density threshold to filaments in the extinction-derived column density maps is infeasible, even with the application of a concave hull algorithm. For the IRDC widths, we applied a column density threshold $\approx 1-2 \sigma$ above the background (see \S \ref{bone_boundaries} for more details on contour application) and picked the longest single continuous component to derive a representative extinction width for the \citet{Zucker_2015} Bones. This isocontour was skeletonized in the same fashion as the Herschel contours and cuts were taken at the same sampling interval (0.5 pc). The masks used to take the perpendicular cuts along the IRDC are available with the rest of the filament masks on the \href{https://dataverse.harvard.edu/dataverse/Galactic-Filaments}{Large-Scale Galactic Filaments Dataverse}.

For these extinction-based Gaussian fits performed on the \citet{Zucker_2015} IRDCs, we choose a representative fitting distance of 1 pc and estimate the background between radii of 1.0 and 1.5 pc. While ideally we would like to match the emission-derived fitting distance and background radii above (2 pc and 3-4 pc) to the extinction-derived fitting distance and background radii (1.0 and 1.0-1.5 pc), the fidelity of the surface mass density maps depends upon tightly fitting an ellipse to the spatial extent of the IRDC to estimate the contribution from a diffuse Galactic background (see Figure 1 in \citet{Battersby_2010}). This prevents us from building a profile out to the same radial distances used for the Herschel maps. Because the extinction widths are several factors narrower than the emission widths, the flattening radius is closer to the spine, so this will have a smaller effect than if the same fitting and background subtraction radii are adopted for the emission widths. 

\clearpage
\begin{turnpage}
\tabletypesize{\scriptsize}
\renewcommand{\arraystretch}{0.9}
\setlength{\tabcolsep}{3pt}
\begin{deluxetable}{cccccccccccccc}
\colnumbers
\tablehead{\colhead{Name} & \colhead{$ p_{3, 3\rightarrow5}$} & \colhead{$ p_{4, 3\rightarrow4}$} & \colhead{$ p_{4, 4\rightarrow5}$} & \colhead{$ p_{5, 3\rightarrow5}$} & \colhead{$ R_{flat \;3, 3\rightarrow5}$} & \colhead{$ R_{flat \;4, 3\rightarrow4}$} & \colhead{$ R_{flat \;4, 4\rightarrow5}$} & \colhead{$ R_{flat \;5, 3\rightarrow5}$} & \colhead{$ FWHM_{2,2\rightarrow3}$} & \colhead{$ FWHM_{1,3\rightarrow4}$} & \colhead{$ FWHM_{2,3\rightarrow4}$} & \colhead{$ FWHM_{0.5,3\rightarrow4}$} & \colhead{$ FWHM_{3,3\rightarrow4}$}\\ \colhead{} & \colhead{} & \colhead{} & \colhead{} & \colhead{} & \colhead{pc} & \colhead{pc} & \colhead{pc} & \colhead{pc} & \colhead{pc} & \colhead{pc} & \colhead{pc} & \colhead{pc} & \colhead{pc}}
\startdata
Fil1 & 3.4 & 3.7 & 3.6 & 3.7 & 0.8 & 0.9 & 0.9 & 0.9 & 1.3 & 1.2 & 1.5 & 0.9 & 1.5 \\
Fil2 & 2.8 & 3.2 & 2.6 & 2.9 & 0.5 & 0.6 & 0.5 & 0.5 & 1.0 & 0.9 & 1.1 & 0.7 & 1.1 \\
Fil3 & 4.0 & 4.2 & 4.3 & 4.1 & 0.8 & 0.9 & 0.9 & 0.9 & 1.3 & 1.2 & 1.3 & 0.8 & 1.3 \\
Fil4 & 2.4 & 3.2 & 2.4 & 3.1 & 0.7 & 1.0 & 0.7 & 1.0 & 1.5 & 1.4 & 1.9 & 0.9 & 2.1 \\
Fil5 & 2.2 & 2.4 & 2.1 & 2.4 & 0.3 & 0.4 & 0.3 & 0.4 & 0.8 & 0.9 & 1.1 & 0.6 & 1.1 \\
Fil6 & 2.8 & 2.9 & 2.8 & 2.9 & 0.4 & 0.4 & 0.4 & 0.4 & 0.9 & 0.9 & 1.0 & 0.6 & 1.0 \\
Fil7 & 2.2 & 2.6 & 2.4 & 2.7 & 0.4 & 0.6 & 0.6 & 0.6 & 1.5 & 1.3 & 1.9 & 0.9 & 2.0 \\
Fil8 & 2.4 & 2.7 & 2.3 & 2.5 & 0.4 & 0.5 & 0.4 & 0.4 & 1.2 & 1.1 & 1.4 & 0.8 & 1.4 \\
Fil9 & 2.4 & 2.6 & 2.4 & 2.6 & 0.4 & 0.5 & 0.4 & 0.5 & 1.2 & 1.1 & 1.5 & 0.9 & 1.5 \\
Fil10 & 2.5 & 2.7 & 2.3 & 2.6 & 0.4 & 0.5 & 0.4 & 0.5 & 1.2 & 1.0 & 1.3 & 0.7 & 1.3 \\
Nessie & 2.5 & 2.3 & 2.5 & 2.4 & 0.3 & 0.3 & 0.3 & 0.3 & 0.9 & 0.8 & 0.9 & 0.6 & 0.9 \\
G24 & 3.6 & 4.6 & 3.9 & 4.5 & 1.2 & 1.5 & 1.3 & 1.5 & 1.6 & 1.5 & 1.9 & 1.2 & 2.0 \\
G26 & 2.9 & 2.9 & 3.4 & 3.1 & 0.5 & 0.5 & 0.6 & 0.5 & 1.1 & 1.0 & 1.2 & 0.8 & 1.2 \\
G28 & 3.5 & 3.9 & 3.3 & 3.7 & 0.7 & 0.8 & 0.7 & 0.8 & 1.1 & 0.9 & 1.1 & 0.9 & 1.1 \\
G29 & 3.1 & 3.9 & 3.1 & 3.7 & 0.8 & 1.0 & 0.9 & 1.0 & 1.3 & 1.2 & 1.6 & 0.6 & 1.6 \\
G47 & 3.8 & 5.3 & 3.5 & 4.8 & 1.3 & 1.7 & 1.2 & 1.6 & 1.4 & 1.3 & 1.7 & 4.6* & 1.8 \\
G49 & 2.7 & 3.1 & 2.5 & 2.5 & 0.8 & 0.9 & 0.7 & 0.7 & 1.4 & 1.2 & 1.5 & 0.8 & 1.6 \\
G64 & 3.1 & 3.2 & 3.2 & 3.2 & 0.5 & 0.5 & 0.5 & 0.5 & 1.0 & 0.9 & 1.0 & 0.6 & 1.0 \\
\enddata
\caption{\label{tab:profilealt} Variation in best-fit Gaussian (deconvolved FWHM) and Plummer (``$p$" and $ R_{flat}$) values given different fitting radii and background estimation radii. In each column header the first subscript value specifies the fitting radius and the second and third subscript values (on either side of the arrow) specify the inner and outer background estimation radii. So for Col (2), the Plummer function was fit out to a radius of 3 pc, and the background was estimated between radii of 3 and 5 pc. The deconvolved FWHM values adopted in this study and used for further analysis come from Col (12) while the ``$p$" values come from Col (3). The other columns are provided for the sake of comparison, to demonstrate the underlying uncertainty in radial profile fitting due to cutoff radii and background estimation choices.}
\tablenotetext{*}{G47 has one of the largest flattening radii in the sample, so fitting out to only half a parsec produces a highly irregular FWHM value}
\end{deluxetable}

\end{turnpage}

\end{document}